\documentclass[a4paper,11pt]{article}
\pdfoutput=1 

\usepackage{jheppub} 
\usepackage{amsmath,amssymb,graphicx,slashed,xcolor,cancel}
\usepackage{dsfont}
\usepackage{stackengine}
\usepackage{accents}

\usepackage{colortbl,colordvi,xcolor,color}
\usepackage{rotating}

\usepackage{array,multirow}
\usepackage{booktabs}
\usepackage{colortbl}
\usepackage{float}

\usepackage[T1]{fontenc} 

\newcommand{\refeq}[1]{(\ref{#1})}

\numberwithin{equation}{section}

\newcommand{\beq}{\begin{equation}}
\newcommand{\eeq}{\end{equation}}
\newcommand{\ba}{\begin{array}}
\newcommand{\ea}{\end{array}}
\newcommand{\bea}{\begin{eqnarray}}
\newcommand{\eea}{\end{eqnarray} }
\newcommand{\bal}{\begin{align}}
\newcommand{\eal}{\end{align}}

\newcommand{\iDslash}{\!~i\!~\cancel{D}\!~}
\renewcommand{\d}{\partial}
\newcommand{\lrD}{~\!\overset{\leftrightarrow}{\hspace{-0.1cm}D}\!}
\newcommand{\lrDa}{~\!\overset{\leftrightarrow}{\hspace{-0.1cm}D}\!^{\!~a}}
\newcommand{\hc}{\mathrm{h.c.}}

\newcommand{\U}[1]{U(#1)}
\newcommand{\SU}[1]{SU(#1)}

\newcommand{\SM}{\mathrm{SM}}
\newcommand{\LBSM}{\mathcal{L}_{\mathrm{BSM}}}
\newcommand{\Lcal}{\mathcal{L}}

\newcommand{\Ocal}{\mathcal{O}}

\newcommand{\scriptmath}[1]{{\scriptsize{\mbox{$#1$}}}}
\newcommand{\ctoprule}{\toprule[0.5mm]}
\newcommand{\cbottomrule}{\bottomrule[0.5mm]}

\newcommand{\cmrule}{\midrule[0.25mm]}
\newcommand{\crowcolor}{\rowcolor[rgb]{0.9,0.9,0.9}}

\makeatletter
\newcommand\xleftrightarrow[2][]{%
  \ext@arrow 9999{\longleftrightarrowfill@}{#1}{#2}}
\newcommand\longleftrightarrowfill@{%
  \arrowfill@\leftarrow\relbar\rightarrow}
\makeatother

\title{\boldmath Effective description of general extensions of the Standard Model: the complete tree-level dictionary}

\author[a,b]{J.\ de Blas,}
\author[c]{J.\ C.\ Criado,}
\author[c,d]{M.\ P\'erez-Victoria}
\author[c]{and J.\ Santiago}

\affiliation[a]{Dipartimento di Fisica e Astronomia ``Galileo Galilei'', Universit\`a di Padova, Via Marzolo 8, I-35131 Padova, Italy}

\affiliation[b]{INFN, Sezione di Padova, Via Marzolo 8, I-35131 Padova, Italy}

\affiliation[c]{CAFPE and Departamento de F\'{\i}sica Te\'orica y del Cosmos,  Universidad de Granada, Campus de Fuentenueva, E-18071, Granada, Spain}

\affiliation[d] {Theoretical Physics Department, CERN, Geneva, Switzerland} 

\emailAdd{Jorge.DeBlasMateo@pd.infn.it}
\emailAdd{jccriadoalamo@ugr.es}
\emailAdd{mpv@ugr.es}
\emailAdd{jsantiago@ugr.es}

\preprint{CERN-TH-2017-251}

\abstract{
We compute all the tree-level contributions to the Wilson coefficients of the dimension-six Standard-Model effective theory in ultraviolet completions with general scalar, spinor and vector field content and arbitrary interactions. No assumption about the renormalizability of the high-energy theory is made. 
This provides a complete ultraviolet/infrared dictionary at the classical level, which can be used to study the low-energy implications of any model of interest, and also to look for explicit completions consistent with low-energy data.
}

\begin{document} 
\maketitle
\flushbottom

\section{Introduction}
\label{sec:intro} 

The discovery of the Higgs boson at the Large Hadron Collider
(LHC)~\cite{Aad:2012tfa,Chatrchyan:2012xdj}  has
represented a monumental step towards a deeper understanding of the
mechanism of electroweak symmetry breaking in Nature. It has also
implied a shift in the main physics goal of the LHC, which has turned its focus onto the search of new physics
beyond the Standard Model (SM). Indeed, the LHC experimental collaborations
have produced an impressive number of new physics searches that cover
almost every possible experimental signature at the LHC. The few gaps still
existing at the available energy are being closed thanks to the intense
collaboration between the theoretical and experimental communities.
Unfortunately, this effort has produced no direct evidence of new physics,
 and the current experimental limits on the masses of new
particles are typically in the TeV range. Even if it
is certainly possible that new physics is waiting for us just around the corner and that it
will be directly produced at the LHC in the near future, it 
seems quite likely that it will stay beyond kinematic reach in the next decades.

If such a scenario, with a significant gap between the mass of
any new particle and the energies probed by experiment, is realized in
Nature, the effective field theories constructed with the SM fields become the most efficient tool to
analyze experimental data. The framework of effective theories provides a smart way of splitting in two
steps the
problem of comparing experimental data with theoretical predictions
to obtain information on possible extensions of the SM. In the first one, experimental
(pseudo)-observables are encoded in terms of the Wilson coefficients
of the effective operators with minimal theoretical bias.
This allows for essentially model-independent parameterizations of
different sets of experimental data~\cite{delAguila:2000aa,Han:2004az,delAguila:2011zs,Ciuchini:2013pca,deBlas:2013qqa,deBlas:2013gla,Pomarol:2013zra,Falkowski:2014tna,Buckley:2015nca,deBlas:2015aea,Falkowski:2015jaa,Berthier:2015gja,Falkowski:2015krw,deBlas:2016ojx,deBlas:2016nqo,Falkowski:2017pss,deBlas:2017wmn}.
Furthermore, this task can be done once and for all, independently of any
choice of new physics models.\footnote{Global fits have to be updated if there
is new experimental data or new theoretical calculations within the
context of the effective theory.}
In a second step, the Wilson coefficients of the
effective operators can be connected to the parameters of specific new-physics models through the process of matching. This reintroduces the model dependence in the
process of comparing experimental data to new physics. 
Both steps can actually be developed simultaneously and almost independently. Put together, they allow us to use experimental data to test theories beyond the SM, even when the new particles they bring about cannot be produced. 

To take full advantage of the model-independence of the low-energy effective theory approach, it would be desirable to 
match it with a completely general class of new physics models.
This task looks hopeless, but effective theories come to our rescue in this too. First, in view of the good agreement so far of data with the SM, it is plausible that any realistic form of new physics can be well described in the multi-TeV regime by a local effective field theory. This is certainly the case for almost all explicit models in the market. So, for phenomenological purposes it is sufficient to consider this intermediate effective theory as the high-energy theory. Second, when this ultraviolet (UV) theory is weakly interacting, its contributions to the infrared (IR) Wilson coefficients can be classified according to the canonical dimension of the corresponding induced operators and to their order in the loop expansion.\footnote{Note that the intermediate effective theory can be weakly interacting even if it arises from a strongly-coupled theory. One example of this is provided by composite Higgs theories in the large-$N$ limit.} As we show in this paper, it turns out that the leading order, given by tree-level contributions to operators up to dimension six, is restrictive enough that
a complete classification of the UV effective theories with contributions at this
order is feasible. Once such a classification is available, 
the Wilson coefficients in the IR effective theory can be computed in terms of the masses and couplings of the UV theory. This information will constitute a complete UV/IR dictionary at the leading order, which provides a direct 
link between experimental data and the parameters of any new physics model that can give such contributions.

The goal of this article is to present the complete tree-level 
UV/IR dictionary up to dimension six for the SM Effective Field Theory (SMEFT), in which the Higgs boson is considered 
to be part of a doublet of the linearly realized $SU(2)_L\times
U(1)_Y$ symmetry (see~\cite{Brivio:2017vri} for a recent
review).\footnote{\label{footnote:Pich}A related effort  
for the case of the electroweak chiral Lagrangian, in which the Higgs boson is a scalar singlet of
the non-linearly realized electroweak symmetry is currently
underway~\cite{Pich:2016lew,Rosell:2017kps}.}
The classification of the new fields that
are relevant for this dictionary was presented before in a series of
papers for new quarks~\cite{delAguila:2000rc},
leptons~\cite{delAguila:2008pw}, vectors~\cite{delAguila:2010mx} and
scalars~\cite{deBlas:2014mba}, respectively. The explicit contributions to the Wilson coefficients were also computed in each case, for renormalizable UV theories. The selected fields that contribute at the leading order have the characteristic property of allowing for linear couplings to the SM fields. Therefore, this classification is also useful beyond its direct application to effective field theories, as it 
provides an exhaustive list of the new particles that can be singly
produced in particle colliders at the classical level and via renormalizable interactions. The couplings that govern single production are
the same as the ones required to generate the corresponding SM effective
operators, which connects direct and indirect
constraints on (singly-produced) particles at colliders. Pair
production of new particles can be similarly related to new particles
that contribute to the effective-theory matching at one loop.

The presence of couplings with positive mass dimension in new-physics models
has two important implications for the IR effective field theory, already at the classical level and for operators of dimension up to six. First, it
allows for mixed contributions in which two or more
heavy particles with different 
spins can simultaneously contribute to certain operators. These mixed
contributions are computed for the first time with complete generality
in the present work. Second, non-renormalizable operators in the new physics
model can also contribute to the SMEFT at tree-level to dimension
six. We also include these non-renormalizable operators in the possible
UV completions of the SM, which again give rise to new contributions.
In an effort to be complete, and once we have given up
renormalizability of the new physics models, we also consider the possibility of
new heavy vectors that do not arise as the gauge bosons of
spontaneously broken gauge symmetries. This introduces a new possible
vector multiplet beyond the ones previously presented in the
literature.

This completes the tree-level 
dictionary to dimension six. To keep it self-contained, we provide here the full
dictionary, including the previous results in renormalizable theories in which only particles of the same spin contribute at a time. Even in that case, the results presented here are not a direct transcription of the ones in the
literature. We provide the results for the first time in a real
dictionary style, listing the contribution to the Wilson coefficients both operator by
operator and field by field. In this way it is trivial to check which
new physics can generate a specific contribution to certain Wilson coefficients and subsequently analyze all the other physical effects of such an extension of the SM. Furthermore we
give all our results in the Warsaw basis~\cite{Grzadkowski:2010es},
following the SM conventions in ref.~\cite{Jenkins:2013zja} for the
relations between redundant operators.\footnote{Our results can be easily translated into other popular bases by using 
publicly available codes~\cite{Falkowski:2015wza}. 
} This allows the direct use of
our results together with the anomalous dimensions computed
in~\cite{Jenkins:2013zja,Jenkins:2013wua,Alonso:2013hga,Alonso:2014zka} (see
also~\cite{Elias-Miro:2013gya,Elias-Miro:2013mua}) to have a proper leading order
calculation with possible large logarithms resummed.\footnote{There
has been an important progress recently towards the automation of
one-loop matching
calculations~\cite{Henning:2014wua,Drozd:2015rsp,delAguila:2016zcb,Henning:2016lyp,Ellis:2016enq,Fuentes-Martin:2016uol,Zhang:2016pja,Ellis:2017jns}
which would allow for 
consistent one-loop calculations in the new models and, eventually,
next-to-leading order ones when the two-loop SMEFT anomalous dimensions are available.} 

The article is organized as follows. We describe our
(minimal) hypotheses and provide the complete list of new particles
that contribute to the tree-level dimension-six dictionary in 
section~\ref{sec:fields}. The general contribution to the tree-level
matching for effective operators up to dimension six is computed in
section~\ref{sec:match}. 
In section~\ref{sec:userguide}, we provide a guide to use our results both in a
bottom-up and in a top-down fashion. Then we give a specific example,
using the recently reported anomalies in certain $B$-meson observables in
section~\ref{sec:example} and we
conclude in section~\ref{sec:conclusions}. Our results, unavoidably long, are given in several appendices for the reader convenience.
In appendix~\ref{sec:lagrangians}, after setting our conventions and notation, we write
down the explicit Lagrangians for all possible extensions of the SM
with new scalars, fermions and vectors that contribute to the dimension-six SMEFT at
the tree-level.  For completeness, we reproduce the operators in the Warsaw
basis in appendix~\ref{sec:d6Basis}. The top-down dictionary is given
in appendix~\ref{sec:Field_Op} and finally the bottom-up one, which collects the expressions of the Wilson coefficients as functions of the UV parameters, is
reproduced in appendix~\ref{sec:results}.

\section{General Extensions of the Standard Model}
\label{sec:fields} 

The SMEFT provides a simple and well-defined model-independent framework to study new physics beyond the SM. Its main limitation is that it is only valid at energies below the threshold of production of any extra degrees of freedom. To study the direct production of new particles, it is mandatory to incorporate into the effective theory the extra fields associated to them. Of course, the problem is that we do not know {\it a priori} which are the particles and fields that are relevant at the energies that can be accessed now and in the near future. So, in order to preserve model independence, we need to consider effective theories with arbitrary field content and arbitrary interactions. This also helps in connecting to particular models and hence in providing a rationale for the values of the low-energy parameters. Such a general space of theories depends on an infinite number of free parameters and looks all but intractable. However, some well-motivated assumptions, together with our aim of matching to the SMEFT at the leading non-trivial order, remarkably reduce it to a manageable subspace of finite dimension. Specifically, we assume here that, at energy scales below a certain cutoff $\Lambda$, nature is well described by a four-dimensional Poincaré-invariant local effective Lagrangian $\Lcal_{\mathrm{BSM}}$ such that
\begin{enumerate}
\item $\Lcal_{\mathrm{BSM}}$ is invariant under the linearly-realized $H \equiv \SU{3}_C\times \SU{2}_L \times \U{1}_Y$ gauge group.
\item $\Lcal_{\mathrm{BSM}}$ contains only fields associated to particles of spin $\leq 1$.
\item $\Lcal_{\mathrm{BSM}}$ includes as a subset all the field multiplets in the SM. In particular, it contains a scalar $\phi$ in the $(1,2)_{1/2}$ representation of the gauge group.
\item The only fermion fields with chiral transformations under the gauge group $H$ are the ones in the SM. In other words, all the extra fermions are vector-like with respect to $H$ or Majorana. This ensures that the symmetry $H$ is non-anomalous.
\end{enumerate}
The first assumption is a requisite for the perturbative unitarity of a theory that contains the SM gauge bosons (see, nevertheless, footnote~\ref{footnote:Pich}). The second one is a restriction we make to avoid subtle consistency issues with interacting particles of spin~$> 1$~\cite{Rahman:2013sta}.\footnote{Local effective field theories involving higher-spin particles are possible, with a restricted region of validity determined by their mass, spin and couplings~\cite{Porrati:2008ha}.}  The third and fourth assumptions are partially justified by the experimental success of the SM, including the discovery of the Higgs boson, precision electroweak data and Higgs data. Importantly for our purposes, the first, third and fourth assumptions ensure that, at energies much smaller than all the (gauge-invariant) masses of the extra particles, the theory is well described by the SMEFT.

The operators of canonical dimension $d > 4$ in $\LBSM$ have dimensionful coefficients, which can be written as $\alpha_i f^{4-d}$, with $f$ some mass scale and $\alpha_i$ dimensionless couplings, which can be related with the cutoff $\Lambda$ by power-counting arguments~\cite{Manohar:1983md,Giudice:2007fh,Jenkins:2013sda,Buchalla:2013eza,Gavela:2016bzc,Buchalla:2016sop,Liu:2016idz}.  
If all the vector bosons in the theory are the additional gauge bosons of an extended gauge symmetry $G \supset H$ (spontaneously broken to $H$) and $\Lcal_{\mathrm{BSM}}$ is invariant under $G$, with no anomalies, then $\Lcal_{\mathrm{BSM}}$ describes a unitary effective quantum field theory that can be used to perform perturbative calculations to arbitrary precision at energies below the cutoff $\Lambda$. However, in agreement with our model-independent spirit, we will consider here general theories with Proca vector bosons without enforcing any gauge invariance beyond $H$.\footnote{Spin-1 particles could alternatively be described by rank-2 antisymmetric tensor fields, which can be related to our vector formulation by a field redefinition, see~\cite{Ecker:1989yg,Pich:2016lew}.} This class of theories contains the ones with extended gauge invariance.  All the covariant derivatives we write are thus understood to be covariant with respect to $H$ only. 

The field content of the theory $\Lcal_{\mathrm{BSM}}$ can be conveniently classified into irreducible representations of the Lorentz and gauge symmetry groups. In this paper, we concentrate on the sector of $\Lcal_{\mathrm{BSM}}$ that can contribute at the classical level to the SMEFT operators of canonical dimension up to six.  As we show in the next section, this sector includes only operators of canonical dimension up to six and only those extra fields that can have gauge-invariant linear interactions with the SM fields of dimension $d\leq 4$. This last requirement strongly restricts the quantum numbers of the extra fields to be considered, as the Lorentz and gauge quantum numbers are given by the ones of the possible bosonic and fermionic operators of dimension 2, 3 and 5/2, respectively, that can be built with SM fields.  All these irreducible representations, together with the notation we use for each of the corresponding fields, are collected in tables~\ref{t:scalars}, \ref{t:fermions} and~\ref{t:vectors}. 


\setlength{\aboverulesep}{0pt}
\setlength{\belowrulesep}{0pt}

\begin{table}[t]
  \begin{center}
    {\small
      \begin{tabular}{lcccccccc}
        \ctoprule
        \crowcolor
        Name &
        ${\cal S}$ &
        ${\cal S}_1$ &
        ${\cal S}_2$ &
        $\varphi$ &
        $\Xi$ &
        $\Xi_1$ &
        $\Theta_1$ &
        $\Theta_3$ \\
        Irrep &
        $\left(1,1\right)_0$ &
        $\left(1,1\right)_1$ &
        $\left(1,1\right)_2$ &
        $\left(1,2\right)_{\frac 12}$ &
        $\left(1,3\right)_0$ &
        $\left(1,3\right)_1$ &
        $\left(1,4\right)_{\frac 12}$ &
        $\left(1,4\right)_{\frac 32}$ \\[1.3mm]
        \cbottomrule
        &&&&&&&\\[-0.4cm]
        \ctoprule
        \crowcolor
        Name &
        ${\omega}_{1}$ &
        ${\omega}_{2}$ &
        ${\omega}_{4}$ &
        $\Pi_1$ &
        $\Pi_7$ &
        $\zeta$ &
        & \\
        Irrep &
        $\left(3,1\right)_{-\frac 13}$ &
        $\left(3,1\right)_{\frac 23}$ &
        $\left(3,1\right)_{-\frac 43}$ &
        $\left(3,2\right)_{\frac 16}$ &
        $\left(3,2\right)_{\frac 76}$ &
        $\left(3,3\right)_{-\frac 13}$ \\[1.3mm]
        \cbottomrule
        &&&&&&&\\[-0.4cm]
        \ctoprule
        \crowcolor
        Name &
        $\Omega_{1}$ &
        $\Omega_{2}$ &
        $\Omega_{4}$ &
        $\Upsilon$ &
        $\Phi$ &
        &
        & \\
        Irrep &
        $\left(6,1\right)_{\frac 13}$ &
        $\left(6,1\right)_{-\frac 23}$ &
        $\left(6,1\right)_{\frac 43}$ &
        $\left(6,3\right)_{\frac 13}$ &
        $\left(8,2\right)_{\frac 12}$ \\[1.3mm]
        \cbottomrule
      \end{tabular}
    }
    \caption{New scalar bosons contributing to the dimension-six SMEFT at tree level.}
    \label{t:scalars}
  \end{center}
\vspace{1cm}
%
%
  \begin{center}
    {\small
      \begin{tabular}{lccccccc}
        \ctoprule
        \crowcolor
        Name &
        $N$ & $E$ & $\Delta_1$ & $\Delta_3$ & $\Sigma$ & $\Sigma_1$ & \\
        Irrep &
        $\left(1, 1\right)_0$ &
        $\left(1, 1\right)_{-1}$ &
        $\left(1, 2\right)_{-\frac{1}{2}}$ &
        $\left(1, 2\right)_{-\frac{3}{2}}$ &
        $\left(1, 3\right)_0$ &
        $\left(1, 3\right)_{-1}$ & \\[1.3mm]
        \cbottomrule
        &&&&&&&\\[-0.4cm]
        \ctoprule
        \crowcolor
        Name &
        $U$ & $D$ & $Q_1$ & $Q_5$ & $Q_7$ & $T_1$ & $T_2$ \\
        Irrep &
        $\left(3, 1\right)_{\frac{2}{3}}$ &
        $\left(3, 1\right)_{-\frac{1}{3}}$ &
        $\left(3, 2\right)_{\frac{1}{6}}$ &
        $\left(3, 2\right)_{-\frac{5}{6}}$ &
        $\left(3, 2\right)_{\frac{7}{6}}$ &
        $\left(3, 3\right)_{-\frac{1}{3}}$ &
        $\left(3, 3\right)_{\frac{2}{3}}$ \\[1.3mm]
        \cbottomrule
      \end{tabular}
    }
    \caption{New vector-like fermions contributing to the dimension-six SMEFT at tree level.}
    \label{t:fermions}
  \end{center}
\vspace{1cm}
%
%
  \begin{center}
    {\small
      \begin{tabular}{lcccccccc} 
        \ctoprule
        \crowcolor
        Name &
        ${\cal B}$ &
        ${\cal B}_1$ &
        ${\cal W}$ &
        ${\cal W}_1$ &
        ${\cal G}$ &
        ${\cal G}_1$ &
        ${\cal H}$ &
        ${\cal L}_1$ \\
        Irrep &
        $\left(1,1\right)_0$ &
        $\left(1,1\right)_1$ &
        $\left(1,3\right)_0$ &
        $\left(1,3\right)_1$ &
        $\left(8,1\right)_0$ &
        $\left(8,1\right)_1$ &
        $\left(8,3\right)_{0}$ &
        $\left(1,2\right)_{\frac 12}$ \\[1.3mm]
        \cbottomrule
        &&&&&&&\\[-0.4cm]
        \ctoprule
        \crowcolor
        Name &
        ${\cal L}_3$ &
        ${\cal U}_2$ &
        ${\cal U}_5$ &
        ${\cal Q}_1$ &
        ${\cal Q}_5$ &
        ${\cal X}$ &
        ${\cal Y}_1$ &
        ${\cal Y}_5$ \\
        Irrep &
        $\left(1,2\right)_{-\frac 32}$ &
        $\left(3,1\right)_{\frac 23}$ &
        $\left(3,1\right)_{\frac 53}$ &
        $\left(3,2\right)_{\frac 16}$ &
        $\left(3,2\right)_{-\frac 56}$ &
        $\left(3,3\right)_{\frac 23}$ &
        $\left(\bar 6,2\right)_{\frac 16}$ &
        $\left(\bar 6,2\right)_{-\frac 56}$ \\[1.3mm]
        \cbottomrule
      \end{tabular}
    }
    \caption{New vector bosons contributing to the dimension-six SMEFT at tree level.}
    \label{t:vectors}
  \end{center}
\end{table}

These new fields with linear couplings have been singled out and studied before, in~\cite{delAguila:2000rc,delAguila:2008pw,delAguila:2010mx,deBlas:2014mba}.\footnote{\label{footnote:L1}There is actually one exception: the vector field $\Lcal_1$ was not included in~\cite{delAguila:2010mx}. There exists only one gauge-invariant operator of dimension $d\leq 4$ that is linear in this vector and has no any other extra field: the super-renormalizable operator $\Lcal^\dagger_{1\mu} D^\mu \phi$, which mixes the longitudinal part of $\Lcal_1$ with the Higgs doublet. Such an operator will not appear, in the unitary gauge, if $\Lcal_1$ is the gauge boson of an extended, spontaneously broken gauge invariance. Therefore, in a complete unitary theory, it will not contribute to the SMEFT operators at the leading order. However, it could appear in other gauges and also in phenomenological models, much as pion-vector resonance mixing is included in certain descriptions of low-energy QCD~\cite{Ecker:1989yg,Pich:2016lew}. In these cases it can be eliminated by a field redefinition, which in general generates local operators of dimension 4, 5 and 6 weighted by the vector mass and the dimensional coefficient of the super-renormalizable operator~\cite{Cirigliano:2006hb}. At the end of the day, as far as low-energy physics is concerned, this is equivalent to integrating the field out, which is our approach here.
} 
Besides the fact that they provide the leading contributions to the SMEFT, and thus to indirect tests, they are also relevant for the resonant production of new particles, as the only new particles that can be singly produced at the classical level in collisions of SM particles are excitations of these fields. In fact, several subsets of the fields in tables~\ref{t:scalars}, \ref{t:fermions} and~\ref{t:vectors}  have appeared in the literature in different contexts, see for instance~\cite{Buchmuller:1986zs,DelNobile:2009st,Han:2010rf,AguilarSaavedra:2011vw,Grinstein:2011dz,Dawson:2017vgm}.

The part of $\Lcal_{\mathrm{BSM}}$ that contributes classically to the effective Lagrangian of dimension six or smaller involves a finite number of fields and a finite number of operators. Therefore, it can be written explicitly and in full generality, as a sum of all the possible independent contributing operators with arbitrary coefficients. The complete Lagrangian can be split in the following way:
\beq
\Lcal_{\mathrm{BSM}} = {\cal L}_0 + \Lcal_{\mathrm{S}} + \Lcal_{\mathrm{F}} + \Lcal_{\mathrm{V}} + \Lcal_{\mathrm{mixed}} + \dots,
\label{LBSMsplit}
\eeq
where ${\cal L}_0$ contains terms of dimension $d\leq6$ with only SM fields, $\Lcal_{\mathrm{S,F,V}}$ contains terms of dimension $d\leq 5$ with extra scalars, fermions and vectors, respectively, but no products of new fields of different spin, and $\Lcal_{\mathrm{mixed}}$ contains terms of dimension $d\leq 4$ involving products of extra fields of different spin. In writing the dimension-five interactions with the heavy particles we remove redundant operators by using the SM equations of motion. The dots indicate terms that do not contribute in our approximation.  

The extra fields can have kinetic or mass mixing with the {\it a priori} SM ones if they share the same quantum numbers. However, field rotations and rescalings can always be performed in such a way that all the kinetic terms in $\Lcal_{\mathrm{BSM}}$ are diagonal and canonical and all the mass terms are diagonal in the electroweak symmetric phase. All our equations are written with this choice of fields (except for the mixing of $\phi$ and possible scalars $\varphi$ with $\Lcal_1$, see footnote~\ref{footnote:L1}).  
Furthermore, we assume that no fields get a non-trivial gauge-invariant vacuum expectation value in the symmetric phase. This can always be achieved by convenient shifts of the scalar singlets. 
To match models written in a different ``field basis'', the shift, diagonalization and canonical normalization must be performed prior to using our formulas.

Working in this ``field basis'' not only fixes the precise meaning of the couplings in $\Lcal_{\mathrm{BSM}}$, but also allows to identify the SM fields that enter in~${\cal L}_0$. The SM fermions and gauge fields are the massless fermion and vector eigenstates, respectively, whereas we identify the Higgs doublet $\phi$ with the $(1,2)_{1/2}$ scalar eigenstate associated to a negative eigenvalue of the squared mass matrix. We assume that this eigenvalue is non-degenerate and that all the other eigenvalues are positive. This is required if we want $\Lcal_{\mathrm{BSM}}$ to be described by the SMEFT at low energies. 
The different pieces that appear in~\refeq{LBSMsplit} are written explicitly in appendix~\ref{sec:lagrangians}.

\vspace{0.15cm}
\section{Effective Lagrangian and Tree-Level Matching}
\label{sec:match} 
\vspace{0.15cm}

In order to study the physics of $\LBSM$ at energy scales much smaller than all the masses of the extra particles, the heavy fields can be integrated out to find the corresponding effective Lagrangian, organized as a power series in the inverse masses: 
\beq
\Lcal_{\mathrm{eff}} = {\cal L}_0 + \sum_{n=2}^\infty \Lcal_{\mathrm{eff}}^{(n)}.
\eeq
$\Lcal_{\mathrm{eff}}^{(n)}$ contain the Lorentz and gauge invariant local operators $\Ocal^{(n)}$ of canonical dimension $n$ that can be constructed with the SM fields,
\beq
\Lcal_{\mathrm{eff}}^{(n)} = \sum_j C^{(n)}_j \Ocal^{(n)}_j .
\eeq
This effective Lagrangian will be a SMEFT with particular Wilson coefficients $C^{(n)}_j$, of mass dimension $4-n$. The dimensions are provided by the masses and other scales in $\LBSM$. 

Not all the operators $\Ocal^{(n)}$ are independent. Making use of algebraic identities and field redefinitions, certain linear combinations can be eliminated from $\Lcal_{\mathrm{eff}}^{(n)}$, at the price of changing $\Lcal_{\mathrm{eff}}^{(>n)}$. Taking this redundance into account, several operator bases have been defined to dimension $n=6$. Here, we employ the Warsaw basis defined in~\cite{Grzadkowski:2010es}. The operators in that basis are collected in~appendix~\ref{sec:d6Basis}. The main purpose of this paper is to calculate the corresponding coefficients $C^{(\leq 6)}$ in the classical approximation, as functions of the couplings and masses in $\LBSM$. 

Note that the generated operators have the same form as the ones in $\Lcal_0$.  The non-trivial contributions we are interested in can be distinguished when there is sufficient information on $\Lcal_0$. This is the case if the coefficients of the non-renormalizable terms in $\Lcal_0$ are suppressed by a scale larger than the masses of the new particles, and also if they are fixed by symmetries or are known functions of the parameters of a given UV completion of $\LBSM$. The requirement of a soft UV behaviour also imposes some constraints~\cite{Ecker:1989yg,Pich:2016lew}.

The individual contributions of heavy fields and the collective contributions of heavy fields with the same spin (except for the ones involving the vector $\Lcal_1$) have been calculated before in~\cite{delAguila:2000rc,delAguila:2008pw,delAguila:2010mx,deBlas:2014mba}. Here, we also incorporate the mixed contributions of heavy particles of different spin, the contributions of $\Lcal_1$ and the contribution of the operators of dimension $d=5$ in $\LBSM$. 

Let us explain the systematics of the integration procedure. With this aim, we first write the part of $\LBSM$ involving new fields as
\beq
\LBSM - {\cal L}_0 = \eta_{(i)} A^\dagger_i  \Delta_{(i)}^{-1} A^i + \sum_{m,n}  A^\dagger_{j_1}\cdots A^\dagger_{j_n} W_{i_1\ldots i_m}^{j_1 \ldots j_n} \,  A^{i_1}\cdots A^{i_m} , \label{LBSMnew}
\eeq
where $A^i$ represent all possible extra fields in $\LBSM$,  
$\Delta_{{(i)}}$ is the covariant propagator for $A^i$ and $W_{i_1\ldots i_m}^{j_1 \ldots j_n}$ are operators constructed with the SM fields, including the identity operator. 
The factor $\eta_{(i)}=1~(1/2)$ yields canonical normalization for complex (real) fields (see appendix \ref{sec:lagrangians}).
Lorentz and Dirac indices are implicit. In general, these operators carry a reducible representation of $H$, but the ones with a single index $i$ belong to the same irreducible representation as $A^i$ or $A_i^\dagger$.
The integration at the classical level can be performed by i) using the equations of motion of $\LBSM$ to eliminate the heavy fields and ii) expanding the propagators of the heavy fields in inverse powers of  $D_{(i)}/M_{(i)}$:
\begin{align}
&\Delta_{(i)}  = - \frac{1}{M_{(i)}^2} \left(1-\frac{D^2_{(i)}}{M_{(i)}^2} \right) + O(1/M^6) ~~~~\mbox{(scalars)} ,\\
&\Delta_{(i)}  =  -\frac{\iDslash_{(i)} + M_{(i)} }{M_{(i)}^2} \left(1-\frac{D^2_{(i)}}{M_{(i)}^2} \right) + O(1/M^5) ~~~~\mbox{(fermions)} ,\\
&\Delta_{(i)}^{\mu\nu}  = \frac{\eta^{\mu\nu}}{M_{(i)}^2}+\frac{D_{(i)}^\nu D_{(i)}^\mu-\eta^{\mu\nu} D_{(i)}^2}{M_{(i)}^4} + O(1/M^6) ~~~~\mbox{(vectors)} .
\end{align}
The result at any finite order in $D_{(i)}/M_{(i)}$ is a local Lagrangian. We have performed the calculations in this algebraic fashion, keeping only the operators of dimension $n\leq6$. To deal in an efficient manner with the large number of terms that appear in this process and minimize the possibility of errors, we have employed the symbolic code \texttt{MatchingTools}~\cite{Criado:2017khh}, where 
we have implemented the algebraic relations and field redefinitions necessary to express our results in terms of the Warsaw-basis operators in~appendix~\ref{sec:d6Basis}.
All the calculations have been double-checked by hand and against previous results in the literature.

\begin{figure}[tbp]
\begin{center}
\includegraphics[width=15cm]{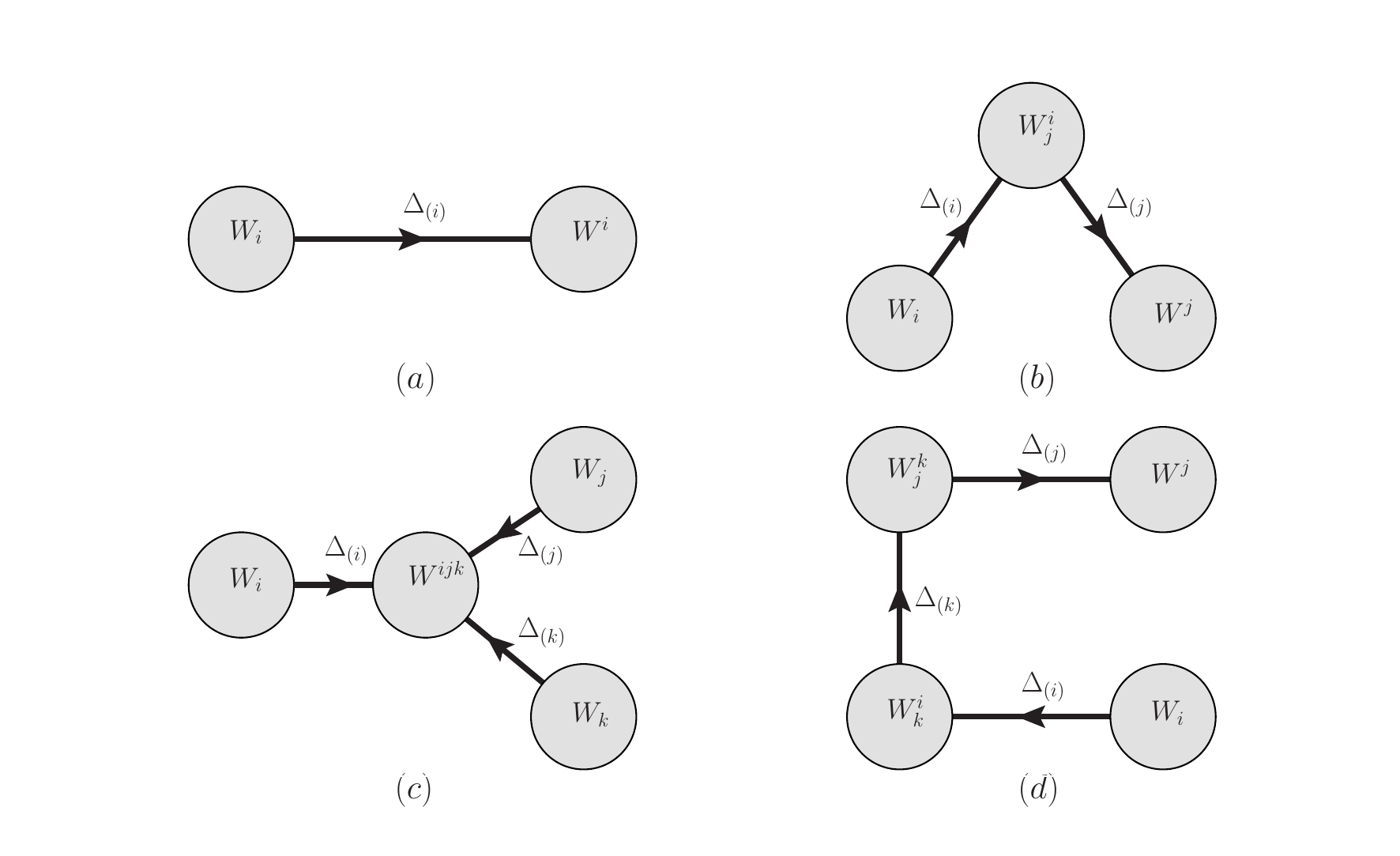}
\end{center}
\caption{
Feynman diagrams contributing to $\Lcal_{\mathrm{eff}}$ to dimension $n=6$. Non-equivalent permutations of the arrow directions shown here should be considered as well.}
\label{fig:diagrams}
\end{figure}

Equivalently, step i) above can be performed in terms of Feynman diagrams. In figure~\ref{fig:diagrams}, we show the tree-level Feynman diagrams with heavy field propagators that contribute to $\Lcal_\mathrm{eff}$ to order $n=6$. The blobs in this figure represent the SM operators $W^{i_1\ldots i_m}_{j_1\dots j_n}$ with $m$ incoming and $n$ outgoing lines, and the arrowed lines represent the covariant propagators $\Delta_{(i)}$. The arrows have no significance for real representations. 
In order to see that these are the only non-trivial tree-level diagrams contributing to $\Lcal_\mathrm{eff}$, note first that the canonical dimension of each term in the expansion of the propagators is non-negative, while the canonical dimension of each blob is equal to the canonical dimension of its corresponding interaction in eq.~(\ref{LBSMnew}) minus the one carried away by the bosonic or fermionic heavy fields. Consider a particular connected tree-level diagram. Let $B^d_f$ be the number of blobs in the diagram with at least one fermionic index and corresponding to interactions of canonical dimension $d$, and $B^d_b$ be the number of blobs in the diagram with no fermionic indices and corresponding to interactions of canonical dimension $d$. Let $L_f $ and $L_b$ be, respectively, the number of fermionic and bosonic propagators in the diagram and let $X_f$ be the number of blocks with uninterrupted heavy-fermion lines. The canonical dimensions $n$ of each term in the diagram, after the propagator expansions, obey
\beq
n\geq \sum_d d (B^d_b+B^d_f) - 2 L_b - 3 L_f. 
\eeq
From the topological relations $L_b + L_f + 1 = \sum_d (B^d_b+B^d_f)$ and $L_f  + X_f = \sum_d B^d_f$, the bound
\beq
n \geq 2 + X_f + \sum_d \left[ (d-2) B^d_b + (d-3) B^d_f \right]   \label{bound}
\eeq
follows.
Using the facts that $B^d_b=0$ if $d<3$ and $B^d_f=0$ if $d<4$, we find in particular that
\beq
n \geq B+2,
\eeq
with $B=\sum_d (B^d_b+B^d_f)$ the total number of blobs. Therefore, only diagrams with 4 blobs or less can contribute to $n\leq 6$. We also see from~\refeq{bound} that only interactions of canonical dimension $d\leq 6$ can contribute to $n\leq 6$. But the operators with $d=6$ only give the trivial contribution of a diagram with one blob and no propagators, which is nothing but the term already present in ${\cal L}_0$. This justifies our restriction to operators with $d\leq 5$ in the explicit expression of~$\LBSM$ written in appendix~\ref{sec:lagrangians}. Finally, we observe that both the operators of dimension $d=5$ and the ones involving more than one heavy field can only contribute to $n\leq 6$ in the presence of super-renormalizable operators of dimension $d=3$, and that operators of dimension $d=5$ with more than one heavy field do not contribute to this order. 
 
Note that in diagrams $(a)$, $(b)$ and $(c)$ of figure~\ref{fig:diagrams}, all the propagators are contracted with one-index operators $W_i$  or $W^i$, which arise from terms in $\LBSM$ with only one heavy field ($A^i$ or $A^\dagger_i$). In diagram $(d)$, on the other hand, the propagator $\Delta_{(k)}$ is attached only to operators with two indices, $W^k_j$ and $W^i_k$. However, upon the covariant-derivative expansion at finite order of the other two propagator, $\Delta_{(i)}$ and $\Delta_{(j)}$, the blobs they connect collapse into one-index local operators $\widetilde{W}^k= W^k_j  [\Delta_{(j)}] W^j$ and $\widetilde{W}_k=W^i_k [\Delta_{(i)}] W_i$, with $[.]$ indicating the derivative expansion. The operators $\widetilde{W}^k$ and $\widetilde{W}_k$ are in the same Lorentz and gauge representation as $W^k$ and $W_k$, respectively. Moreover, to allow for a dimension-six contribution, both of them must have canonical dimension $d=4$.  Hence, the fields $A^k$ ($A^\dagger_k$) associated to $\widetilde{W}_k$ ($\widetilde{W}^k$) must also belong to a representation that can couple linearly to the SM fields to give a scalar gauge-invariant operator of dimension $\leq 4$. We conclude that, as promised, only the heavy fields in the irreducible representations of tables~\ref{t:scalars}, \ref{t:fermions} and~\ref{t:vectors} contribute at the tree level to the effective Lagrangian to dimension six.

We can draw another interesting corollary from this discussion. Let us define {\em tree-level operators} as those for which there exists a renormalizable UV theory that induces them at the tree-level, when the effective Lagrangian is written in the Warsaw basis, and {\em loop operators} as those for which no such theory exists.\footnote{The requirement of renormalizability is crucial to make the distinction. Without constraints on the dimension of the interactions, any gauge-invariant operator could be trivially induced at the tree level by directly including it in the UV theory. Considering a complete basis gives definite physical meaning to each operator. Of course, which operators are potentially generated at tree or loop level depends on the particular choice of basis, but the implications for physical observables remain unchanged.} As we have just argued, tree-level operators of dimension six can only be generated by the diagrams in figure~\ref{fig:diagrams} and only by extra fields that allow for linear couplings to SM operators. This is also true if, instead of using the effective theory $\LBSM$ as a starting point, we directly integrate out at the classical level all the fields beyond the SM in a renormalizable completion of $\LBSM$. Therefore, our results in appendix~\ref{sec:results} explicitly show which operators are tree-level: those that (potentially) receive contributions in the absence of non-renormalizable interactions, that is, when $f\to \infty$ and $\gamma_{\mathcal{L}_1}\to 0$. Conversely, the operators that can {\em only} have, at most, $1/f$ or $\gamma_{\mathcal{L}_1}$ contributions are loop operators.\footnote{The possibility of generating operators of this type with tree-level diagrams involving higher-dimensional interactions was pointed out and emphasized in~\cite{Jenkins:2013fya}.} Even if the latter are connected to $\LBSM$ by tree-level diagrams, they cannot be generated at the tree level in any renormalizable completion of it. That is, the necessary dimension-five interactions are only generated by loop diagrams in any such UV completion. If this completion is weakly coupled, their coefficients will have a loop suppression that carries over to the Wilson coefficients in the SMEFT. Of course, such a suppression will not occur if the UV completion is strongly coupled. 
This classification agrees with the one in~\cite{Arzt:1994gp}, as it should, since we employ the same criteria.

\section{Results of the Matching: User Guide}
\label{sec:userguide}

The tree-level integration of the 48 fields of spin 0, 1/2 and 1 that can contribute to the dimension-six SMEFT, via the interactions in eqs.~(\ref{eq:LS4})-(\ref{eq:LVF}), generates all the effective operators in the basis of ref.~\cite{Grzadkowski:2010es}, with the exception of the four operators ${\cal O}_{G,\tilde{G},W,\tilde{W}}$. The explicit expressions of the contributions to the different Wilson coefficients are collected in~appendix~\ref{sec:results}. In this section we offer a basic guidance so that users can quickly find  the required entries of the UV/IR dictionary inside our long and numerous equations. 

We present our results by writing, for each operator, all the possible contributions of all the multiplets to its Wilson coefficient. 
The results for the different operators have been organized in the following way:
\begin{itemize}
  {\item Pure four-fermion operators (appendix \ref{sec:d64F}), classified according to the structure of chiralities of the fields in the operator, i.e.
    $\left(\bar{L}L\right)\left(\bar{L}L\right)$,
    $\left(\bar{R}R\right)\left(\bar{R}R\right)$,
    $\left(\bar{L}L\right)\left(\bar{R}R\right)$,
    $\left(\bar{L}R\right)\left(\bar{R}L\right)$,
    $\left(\bar{L}R\right)\left(\bar{L}R\right)$,
    and, separately, the baryon-number (B) violating interactions.}
{\item Pure bosonic interactions (appendix \ref{sec:d6Bosonic}). We follow the classification of ref.~\cite{Grzadkowski:2010es} and include here the operators of the form $\phi^6$, $\phi^4 D^2$ and $X^2 \phi^2$, where $X$ refers to a field-strength tensor.} 
{\item Interactions between bosons and fermions (appendix \ref{sec:d6BosonFermion}). We again follow the classification of ref.~\cite{Grzadkowski:2010es}, and separate the operators of the form $\psi^2\phi^3$, $X \psi^2 \phi$ and $\psi^2 \phi^2 D$.}
\end{itemize}
Unless otherwise stated, for each Wilson coefficient, the contributions of the different types of fields are ordered in the following way:
\begin{equation}
C_i= C_i^{\mathrm{Scalars}} + C_i^{\mathrm{Fermions}} + C_i^{\mathrm{Vectors}} + C_i^{\mathrm{Mixed}}+ \frac{1}{f}  C_i^{\mathrm{dim~\!5}},
\label{eq:Ci_contrib}
\end{equation}
where $C_i^{P}$, $P=$ Scalars, Fermions, Vectors, contains the information from the integration of only one type of spin, in the same order as presented in tables \ref{t:scalars}, \ref{t:fermions} and \ref{t:vectors}, respectively. Each of these are further separated, with the contributions from one type of particle appearing first, and mixing between particles of same spin, afterwards:
\begin{equation}
C_i^{P}=\sum_{m \in P}  C_i^{m} + \sum_{m,n \in P} C_i^{mn} + \sum_{m,n,p \in P} C_i^{mnp}.
\label{eq:Ci_contrib_2}
\end{equation}
The  contributions coming from Lagrangian interactions between particles of different spin, eqs.~(\ref{eq:LSF})-(\ref{eq:LVF}), are contained in $C_i^{\mathrm{Mixed}}$. The coefficient $C_i^{\mathrm{dim~\!5}}$ includes the dimension-six interactions generated by the non-renomalizable couplings in eqs.~(\ref{eq:LS5}), (\ref{eq:Lleptons5}), (\ref{eq:Lquarks5}) and (\ref{eq:LV5}). These can be easily distinguished noting the prefactor $1/f$. Finally, some of the new particles induce modifications on the kinetic term of the SM Higgs doublet in the effective theory. Our results are given in a basis where all fields are canonically normalized, and we include such corrections into a renormalization of the Higgs doublet $\phi \rightarrow Z_\phi^{-\frac 12}\phi$, with $Z_\phi^{-\frac 12}$ given in eq.~(\ref{eq:Zphi}). The corresponding factors of $Z_\phi^{-\frac {n_\phi}{2}}$ renormalizing operators with $n_\phi$ scalar doublets are shown explicitly in the coefficients.

Finally, for those operators that are non-hermitian we only report the coefficient of the interaction in tables~\ref{tab:dim45Basis},~\ref{tab:dim6Basis4F} and \ref{tab:dim6BasisBF}. The corresponding contributions to the coefficients of the hermitian conjugates can be obtained by complex conjugation.

The results of the matching can be employed in both directions:

\subsection*{Top-down}
\label{sec:topdown}

To facilitate the matching of particular models with the SMEFT---for instance to profit from the abundant model-independent constraints phrased in this language (see, e.g. \cite{delAguila:2000aa,Han:2004az,delAguila:2011zs,Ciuchini:2013pca,deBlas:2013qqa,deBlas:2013gla,Pomarol:2013zra,Falkowski:2014tna,Buckley:2015nca,deBlas:2015aea,Falkowski:2015jaa,Berthier:2015gja,Falkowski:2015krw,deBlas:2016ojx,deBlas:2016nqo,Falkowski:2017pss,deBlas:2017wmn})---we have collected in tables~\ref{tab:topdown_scalars}, \ref{tab:topdown_fermions} and \ref{tab:topdown_vectors}, in appendix~\ref{sec:Field_Op}, the different operators resulting from the integration of each of the scalar, fermion and vector multiplets, respectively. It turns out that all the operators that receive contributions involving couplings between different types of extra fields (with the same or different spin) can always be generated as well by at least one of the particles entering in the interaction {\it individually}. Therefore, tables~\ref{tab:topdown_scalars}-\ref{tab:topdown_vectors} contain all the information necessary to identify which operators can be generated in any scenario.

In this way, these tables show all the operators that can be generated given the field content of the model. One can then look at the corresponding Wilson coefficients in appendix~\ref{sec:results} and use eqs.~(\ref{eq:Ci_contrib}) and (\ref{eq:Ci_contrib_2}) to find the explicit contributions in terms of the masses and couplings of the new particles.

\subsection*{Bottom-up}

Our results can also be used in a bottom-up fashion, to find the explicit SM extensions that can give rise to a given set of effective interactions. To identify which multiplets contribute to each dimension-six operator in the effective theory, one simply needs to look at the labels of the masses in the denominators of each term in the expression of the Wilson coefficient. 
For operators involving the SM scalar doublet, one must also take into account that ${\cal L}_1$ and $\varphi$ can contribute to the renormalization of the scalar doublet $Z_\phi$. 
Finally, upon integration of the ${\cal L}_1$ vector field, the effects of its interactions with the vectors ${\cal B}$, ${\cal B}_{1}$,  ${\cal W}$ and ${\cal W}_{1}$ ---parameterized by the $\zeta_{{\cal L}_1 V}$ couplings in the Lagrangian (\ref{eq:LV4})--- can be described in a compact form by using modified couplings of ${\cal B}$, ${\cal B}_{1}$,  ${\cal W}$ and ${\cal W}_{1}$ to the corresponding SM scalar currents. Explicitly, they can be described by replacing
\begin{eqnarray}
 ( g^\phi_{{\cal B}})_r &
  \to &
  (\hat{g}^\phi_{{\cal B}})_r \equiv
  (g^\phi_{{\cal B}})_r
  - i \frac{(\zeta_{\mathcal{L}_1 \cal B})^*_{sr} (\gamma_{{\cal L}_1})_s}{M_{{\cal L}_{1s}}^2},
  \label{eq:ghat_first}\\
 ( g^\phi_{{\cal W}})_r &
  \to &
  (\hat{g}^\phi_{{\cal W}})_r \equiv
  (g^\phi_{{\cal W}})_r
  - 2 i \frac{(\zeta_{\mathcal{L}_1 \cal W})^*_{sr} (\gamma_{{\cal L}_1})_s}{M_{{\cal L}_{1s}}^2}, \\
 ( g^\phi_{{\cal B}_{1}})_r &
  \to &
  (\hat{g}^\phi_{{\cal B}_{1}})_r \equiv
  (g^\phi_{{\cal B}_{1}})_r
  + i \frac{(\zeta_{{\mathcal{L}_1 \cal B}_1})_{sr} (\gamma_{{\cal L}_1})_s}{M_{{\cal L}_{1s}}^2},\\
 ( g^\phi_{{\cal W}_{1}})_r &
  \to &
  (\hat{g}^\phi_{{\cal W}_{1}})_r \equiv
  (g^\phi_{{\cal W}_{1}})_r
  + 2 i \frac{(\zeta_{{\mathcal{L}_1 \cal W}_1})_{sr} (\gamma_{{\cal L}_1})_s}{M_{{\cal L}_{1s}}^2}.
  \label{eq:ghat_last}
\end{eqnarray}
Writing the solution in terms of the $\hat{g}_V^\phi$ couplings has the advantage of simplifying significantly many of the expressions, but obscures a bit the origin of the contribution. So, besides looking at the explicit masses, one should take into account that any $\hat{g}_V^\phi$ coupling implicitly involves a dependence on the couplings and mass of the field(s) ${\cal L}_1$. For instance,
\begin{equation}
\begin{array}{c c c}
  \!& \scriptmath{
 \!\!\!\!\!\!\!\!\!\!\!\!  (\hat{g}^\phi_{{\cal B}})_r \equiv
    (g^\phi_{{\cal B}})_r
    - i \frac{(\zeta_{\mathcal{L}_1 \cal B})^*_{rs} (\gamma_{{\cal L}_1})_s}{M_{{\cal L}_{1s}}^2}}~~~& \\
  &\downarrow~~~~~~~~~~~~~~~~~~~~~~~~~~&\\
  \!~~~\Delta C= &\!\!\!\!\!\!\!\!\!\!\!\!\!\!\!\!\!\!\!\!\!\!\!\!\!\!\!\!\!\!\!\!\!\!\!\!\!
  \frac{(\hat{g}_{{\cal B}}^\phi)_r^2}{M_{{\cal B}_r}^2}~\longrightarrow &
  \!\!\!\!\!\!\!\!\!\!\!\!\!\!\!\!\!\!\!\!\!\!\!\!\!\!\!\!\!\!\!\!\!\Delta C = 
  \frac{(g_{{\cal B}}^\phi)_r^2}{M_{{\cal B}_r}^2}
  - 2i \frac{
    (g_{{\cal B}}^\phi)_r (\zeta_{\mathcal{L}_1 \cal B})^*_{sr} (\gamma_{{\cal L}_1})_s}{
    M_{{\cal B}_r}^2 M_{{\cal L}_{1s}}^2}
  - \frac{
    (\zeta_{\mathcal{L}_1 \cal B})^*_{sr} (\gamma_{{\cal L}_1})_s
    (\zeta_{\mathcal{L}_1 \cal B})^*_{tr} (\gamma_{{\cal L}_1})_t}{
    M_{{\cal B}_r}^2 M_{{\cal L}_{1s}}^2 M_{{\cal L}_{1t}}^2}.
\end{array}
\end{equation}
Remember, nevertheless, that the vector multiplets~$\mathcal{L}_1$ will not contribute at all if they are the gauge bosons of an extended gauge invariance.

Similarly, the tree-level matching leads to a redefinition of the coefficients of the SM operators, see section~\ref{sec:redefinitions}. Then there are indirect effects in the dimension-six coefficients when the original SM couplings, which wear a {\em hat}, are written in terms of the redefined ones, without {\em hat}, as specified in eqs.~\refeq{eq:y}, \refeq{eq:lambda} and~\refeq{eq:mu}. Moreover, the covariant kinetic term of the Higgs doublet is modified in the presence of $\gamma_{\mathcal{L}_1}$, which leads to the Higgs-field renormalization in eq.~\refeq{eq:Zphi}. Therefore, one should also keep track of the Yukawa couplings $\hat{y}^{e,u,d}$ and the quartic coupling $\hat{\lambda}_\phi$ in order to check which fields can contribute to the Wilson coefficients. 

We include reminders of all these implicit dependences, where appropriate, in appendix~\ref{sec:results}.

\section{Example: Interpretation of LHCb Anomalies}
\label{sec:example}

Our UV/IR dictionary is a tool that can be used for different phenomenological purposes, such as finding indirect limits on the parameters of explicit models, constructing BSM models consistent with existing data or analyzing deviations with respect to the SM in terms of new physics. In this section we illustrate the latter application with a particular example: explaining the recent hints in LHCb data of a violation of lepton flavor universality (LFU) in $B$-meson decays~\cite{Aaij:2014ora,Aaij:2017vbb}. We will first identify which heavy multiplets can generate the necessary operators and then look at correlated effects that could constrain or test the different possibilities. Our schematic analysis is just intended as an illustration. Most of the results in this section have in fact already appeared in the literature.

The measurement of the observables $R_{K}\equiv \mathrm{Br}(B^+\to K^+ \mu^+\mu^-)/\mathrm{Br}(B^+\to K^+ e^+ e^-)$ and $R_{K^*}\equiv \mathrm{Br}(B\to K^* \mu^+\mu^-)/\mathrm{Br}(B\to K^* e^+ e^-)$ provides a particularly clean test of LFU of the gauge interactions, since a large component of the SM theory uncertainties cancel in the ratio. The LHCb collaboration has recently presented measurements of these ratios, both of which deviate from the SM predictions by $\sim 2.6~\sigma$~\cite{Aaij:2014ora} and $\sim 2.4~\sigma$~\cite{Aaij:2017vbb}, respectively. 
These are not the only anomalies in $b\to s\ell^+ \ell^-$ processes, with some discrepancies also in the angular distributions of $B\to K^* \mu^+\mu^-$~\cite{Aaij:2014pli,Aaij:2013qta,Aaij:2015oid},  or in the differential branching fractions of $B\to K \mu^+ \mu^-$~\cite{Aaij:2014pli} and $B_s\to \phi \mu^+ \mu^-$~\cite{Aaij:2015esa}. 
At present, the different deviations follow a pattern that can be consistently explained by the presence of new physics. 
Altogether, the global fit to all flavour anomalies points to a deviation with respect to the SM hypotheses of $\sim 3$-$5~\sigma$, depending on the estimates assumed for the SM hadronic uncertainties in some of the observables \cite{Capdevila:2017bsm,DAmico:2017mtc,Altmannshofer:2017yso,Geng:2017svp,Ciuchini:2017mik,Celis:2017doq}.

The observed deviations from LFU in $B$ decays are well described by the following four-fermion effective Hamiltonian, valid at energies $E \ll M_W$, 
\begin{equation}
{\cal H}_{\mathrm Eff}^{b\to s \ell\ell}=-V_{tb}V_{ts}^* \frac{\alpha_{\mathrm{em}}}{4\pi} \frac{4 G_F}{\sqrt{2}}\sum C_{ij}^{\ell} {\cal O}_{ij}^\ell +\mathrm{h.c.}, 
\label{eq:Heff}
\end{equation}
where
\begin{equation}
{\cal O}_{ij}^\ell=(\bar{s}\gamma^\mu P_i b)(\bar{\ell}\gamma_\mu P_j \ell)
\end{equation}
are the different chiral four-fermion operators that can be obtained from the product of two vector currents, with $P_{L,R}=\frac 12 (1\mp \gamma_5)$. The fit to $R_{K,K*}$ favors an explanation where new physics is present in left-handed leptons and, in particular, points to a sizable deviation from the SM hypotheses in $C_{LL}^\ell$. 
For the purpose of this example, we focus the discussion around these interactions.
They can be either $C_{LL}^\mu<0$ or  $C_{LL}^e>0$, although a global fit to all $B$ anomalies prefers new physics in the muon sector, with $C_{LL}^\mu \approx -1.2\pm 0.3$~\cite{Capdevila:2017bsm,DAmico:2017mtc,Altmannshofer:2017yso,Geng:2017svp,Ciuchini:2017mik,Celis:2017doq}. 

Matching ${\cal O}_{LL}^\ell$ with the dimension-six SMEFT at the tree level results in the following four-fermion contributions to $C_{LL}^{\ell}$:
\begin{equation}
C_{LL}^\ell = \lambda_t^{-1} \left( C_{lq}^{(1)} + C_{lq}^{(3)} \right)_{\ell\ell 23},
\label{eq:matchingHeff}
\end{equation}
where $\lambda_t\equiv V_{tb}V_{ts}^* \frac{\alpha_{\mathrm{em}}}{4\pi} \frac{4 G_F}{\sqrt{2}}$, and we are working in a fermion basis with diagonal Yukawa interactions for the down-type quarks. The operators $\Ocal^{(1,3)}_{\phi q}$ and $\Ocal^{(1,3)}_{\phi l}$ also contribute, via a modification of the couplings of the $Z$ boson to the relevant quarks and leptons. However, such non-universal anomalous couplings are strongly bounded by LEP data, so we concentrate on the operators ${\cal O}_{lq}^{(1)}$ and ${\cal O}_{lq}^{(3)}$.

The relevant entries of the UV/IR dictionary are eqs.~(\ref{eq:Clq1}) and (\ref{eq:Clq3}). A look at the masses in the denominators of each term allows us to easily identify all the types of multiplets that can contribute to ${C}_{lq}^{(1)}$ and ${C}_{lq}^{(3)}$ at the tree level: 
\begin{equation}
\begin{array}{ c c c c c c c }
 &\scriptmath{(3,3)_{-\frac 13}}&\scriptmath{(1,1)_{0}}  & \scriptmath{(1,3)_{0} } & \scriptmath{(3,1)_{\frac 23} } & \scriptmath{(3,3)_{\frac 23} } & \\
\left\{\right.&\zeta,& {\cal B},&{\cal W},&{\cal U}_2,&{\cal X}&\left.\right\}.
\end{array}
\label{eq:Pbsll}
\end{equation}
Note that for $\omega_1$, $C_{lq}^{(1)}=-C_{lq}^{(3)}$ and therefore $C_{LL}^\ell=0$. 
This list with one scalar and four vector multiplets agrees with the classification in other studies, see, e.g.~\cite{DAmico:2017mtc,DiLuzio:2017chi,Crivellin:2017dsk,Buttazzo:2017ixm}. 
From eqs.~(\ref{eq:Clq1}) and (\ref{eq:Clq3}) we also see that there is no collective contribution with several heavy propagators in the same diagram. Most importantly, we can pinpoint the relevant couplings in~$\LBSM$. This is a simple example of looking at an IR entry of the dictionary to find its UV translation.

For instance, we can readily check in eqs.~(\ref{eq:Clq1}) and (\ref{eq:Clq3}) that a product of lepto-quark couplings is involved in the case of the scalar $\zeta$ and the vector bosons $\mathcal{X}$ and $\mathcal{U}^2$, while the vectors $\mathcal{B}$ and $\mathcal{W}$ contribute through a product of a diagonal lepton coupling and a flavor-changing quark coupling.

With this information, one can proceed to investigate in a systematic way all the different constraints (or signals) arising from other processes that involve the same couplings and particles. Processes involving other couplings will also be of great interest if the anomalies are confirmed. Direct searches with resonant production can be very relevant, but here we focus mostly on indirect searches. They reduce essentially to an analysis of the different operators, besides ${\cal O}_{lq}^{(1)}$ and ${\cal O}_{lq}^{(3)}$, that are generated when the heavy particles are integrated out. We can distinguish three kinds of contributions to the Wilson coefficients of the other induced operators:
\begin{description}
\item{Type I:} Contributions that depend {\em only} on couplings that enter in $C_ {LL}^{\ell}$. The corresponding observable effects are then correlated with the ones entering in $b\to s\ell^+ \ell^-$, and can be used to constrain or probe a given solution to the $B$-meson anomalies.
\item{Type II:} Contributions that depend on these couplings but can be made arbitrarily small by tuning an interaction not entering in $C_ {LL}^{\ell}$. In this case, the correlations require extra information on that coupling.
\item{Type III:} Contributions that do not depend on the couplings that appear in $C_ {LL}^{\ell}$. These are completely uncorrelated.
\end{description}
In this classification it is of course crucial to take flavor indices into account.
Even if contributions of type I are more relevant, an observation of the effects of contributions of type II and III could also be used to support the new physics interpretation and for model discrimination.

Let us examine along these lines the multiplets $\zeta$, $\mathcal{X}$ and $\mathcal{W}$, which have the compelling feature of allowing only for the required left-handed couplings. 
In this case, we will use the dictionary in the UV to IR direction. Tables~\ref{tab:topdown_scalars} and~\ref{tab:topdown_vectors} prove handy for this task, as they list the operators we need to look at for each assumed multiplet. 

\subsection*{Scalar leptoquark $\zeta$}

The interactions of $\zeta$ can be found in eq.~(\ref{eq:LS4}).  We see that the scalar $\zeta$ has, up to flavor indices, two couplings (besides the gauge couplings, determined by quantum numbers): the lepto-quark coupling $y_{\zeta}^{ql}$ and the coupling to quarks $y_{\zeta}^{qq}$. 
A glimpse at table~\ref{tab:topdown_scalars} tells us that the following operators are induced: ${\cal O}_{lq}^{(1,3)}$, ${\cal O}_{qq}^{(1,3)}$ and $\Ocal_{qqq}$. Then, we read the precise contributions to their Wilson coefficients from eqs.~(\ref{eq:Cqq1})-(\ref{eq:Clq3}) and (\ref{eq:Cqqq}). Assuming only one $\zeta$ multiplet,
\begin{align}
&(C_{lq}^{(1)})_{ijkl}=3 (C_{lq}^{(3)})_{ijkl}=\frac 34\frac{(y_{\zeta}^{ql})_{lj} (y_{\zeta}^{ql})_{ki}^*}{M_{\zeta}^2}, \label{eq:Clq13Zeta}\\
&(C_{qq}^{(1)})_{ijkl}=- 3 (C_{qq}^{(3)})_{ilkj}=\frac{
    3   (y^{qq}_{\zeta})_{ki}
    (y^{qq}_{\zeta})^*_{lj}}{
    2  M_{\zeta}^{2}}, 
\label{eq:Cqq13Zeta}    \\
&(C_{qqq})_{ijkl}=- \frac{
    2   (y^{qq}_{\zeta})^*_{ij}
    (y^{ql}_{\zeta})_{kl}}{
    M_{\zeta}^{2}}.
\label{eq:ClqZeta}
\end{align}
Looking at the flavor structure of~\refeq{eq:Clq13Zeta}, we see that we need sizable couplings $(y_{\zeta}^{ql})_{2\ell}$ and $(y_{\zeta}^{ql})_{3\ell}$ to explain the anomalies. For sufficiently low mass $M_{\zeta}$, these couplings can be probed by analyses of single and pair production of $\zeta$ at the LHC~\cite{Dorsner:2014axa}. The very same couplings also contribute to other components of $C_{qq}^{(1,3)}$, and we conclude that
\begin{equation}
C_{LL}^l \not=0 \longrightarrow \left\{
\begin{array}{c}
(C_{lq}^{(1)})_{\ell\ell 33}=3(C_{lq}^{(3)})_{\ell\ell 33}=\frac{3\left|(y_{\zeta}^{ql})_{3\ell}\right|^2}{4M_{\zeta}^2}\not = 0,\\
(C_{lq}^{(1)})_{\ell\ell 22}=3(C_{lq}^{(3)})_{\ell\ell 22}=\frac{3\left|(y_{\zeta}^{ql})_{2\ell}\right|^2}{4M_{\zeta}^2}\not = 0.
\end{array}
\right.
\label{eq:CorrZeta}
\end{equation}
These are contributions of type I. The corresponding effects in hadronic-flavor-preserving processes are correlated with the $B$ anomalies. From~\refeq{eq:CorrZeta} it is also clear that in these processes each of the two couplings can be measured, in principle, independently. Both the flavor-preserving and flavor-violating effects in an electron explanation of the anomalies can be tested in $e^+ e^-$ colliders. The observed values of $R_{K,K^*}$ can be reproduced with $C_{LL}^{\ell}\sim O(1)$, which corresponds to a new physics interaction scale of about 35 TeV, well above the sensitivity of LEP2. Therefore, current $e^+ e^-$ data do not provide significant constraints on the relevant couplings. However, they could be tested at future lepton colliders. Any other combination of flavor indices gives contributions of type III, with effects that are uncorrelated with the anomalies. The same holds for the contributions to the operators $\Ocal_{qq}^{(1,3)}$, which involve the quark couplings $y^{qq}_\zeta$. Finally, the baryon-number violating operator $\Ocal_{qqq}$ receives contributions of type~II or type~III, depending on the flavor indices. Note in particular that the quark couplings for the first family are strongly constrained by the non-observation of proton decay. 

\subsection*{Vector leptoquark $\mathcal{X}$}

The analysis of the vector multiplet $\mathcal{X}$ is similar, but as we can see in eq.~(\ref{eq:LV4}) in this case there is only one non-gauge coupling (up to flavor indices): the lepto-quark coupling $g_{\cal X}$. In table~\ref{tab:topdown_vectors} we see that only the operators ${\cal O}_{lq}^{(1,3)}$ are generated in the effective theory below the mass $M_\mathcal{X}$. Assuming only one replica of $\mathcal{X}$, eq.~(\ref{eq:LS4}) gives
\begin{equation}
(C_{lq}^{(1)})_{ijkl}=-3(C_{lq}^{(3)})_{ijkl}=-\frac{3(g_{\cal X})_{jk}^* (g_{\cal X})_{il}}{8M_{{\cal X}}^2}.
\label{eq:ClqX}
\end{equation}
We see that the contribution of $\mathcal{X}$ to $C_{LL}^\ell$ is proportional to the product of $(g_{\cal X})^*_{\ell 2}$ and $(g_{\cal X})_{\ell 3}$. Again, there are correlations with the coefficients of the corresponding hadronic flavor-conserving operators:
\begin{equation}
C_{LL}^l \not= 0 \longrightarrow \left\{
\begin{array}{c}
(C_{lq}^{(1)})_{\ell\ell 33}=-3(C_{lq}^{(3)})_{\ell\ell 33}=-\frac{3\left|(g_{\cal X})_{\ell 3}\right|^2}{8M_{\cal X}^2}\not = 0,\\
(C_{lq}^{(1)})_{\ell\ell 22}=-3(C_{lq}^{(3)})_{\ell\ell 22}=-\frac{3\left|g_{\cal X})_{\ell 2}\right|^2}{8M_{\cal X}^2}\not = 0.
\end{array}
\right.
\label{eq:CorrX}
\end{equation}
The same discussion in the paragraph below eq.~\refeq{eq:CorrZeta} applies to this case, except for the fact that now there are no purely-hadronic couplings.

\subsection*{Vector iso-triplet $\mathcal{W}$}

As we can check in eq.~(\ref{eq:LV4}), the vector iso-triplet ${\cal W}$ has couplings $g_{\mathcal{W}}^l$ and $g_{\mathcal{W}}^q$ to left-handed fermions and $g_{\mathcal{W}}^\phi$ to the Higgs doublet. The latter induces a mixing of the $Z^\prime$ and $W^\prime$ components with the $Z$ and $W$ bosons, respectively. There are also couplings involving a possible vector doublet $\mathcal{L}_1$, which we shall not consider. For masses $M_{\mathcal{W}}$ light enough, the $Z^\prime$ and $W^\prime$ bosons in $\mathcal{W}$ can be produced at hadron colliders if the light-quark couplings are not too small. They then decay into di-leptons (including lepton + MET)~\cite{deBlas:2012qp} and di-bosons~\cite{Pappadopulo:2014qza} through the couplings to leptons and to the Higgs, respectively. Regarding indirect effects, the operators that can be induced are listed in the $\mathcal{W}$ entry of table~\ref{tab:topdown_vectors}. The most relevant ones in the context of the $B$ anomalies are $\Ocal^{(3)}_{lq}$, $\Ocal_{ll}$ and $\Ocal_{qq}^{(3)}$, with Wilson coefficients given by (see eqs.~\refeq{eq:Clq3},~\refeq{eq:Cll} and~\refeq{eq:Cqq3})
\begin{align}
&(C_{lq}^{(3)})_{ijkl}=-\frac{(g_{\mathcal{W}}^l)_{ij} (g_{\mathcal{W}}^q)_{kl}}{4 M_{\mathcal{W}}^2}, \label{eq:Clq3W}\\
&(C_{ll})_{ijkl}= - \frac{ (g^{l}_{\mathcal{W}})_{ij}  (g^{l}_{\mathcal{W}})_{kl}}{ 8  M_{\mathcal{W}}^2}  , \label{eq:Cll3W}    \\
&(C_{qq}^{(3)})_{ijkl}= - \frac{ (g^{q}_{\mathcal{W}})_{ij}  (g^{q}_{\mathcal{W}})_{kl}}{ 8  M_{\mathcal{W}}^2}  . \label{eq:Cqq3W}
\end{align}
We see that to get the necessary $C_{LL}^{\ell}$ we need sizable couplings $(g^l_\mathcal{W})_{\ell\ell}$ and $(g^q_\mathcal{W})_{23}$. The first one must be non-universal, while the second one is explicitly flavor-changing. Schematically, we have the following correlations:
\begin{equation}
C_{LL}^l \not=0 \longrightarrow \left\{
\begin{array}{l c}
(C_{ll})_{\ell\ell \ell\ell}=-\frac{(g_{\cal W}^l)_{\ell\ell}^2}{8M_{\cal W}^2}\not = 0,&\\
(C_{qq}^{(3)})_{2323}=-\frac{(g_{\cal W}^q)_{23}^2}{8M_{\cal W}^2}\not = 0,&\\
(C_{qq}^{(3)})_{2332}=-\frac{|(g_{\cal W}^q)_{23}|^2}{8M_{\cal B}^2}\not = 0.
\end{array}
\right.
\end{equation}
Of particular importance is the contribution to
$(C_{qq}^{(3)})_{2323}$, as it generates contributions to
$B_s\!-\!\bar{B}_s$ mixing amplitudes. Such contributions are tightly
constrained, pushing the new physical interaction scale to values of
$O(100)$ TeV~\cite{Bevan:2014cya,Carrasco:2013zta}.\footnote{These
bounds, together with the ones discussed below, can be relaxed by
reducing the $(g_\mathcal{W}^q)_{23}$ and $g^\phi_{\cal W}$ 
couplings at the expense of increasing the corresponding
$(g_\mathcal{W}^{l})_{\ell \ell}$ ones~\cite{Boucenna:2016wpr,Boucenna:2016qad}. A similar comment applies to
the case of $\mathcal{B}$. Such leptophilic vector
bosons can be probed at colliders in multi-lepton searches~\cite{delAguila:2014soa,delAguila:2015vza}.}
 This case shows that, although $\Delta F=1$ and $\Delta F=2$ bounds are uncorrelated in a low-energy operator analysis, correlations may exist and be crucial in specific explanations of the $B$ anomalies. Similar considerations apply often to processes that may not appear to be connected in an effective low-energy description. 
Of course, the correlations become weaker as more particles (with the same or different quantum numbers) are included, but some of them are unavoidable~\cite{delAguila:2011yd}. 

Again, other combinations of flavor indices give contributions of type II and~III. 
The contributions of $\mathcal{W}$ to $\psi^2 \phi^2 D$ operators,
\begin{align}
&(C_{\phi l}^{(3)})_{ij}=-\frac{\mbox{Re}\left\{(g_{\cal W}^l)_{ij}~g_{\cal W}^\phi\right\}}{4 M_{\cal W}^2},\\
&(C_{\phi q}^{(3)})_{ij}=-\frac{\mbox{Re}\left\{(g_{\cal W}^q)_{ij}~g_{\cal W}^\phi\right\}}{4 M_{\cal W}^2},
\label{eq:CHf2W}
\end{align}
are of type~II for $ij=\ell\ell$ and $ij=23,32$, respectively, and of type~III otherwise. These operators modify the $Z$ and $W$ couplings to leptons and quarks, so they are constrained by electroweak precision data, by observables sensitive to flavor-changing decays of the $Z$ boson, $B_s\!-\!\bar{B}_s$ mixing and by non-resonant processes with di-lepton and di-jet final states at the LHC. But these limits can always be made compatible with the lepton and quark couplings that explain the anomalies by tuning the Higgs coupling $g_\mathcal{W}^\phi$ to be small. This coupling also induces type-III effects in Higgs  physics, via the operators $\Ocal_{\phi D}$, $\Ocal_{\phi}$, $\Ocal_{\phi \Box}$ and $\Ocal_{f \phi}$ ($f=e,d,u$), with Wilson coefficients
\begin{eqnarray}
C_{\phi D}=-\frac{\mbox{Im}\left\{(g_{\cal W}^{\phi})\right\}^2}{2M_{\cal W}^2},&
C_\phi=-\frac{\lambda_{\phi} (g_{\cal W}^{\phi})^2}{M_{\cal W}^2},&
C_{\phi\Box}=-\frac{|g_{\cal W}^{\phi}|^2}{4M_{\cal W}^2},\nonumber\\
(C_{e\phi (d\phi)})_{ij}= y_{ji}^{e (d)*} a,&
~~~~~(C_{u\phi})_{ij}=-2 y_{ji}^{u*} a^*,&
a\equiv -\frac{2 |g_{\cal W}^{\phi}|^2 + i~\!\mathrm{Im}((g_{\cal W}^{\phi})^2)}{8M_{\cal W}^2}.
\end{eqnarray}
(Note that we have replaced $\hat{\lambda}_\phi$ and $\hat{y}^{e,d,u}$ by $\lambda_\phi$ and $y^{e,d,u}$, respectively, as in the extension we are considering there are no contributions to dimension-four operators.)

Before finishing this section let us point out another possible usage of the UV/IR dictionary for model building. Say we are interested in a given class of models, including one or more of the multiplets that contribute at the tree level to the dimension-six effective Lagrangian. Then we can relax the indirect limits on the corresponding couplings by including other multiplets that (partially) cancel the contributions to the Wilson coefficients of interest. The different possibilities can be easily determined by a scan of our results in appendix~\ref{sec:results}. For instance, it is easy to see that the contributions of $\mathcal{W}$ to $(C_{ll})_{1111}$, which could be tested at future $e^+ e^-\rightarrow e^+ e^-$ colliders, can be (partially) cancelled, with some tuning, against the ones of a hypercharge 1 scalar singlet $\mathcal{S}_1$ or triplet $\Xi_1$~\cite{delAguila:2011yd}. 

\section{Conclusions}
\label{sec:conclusions}

The quest for new physics beyond the SM often takes the form of a detailed study of explicit models, or classes of models, which can be motivated by experimental, theoretical or aesthetical puzzles, or designed to give new signatures in current experiments. The models typically predict new particles, which could in principle be observed at colliders, and deviations in different observables from the indirect effects of those new particles. This approach has many advantages, but it suffers from one obvious drawback: by definition we do not know a priori which model has been chosen by Nature, so the possibilities are infinite. At energies sufficiently lower than the production threshold of any new particle, a model-independent and rigorous framework exists to study possible deviations of experimental observations with respect to the SM
theoretical predictions: effective field theories built with the SM degrees of freedom. Their underlying power counting also provides a rationale for the expected size of eventual corrections. In this case, the disadvantages are the limitation in energy---which make them invalid to study direct production of the new particles, the number of free parameters and the lack of obvious physical insight about the nature of the new physics. These two approaches are complementary and relating one to each other is essential to take advantage of their synergy. Actually, integrating the heavy degrees of freedom of particular models to find the corresponding low-energy effective Lagrangian is common practice. But once again, it seems at first sight that this task must be done model by model.

In this work we have shown, however, that the matching between the IR and UV descriptions can be performed once and for all at the leading order, namely for operators of canonical dimension up to six and at the classical level. 
The idea is to map the model-independent low-energy effective theory approach to arbitrary models of new physics.
With this purpose, we have considered a completely general extension of the SM, subject only to a few mild assumptions. This extension has an arbitrary number of new scalars, fermions and vectors, with no restrictions on their gauge quantum numbers nor on their possible interactions. In particular, we
have made no assumption about renormalizability.  

In order to construct explicitly the relevant part of this general extension of the SM, we have first examined which new particles and which couplings can contribute to
the SMEFT at the tree level. The result is that we only have to
consider new fields that can have linear interactions with the SM fields and no other extra field. The origin of this property lies in the fact that tree-level diagrams without external legs of heavy particles always end in two or more vertices of this type. With our restriction to contributions to operators of dimension up to six, power counting further shows that these linear interactions must be of dimension $\leq 4$. The complete list of these selected extra fields comprises
19 scalar, 13 fermion and 16 vector irreducible representations of the SM gauge group. These field multiplets are collected in
tables~\ref{t:scalars}, \ref{t:fermions}
and \ref{t:vectors}. After determining the relevant field content, we have proceeded to write all the possible gauge-invariant terms that can be constructed with these fields and the SM ones and that affect the tree-level matching.  The resulting Lagrangian provides a general parametrization, in terms of masses and coupling constants, of essentially any kind of new physics with unsuppressed impact at low energies. 

We emphasize that the fact that the new particles must have quantum numbers that allow for such linear interactions does not mean
that these are the only relevant couplings. Couplings
involving more than one heavy field can actually result in non-trivial
contributions to the SMEFT. Also non-renormalizable interactions
involving one new particle in interactions up to canonical dimension five may
be relevant. We have classified all these couplings and reported
them in appendix~\ref{sec:lagrangians}. 

We have then integrated out the new heavy particles in this completely
general extension of the SM at the tree-level and have computed the
Wilson coefficients of the corresponding SMEFT operators of dimension up to six in the
Warsaw basis. This is the main contribution of this work. We report
our results in the form of a UV/IR dictionary. A top-down approach to the analysis of new physics would first use our appendix~\ref{sec:Field_Op}, where we list all the operators that are
generated for specific new particles. In appendix~\ref{sec:results}, on the other hand, we
give our results organized from the bottom-up point of view, by writing the contribution
to each Wilson coefficient from an arbitrary number of new particles. This dictionary greatly simplifies the task of analyzing the low-energy implications of explicit models and obtaining the corresponding bounds on their parameters. It also helps
disentangle the origin of possible anomalies eventually
observed in experiments. We have included a short section to guide the
reader through our results and have provided a simple example to illustrate the
use of this dictionary.

It is interesting that all operators in the Warsaw
basis, except for the ones involving three field strength tensors, are
generated in our tree-level integration. This would naively seem to
contradict the arguments in ref.~\cite{Arzt:1994gp}, which, up to the
presence of $\mathcal{L}_1$, share our assumptions. 
In fact there is no contradiction since, as
we have shown, tree level contributions to operators that are
classified as ``loop generated'' in~\cite{Arzt:1994gp} only arise due
to non-renormalizable, dimension-five operators in our SM extension, which can only be generated in turn at the loop level in
any weakly-coupled renormalizable UV
completion of that theory. (See~\cite{Jenkins:2013fya} for a 
related discussion.)  However, we have included these operators in our dictionary because they could be unsuppressed in strongly-coupled completions.

We conclude by emphasizing that we have provided a complete
classification of all possible extensions of the SM (with new
particles up to spin 1) with low-energy implications at the leading order. These implications are
encoded in tree-level contributions to the Wilson coefficients of the dimension-six
operators in the SMEFT, which we have computed explicitly in terms of the masses and couplings of the new particles. This result can in principle be extended to
operators of higher dimension: as long as the classical approximation is used, the number of extra fields and extra couplings to be considered will be finite (even if huge). On the other hand, at the loop level this endeavor faces an additional problem: there are infinitely-many types of extra fields that can contribute, already at one loop, to dimension-six operators. The reason is that fields without linear couplings to the SM need also be considered in this case. So,  a complete matching to general extensions  beyond the classical approximation will need to deal with this difficulty.

\acknowledgments

We thank Nuria Rius and Arcadi Santamaria for an interesting discussion that motivated this work.
We also thank Paco del {\'A}guila and Toni Pich for useful comments.
The work of J.C.C., M.P.V. and J.S. has been supported by the Spanish MICINN project FPA2013- 47836-C3-2-P, 
the MINECO project FPA2016-78220-C3-1-P (Fondos FEDER) and the Junta de Andaluc{\'i}a grant FQM101.
The work of J.C.C. has also been supported by the Spanish MECD grant FPU14.
The work of M.P.V. and J.S. has also been supported by the European Commission through the contract PITN-GA-2012-316704 (HIGGSTOOLS).
J.C.C. is grateful for the hospitality of the Dipartimento di Fisica e Astronomia ``Galileo Galilei'' of the University of Padova during part of this work.
J.S. would like to thank the Mainz Institute for Theoretical Physics (MITP)  for  its  hospitality  and  partial  support  during the completion of this work.
 

\newpage

\appendix

\clearpage

\section{Explicit BSM Effective Lagrangian}
\label{sec:lagrangians}

In this appendix we present the explicit form of the different terms contributing to the BSM Lagrangian in eq.~(\ref{LBSMsplit}).
In these and the rest of the results in this paper we use a notation where
color indices
are labeled by capital letters, $A, B, C$, running over
the dimensionality of the corresponding $SU(3)_c$ representation. 
Whenever possible, objects in the fundamental representations of $SU(2)_L$ and $SU(3)_c$
have been written
as row or column vectors, with matrix products implied. 
The superscript symbol
``$^\mathrm{T}$'' indicates transposition of the $SU(2)_L$ indices exclusively.
When showing these indices explicitly, we use the following different labels, depending on the $SU(2)_L$ representation:
$\alpha,\beta=\frac 12,-\frac 12$ for $SU(2)_L$ doublets; 
$a,b,c = 1, 2, 3$ for the components of $SU(2)_L$ adjoints/triplets in Cartesian coordinates;
and $I,J,K = \frac 32, \frac 12,-\frac 12,-\frac 32$ for the components of the $SU(2)_L$ quadruplets.

The symbols $T_A=\frac 12 \lambda_A$ and
$f_{ABC}$, $A,B,C=1,\ldots,8$, denote the $SU(3)_c$ generators and structure
constants, respectively, with $\lambda_A$ the Gell-Mann matrices. 
$\epsilon_{ABC}$ ($\varepsilon_{abc}$) , $A,B,C=1,2,3$ ($a,b,c=1,2,3$) is the
totally antisymmetric tensor in color (weak isospin) indices; $\sigma_a$ or $\sigma^a$,
$a=1,2,3$ are the Pauli matrices; $\sigma_{\mu\nu}=\frac {i}{2}[\gamma_\mu,\gamma_\nu]$; and
$\tilde{A}_{\mu\nu}=\frac12\varepsilon_{\mu\nu\rho\sigma}A^{\rho\sigma}$ is the
Hodge-dual of the field strength $A_{\mu\nu}$. 
 
In the construction of the different $SU(2)_L$ invariants we also use the following:
\begin{itemize}
{\item The isospin-1 product of two triplets is obtained through:
  \beq
  f_{abc}=\frac{i}{\sqrt{2}}\varepsilon_{abc}.\nonumber
  \eeq
}
{\item Quadruplets are obtained from the product of an isospin-1 field and a doublet by means of 
  \bea
  C^{3/2}_{a\beta} = \frac{1}{\sqrt{2}}\left(
    \begin{array}{rr}
      1 & 0 \\
      -i & 0 \\
      0 & 0
    \end{array}
  \right),
  &
  ~~~~~~C^{1/2}_{a\beta} = \frac{1}{\sqrt{6}}\left(
    \begin{array}{rr}
      0 & 1 \\
      0 & -i \\
      -2 & 0
    \end{array}
  \right),\nonumber
  \eea
  \bea
  C^{-1/2}_{a\beta} = -\frac{1}{\sqrt{6}}\left(
    \begin{array}{rr}
      1 & 0 \\
      i & 0 \\
      0 & 2
    \end{array}
  \right),
  &
  ~~~~C^{-3/2}_{a\beta} = -\frac{1}{\sqrt{2}}\left(
    \begin{array}{rr}
      0 & 1 \\
      0 & i \\
      0 & 0
    \end{array}
  \right).\nonumber
  \eea
}
{\item The singlet product of two quadruplets is obtained through the $SU(2)$
  product
  \beq
  \epsilon_{IJ} = \frac 12 \left(
    \begin{array}{cccc}
      0 & 0 & 0 & 1 \\
      0 & 0 & -1 & 0 \\
      0 & 1 & 0 & 0 \\
      -1 & 0 & 0 & 0
    \end{array}\right).\nonumber
  \eeq
}
\end{itemize}
Finally, for $SU(3)_c$ indices, we use the following notation for the
symmetric product of colored fields:
$$
\psi_1^{\left(A\right|}\ldots \psi_2^{\left|B\right)}\equiv
\frac 12\left(\psi_1^{A}\ldots \psi_2^{B} + \psi_1^{B}\ldots \psi_2^{A}\right).
$$

\newpage

\subsection{Standard Model Lagrangian}

The renormalizable part of ${\cal L}_0$ in eq.~(\ref{LBSMsplit}) is just the SM Lagrangian
$\Lcal_{\mathrm{SM}}$. Let us write it explicitly (up to total derivatives). In standard
notation,\footnote{Latin indices $i,j,k$
  are used to label different generations. $L,R$ indicate the chiral
  components of spinors, written in Dirac's four-component notation.
  We use the notation $\hat{y}^{e,u,d}$, $\hat{\mu}_\phi$ and $\hat{\lambda}_\phi$ to denote couplings that will be renomalized by the effects of the heavy particles (see section~\ref{sec:redefinitions}).} it reads
\begin{align}
  {\cal L}_{\SM} = &
  -\frac 14 G^A_{\mu\nu} G^{A\ \mu\nu}
  -\frac 14 W^a_{\mu\nu}W^{a\ \mu\nu}
  -\frac 14 B_{\mu\nu}B^{\mu\nu} +
  \nonumber \\ &
  +\bar{l}_{Li} \iDslash  l_{Li}
  +\bar{q}_{Li} \iDslash  q_{Li}
  +\bar{e}_{Ri} \iDslash e_{Ri}
  +\bar{u}_{Ri} \iDslash u_{Ri}
  +\bar{d}_{Ri} \iDslash d_{Ri} +
  \nonumber \\ &
  +\left(D_\mu \phi\right)^\dagger D^\mu \phi
  -V\left(\phi\right)
  -\left(
    \hat{y}^e_{ij} ~ \bar{e}_{Ri} \phi l_{Lj}
    +\hat{y}^d_{ij} ~ \bar{d}_{Ri} \phi q_{Lj}
    +\hat{y}^u_{ij} ~ \bar{u}_{Ri} \tilde{\phi}^\dagger q_{Lj}
    +\hc
  \right).
\label{SMLag}
\end{align}
As usual,  ${\tilde \phi}=i\sigma_2 \phi^*$ denotes the iso-doublet of hypercharge $-1/2$. Here and below, the covariant derivatives acting on a field or operator $P$ in the representation $(C,I)_Y$ are 
\begin{equation}
  D_\mu P =
  \left(
    \d_\mu
    + i g_3 G^A_\mu \mathbf{T}^A_{C}
    + i g_2 W^a_\mu \mathrm{T}^a_{I}
    + i g_1 Y B_\mu
  \right) P,
\end{equation}
with $\mathbf{T}^A_{C}$ the \SU{3} generators in the $C$ representation and $\mathrm{T}^a_{I}$ the \SU{2} generators in the $I$ representation. Our normalization of the hypercharge is such that $Y=Q+\mathrm{T}^3_I$, with $Q$ the electric charge. 
The Higgs scalar potential is
\begin{equation}
  V\left(\phi\right) =
  -\hat{\mu}_\phi^2 \left|\phi \right|^2
  +\hat{\lambda}_\phi \left|\phi\right|^4.
\label{Scalar potential 1}
\end{equation}
We will not need to write explicitly the non-renormalizable part of ${\cal L}_0$.

\subsection{New Scalars}

The Lagrangian $\Lcal_{\mathrm{S}}$ can be written as the sum of two pieces:
\beq
\Lcal_{\mathrm{S}} = \Lcal_{\mathrm{S}}^{\mathrm{quad}}+\Lcal_{\mathrm{S}}^{\mathrm{int}}. \label{LSsplit}
\eeq
The first one contains the kinetic terms (with covariant derivatives) and mass terms of the new scalars:
\beq
\Lcal_{\mathrm{S}}^{\mathrm{quad}} = \sum_\sigma \eta_{\sigma} \left[\left(D_\mu \sigma\right)^\dagger D^\mu \sigma-M_{\sigma}^2\sigma^\dagger \sigma \right].
 \label{LSquad}
\eeq
Here, $\sigma$ are the different scalar fields in~table~\ref{t:scalars}. More than one scalar field in each representation is allowed.  The prefactor  $\eta_{\sigma}$ takes the value $1$ ($\frac 12$) when $\sigma$ is in a complex (real) representation of the gauge group. 
The second piece in~\refeq{LSsplit} contains the general interactions of the new scalars with the SM fields and among themselves. We distinguish the terms of dimension $d\leq 4$ and the ones of dimension $d=5$:
\begin{equation}
  \Lcal^{\mathrm{int}}_{\mathrm{S}} =
  \Lcal^{(\leq 4)}_{\mathrm{S}} + \Lcal^{(5)}_{\mathrm{S}},
\end{equation}
where
\begin{align}
  -\Lcal^{(\leq4)}_{\mathrm{S}} = & \;
  (\kappa_{\mathcal{S}})_r \mathcal{S}_r \phi^\dagger \phi
  + (\lambda_{\mathcal{S}})_{rs} \mathcal{S}_r \mathcal{S}_s \phi^\dagger \phi
  + (\kappa_{\mathcal{S}^3})_{rst} \mathcal{S}_r \mathcal{S}_s \mathcal{S}_t
  \nonumber \\ &
  + \left\{
    (y_{\mathcal{S}_1})_{rij}
    \mathcal{S}_{1r}^\dagger \bar{l}_{Li} i\sigma_2 l^c_{Lj}
    + \hc
  \right\}
  \nonumber \\ &
  + \left\{
    (y_{\mathcal{S}_2})_{rij} \mathcal{S}_{2k}^\dagger \bar{e}_{Ri} e^c_{Rj}
    + \hc
  \right\}
  \nonumber \\ &
  + \left\{
    (y^e_{\varphi})_{rij} \varphi_r^\dagger \bar{e}_{Ri} l_{Lj}
    + (y^d_{\varphi})_{rij} \varphi_r^\dagger \bar{d}_{Ri} q_{Lj}
    + (y^u_{\varphi})_{rij} \varphi_r^\dagger i\sigma_2 \bar{q}^T_{Li} u_{Rj}
  \right.
  \nonumber \\ &
  \quad
  \left.
    + (\lambda_{\varphi})_r
    \left(\varphi^\dagger_r \phi \right) \left(\phi^\dagger \phi\right)
    + \hc
  \right\}
  \nonumber \\ &
  + (\kappa_{\Xi})_r \phi^\dagger \Xi^a_r \sigma^a \phi
  + (\lambda_{\Xi})_{rs}
  \left(\Xi^a_r \Xi^a_s\right) \left(\phi^\dagger \phi\right)
  \nonumber \\ &
  + \frac{1}{2} (\lambda_{\Xi_1})_{rs}
  \left(\Xi^{a\dagger}_{1r} \Xi^a_{1s}\right)
  \left(\phi^\dagger\phi\right)
  + \frac{1}{2} (\lambda'_{\Xi_1})_{rs}
  f_{abc} \left(\Xi_{1r}^{a\dagger} \Xi_{1s}^b\right)
  \left(\phi^\dagger \sigma^c \phi\right)
  \nonumber \\ &
  +\left\{
    (y_{\Xi_1})_{rij}
    \Xi^{a\dagger}_{1r} \bar{l}_{Li} \sigma^a i\sigma_2 l^c_{Lj}
    + (\kappa_{\Xi_1})_r
    \Xi_{1r}^{a\dagger} \left(\tilde{\phi}^\dagger \sigma^a \phi\right)
    + \hc
  \right\}
  \nonumber \\ &
  + \left\{
    (\lambda_{\Theta_1})_r
    \left(\phi^\dagger \sigma^a \phi\right) C^I_{a\beta}
    \tilde{\phi}_\beta \epsilon_{IJ} \Theta_{1r}^{J}
    + \hc
  \right\}
  \nonumber \\ &
  + \left\{
    (\lambda_{\Theta_3})_r
    \left(\phi^\dagger \sigma^a \tilde{\phi}\right) C^I_{a\beta}
    \tilde{\phi}_\beta \epsilon_{IJ} \Theta_{3r}^{J}
    + \hc
  \right\}
  \nonumber \\ &
  + \left\{
    (y^{ql}_{\omega_1})_{rij}
    \omega_{1r}^\dagger \bar{q}^c_{Li} i\sigma_2 l_{Lj}
    +(y^{qq}_{\omega_1})_{rij}
    \omega_{1r}^{A\dagger} \epsilon_{ABC} \bar{q}^B_{Li} i\sigma_2 q^{c\,C}_{Lj}
  \right.
  \nonumber \\ &
  \quad
  \left.
    +(y^{eu}_{\omega_1})_{rij}
    \omega_{1r}^\dagger \bar{e}^c_{Ri} u_{Rj}
    + (y^{du}_{\omega_1})_{rij}
    \omega_{1r}^{A\dagger} \epsilon_{ABC} \bar{d}^B_{Ri} u^{c\,C}_{Rj}
    + \hc
  \right\}
  \nonumber \\ &
  + \left\{
    (y_{\omega_2})_{rij}
    \omega^{A\dagger}_{2r}\epsilon_{ABC} \bar{d}^B_{Ri} d^{c\,C}_{Rj}
    + \hc
  \right\}
  \nonumber \\ &
  + \left\{
    (y^{ed}_{\omega_4})_{rij} \omega^{A\dagger}_{4r} \bar{e}^c_{Ri} d_{Rj}
    + (y^{uu}_{\omega_4})_{rij}
    \omega^{A\dagger}_{4r} \epsilon_{ABC} \bar{u}^B_{Ri} u^{c\,C}_{Rj}
    + \hc
  \right\}
  \nonumber \\ &
  + \left\{
    (y_{\Pi_1})_{rij} \Pi_{1r}^\dagger i\sigma_2 \bar{l}^T_{Li} d_{Rj}
    + \hc
  \right\}
  \nonumber \\ &
  + \left\{
    (y^{lu}_{\Pi_7})_{rij} \Pi_{7r}^\dagger i\sigma_2 \bar{l}^T_{Li} u_{Rj}
    + (y^{eq}_{\Pi_7})_{rij} \Pi_{7r}^\dagger \bar{e}_{Ri} q_{Lj}
    + \hc
  \right\}
  \nonumber \\ &
  + \left\{
    (y^{ql}_{\zeta})_{rij}
    \zeta_r^{a\dagger} \bar{q}^c_{Li} i\sigma_2 \sigma^a l_{Lj}
    + (y^{qq}_{\zeta})_{rij} \zeta^{a\dagger}_r \epsilon_{ABC}
    \bar{q}^B_{Li}\sigma^a i\sigma_2 q^{c\,C}_{Lj}
    + \hc
  \right\}
  \nonumber \\ &
  + \left\{
    (y^{ud}_{\Omega_1})_{rij}
    \Omega_{1r}^{AB\dagger} \bar{u}^{c(A|}_{Ri} d^{|B)}_{Rj}
    + (y^{qq}_{\Omega_1})_{rij}
    \Omega_{1r}^{AB\dagger} \bar{q}^{c(A|}_{Li} i\sigma_2 q^{|B)}_{Lj}
    + \hc
  \right\}
  \nonumber \\ &
  + \left\{
    (y_{\Omega_2})_{rij}
    \Omega_{2r}^{AB\dagger} \bar{d}^{c(A|}_{Ri} d^{|B)}_{Rj}
    + \hc
  \right\}
  \nonumber \\ &
  + \left\{
    (y_{\Omega_4})_{rij}
    \Omega_{4r}^{AB\dagger} \bar{u}^{c(A|}_{Ri} u^{|B)}_{Rj}
    + \hc
  \right\}
  \nonumber \\ &
  + \left\{
    (y_{\Upsilon})_{rij}\Upsilon^{AB\dagger}_r
    \bar{q}^{c(A|}_{Li} i\sigma_2 \sigma^a q^{|B)}_{Lj}
    + \hc
  \right\}
  \nonumber \\ &
  + \left\{
    (y^{qu}_{\Phi})_{rij} \Phi^{A\dagger}_r i\sigma_2 \bar{q}^T_{Li} T_A u_{Rj}
    + (y^{dq}_{\Phi})_{rij} \Phi^{A\dagger}_r \bar{d}_{Ri} T_A q_{Lj}
    + \hc
  \right\}
  \nonumber \\ &
  + (\lambda_{\mathcal{S}\Xi})_{rs}
  \mathcal{S}_r \Xi^a_s \left(\phi^\dagger \sigma^a \phi\right)
  + (\kappa_{\mathcal{S}\Xi})_{rst} \mathcal{S}_r \Xi^a_s \Xi^a_t
  \nonumber \\ &
  + (\kappa_{\mathcal{S}\Xi_1})_{rst} \mathcal{S}_r \Xi^{a\dagger}_{1s} \Xi^a_{1t}
  + \left\{
    (\lambda_{\mathcal{S}\Xi_1})_{rs}
    \mathcal{S}_r \Xi^{a\dagger}_{1s}
    \left(\tilde{\phi}^\dagger \sigma^a \phi\right)
    + \hc
  \right\}
  \nonumber \\ &
  + \left\{
    (\kappa_{\mathcal{S}\varphi})_{rs} \mathcal{S}_r \varphi^\dagger_s \phi
    + (\kappa_{\Xi\varphi})_{rs} \Xi^a_r (\varphi^\dagger_s \sigma^a \phi)
    + (\kappa_{\Xi_1\varphi})_{rs}
    \Xi^{a\dagger}_{1r} \left(\tilde{\varphi}^\dagger_s \sigma^a \phi\right)
    + \hc
  \right\}
  \nonumber \\ &
  + (\kappa_{\Xi\Xi_1})_{rst} f_{abc} \Xi^a_r \Xi^{b\dagger}_{1s} \Xi^b_{1t}
  + \left\{
    (\lambda_{\Xi_1\Xi})_{rs}
    f_{abc} \Xi^{a\dagger}_{1r} \Xi^b_s
    \left(\tilde{\phi}^\dagger \sigma^c \phi\right)
    + \hc
  \right\}
  \nonumber \\ &
  + \Big\{
  (\kappa_{\Xi\Theta_1})_{rs}
  \Xi^a_r C^I_{a\beta} \tilde{\phi}_\beta \epsilon_{IJ} \Theta^J_{1s}
  + (\kappa_{\Xi_1\Theta_1})_{rs}
  \Xi^{a\dagger}_{1r} C^I_{a\beta} \phi_\beta \epsilon_{IJ}\Theta^J_{1s}
  \nonumber \\ &
  \quad
  + (\kappa_{\Xi_1\Theta_3})_{rs}
  \Xi^{a\dagger}_{1r} C^I_{a\beta} \tilde{\phi}_\beta \epsilon_{IJ} \Theta^J_{3s}
  +\hc
  \Big\},
  \label{eq:LS4}
\end{align}

\newpage

and
\begin{align}
  -\Lcal^{(5)}_{\mathrm{S}} = & \;
  \frac{1}{f} \Bigg[
    (\tilde{k}^\phi_{\mathcal{S}})_r \mathcal{S}_r D_\mu \phi^\dagger D^\mu \phi
    + (\tilde{\lambda}_{\mathcal{S}})_r \mathcal{S}_r |\phi|^4
  \nonumber \\ &
  \quad
  + (\tilde{k}^B_{\mathcal{S}})_r \mathcal{S}_r B_{\mu\nu} B^{\mu\nu}
  + (\tilde{k}^W_{\mathcal{S}})_r \mathcal{S}_r W^a_{\mu\nu} W^{a\,\mu\nu}
  + (\tilde{k}^G_{\mathcal{S}})_r \mathcal{S}_r G^A_{\mu\nu} G^{A\,\mu\nu}
  \nonumber \\ &
  \quad
  + (\tilde{k}^{\tilde{B}}_{\mathcal{S}})_r
  \mathcal{S}_r B_{\mu\nu} \tilde{B}^{\mu\nu}
  + (\tilde{k}^{\tilde{W}}_{\mathcal{S}})_r
  \mathcal{S}_r W^a_{\mu\nu} \tilde{W}^{a\,\mu\nu}
  + (\tilde{k}^{\tilde{G}}_{\mathcal{S}})_r
  \mathcal{S}_r G^A_{\mu\nu} \tilde{G}^{A\,\mu\nu}
  \nonumber \\ &
  \quad
  + \left\{
    (\tilde{y}^e_{\mathcal{S}})_{rij}
    \mathcal{S}_r \bar{e}_{Ri} \phi^\dagger l_{Lj}
    + (\tilde{y}^d_{\mathcal{S}})_{rij}
    \mathcal{S}_r \bar{d}_{Ri} \phi^\dagger q_{Lj}
    + (\tilde{y}^u_{\mathcal{S}})_{rij}
    \mathcal{S}_r \bar{u}_{Ri} \tilde{\phi}^\dagger q_{Lj}
    + \hc
  \right\}
  \nonumber \\ &
  \quad
  + (\tilde{k}^\phi_{\Xi})_r \Xi^a_r D_\mu \phi^\dagger \sigma^a D^\mu \phi
  + (\tilde{\lambda}_{\Xi})_r \Xi^a_r |\phi|^2 \phi^\dagger \sigma^a \phi
  \nonumber \\ &
  \quad
  + (\tilde{k}^{WB}_{\Xi})_r \Xi^a_r W^a_{\mu\nu} B^{\mu\nu}
  + (\tilde{k}^{W\tilde{B}}_{\Xi})_r \Xi^a_r W^a_{\mu\nu} \tilde{B}^{\mu\nu}
  \nonumber \\ &
  \quad
  + \left\{
    (\tilde{y}^e_{\Xi})_{rij}
    \Xi^a_r \bar{e}_{Ri} \phi^\dagger \sigma^a l_{Lj}
    + (\tilde{y}^d_{\Xi})_{rij}
    \Xi^a_r \bar{d}_{Ri} \phi^\dagger \sigma^a q_{Lj}
    + (\tilde{y}^u_{\Xi})_{rij}
    \Xi^a_r \bar{u}_{Ri} \tilde{\phi}^\dagger \sigma^a q_{Lj}
    + \hc
  \right\}
  \nonumber \\ &
  \quad 
  + \left\{
    (\tilde{k}_{\Xi_1})_r
    \Xi^{a\dagger}_{1r} D_\mu \tilde{\phi}^\dagger \sigma^a D^\mu \phi
    + (\tilde{\lambda}_{\Xi_1})_r
    \Xi^{a\dagger}_{1r} |\phi|^2 \tilde{\phi}^\dagger \sigma^a \phi
    + (\tilde{y}^e_{\Xi_1})_{rij}
    \Xi^{a\dagger}_{1r} \bar{e}_{Ri} \tilde{\phi}^\dagger \sigma^a l_{Lj}
  \right.
  \nonumber \\ &
  \quad \quad
  \left.
    + (\tilde{y}^d_{\Xi_1})_{rij}
    \Xi^{a\dagger}_{1r} \bar{d}_{Ri} \tilde{\phi}^\dagger \sigma^a q_{Lj}
    + (\tilde{y}^u_{\Xi_1})_{rij}
    \Xi^{a\dagger}_{1r} \bar{q}_{Li} \sigma^a \phi u_{Rj}
    + \hc
  \right\}
  \Bigg] .
  \label{eq:LS5}
\end{align}

\subsection{New Fermions}
As indicated in section~\ref{sec:fields}, we exclude the possibility of extra fermions with chiral transformations under the gauge group $H$. Then, in the massive fermion sector, the complex irreducible representations of $H$ are carried by vector-like Dirac spinors, while the real irreducible representations are carried by Majorana spinors $\psi$, with $\psi_L = (\psi_R)^c \equiv \psi_R^c$. The only instances of the latter possibility are the extra leptons $N$ and $\Sigma$ in table~\ref{t:fermions}. In our ``field basis'', the diagonal mass matrices are given by sums of Dirac mass terms (for the complex representations) and Majorana mass terms (for the real representations).\footnote{Note that the particular case of a Dirac fermion $\Psi$ of mass $M_\Psi$ in a real representation of $H$ is equivalent to our description with two degenerate Majorana fields $\psi_1$ and $\psi_2$ of mass $M_\Psi$, with $\Psi_{R} = 1/\sqrt{2} \left(\psi_{1R} + i \psi_{2R} \right)$ and  $\Psi_{L} = 1/\sqrt{2} \left(\psi_{1R}^c + i \psi_{2R}^c \right)$.}

The general Lagrangian $\Lcal_{\mathrm{F}}$ is given by
\beq
\Lcal_{\mathrm{F}} = \Lcal_{\mathrm{F}}^{\mathrm{quad}} + \Lcal_{\mathrm{F}}^{\mathrm{int}},
\eeq 
where
\beq
\Lcal_{\mathrm{F}}^{\mathrm{quad}} =  \sum_\psi  \eta_{\psi} \left[ \bar{\psi} \iDslash \psi - M_{\psi} \bar{\psi} \psi \right] ,
\eeq
with $\psi$ labelling the different fields in table~\ref{t:fermions}, with an arbitrary number of fields in each irreducible representation, and $\eta_{\psi}=1$ ($\eta_{\psi}=1/2$) when $\psi$ is Dirac (Majorana), and
\begin{equation}
  \Lcal^{\mathrm{int}}_{\mathrm{F}} =
  \Lcal^{(4)}_{\mathrm{leptons}} + \Lcal^{(4)}_{\mathrm{quarks}}
  + \Lcal^{(5)}_{\mathrm{leptons}} + \Lcal^{(5)}_{\mathrm{quarks}} ,
\end{equation}
where
\begin{align}
  -\Lcal^{(4)}_{\mathrm{leptons}} = & \;
  (\lambda_N)_{ri} \bar{N}_{Rr} \tilde{\phi}^\dagger l_{Li}
  + (\lambda_E)_{ri} \bar{E}_{Rr} \phi^\dagger l_{Li}
  \nonumber \\ &
  + (\lambda_{\Delta_1})_{ri} \bar{\Delta}_{1Lr} \phi e_{Ri}
  + (\lambda_{\Delta_3})_{ri} \bar{\Delta}_{3Lr} \tilde{\phi} e_{Ri}
  \nonumber \\ &
  + \frac{1}{2} (\lambda_{\Sigma})_{ri}
  \bar{\Sigma}^a_{Rr} \tilde{\phi}^\dagger \sigma^a l_{Li}
  + \frac{1}{2} (\lambda_{\Sigma_1})_{ri}
  \bar{\Sigma}^a_{1Rr} \phi^\dagger \sigma^a l_{Li}
  \nonumber \\ &
  + (\lambda_{N \Delta_1})_{rs} \bar{N}^c_{Rr} \phi^\dagger \Delta_{1Rs}
  + (\lambda_{E \Delta_1})_{rs} \bar{E}_{Lr} \phi^\dagger \Delta_{1Rs}
  \nonumber \\ &
  + (\lambda_{E \Delta_3})_{rs} \bar{E}_{Lr} \tilde{\phi}^\dagger \Delta_{3Rs}
  + \frac{1}{2} (\lambda_{\Sigma \Delta_1})_{rs}
  \bar{\Sigma}^{c\,a}_{Rr} \tilde{\phi}^\dagger \sigma^a \Delta_{1Rs}
  \nonumber \\ &
  + \frac{1}{2} (\lambda_{\Sigma_1 \Delta_1})_{rs}
  \bar{\Sigma}^a_{1Lr} \phi^\dagger \sigma^a \Delta_{1Rs}
  + \frac{1}{2} (\lambda_{\Sigma_1 \Delta_3})_{rs}
  \bar{\Sigma}^a_{1Lr} \tilde{\phi}^\dagger \sigma^a \Delta_{3Rs}
  + \hc,
  \label{eq:Lleptons4} \\[5mm]
  -\Lcal^{(4)}_{\mathrm{quarks}} = & \;
  (\lambda_U)_{ri} \bar{U}_{Rr} \tilde{\phi}^\dagger q_{Li}
  + (\lambda_D)_{ri} \bar{D}_{Rr} \phi^\dagger q_{Li}
  \nonumber \\ &
  + (\lambda^u_{Q_1})_{ri} \bar{Q}_{1Lr} \tilde{\phi} u_{Ri}
  + (\lambda^d_{Q_1})_{ri} \bar{Q}_{1Lr} \phi d_{Ri}
  \nonumber \\ &
  + (\lambda_{Q_5})_{ri} \bar{Q}_{5Lr} \tilde{\phi} d_{Ri}
  + (\lambda_{Q_7})_{ri} \bar{Q}_{7Lr} \phi u_{Rj}
  \nonumber \\ &
  + \frac{1}{2} (\lambda_{T_1})_{ri}
  \bar{T}^a_{1Rr} \phi^\dagger \sigma^a q_{Li}
  + \frac{1}{2} (\lambda_{T_2})_{ri}
  \bar{T}^a_{2Rr} \tilde{\phi}^\dagger \sigma^a q_{Li}
  \nonumber \\ &
  + (\lambda_{U Q_1})_{rs} \bar{U}_{Lr} \tilde{\phi}^\dagger Q_{1Rs}
  + (\lambda_{U Q_7})_{rs} \bar{U}_{Lr} \phi^\dagger Q_{7Rs}
  \nonumber \\ &
  + (\lambda_{D Q_1})_{rs} \bar{D}_{Lr} \phi^\dagger Q_{1Rs}
  + (\lambda_{D Q_5})_{rs} \bar{D}_{Lr} \tilde{\phi}^\dagger Q_{5Rs}
  \nonumber \\ &
  + \frac{1}{2} (\lambda_{T_1 Q_1})_{rs} \bar{T}^a_{1Lr} \phi^\dagger \sigma^a Q_{1Rs}
  + \frac{1}{2} (\lambda_{T_1 Q_5})_{rs}
  \bar{T}^a_{1Lr} \tilde{\phi}^\dagger \sigma^a Q_{5Rs}
  \nonumber \\ &
  + \frac{1}{2} (\lambda_{T_2 Q_1})_{rs}
  \bar{T}^a_{2Lr} \tilde{\phi}^\dagger \sigma^a Q_{1Rs}
  + \frac{1}{2} (\lambda_{T_2 Q_7})_{rs} \bar{T}^a_{2Lr} \phi^\dagger \sigma^a Q_{7Rs}
  + \hc,
  \label{eq:Lquarks4} \\[5mm]
  - \Lcal^{(5)}_{\mathrm{leptons}} = & \; \frac{1}{f} \Bigg[
  (\tilde{\lambda}_{N})_{ri} \bar{N}^c_{Rr}
  \gamma^\mu \left(D_\mu \tilde{\phi}\right)^\dagger l_{Li}
  \nonumber \\ &
  \quad
  + (\tilde{\lambda}^l_{E})_{ri} \bar{E}_{Lr}
  \gamma^\mu \left(D_\mu \phi\right)^\dagger l_{Li}
  + (\tilde{\lambda}^B_{E})_{ri} \bar{E}_{Lr} \sigma^{\mu\nu} e_{Ri} B_{\mu\nu}
  + (\tilde{\lambda}^e_{E})_{ri} \bar{E}_{Lr} \phi^\dagger \phi e_{Ri}
  \nonumber \\ &
  \quad
  + (\tilde{\lambda}^e_{\Delta_1})_{ri} \bar{\Delta}_{1Rr}
  \slashed{D} \phi e_{Ri}
  + (\tilde{\lambda}^l_{\Delta_1})_{ri} \left(\bar{\Delta}_{1Rr} 
    \phi\right) \left(\phi^\dagger l_{Li}\right)
  + (\tilde{\lambda}^{l\prime}_{\Delta_1})_{ri} \left(\bar{\Delta}_{1Rr} 
    l_{Li}\right) \left(\phi^\dagger \phi\right)
  \nonumber \\ &
  \quad
  + (\tilde{\lambda}^B_{\Delta_1})_{ri} \bar{\Delta}_{1Rr}
  \sigma^{\mu\nu} l_{Li} B_{\mu\nu}
  + (\tilde{\lambda}^W_{\Delta_1})_{ri} \bar{\Delta}_{1Rr}
  \sigma^{\mu\nu} \sigma^a l_{Li} W^a_{\mu\nu}
  \nonumber \\ &
  \quad
  + (\tilde{\lambda}^e_{\Delta_3})_{ri} \bar{\Delta}_{3Rr}
  \slashed{D} \tilde{\phi} e_{Ri}
  + (\tilde{\lambda}^l_{\Delta_3})_{ri} \left(\bar{\Delta}_{3Rr} 
    \tilde{\phi}\right) \left(\phi^\dagger l_{Li}\right)
  \nonumber \\ &
  \quad
  + (\tilde{\lambda}^l_{\Sigma})_{ri} \bar{\Sigma}^{c\,a}_{Rr}
  \gamma^\mu \left(D_\mu \tilde{\phi}\right)^\dagger \sigma^a l_{Li}
  + (\tilde{\lambda}^e_{\Sigma})_{ri} \bar{\Sigma}^{c\,a}_{Rr}
  \tilde{\phi}^\dagger \sigma^a \phi e_{Ri}
  \nonumber \\ &
  \quad
  + (\tilde{\lambda}^l_{\Sigma_1})_{ri} \bar{\Sigma}^a_{1Lr}
  \gamma^\mu \left(D_\mu \phi\right)^\dagger \sigma^a l_{Li}
  + (\tilde{\lambda}^e_{\Sigma_1})_{ri} \bar{\Sigma}^a_{1Lr}
  \phi^\dagger \sigma^a \phi e_{Ri}
  \nonumber \\ &
  \quad
  + (\tilde{\lambda}^W_{\Sigma_1})_{ri} \bar{\Sigma}^a_{1Lr}
  \sigma^{\mu\nu} e_{Ri} W^a_{\mu\nu}
  \Bigg] + \hc,
  \label{eq:Lleptons5}
\end{align}

\newpage

\begin{align}
  - \Lcal^{(5)}_{\mathrm{quarks}} = & \; \frac{1}{f} \Bigg[
  (\tilde{\lambda}^q_{U})_{ri} \bar{U}_{Lr}
  \gamma^\mu \left(D_\mu \tilde{\phi}\right)^\dagger q_{Li}
  + (\tilde{\lambda}^u_{U})_{ri} \bar{U}_{Lr}
  \phi^\dagger \phi u_{Ri}
  \nonumber \\ &
  \quad
  + (\tilde{\lambda}^B_{U})_{ri} \bar{U}_{Lr}
  \sigma^{\mu\nu} u_{Ri} B_{\mu\nu}
  + (\tilde{\lambda}^G_{U})_{ri} \bar{U}_{Lr}
  T_A \sigma^{\mu\nu} u_{Ri} G^A_{\mu\nu}
  \nonumber \\ &
  \quad
  + (\tilde{\lambda}^q_{D})_{ri} \bar{D}_{Lr}
  \gamma^\mu \left(D_\mu \phi\right)^\dagger q_{Li}
  + (\tilde{\lambda}^d_{D})_{ri} \bar{D}_{Lr}
  \phi^\dagger \phi d_{Ri}
  \nonumber \\ &
  \quad
  + (\tilde{\lambda}^B_{D})_{ri} \bar{D}_{Lr}
  \sigma^{\mu\nu} d_{Ri} B_{\mu\nu}
  + (\tilde{\lambda}^G_{D})_{ri} \bar{D}_{Lr}
  T_A \sigma^{\mu\nu} d_{Ri} G^A_{\mu\nu}
  \nonumber \\ &
  \quad
  + (\tilde{\lambda}^u_{Q_1})_{ri} \bar{Q}_{1Rr}
  \slashed{D} \tilde{\phi} u_{Ri}
  + (\tilde{\lambda}^d_{Q_1})_{ri} \bar{Q}_{1Rr}
  \slashed{D} \phi d_{Ri}
  \nonumber \\ &
  \quad
  + (\tilde{\lambda}^q_{Q_1})_{ri} \left(\bar{Q}_{1Rr}
    \phi\right) \left(\phi^\dagger q_{Li}\right)
  + (\tilde{\lambda}^{q\prime}_{Q_1})_{ri} \left(\bar{Q}_{1Rr}
    q_{Li}\right) \left(\phi^\dagger \phi\right)
  \nonumber \\ &
  \quad
  + (\tilde{\lambda}^B_{Q_1})_{ri} \bar{Q}_{1Rr}
  \sigma^{\mu\nu} q_{Li} B_{\mu\nu}
  + (\tilde{\lambda}^W_{Q_1})_{ri} \bar{Q}_{1Rr}
  \sigma^{\mu\nu} \sigma^a q_{Li} W^a_{\mu\nu}
  \nonumber \\ &
  \quad
  + (\tilde{\lambda}^G_{Q_1})_{ri} \bar{Q}_{1Rr}
  \sigma^{\mu\nu} T^A q_{Li} G^A_{\mu\nu}
  \nonumber \\ &
  \quad
  + (\tilde{\lambda}^d_{Q_5})_{ri} \bar{Q}_{5Rr}
  \slashed{D} \tilde{\phi} d_{Ri}
  + (\tilde{\lambda}^q_{Q_5})_{ri} \left(\bar{Q}_{5Rr}
    \tilde{\phi}\right) \left(\phi^\dagger q_{Li}\right)
  \nonumber \\ &
  \quad
  + (\tilde{\lambda}^u_{Q_7})_{ri} \bar{Q}_{7Rr}
  \slashed{D} \phi u_{Ri}
  + (\tilde{\lambda}^q_{Q_7})_{ri} \left(\bar{Q}_{7Rr}
    \phi\right) \left(\tilde{\phi}^\dagger q_{Li}\right)
  \nonumber \\ &
  \quad
  + (\tilde{\lambda}^q_{T_1})_{ri} \bar{T}^a_{1Lr}
  \gamma^\mu \left(D_\mu \phi\right)^\dagger \sigma^a q_{Li}
  + (\tilde{\lambda}^u_{T_1})_{ri} \bar{T}^a_{1Lr}
  \phi^\dagger \sigma^a \tilde{\phi} u_{Ri}
  \nonumber \\ &
  \quad
  + (\tilde{\lambda}^d_{T_1})_{ri} \bar{T}^a_{1Lr}
  \phi^\dagger \sigma^a \phi d_{Ri}
  + (\tilde{\lambda}^W_{T_1})_{ri} \bar{T}^a_{1Lr}
  \sigma^{\mu\nu} d_{Ri} W^a_{\mu\nu}
  \nonumber \\ &
  \quad
  + (\tilde{\lambda}^q_{T_2})_{ri} \bar{T}^a_{2Lr}
  \gamma^\mu \left(D_\mu \tilde{\phi}\right)^\dagger \sigma^a q_{Li}
  + (\tilde{\lambda}^u_{T_2})_{ri} \bar{T}^a_{2Lr}
  \phi^\dagger \sigma^a \phi u_{Ri}
  \nonumber \\ &
  \quad
  + (\tilde{\lambda}^d_{T_2})_{ri} \bar{T}^a_{2Lr}
  \tilde{\phi}^\dagger \sigma^a \phi d_{Ri}
  + (\tilde{\lambda}^W_{T_2})_{ri} \bar{T}^a_{2Lr}
  \sigma^{\mu\nu} u_{Ri} W^a_{\mu\nu}
  \Bigg]
  +\hc~\!.
  \label{eq:Lquarks5}
\end{align}

\subsection{New Vectors}
For the extra vectors, we write
\beq
\Lcal_{\mathrm{V}} = \Lcal_{\mathrm{V}}^{\mathrm{quad}} + \Lcal_{\mathrm{V}}^{\mathrm{int}},
\eeq
where\footnote{For each $V$, this covariant Proca Lagrangian describes a particle of spin 1 coupled to the SM gauge fields. Other choices of the kinetic term would give rise to ghosts.}
\beq
 \Lcal_{\mathrm{V}}^{\mathrm{quad}}  = \sum_V \eta_V \left( D_\mu V_\nu^\dagger D^\nu V^\mu - D_\mu V_\nu^\dagger D^\mu V^\nu + M_V^2 V_\mu^\dagger V^\mu \right),
\eeq
with $V$ on the right-hand side labelling the different fields in table~\ref{t:vectors}, with an arbitrary number of fields in each irreducible representation, and $\eta_{V}=1$ ($\eta_{V}=1/2$) when $V$ is in a complex (real) representation of $H$, and
\begin{equation}
  \Lcal^{\mathrm{int}}_{\mathrm{V}} =
  \Lcal^{(\leq4)}_{\mathrm{V}} + \Lcal^{(5)}_{\mathrm{V}},
\end{equation}
where
\begin{align}
  -\Lcal^{(\leq 4)}_{\mathrm{V}} = & \;
  (g^l_{\mathcal{B}})_{rij} \mathcal{B}^\mu_r \bar{l}_{Li} \gamma_\mu l_{Lj}
  + (g^q_{\mathcal{B}})_{rij} \mathcal{B}^\mu_r \bar{q}_{Li} \gamma_\mu q_{Lj}
  + (g^e_{\mathcal{B}})_{rij} \mathcal{B}^\mu_r \bar{e}_{Li} \gamma_\mu e_{Lj}
  \nonumber \\ &
  + (g^d_{\mathcal{B}})_{rij} \mathcal{B}^\mu_r \bar{d}_{Li} \gamma_\mu d_{Lj}
  + (g^u_{\mathcal{B}})_{rij} \mathcal{B}^\mu_r \bar{u}_{Li} \gamma_\mu u_{Lj}
  + \left\{
    (g^\phi_{\mathcal{B}})_r \mathcal{B}^\mu_r \phi^\dagger iD_\mu \phi + \hc
  \right\}
  \nonumber \\ &
  + \left\{
    (g^{du}_{\mathcal{B}_1})_{rij}
    \mathcal{B}^{\mu\dagger}_{1r} \bar{d}_{Ri}\gamma_\mu u_{Rj}
    + (g^\phi_{\mathcal{B}_1})_r
    \mathcal{B}^{\mu\dagger}_{1r} i D_\mu \phi^T i\sigma_2 \phi
    + \hc
  \right\}
  \nonumber \\ &
  + \frac{1}{2} (g^l_{\mathcal{W}})_{rij}
  \mathcal{W}^{\mu a}_r \bar{l}_{Li} \sigma^a \gamma_\mu l_{Lj}
  + \frac{1}{2} (g^q_{\mathcal{W}})_{rij}
  \mathcal{W}^{\mu a}_r \bar{q}_{Li} \sigma^a \gamma_\mu q_{Lj}
  \nonumber \\ &
  + \left\{
    \frac{1}{2} (g^\phi_{\mathcal{W}})_r
    \mathcal{W}^{\mu a}_r \phi^\dagger \sigma^a iD_\mu \phi + \hc
  \right\}
  \nonumber \\ &
  + \left\{
    \frac{1}{2} (g_{\mathcal{W}_1})_r
    \mathcal{W}^{\mu a\dagger}_{1r} i D_\mu\phi^T i\sigma_2 \sigma^a \phi
    + \hc
  \right\}
  \nonumber \\ &
  + (g^q_{\mathcal{G}})_{rij}
  \mathcal{G}^{\mu A}_r \bar{q}_{Li} \gamma_\mu T_A q_{Lj}
  +  (g^u_{\mathcal{G}})_{rij}
  \mathcal{G}^{\mu A}_r \bar{u}_{Li} \gamma_\mu T_A u_{Rj}
  + (g^d_{\mathcal{G}})_{rij}
  \mathcal{G}^{\mu A}_r \bar{d}_{Ri} \gamma_\mu T_A d_{Rj}
  \nonumber \\ &
  + \left\{
    (g_{\mathcal{G}_1})_{rij}
    \mathcal{G}^{A\mu\dagger}_{1r} \bar{d}_{Ri} T_A \gamma_\mu u_{Rj}
    + \hc
  \right\}
  \nonumber \\ &
  + \frac{1}{2} (g_{\mathcal{H}})_{rij}
  \mathcal{H}^{\mu a A}_r \bar{q}_{Li} \gamma_\mu \sigma^a T_A q_{Lj}
  \nonumber \\ &
  + \left\{
    (\gamma_{\mathcal{L}_1})_r \mathcal{L}_{1r\mu}^{\dagger} D^{\mu} \phi
    + \hc
  \right\}
  \nonumber \\ &
  + i (g^B_{\mathcal{L}_1})_{rs}
  \mathcal{L}_{1r\mu}^{\dagger} \mathcal{L}_{1s\nu} B^{\mu\nu}
  + i (g^W_{\mathcal{L}_1})_{rs}
  \mathcal{L}_{1i\mu}^{\dagger} \sigma^a \mathcal{L}_{1j\nu} W^{a\,\mu\nu}
  \nonumber \\ &
  + i (g^{\tilde{B}}_{\mathcal{L}_1})_{rs}
  \mathcal{L}_{1r\mu}^{\dagger} \mathcal{L}_{1s\nu} \tilde{B}^{\mu\nu} 
  + i (g^{\tilde{W}}_{\mathcal{L}_1})_{rs}
  \mathcal{L}_{1r\mu}^{\dagger}
  \sigma^{a} \mathcal{L}_{1s\nu} \tilde{W}^{a\,\mu\nu}
  \nonumber \\ &
  + (h^{(1)}_{\mathcal{L}_1})_{rs}
  \left(\mathcal{L}_{1r\mu}^{\dagger} \mathcal{L}^{\mu}_{1s}\right)
  \left(\phi^{\dagger} \phi\right)
  + (h^{(2)}_{\mathcal{L}_1})_{rs}
  \left(\mathcal{L}_{1r\mu}^{\dagger} \phi\right)
  \left(\phi^{\dagger} \mathcal{L}^{\mu}_{1s}\right)
  \nonumber \\ &
  + \left\{
    (h^{(3)}_{\mathcal{L}_1})_{rs}
    \left( \mathcal{L}_{1r\mu}^{1\dagger} \phi\right)
    \left(\mathcal{L}^{\dagger\mu}_{1s} \phi\right)
    + \hc
  \right\}
  \nonumber \\ &
  + \left\{
    (g_{\mathcal{L}_3})_{rij}
    \mathcal{L}^{\mu\dagger}_{3r} \bar{e}^c_{Ri} \gamma_\mu l_{Lj}
    + \hc
  \right\}
  \nonumber \\ &
  + \left\{
    (g^{ed}_{\mathcal{U}_2})_{rij}
    \mathcal{U}^{\mu\dagger}_{2r} \bar{e}_{Ri} \gamma_\mu d_{Rj}
    + (g^{lq}_{\mathcal{U}_2})_{rij}
    \mathcal{U}^{\mu\dagger}_{2r} \bar{l}_{Li} \gamma_\mu q_{Lj}
    + \hc
  \right\}
  \nonumber \\ &
  + \left\{
    (g_{\mathcal{U}_5})_{rij}
    \mathcal{U}^{\mu\dagger}_{5r} \bar{e}_{Ri} \gamma_\mu u_{Rj}
    + \hc
  \right\}
  \nonumber \\ &
  + \left\{
    (g^{ul}_{\mathcal{Q}_1})_{rij}
    \mathcal{Q}^{\mu\dagger}_{1r} \bar{u}^c_{Ri} \gamma_\mu l_{Lj}
    + (g^{dq}_{\mathcal{Q}_1})_{rij}
    \mathcal{Q}^{A\mu\dagger}_{1r}
    \epsilon_{ABC} \bar{d}^B_{Ri} \gamma_\mu i\sigma_2 q^{c\,C}_{Lj}
    + \hc
  \right\}
  \nonumber \\ &
  + \left\{
    (g^{dl}_{\mathcal{Q}_5})_{rij}
    \mathcal{Q}^{\mu\dagger}_{5r} \bar{d}^c_{Ri} \gamma_\mu l_{Lj}
    + (g^{eq}_{\mathcal{Q}_5})_{rij}
    \mathcal{Q}^{\mu\dagger}_{5r}
    \bar{e}^c_{Ri} \gamma_\mu q_{Lj}
  \right.
  \nonumber \\ &
  \quad
  \left.
    + (g^{uq}_{\mathcal{Q}_5})_{rij}
    \mathcal{Q}^{A\mu\dagger}_{5r}
    \epsilon_{ABC} \bar{u}^B_{Ri} \gamma_\mu q^{c\,C}_{Lj}
    + \hc
  \right\}
  \nonumber \\ &
  + \left\{
    \frac{1}{2} (g_{\mathcal{X}})_{rij}
    \mathcal{X}^{a\mu\dagger}_r
    \bar{l}_{Li} \gamma_\mu \sigma^a q_{Lj}
    + \hc
  \right\}
  \nonumber \\ &
  + \left\{
    \frac{1}{2} (g_{\mathcal{Y}_1})_{rij}
    \mathcal{Y}^{AB\mu\dagger}_{1r}
    \bar{d}^{(A|}_{Ri} \gamma_\mu i\sigma_2 q^{c|B)}_{Lj}
    + \hc
  \right\}
  \nonumber \\ &
  + \left\{
    \frac{1}{2} (g_{\mathcal{Y}_5})_{rij}
    \mathcal{Y}^{AB\mu\dagger}_{5r}
    \bar{u}^{(A|}_{Ri} \gamma_\mu i\sigma_2 q^{c|B)}_{Lj}
    + \hc
  \right\}
  \nonumber \\ &
  +\left\{
    (\zeta_{\mathcal{L}_1\mathcal{B}})_{rs}
    \left(\mathcal{L}_{1r\mu}^{\dagger}\phi\right) \mathcal{B}^{\mu}_s
    + (\zeta_{\mathcal{L}_1\mathcal{B}_1})_{rs}
    \tilde{\mathcal{L}}_{1r\mu}^{\dagger} \phi \mathcal{B}_{1s}^{\mu\dagger}
  \right.
  \nonumber \\ &
  \quad
  \left.
    + (\zeta_{\mathcal{L}_1\mathcal{W}})_{rs}
    \left(\mathcal{L}_{1r\mu}^{\dagger} \sigma^{a} \phi \right)
    \mathcal{W}^{a\mu}_s
    + (\zeta_{\mathcal{L}_1\mathcal{W}_1})_{rs}
    \tilde{\mathcal{L}}_{1r\mu}^{\dagger}
    \sigma^{a}\phi\mathcal{W}^{a\mu\dagger}_{1s}
    + \hc
  \right\},
  \label{eq:LV4}
\end{align}

\newpage

and
\begin{align}
  -\Lcal^{(5)}_{\mathrm{V}} = & \;
  \frac{1}{f} \mathcal{L}^{\mu\dagger}_{1r} \bigg[
  (\tilde{\gamma}^{(1)}_{\mathcal{L}_1})_r
  \left(\phi^\dagger D_\mu \phi\right) \phi
  + (\tilde{\gamma}^{(2)}_{\mathcal{L}_1})_r
  \left(D_\mu\phi^\dagger \phi\right) \phi
  + (\tilde{\gamma}^{(3)}_{\mathcal{L}_1})_r
  \left(\phi^\dagger \phi\right) D_\mu \phi
  \nonumber \\ &
  \quad \qquad
  + (\tilde{\gamma}^B_{\mathcal{L}_1})_r B_{\mu\nu} D^\nu \phi
  + (\tilde{\gamma}^{\tilde{B}}_{\mathcal{L}_1})_r \tilde{B}_{\mu\nu} D^\nu \phi
  \nonumber \\ &
  \quad \qquad
  + (\tilde{\gamma}^W_{\mathcal{L}_1})_r W^a_{\mu\nu} \sigma^a D^\nu \phi
  + (\tilde{\gamma}^{\tilde{W}}_{\mathcal{L}_1})_r
  \tilde{W}^a_{\mu\nu} \sigma^a D^\nu \phi
  \nonumber \\ &
  \quad \qquad
  + (\tilde{g}^{eDl}_{\mathcal{L}_1})_{rij} \bar{e}_{Ri} D_\mu l_{Lj}
  + (\tilde{g}^{Del}_{\mathcal{L}_1})_{rij} D_\mu \bar{e}_{Ri} l_{Lj}
  + (\tilde{g}^{dDq}_{\mathcal{L}_1})_{rij} \bar{d}_{Ri} D_\mu q_{Lj}
  \nonumber \\ &
  \quad \qquad
  + (\tilde{g}^{Ddq}_{\mathcal{L}_1})_{rij} D_\mu \bar{d}_{Ri} q_{Lj}
  + (\tilde{g}^{qDu}_{\mathcal{L}_1})_{rij} i\sigma_2 \bar{q}^T_{Li} D_\mu u_{Rj} 
  + (\tilde{g}^{Dqu}_{\mathcal{L}_1})_{rij} i\sigma_2 D_\mu \bar{q}^T_{Li} u_{Rj}
  \nonumber \\ &
  \quad \qquad
  + (\tilde{g}^{du}_{\mathcal{L}_1})_{rij}
  \tilde{\phi} \bar{d}_{Ri} \gamma_\mu u_{Rj}
  + (\tilde{g}^e_{\mathcal{L}_1})_{rij} \phi \bar{e}_{Ri} \gamma_\mu e_{Rj}
  + (\tilde{g}^d_{\mathcal{L}_1})_{rij} \phi \bar{d}_{Ri} \gamma_\mu d_{Rj}
  \nonumber \\ &
  \quad \qquad
  + (\tilde{g}^u_{\mathcal{L}_1})_{rij} \phi \bar{u}_{Ri} \gamma_\mu u_{Rj}
  + (\tilde{g}^{l}_{\mathcal{L}_1})_{rij} \phi \bar{l}_{Ri} \gamma_\mu l_{Lj}
  + (\tilde{g}^{l\prime}_{\mathcal{L}_1})_{rij} \left(\sigma^a \phi\right)
  \left(\bar{l}_{Li} \gamma_\mu \sigma^a l_{Lj}\right)
  \nonumber \\ &
  \quad \qquad
  + (\tilde{g}^{q}_{\mathcal{L}_1})_{rij} \phi \bar{q}_{Li} \gamma_\mu q_{Lj}
  + (\tilde{g}^{q\prime}_{\mathcal{L}_1})_{rij} \left(\sigma^a \phi\right)
  \left(\bar{q}_{Li} \gamma_\mu \sigma^a q_{Lj}\right)
  \bigg]
  + \hc~\!.
  \label{eq:LV5}
\end{align}

\subsection{Mixed Terms}
$\Lcal_{\mathrm{mixed}}$ can be further decomposed as
\beq
\Lcal_{\mathrm{mixed}} = \Lcal_{\mathrm{SF}} + \Lcal_{\mathrm{SV}} + \Lcal_{\mathrm{VF}},
\eeq
where the different pieces are given by
\begin{align}
  -\Lcal_{\mathrm{SF}} = & \;
  (\lambda_{\mathcal{S}E})_{rsi} \mathcal{S}_r \bar{E}_{Ls} e_{Ri}
  + (\lambda_{\mathcal{S}\Delta_1})_{rsi} \mathcal{S}_r \bar{\Delta}_{1Rs} l_{Li}
  \nonumber \\ &
  + (\lambda_{\mathcal{S}U})_{rsi} \mathcal{S}_r \bar{U}_{Ls} u_{Ri}
  + (\lambda_{\mathcal{S}D})_{rsi} \mathcal{S}_r \bar{D}_{Ls} d_{Ri}
  + (\lambda_{\mathcal{S}Q_1})_{rsi} \mathcal{S}_r \bar{Q}_{1Rs} q_{Li}
  \nonumber \\ &
  + (\lambda_{\Xi \Delta_1})_{rsi} \Xi_r^a \bar{\Delta}_{1Rs} \sigma^a l_{Li}
  + (\lambda_{\Xi \Sigma_1})_{rsi} \Xi_r^a \bar{\Sigma}_{1Ls}^a e_{Ri}
  \nonumber \\ &
  + (\lambda_{\Xi Q_1})_{rsi} \Xi_r^a \bar{Q}_{1Rs} \sigma^a q_{Li}
  + (\lambda_{\Xi T_1})_{rsi} \Xi_r^a \bar{T}_{1Ls}^a d_{Ri}
  + (\lambda_{\Xi T_2})_{rsi} \Xi_r^a \bar{T}_{2Ls}^a u_{Ri}
  \nonumber \\ &
  + (\lambda_{\Xi_{1}\Delta_3})_{rsi}
  \Xi_{1r}^{a\dagger} \bar{\Delta}_{3Rs} \sigma^a l_{Li}
  + (\lambda_{\Xi_{1}\Sigma})_{rsi}
  \Xi_{1r}^{a\dagger} \bar{\Sigma}_{Rs}^{c\ a} e_{Ri}^{c}
  \nonumber \\ &
  + (\lambda_{\Xi_{1}Q_5})_{rsi} \Xi_{1r}^{a\dagger} \bar{Q}_{5Rs} \sigma^a q_{Li}
  + (\lambda_{\Xi_{1}Q_7})_{rsi} \Xi_{1r}^{a} \bar{Q}_{7Rs} \sigma^a q_{Li}
  \nonumber \\ &
  + (\lambda_{\Xi_{1}T_1})_{rsi} \Xi_{1r}^{a\dagger} \bar{T}_{1Ls}^a u_{Ri}
  + (\lambda_{\Xi_{1}T_2})_{rsi} \Xi_{1r}^a \bar{T}_{2Ls}^a d_{Ri}
  + \hc~\!,
  \label{eq:LSF} \\[5mm]
  -\Lcal_{\mathrm{SV}} = & \;
  (\delta_{\mathcal{B}\mathcal{S}})_{rs}
  \mathcal{B}_{r\mu} D^{\mu} \mathcal{S}_s
  + (\delta_{\mathcal{W}\Xi})_{rs}
  \mathcal{W}_{r,\mu} D^{\mu} \Xi_s
  \nonumber \\ &
  +\left\{
    (\delta_{\mathcal{L}^1\varphi})_{rs}
    \mathcal{L}_{1r\mu}^{1\dagger} D^{\mu} \varphi_s
    + (\delta_{\mathcal{W}^1\Xi_1})_{rs}
    \mathcal{W}_{1r\mu}^{1\dagger} D^{\mu} \Xi_{1s}
    + \hc
  \right\}
  \nonumber \\ &
  + (\varepsilon_{\mathcal{S}\mathcal{L}_1})_{rst}
  \mathcal{S}_r \mathcal{L}_{1s\mu}^{\dagger}\mathcal{L}_{1t}^{\mu}
  + (\varepsilon_{\Xi \mathcal{L}_1})_{rst}
  \Xi_r^{a} \mathcal{L}_{1s\mu}^{\dagger}\sigma^{a}\mathcal{L}_{1t}^{\mu}
  \nonumber \\ &
  +\left\{
    (\varepsilon_{\Xi_{1}\mathcal{L}_1})_{rst}
    \Xi_{1i}^a \mathcal{L}_{1s\mu}^{\dagger} \sigma^a \tilde{\mathcal{L}}_{1t}^{\mu}
    + \hc
  \right\}
  \nonumber \\ &
  + \Big\{
  (g_{\mathcal{S}\mathcal{L}_1})_{rs}
  \phi^{\dagger} \left(D_\mu \mathcal{S}_r\right) \mathcal{L}_{1s}^{\mu}
  + (g'_{\mathcal{S}\mathcal{L}_1})_{rs}
  \left(D_\mu \phi\right)^{\dagger} \mathcal{S}_r \mathcal{L}_{1s}^{\mu}
  \nonumber \\ &
  \quad
  + (g_{\Xi\mathcal{L}_1})_{rs}
  \phi^{\dagger} \sigma^a \left(D_\mu \Xi_r^a\right) \mathcal{L}_{1s}^{\mu}
  + (g'_{\Xi\mathcal{L}_1})_{rs}
  \left(D_\mu \phi\right)^{\dagger} \sigma^{a} \Xi_r^{a} \mathcal{L}_{1s}^{\mu}
  \nonumber \\ &
  \quad
  + (g_{\Xi_1\mathcal{L}_1})_{rs}
  \tilde{\phi}^{\dagger} \sigma^a
  \left(D_\mu \Xi_{1r}^a\right)^{\dagger} \mathcal{L}_{1s}^{\mu}
  + (g'_{\Xi_1\mathcal{L}_1})_{rs}
  \left(D_\mu \tilde{\phi}\right)^{\dagger} \sigma^a
  \Xi_{1r}^{a\dagger} \mathcal{L}_{1s}^{\mu}
  + \hc
  \Big\}~\!,
  \label{eq:LVS}
\end{align}
and
\begin{align}
  -\Lcal_{\mathrm{VF}} = & \;
  (z_{N\mathcal{L}_1})_{rsi}
  \bar{N}_{Rr}^c\gamma^{\mu}\tilde{\mathcal{L}}_{1s\mu}^{\dagger}l_{Li}
  + (z_{E\mathcal{L}_1})_{rsi}
  \bar{E}_{Lr}\gamma^{\mu}\mathcal{L}_{1s\mu}^{\dagger}l_{Li}
  \nonumber \\ &
  + (z_{\Delta_{1}\mathcal{L}_1})_{rsi}
  \bar{\Delta}_{1Rr}\gamma^{\mu}\mathcal{L}_{1s\mu}e_{Ri}
  + (z_{\Delta_{3}\mathcal{L}_1})_{rsi}
  \bar{\Delta}_{3Rr}\gamma^{\mu}\tilde{\mathcal{L}}_{1s\mu}e_{Ri}
  \nonumber \\ &
  + (z_{\Sigma\mathcal{L}_1})_{rsi}
  \bar{\Sigma}_{Rr}^{c\ a}\gamma^{\mu}
  \tilde{\mathcal{L}}_{1s\mu}^{\dagger}\sigma^{a}l_{Li}
  + (z_{\Sigma_{1}\mathcal{L}_1})_{rsi}
  \bar{\Sigma}_{1Lr}^{a}\gamma^{\mu}\mathcal{L}_{1s\mu}^{\dagger}\sigma^{a}l_{Li}
  \nonumber \\ &
  +(z_{U\mathcal{L}_1})_{rsi}
  \bar{U}_{Lr}\gamma^{\mu}\tilde{\mathcal{L}}_{1s\mu}^{\dagger}q_{Li}
  +(z_{D\mathcal{L}_1})_{rsi}
  \bar{D}_{Lr}\gamma^{\mu}\mathcal{L}_{1s\mu}^{\dagger}q_{Li}
  \nonumber \\ &
  +(z^u_{Q_1\mathcal{L}_1})_{rsi}
  \bar{Q}_{1Rr}\gamma^{\mu}\tilde{\mathcal{L}}_{1s\mu}u_{Ri}
  +(z^d_{Q_1\mathcal{L}_1})_{rsi}
  \bar{Q}_{1Rr}\gamma^{\mu}\mathcal{L}_{1s\mu}d_{Ri}
  \nonumber \\ &
  +(z_{Q_5\mathcal{L}_1})_{rsi}
  \bar{Q}_{5Rr}\gamma^{\mu}\tilde{\mathcal{L}}_{1s\mu}d_{Ri}
  +(z_{Q_7\mathcal{L}_1})_{rsi}\bar{Q}_{7Rr}
  \gamma^{\mu}\mathcal{L}_{1s\mu}u_{Ri}
  \nonumber \\ &
  + (z_{T_1\mathcal{L}_1})_{rsi}
  \bar{T}_{1Lr}^{a}\gamma^{\mu}\mathcal{L}_{1s\mu}^{\dagger}\sigma^{a}q_{Li}
  + (z_{T_2\mathcal{L}_1})_{rsi}
  \bar{T}_{2Lr}^{a}\gamma^{\mu}\tilde{\mathcal{L}}_{1s\mu}^{\dagger}\sigma^{a}q_{Li}
  +\hc~\!.
  \label{eq:LVF}
\end{align}

No renormalizable operators exist that contain extra scalars, fermions and vectors simultaneously.

Finally, in order to keep track of the dimensionality of the different contributions to the operators in the effective Lagrangian presented in appendix~\ref{sec:results} we collect here the mass dimensions of the different types of couplings appearing in the new physics Lagrangians introduced above:
\begin{equation}
[\kappa]=1,~~[\lambda]=[\lambda^\prime]=0,~~[y]=0,
\end{equation}
\begin{equation}
[\tilde{k}]=0,~~[\tilde{\lambda}]=0,~~[\tilde{y}]=0,
\end{equation}
\begin{equation}
[g]=[g^\prime]=0,~~[\gamma]=1,~~[h]=0,~~[\zeta]=1,
\end{equation}
\begin{equation}
[\tilde{g}]=0,~~[\tilde{\gamma}]=0, 
\end{equation}
\begin{equation}
[\delta]=1,~~[\varepsilon]=1,~~[z]=0.
\end{equation}

\newpage

\section{Dimension-Six Basis}
\label{sec:d6Basis}

In this appendix we present the complete basis of gauge-invariant operators ${\cal O}_i$ that we use in this paper in the analysis of the general SM effective Lagrangian to dimension six.

Table~\ref{tab:dim45Basis} defines our notation for those operators of mass dimension four that appear in the integration of the heavy particles. These renormalize the SM interactions. The table also presents the only possible dimension-five interaction: the Weinberg operator, which gives Majorana masses to the SM neutrinos. Tables~\ref{tab:dim6Basis4F} and \ref{tab:dim6BasisBF} contain the basis of dimension-six operators as introduced in ref.~\cite{Grzadkowski:2010es}.

The notation used in the tables is defined in appendix~\ref{sec:lagrangians}. 
Flavor indices of the operators and their coefficients are defined to appear
in the same order as the corresponding fermion fields inside the operator.
Finally, the hermitian derivatives $\lrD$ and $\lrDa$ appearing in the ${\cal O}_{\phi \psi}^{(1)}$ and ${\cal O}_{\phi \psi}^{(3)}$ operators in table~\ref{tab:dim6BasisBF} are defined by: 
\begin{eqnarray}
\lrD_\mu &\equiv& {D}_\mu - \overset{\leftarrow}{D}_\mu,\nonumber\\
\lrDa_\mu &\equiv& \sigma_a {D}_\mu - \overset{\leftarrow}{D}_\mu \sigma_a.\nonumber
\end{eqnarray}

\vspace{2cm}

\begin{table}[H]
  \begin{center}
    \begin{tabular}{ccl}
      \ctoprule
      & Operator &  Notation \\
      \cmidrule{2-3}
      \multirow{5}{*}
      {Dim. 4} &
      & \\[-5mm]
      &
      $\left(\phi^{\dagger} \phi\right)^2$ &
      $\mathcal{O}_{\phi 4}$ \\
      &
      $\bar{e}_R \phi^\dagger l_L$ &
      $\mathcal{O}_{y^e}$ \\
      &
      $\bar{d}_R \phi^\dagger q_L$ &
      $\mathcal{O}_{y^d}$ \\
      &
      $\bar{u}_R \tilde{\phi}^\dagger q_L$ &
      $\mathcal{O}_{y^u}$ \\[1mm]
      \cmrule
      Dim. 5 &
      $\overline{l_L^c} \tilde{\phi}^* \tilde{\phi}^\dagger l_L$ &
      $\mathcal{O}_{5}$ \\[1mm]
      \cbottomrule
    \end{tabular}
    \caption{Operators of dimension four and five.\label{tab:dim45Basis}}
  \end{center}
\end{table}

\newpage

\begin{table}[H]
  \begin{center}
    \begin{tabular}{cclccl}
      \ctoprule
      & Operator & Notation & & Operator & Notation \\
       \cmidrule{2-3} \cmidrule{5-6}
      %
      \multirow{3}{*}
      {$\left(\bar{L}L\right)\left(\bar{L}L\right)$} &
      $\left(\bar{l}_L \gamma_\mu l_L\right)
      \left(\bar{l}_L \gamma^\mu l_L\right)$ &
      $\mathcal{O}_{ll}$ &
      & & \\
      &
      $\left(\bar{q}_L \gamma_\mu q_L\right)
      \left(\bar{q}_L \gamma^\mu q_L\right)$ &
      $\mathcal{O}_{qq}^{(1)}$ &
      &
      $\left(\bar{q}_L \gamma_\mu \sigma_a q_L\right)
      \left(\bar{q}_L \gamma^\mu \sigma_a q_L\right)$ &
      $\mathcal{O}_{qq}^{(3)}$ \\
      &
      $\left(\bar{l}_L\gamma_\mu l_L\right)
      \left(\bar{q}_L \gamma^\mu q_L\right)$ &
      $\mathcal{O}_{lq}^{(1)}$ &
      &
      $\left(\bar{l}_L \gamma_\mu\sigma_a l_L\right)
      \left(\bar{q}_L \gamma^\mu\sigma_a q_L\right)$ &
      $\mathcal{O}_{lq}^{(3)}$ \\[1mm]
      \cmrule
      \multirow{4}{*}
      {$\left(\bar{R}R\right)\left(\bar{R}R\right)$} &
      $\left(\bar{e}_R \gamma_\mu e_R\right)
      \left(\bar{e}_R \gamma^\mu e_R\right)$ &
      $\mathcal{O}_{ee}$ &
      & \\
      &
      $\left(\bar{u}_R \gamma_\mu u_R\right)
      \left(\bar{u}_R \gamma^\mu u_R\right)$ &
      $\mathcal{O}_{uu}$ &
      &
      $\left(\bar{d}_R \gamma_\mu d_R\right)
      \left(\bar{d}_R \gamma^\mu d_R\right)$ &
      $\mathcal{O}_{dd}$ \\
      &
      $\left(\bar{u}_R \gamma_\mu u_R\right)
      \left(\bar{d}_R \gamma^\mu d_R\right)$ &
      $\mathcal{O}_{ud}^{(1)}$ &
      &
      $\left(\bar{u}_R \gamma_\mu T_A u_R\right)
      \left(\bar{d}_R \gamma^\mu T_A d_R\right)$ &
      $\mathcal{O}_{ud}^{(8)}$ \\
      &
      $\left(\bar{e}_R \gamma_\mu e_R\right)
      \left(\bar{u}_R \gamma^\mu u_R\right)$ &
      $\mathcal{O}_{eu}$ &
      &
      $\left(\bar{e}_R \gamma_\mu e_R\right)
      \left(\bar{d}_R \gamma^\mu d_R\right)$ &
      $\mathcal{O}_{ed}$ \\[1mm]
      \cmrule
      \multirow{4}{*}
      {$\left(\bar{L}L\right)\left(\bar{R}R\right)$} &
      $\left(\bar{l}_L \gamma_\mu l_L\right)
      \left(\bar{e}_R \gamma^\mu e_R\right)$ &
      $\mathcal{O}_{le}$ &
      &
      $\left(\bar{q}_L \gamma_\mu q_L\right)
      \left(\bar{e}_R \gamma^\mu e_R\right)$
      &
      $\mathcal{O}_{qe}$ \\
      &
      $\left(\bar{l}_L \gamma_\mu l_L\right)
      \left(\bar{u}_R \gamma^\mu u_R\right)$ &
      $\mathcal{O}_{lu}$ &
      &
      $\left(\bar{l}_L \gamma_\mu l_L\right)
      \left(\bar{d}_R \gamma^\mu d_R\right)$ &
      $\mathcal{O}_{ld}$ \\
      &
      $\left(\bar{q}_L \gamma_\mu q_L\right)
      \left(\bar{u}_R \gamma^\mu u_R\right)$ &
      $\mathcal{O}_{qu}^{(1)}$ &
      &
      $\left(\bar{q}_L \gamma_\mu T_A q_L\right)
      \left(\bar{u}_R \gamma^\mu T_A u_R\right)$ &
      $\mathcal{O}_{qu}^{(8)}$ \\
      &
      $\left(\bar{q}_L \gamma_\mu q_L\right)
      \left(\bar{d}_R \gamma^\mu d_R\right)$ &
      $\mathcal{O}_{qd}^{(1)}$ &
      &
      $\left(\bar{q}_L \gamma_\mu T_A q_L\right)
      \left(\bar{d}_R \gamma^\mu T_A d_R\right)$ &
      $\mathcal{O}_{qd}^{(8)}$ \\[1mm]
      \cmrule
      $\left(\bar{L}R\right)\left(\bar{R}L\right)$ &
      $\left(\bar{l}_L e_R\right)
      \left(\bar{d}_R q_L\right)$ &
      $\mathcal{O}_{ledq}$ &
      & & \\[1mm]
      \cmrule
      \multirow{2}{*}
      {$\left(\bar{L}R\right)\left(\bar{L}R\right)$} &
      $\left(\bar{q}_L  u_R\right) i\sigma_2
      \left(\bar{q}_L d_R\right)^{\mathrm{T}}$ &
      $\mathcal{O}_{quqd}^{(1)}$ &
      &
      $\left(\bar{q}_L T_A u_R\right) i\sigma_2
      \left(\bar{q}_L T_A d_R\right)^{\mathrm{T}}$ &
      $\mathcal{O}_{quqd}^{(8)}$ \\
      &
      $\left(\bar{l}_L  e_R\right) i\sigma_2
      \left(\bar{q}_L u_R\right)^{\mathrm{T}}$ &
      $\mathcal{O}^{(1)}_{lequ}$ &
      &
      $\left(\bar{l}_L  \sigma_{\mu\nu} e_R\right) i\sigma_2
      \left(\bar{q}_L \sigma^{\mu\nu} u_R\right)^{\mathrm{T}}$ &
      $\mathcal{O}^{(3)}_{lequ}$ \\[1mm]
      \cmrule
      \multirow{4}{*}
      {B-violating} &
      \multicolumn{4}{r}{
        $\epsilon_{ABC} \left(\bar{d}^{c\,A}_R u^B_R\right)
        \left(\bar{q}^{c\,C}_L i\sigma_2 l_L\right)$} &
      $\mathcal{O}_{duq}$ \\
      &
      \multicolumn{4}{r}{
        $\epsilon_{ABC} \left(\bar{q}^{c\,A}_L i\sigma_2 q^B_L\right)
        \left(\bar{u}^{c\,C}_R e_R\right)$} &
      $\mathcal{O}_{qqu}$ \\
      &
      \multicolumn{4}{r}{
        $\epsilon_{ABC} \left(\bar{d}^{c\,A}_R u^B_R\right)
        \left(\bar{u}^{c\,C}_R e_R\right)$} &
      $\mathcal{O}_{duu}$ \\
      &
      \multicolumn{4}{r}{
        $\epsilon_{ABC} (i\sigma_2)_{\alpha\delta} (i\sigma_2)_{\beta\gamma}
      \left(\bar{q}^{c\,A\alpha}_L q^{B\beta}_L\right)
      \left(\bar{q}^{c\,C\gamma}_L l^{\delta}_L\right)$} &
      $\mathcal{O}_{qqq}$ \\[1mm]
      \cbottomrule
    \end{tabular}
    \caption{Basis of dimension-six operators: four-fermion interactions.
      Flavor indices are omitted.
      \label{tab:dim6Basis4F}}
  \end{center}
\end{table}
%

\begin{table}[H]
  \begin{center}
    \begin{tabular}{cclccl}
      \ctoprule
      & Operator & Notation & & Operator & Notation \\
      \cmidrule{2-3} \cmidrule{5-6}
      \multirow{2}{*}{$X^3$}
      &
      $\varepsilon_{abc} W^{a\,\nu}_\mu W^{b\,\rho}_\nu W^{c\,\mu}_\rho$ &
      $\mathcal{O}_{W}$ &
      &
      $\varepsilon_{abc} \tilde{W}^{a\,\nu}_\mu W^{b\,\rho}_\nu W^{c\,\mu}_\rho$ &
      $\mathcal{O}_{\tilde{W}}$ \\
      &
      $f_{ABC} G^{A\,\nu}_\mu G^{B\,\rho}_\nu G^{C\,\mu}_\rho$ &
      $\mathcal{O}_{G}$ &
      &
      $f_{ABC} \tilde{G}^{A\,\nu}_\mu G^{B\,\rho}_\nu G^{C\,\mu}_\rho$ &
      $\mathcal{O}_{\tilde{G}}$ \\[1mm]
      \cmrule
      \vspace{-5mm} & & & & \\
      $\phi^6$ &
      $\left(\phi^{\dagger} \phi\right)^3$ &
      $\mathcal{O}_{\phi}$ &
      & & \\[1mm]
      \cmrule
      \vspace{-5mm} & & & & \\
      $\phi^4 D^2$ &
      $\left(\phi^{\dagger} \phi\right)\square
      \left(\phi^{\dagger} \phi\right)$ &
      $\mathcal{O}_{\phi \square}$ &
      &
      $\left(\phi^{\dagger}D_\mu \phi\right)
      (\left(D^\mu \phi\right)^{\dagger}\phi)$ &
      $\mathcal{O}_{\phi D}$ \\[1mm]
      \cmrule
      \multirow{2}{*}{$\psi^2 \phi^2$} &
      $\left(\phi^{\dagger} \phi\right)
      \left(\bar{l}_L \phi e_R\right)$ &
      $\mathcal{O}_{e \phi}$ & & \\
      &
      $\left(\phi^{\dagger} \phi\right)
      \left(\bar{q}_L \phi d_R\right)$ &
      $\mathcal{O}_{d \phi }$ &
      & 
      $\left(\phi^{\dagger} \phi\right)
      \left(\bar{q}_L \tilde{\phi} u_R\right)$ &
      $\mathcal{O}_{u \phi}$ \\[1mm]
      \cmrule
      \multirow{4}{*}{$X^2 \phi^2$} &
      $\phi^\dagger \phi B_{\mu\nu} B^{\mu\nu} $ &
      $\mathcal{O}_{\phi B}$ &
      &
      $\phi^\dagger \phi \tilde{B}_{\mu\nu} B^{\mu\nu}$ &
      $\mathcal{O}_{\phi \tilde{B}}$ \\
      &
      $\phi^\dagger \phi W_{\mu\nu}^a W^{a\,\mu\nu} $ &
      $\mathcal{O}_{\phi W}$ &
      &
      $\phi^\dagger \phi \tilde{W}_{\mu\nu}^a W^{a\,\mu\nu}$ &
      $\mathcal{O}_{\phi \tilde{W}}$ \\
      &
      $\phi^\dagger \sigma_a \phi W^a_{\mu\nu} B^{\mu\nu}$ &
      $\mathcal{O}_{\phi WB}$ &
      &
      $\phi^\dagger \sigma_a \phi \tilde{W}^a_{\mu\nu} B^{\mu\nu}$ &
      $\mathcal{O}_{\phi\tilde{W}B}$ \\
      &
      $\phi^\dagger \phi G_{\mu\nu}^A G^{A\,\mu\nu} $ &
      $\mathcal{O}_{\phi G}$ &
      &
      $\phi^\dagger \phi \tilde{G}_{\mu\nu}^A G^{A\,\mu\nu}$ &
      $\mathcal{O}_{\phi\tilde{G}}$ \\[1mm]
      \cmrule
      \multirow{4}{*}{$\psi^2 X \phi$} &
      $\left(\bar{l}_L \sigma^{\mu\nu} e_R\right)
      \phi B_{\mu\nu}$ &
      $\mathcal{O}_{eB}$ &
      &
      $\left(\bar{l}_L \sigma^{\mu\nu} e_R\right)
      \sigma^a \phi W_{\mu\nu}^a$ &
      $\mathcal{O}_{eW}$ \\
      & $\left(\bar{q}_L \sigma^{\mu\nu} u_R\right)
      \tilde{\phi} B_{\mu\nu}$ &
      $\mathcal{O}_{uB}$ &
      &
      $\left(\bar{q}_L \sigma^{\mu\nu} u_R\right)
      \sigma^a \tilde{\phi} W_{\mu\nu}^a$ &
      $\mathcal{O}_{uW}$ \\
      &
      $\left(\bar{q}_L \sigma^{\mu\nu} d_R\right)
      \phi B_{\mu\nu}$ &
      $\mathcal{O}_{dB}$ &
      &
      $\left(\bar{q}_L \sigma^{\mu\nu} d_R\right)
      \sigma^a \phi W_{\mu\nu}^a$ &
      $\mathcal{O}_{dW}$ \\
      &
      $\left(\bar{q}_L \sigma^{\mu\nu} T_A u_R\right)
      \tilde{\phi} G_{\mu\nu}^A$ &
      $\mathcal{O}_{uG}$ &
      &
      $\left(\bar{q}_L \sigma^{\mu\nu} T_A d_R\right)
      \phi G^A_{\mu\nu}$ &
      $\mathcal{O}_{dG}$ \\[1mm]
      \cmrule
      \multirow{5}{*}{$\psi^2 \phi^2 D$} &
      $(\phi^{\dagger} i\overset{\leftrightarrow}{D}_\mu \phi)
      \left(\bar{l}_L \gamma^\mu l_L\right)$ &
      $\mathcal{O}_{\phi l}^{(1)}$ &
      &
      $(\phi^{\dagger} i\lrDa_\mu \phi)
      \left(\bar{l}_L \gamma^\mu \sigma_a l_L\right)$ &
      $\mathcal{O}_{\phi l}^{(3)}$ \\
      &
      $(\phi^{\dagger} i\overset{\leftrightarrow}{D}_\mu \phi)
      \left(\bar{e}_R \gamma^\mu e_R\right)$ &
      $\mathcal{O}_{\phi e}$ &
      & & \\
      &
      $(\phi^{\dagger} i\overset{\leftrightarrow}{D}_\mu \phi)
      \left(\bar{q}_L \gamma^\mu q_L\right)$ &
      $\mathcal{O}_{\phi q}^{(1)}$ &
      &
      $(\phi^{\dagger} i \lrDa_\mu \phi)
      \left(\bar{q}_L \gamma^\mu\sigma_a q_L\right)$ &
      $\mathcal{O}_{\phi q}^{(3)}$ \\
      &
      $(\phi^{\dagger} i\overset{\leftrightarrow}{D}_\mu \phi)
      \left(\bar{u}_R \gamma^\mu u_R\right)$ &
      $\mathcal{O}_{\phi u}$ &
      &
      $(\phi^{\dagger} i\overset{\leftrightarrow}{D}_\mu \phi)
      \left(\bar{d}_R \gamma^\mu d_R\right)$ &
      $\mathcal{O}_{\phi d}$ \\
      &
      $(\tilde{\phi}^{\dagger} iD_\mu \phi)
      \left(\bar{u}_R \gamma^\mu d_R\right)$ &
      $\mathcal{O}_{\phi ud}$ &
      & & \\[1mm]
      \cbottomrule
    \end{tabular}
    \caption{Basis of dimension-six operators: operators other than four-fermion
      interactions. Flavor indices are omitted. \label{tab:dim6BasisBF}}
  \end{center}
\end{table}

\newpage
\section{Operators Generated by Each Field Multiplet}
\label{sec:Field_Op}

In this appendix we provide the representation of each heavy multiplet introduced in section~\ref{sec:fields} in terms of operators of dimension $n\leq6$ in the low energy effective Lagrangian. The results for the corresponding coefficients are given in appendix~\ref{sec:results}. See section~\ref{sec:userguide} for details.

\begin{table}[H]
  \centering
  \begin{tabular}{cl}
  \ctoprule
    Fields & Operators \\
    \cmrule
    $\mathcal{S}$ &
    $\mathcal{O}_{\phi 4}$, $\mathcal{O}_{\phi}$,
    $\mathcal{O}_{\phi \square}$, $\mathcal{O}_{\phi B}$,
    $\mathcal{O}_{\phi \tilde{B}}$, $\mathcal{O}_{\phi W}$,
    $\mathcal{O}_{\phi \tilde{W}}$, $\mathcal{O}_{\phi G}$,
    $\mathcal{O}_{\phi \tilde{G}}$, $\mathcal{O}_{e\phi}$,
    $\mathcal{O}_{d\phi}$, $\mathcal{O}_{u\phi}$ \\
    $\mathcal{S}_1$ &
    $\mathcal{O}_{ll}$ \\
    $\mathcal{S}_2$ &
    $\mathcal{O}_{ee}$ \\
    $\varphi$ &
    $\mathcal{O}_{le}$, $\mathcal{O}^{(1)}_{qu}$,
    $\mathcal{O}^{(8)}_{qu}$, $\mathcal{O}^{(1)}_{qd}$,
    $\mathcal{O}^{(8)}_{qd}$, $\mathcal{O}_{ledq}$,
    $\mathcal{O}^{(1)}_{quqd}$, $\mathcal{O}^{(1)}_{lequ}$,
    $\mathcal{O}_{\phi}$, $\mathcal{O}_{e\phi}$,
    $\mathcal{O}_{d\phi}$, $\mathcal{O}_{u\phi}$ \\
    $\Xi$ &
    $\mathcal{O}_{\phi 4}$, $\mathcal{O}_{\phi}$,
    $\mathcal{O}_{\phi D}$, $\mathcal{O}_{\phi \square}$,
    $\mathcal{O}_{\phi WB}$, $\mathcal{O}_{\phi W\tilde{B}}$,
    $\mathcal{O}_{e\phi}$, $\mathcal{O}_{d\phi}$,
    $\mathcal{O}_{u\phi}$ \\
    $\Xi_1$ & 
    $\mathcal{O}_{\phi 4}$, $\mathcal{O}_5$,
    $\mathcal{O}_{ll}$, $\mathcal{O}_{\phi}$,
    $\mathcal{O}_{\phi D}$, $\mathcal{O}_{\phi \square}$,
    $\mathcal{O}_{e\phi}$, $\mathcal{O}_{d\phi}$,
    $\mathcal{O}_{u\phi}$ \\
    $\Theta_1$ & 
    $\mathcal{O}_{\phi}$ \\
    $\Theta_3$ & 
    $\mathcal{O}_{\phi}$ \\
    $\omega_{1}$ & 
    $\mathcal{O}^{(1)}_{qq}$, $\mathcal{O}^{(3)}_{qq}$,
    $\mathcal{O}^{(1)}_{lq}$, $\mathcal{O}^{(3)}_{lq}$,
    $\mathcal{O}_{eu}$, $\mathcal{O}^{(1)}_{ud}$,
    $\mathcal{O}^{(8)}_{ud}$, $\mathcal{O}^{(1)}_{quqd}$,
    $\mathcal{O}^{(8)}_{quqd}$, \\
    &
    $\mathcal{O}^{(1)}_{lequ}$
    $\mathcal{O}^{(3)}_{lequ}$, $\mathcal{O}_{duq}$,
    $\mathcal{O}_{qqu}$, $\mathcal{O}_{qqq}$,
    $\mathcal{O}_{duu}$ \\
    $\omega_{2}$ & 
    $\mathcal{O}_{dd}$ \\
    $\omega_{4}$ & 
    $\mathcal{O}_{uu}$, $\mathcal{O}_{ed}$,
    $\mathcal{O}_{duu}$ \\
    $\Pi_1$ & 
    $\mathcal{O}_{ld}$ \\
    $\Pi_7$ & 
    $\mathcal{O}_{lu}$, $\mathcal{O}_{qe}$,
    $\mathcal{O}^{(1)}_{lequ}$, $\mathcal{O}^{(3)}_{lequ}$ \\
    $\zeta$ &
    $\mathcal{O}^{(1)}_{qq}$, $\mathcal{O}^{(3)}_{qq}$,
    $\mathcal{O}^{(1)}_{lq}$, $\mathcal{O}^{(3)}_{lq}$,
    $\mathcal{O}_{qqq}$ \\
    $\Omega_{1}$ & 
    $\mathcal{O}^{(1)}_{qq}$, $\mathcal{O}^{(3)}_{qq}$,
    $\mathcal{O}^{(1)}_{ud}$, $\mathcal{O}^{(8)}_{ud}$,
    $\mathcal{O}^{(1)}_{quqd}$, $\mathcal{O}^{(8)}_{quqd}$ \\
    $\Omega_{2}$ & 
    $\mathcal{O}_{dd}$ \\
    $\Omega_{4}$ & 
    $\mathcal{O}_{uu}$ \\
    $\Upsilon$ & 
    $\mathcal{O}^{(1)}_{qq}$, $\mathcal{O}^{(3)}_{qq}$ \\
    $\Phi$ &
    $\mathcal{O}^{(1)}_{qu}$, $\mathcal{O}^{(8)}_{qu}$,
    $\mathcal{O}^{(1)}_{qd}$, $\mathcal{O}^{(8)}_{qd}$,
    $\mathcal{O}^{(8)}_{quqd}$\\
    & \\[-0.4cm]
    \cbottomrule
  \end{tabular}
  \caption{Operators generated by the heavy scalar fields introduced in table~\ref{t:scalars}.}
  \label{tab:topdown_scalars}
\end{table}

\newpage

\begin{table}[H]
  \centering
  \begin{tabular}{cl}
  \ctoprule
    Fields & Operators 
    \\
    \cmrule
    $N$ & 
    $\mathcal{O}_5$, $\mathcal{O}^{(1)}_{\phi l}$,
    $\mathcal{O}^{(3)}_{\phi l}$ \\
    $E$ & 
    $\mathcal{O}_{e\phi}$, $\mathcal{O}_{eB}$,
    $\mathcal{O}^{(1)}_{\phi l}$, $\mathcal{O}^{(3)}_{\phi l}$ \\
    $\Delta_1$ & 
    $\mathcal{O}_{e\phi}$, $\mathcal{O}_{eB}$,
    $\mathcal{O}_{eW}$, $\mathcal{O}_{\phi e}$ \\
    $\Delta_3$ & 
    $\mathcal{O}_{e\phi}$, $\mathcal{O}_{\phi e}$ \\
    $\Sigma$ & 
    $\mathcal{O}_5$, $\mathcal{O}_{e\phi}$,
    $\mathcal{O}^{(1)}_{\phi l}$, $\mathcal{O}^{(3)}_{\phi l}$ \\
    $\Sigma_1$ & 
    $\mathcal{O}_{e\phi}$, $\mathcal{O}_{eW}$,
    $\mathcal{O}^{(1)}_{\phi l}$, $\mathcal{O}^{(3)}_{\phi l}$ \\
    $U$ & 
    $\mathcal{O}_{u\phi}$, $\mathcal{O}_{uB}$,
    $\mathcal{O}_{uG}$, $\mathcal{O}^{(1)}_{\phi q}$,
    $\mathcal{O}^{(3)}_{\phi q}$ \\
    $D$ & 
    $\mathcal{O}_{d\phi}$, $\mathcal{O}_{dB}$,
    $\mathcal{O}_{dG}$, $\mathcal{O}^{(1)}_{\phi q}$,
    $\mathcal{O}^{(3)}_{\phi q}$ \\
    $Q_1$ & 
    $\mathcal{O}_{d\phi}$, $\mathcal{O}_{u\phi}$,
    $\mathcal{O}_{dB}$,  $\mathcal{O}_{dW}$,
    $\mathcal{O}_{dG}$, $\mathcal{O}_{uB}$,
    $\mathcal{O}_{uW}$, $\mathcal{O}_{uG}$,
    $\mathcal{O}_{\phi d}$, $\mathcal{O}_{\phi u}$,
    $\mathcal{O}_{\phi ud}$ \\
    $Q_5$ & 
    $\mathcal{O}_{d\phi}$, $\mathcal{O}_{\phi d}$ \\
    $Q_7$ & 
    $\mathcal{O}_{u\phi}$, $\mathcal{O}_{\phi u}$ \\
    $T_1$ & 
    $\mathcal{O}_{d\phi}$, $\mathcal{O}_{u\phi}$,
    $\mathcal{O}_{dW}$, $\mathcal{O}^{(1)}_{\phi q}$,
    $\mathcal{O}^{(3)}_{\phi q}$ \\
    $T_2$ &
    $\mathcal{O}_{d\phi}$, $\mathcal{O}_{u\phi}$,
    $\mathcal{O}_{uW}$, $\mathcal{O}^{(1)}_{\phi q}$,
    $\mathcal{O}^{(3)}_{\phi q}$\\
    & \\[-0.5cm]
    \cbottomrule
  \end{tabular}
  \caption{Operators generated by the heavy vector-like fermions in table~\ref{t:fermions}.}
  \label{tab:topdown_fermions}
\vspace{1cm}
  \centering
  \begin{tabular}{cl}
  \ctoprule
    Fields & Operators 
    \\
    \cmrule
    $\mathcal{B}$ &
    $\mathcal{O}_{ll}$, $\mathcal{O}^{(1)}_{qq}$,
    $\mathcal{O}^{(1)}_{lq}$, $\mathcal{O}_{ee}$,
    $\mathcal{O}_{dd}$, $\mathcal{O}_{uu}$,
    $\mathcal{O}_{ed}$, $\mathcal{O}_{eu}$,
    $\mathcal{O}^{(1)}_{ud}$, $\mathcal{O}_{le}$,
    $\mathcal{O}_{ld}$, $\mathcal{O}_{lu}$,
    $\mathcal{O}_{qe}$, $\mathcal{O}^{(1)}_{qu}$,
    $\mathcal{O}^{(1)}_{qd}$, \\
    &
    $\mathcal{O}_{\phi D}$, $\mathcal{O}_{\phi \square}$,
    $\mathcal{O}_{e\phi}$, $\mathcal{O}_{d\phi}$,
    $\mathcal{O}_{u\phi}$, $\mathcal{O}^{(1)}_{\phi l}$,
    $\mathcal{O}^{(1)}_{\phi q}$, $\mathcal{O}_{\phi e}$,
    $\mathcal{O}_{\phi d}$, $\mathcal{O}_{\phi u}$ \\
    $\mathcal{B}_1$ &
    $\mathcal{O}_{\phi 4}$, $\mathcal{O}^{(1)}_{ud}$,
    $\mathcal{O}^{(8)}_{ud}$, $\mathcal{O}_{\phi}$,
    $\mathcal{O}_{\phi D}$, $\mathcal{O}_{\phi \square}$,
    $\mathcal{O}_{e\phi}$, $\mathcal{O}_{d\phi}$,
    $\mathcal{O}_{u\phi}$, $\mathcal{O}_{\phi ud}$ \\
    $\mathcal{W}$ &
    $\mathcal{O}_{\phi 4}$, $\mathcal{O}_{ll}$,
    $\mathcal{O}^{(3)}_{qq}$, $\mathcal{O}^{(3)}_{lq}$,
    $\mathcal{O}_{\phi}$, $\mathcal{O}_{\phi D}$,
    $\mathcal{O}_{\phi \square}$, $\mathcal{O}_{e\phi}$,
    $\mathcal{O}_{d\phi}$, $\mathcal{O}_{u\phi}$,
    $\mathcal{O}^{(3)}_{\phi l}$, $\mathcal{O}^{(3)}_{\phi q}$ \\
    $\mathcal{W}_1$ & 
    $\mathcal{O}_{\phi 4}$, $\mathcal{O}_{\phi}$,
    $\mathcal{O}_{\phi D}$, $\mathcal{O}_{\phi \square}$,
    $\mathcal{O}_{e\phi}$, $\mathcal{O}_{d\phi}$,
    $\mathcal{O}_{u\phi}$ \\
    $\mathcal{G}$ & 
    $\mathcal{O}^{(1)}_{qq}$, $\mathcal{O}^{(3)}_{qq}$,
    $\mathcal{O}_{dd}$, $\mathcal{O}_{uu}$,
    $\mathcal{O}^{(8)}_{ud}$, $\mathcal{O}^{(8)}_{qu}$,
    $\mathcal{O}^{(8)}_{qd}$ \\
    $\mathcal{G}_1$ & 
    $\mathcal{O}^{(1)}_{ud}$, $\mathcal{O}^{(8)}_{ud}$ \\
    $\mathcal{H}$ & 
    $\mathcal{O}^{(1)}_{qq}$, $\mathcal{O}^{(3)}_{qq}$ \\
    $\mathcal{L}_1$ &
    $\mathcal{O}_{\phi 4}$,
    $\mathcal{O}_{y^e}$, $\mathcal{O}_{y^d}$,
    $\mathcal{O}_{y^u}$, $\mathcal{O}_{le}$,
    $\mathcal{O}^{(1)}_{qu}$, $\mathcal{O}^{(8)}_{qu}$,
    $\mathcal{O}^{(1)}_{qd}$, $\mathcal{O}^{(8)}_{qd}$,
    $\mathcal{O}_{ledq}$, $\mathcal{O}^{(1)}_{quqd}$,
    $\mathcal{O}^{(1)}_{lequ}$, \\
    &
    $\mathcal{O}_{\phi}$, $\mathcal{O}_{\phi D}$,
    $\mathcal{O}_{\phi \square}$, $\mathcal{O}_{\phi B}$,
    $\mathcal{O}_{\phi \tilde{B}}$, $\mathcal{O}_{\phi W}$,
    $\mathcal{O}_{\phi \tilde{W}}$, $\mathcal{O}_{\phi WB}$,
    $\mathcal{O}_{\phi W\tilde{B}}$, $\mathcal{O}_{e\phi}$,
    $\mathcal{O}_{d\phi}$, $\mathcal{O}_{u\phi}$, \\
    &
    $\mathcal{O}_{eB}$, $\mathcal{O}_{eW}$,
    $\mathcal{O}_{dB}$, $\mathcal{O}_{dW}$,
    $\mathcal{O}_{uB}$, $\mathcal{O}_{uW}$,
    $\mathcal{O}^{(1)}_{\phi l}$, $\mathcal{O}^{(3)}_{\phi l}$,
    $\mathcal{O}^{(1)}_{\phi q}$, $\mathcal{O}^{(3)}_{\phi q}$,
    $\mathcal{O}_{\phi e}$, $\mathcal{O}_{\phi d}$,
    $\mathcal{O}_{\phi u}$ \\
    $\mathcal{L}_3$ & 
    $\mathcal{O}_{le}$ \\
    $\mathcal{U}_2$ &
    $\mathcal{O}^{(1)}_{lq}$, $\mathcal{O}^{(3)}_{lq}$,
    $\mathcal{O}_{ed}$, $\mathcal{O}_{ledq}$ \\
    $\mathcal{U}_5$ & 
    $\mathcal{O}_{eu}$ \\
    $\mathcal{Q}_1$ & 
    $\mathcal{O}_{lu}$, $\mathcal{O}^{(1)}_{qd}$,
    $\mathcal{O}^{(8)}_{qd}$, $\mathcal{O}_{duq}$ \\
    $\mathcal{Q}_5$ & 
    $\mathcal{O}_{qe}$, $\mathcal{O}^{(1)}_{qu}$,
    $\mathcal{O}^{(8)}_{qu}$, $\mathcal{O}_{ledq}$,
    $\mathcal{O}_{duq}$, $\mathcal{O}_{qqu}$ \\
    $\mathcal{X}$ &
    $\mathcal{O}^{(1)}_{lq}$, $\mathcal{O}^{(3)}_{lq}$ \\
    $\mathcal{Y}_1$ & 
    $\mathcal{O}^{(1)}_{qd}$, $\mathcal{O}^{(8)}_{qd}$ \\
    $\mathcal{Y}_5$ &
    $\mathcal{O}^{(1)}_{qu}$, $\mathcal{O}^{(8)}_{qu}$\\
    & \\[-0.5cm]
    \cbottomrule
  \end{tabular}
  \caption{Operators generated by the heavy vector bosons presented in table~\ref{t:vectors}.}
  \label{tab:topdown_vectors}
\end{table}

\newpage
\section{Complete Contributions to Wilson Coefficients}
\label{sec:results}

In this appendix we present the contributions to the dimension-six SMEFT 
induced by the heavy scalars, fermions and vectors introduced in section~\ref{sec:fields}. 
See section~\ref{sec:userguide} for details.

\subsection{Redefinitions of Standard Model interactions}
\label{sec:redefinitions}

Upon integrating the heavy fields ${\cal L}_1$ and $\varphi$ out, the kinetic term of the SM Higgs doublet receives extra contributions, yielding a non-canonically normalized field:
\begin{equation}
{\cal L}_{\mathrm{kin},\phi}=Z_\phi D_\mu\phi^\dagger D^\mu \phi,
\end{equation}
where
\begin{equation}
Z_\phi \equiv 1 - \frac{
    (\gamma_{\mathcal{L}_1})^*_{r}
    (\gamma_{\mathcal{L}_1})_{r}}{
    M_{\mathcal{L}_{1r}}^{2}}
   - \frac{
    \hat{\mu}^2_{\phi}
    (\delta_{\mathcal{L}_1\varphi})^*_{ts}
    (\gamma_{\mathcal{L}_1})_{t}
    (\delta_{\mathcal{L}_1\varphi})_{rs}
    (\gamma_{\mathcal{L}_1})^*_{r}}{
    M_{\mathcal{L}_{1r}}^{2}
    M_{\varphi_s}^{2}
    M_{\mathcal{L}_{1t}}^{2}}.
\label{eq:Zphi}
\end{equation}
In what follows, we renormalize $\phi \rightarrow Z_\phi^{-1/2} \phi$ and present our results in a basis where all fields have canonical kinetic terms (in the electroweak exact phase). The operators with $n_\phi$ doublets are therefore renormalized with $Z_\phi^{-n_\phi/2}$. (This includes also the operators in ${\cal L}_0$.) We will show these factors explicitly wherever they are needed, such that all the Wilson coefficients $C_i$ in this appendix are defined as the coefficients multiplying the corresponding operators with canonical fields in the effective Lagrangian. Let us make two observations about $Z_\phi$. First, the effect of the second term in eq.~\refeq{eq:Zphi} on the Wilson coefficients of dimension-six operators will have have an extra suppression of the form $\hat{\mu}_\phi^2/M^2$, with $M$ a heavy mass scale, comparable to the one of the typical Wilson coefficients of dimension-eight operators with respect to the dimension-six ones. Hence, even if we include it for completeness of the dimension-six results, for most practical purposes this second term can be neglected. The first term, on the other hand, will not be suppressed if the dimensionful coupling $\gamma_{\mathcal{L}_1}$ is of order $M_{\mathcal{L}_1}$. Second, $Z_\phi$ is non-trivial only when $\gamma_{\mathcal{L}_1} \neq 0$, so it can be ignored in perturbative unitary extensions of the SM. 

The contributions to the renormalizable SM interactions in table~\ref{tab:dim45Basis} are given by
\begin{align}
  Z_\phi^{\frac {1}{2}}\left(C_{y^e}\right)_{ij}= &
  - \frac{
    \hat{\mu}^2_{\phi}
    (\delta_{\mathcal{L}_1\varphi})_{sr}
    (\gamma_{\mathcal{L}_1})^*_{s}
    (y^e_{\varphi})_{rij}}{
    M_{\varphi_r}^{2}
    M_{\mathcal{L}_{1s}}^{2}}
  - \frac{
    \hat{\mu}^2_{\phi}
    \hat{y}^e_{ij}
    (\delta_{\mathcal{L}_1\varphi})^*_{ts}
    (\gamma_{\mathcal{L}_1})_{t}
    (\delta_{\mathcal{L}_1\varphi})_{rs}
    (\gamma_{\mathcal{L}_1})^*_{r}}{
    2 M_{\mathcal{L}_{1r}}^{2}
    M_{\varphi_s}^{2}
    M_{\mathcal{L}_{1t}}^{2}}
  \nonumber \\ &
  + \frac{1}{f} \Bigg\{
  \frac{
    \hat{\mu}^2_{\phi}
    (\tilde{g}^{eDl}_{\mathcal{L}_1})_{rij}
    (\gamma_{\mathcal{L}_1})^*_{r}}{
    2 M_{\mathcal{L}_{1r}}^{2}}
  + \frac{
    \hat{\mu}^2_{\phi}
    (\tilde{g}^{Del}_{\mathcal{L}_1})_{rij}
    (\gamma_{\mathcal{L}_1})^*_{r}}{
    2 M_{\mathcal{L}_{1r}}^{2}}
  \Bigg\},
  \\[5mm]
  Z_\phi^{\frac {1}{2}}\left(C_{y^d}\right)_{ij}= &
  - \frac{
    \hat{\mu}^2_{\phi}
    (\delta_{\mathcal{L}_1\varphi})_{sr}
    (\gamma_{\mathcal{L}_1})^*_{s}
    (y^d_{\varphi})_{rij}}{
    M_{\varphi_r}^{2}
    M_{\mathcal{L}_{1s}}^{2}}
  - \frac{
    \hat{\mu}^2_{\phi}
    \hat{y}^d_{ij}
    (\delta_{\mathcal{L}_1\varphi})^*_{ts}
    (\gamma_{\mathcal{L}_1})_{t}
    (\delta_{\mathcal{L}_1\varphi})_{rs}
    (\gamma_{\mathcal{L}_1})^*_{r}}{
    2 M_{\mathcal{L}_{1r}}^{2}
    M_{\varphi_s}^{2}
    M_{\mathcal{L}_{1t}}^{2}}
  \nonumber \\ &
  + \frac{1}{f} \Bigg\{
  \frac{
    \hat{\mu}^2_{\phi}
    (\tilde{g}^{dDq}_{\mathcal{L}_1})_{rij}
    (\gamma_{\mathcal{L}_1})^*_{r}}{
    2 M_{\mathcal{L}_{1r}}^{2}}
  + \frac{
    \hat{\mu}^2_{\phi}
    (\tilde{g}^{Ddq}_{\mathcal{L}_1})_{rij}
    (\gamma_{\mathcal{L}_1})^*_{r}}{
    2 M_{\mathcal{L}_{1r}}^{2}}
  \Bigg\},
\end{align}

\begin{align}
  Z_\phi^{\frac {1}{2}}\left(C_{y^u}\right)_{ij}= &
  \frac{
    \hat{\mu}^2_{\phi}
    (\delta_{\mathcal{L}_1\varphi})^*_{sr}
    (\gamma_{\mathcal{L}_1})_{s}
    (y^u_{\varphi})^*_{rji}}{
    M_{\varphi_r}^{2}
    M_{\mathcal{L}_{1s}}^{2}}
  - \frac{
    \hat{\mu}^2_{\phi}
    \hat{y}^u_{ij}
    (\delta_{\mathcal{L}_1\varphi})^*_{ts}
    (\gamma_{\mathcal{L}_1})_{t}
    (\delta_{\mathcal{L}_1\varphi})_{rs}
    (\gamma_{\mathcal{L}_1})^*_{r}}{
    2 M_{\mathcal{L}_{1r}}^{2}
    M_{\varphi_s}^{2}
    M_{\mathcal{L}_{1t}}^{2}}
  \nonumber \\ &
  + \frac{1}{f} \Bigg\{
  - \frac{
    \hat{\mu}^2_{\phi}
    (\tilde{g}^{qDu}_{\mathcal{L}_1})^*_{rij}
    (\gamma_{\mathcal{L}_1})_{r}}{
    2 M_{\mathcal{L}_{1r}}^{2}}
  - \frac{
    \hat{\mu}^2_{\phi}
    (\tilde{g}^{Dqu}_{\mathcal{L}_1})^*_{rij}
    (\gamma_{\mathcal{L}_1})_{r}}{
    2 M_{\mathcal{L}_{1r}}^{2}}
  \Bigg\},
  \\[10mm]
  Z_\phi^{2} C_{\phi 4}= & 
  \frac{
    (\kappa_{\mathcal{S}})_{r}
    (\kappa_{\mathcal{S}})_{r}}{
    2  M_{\mathcal{S}_r}^{2}}
  + \frac{
    (\kappa_{\Xi})_{r}
    (\kappa_{\Xi})_{r}}{
    2  M_{\Xi_r}^{2}}
   - \frac{
    2   \hat{\mu}^2_{\phi}
    (\kappa_{\Xi})_{r}
    (\kappa_{\Xi})_{r}}{
    M_{\Xi_r}^{4}}
  + \frac{
    2   (\kappa_{\Xi_1})^*_{r}
    (\kappa_{\Xi_1})_{r}}{
    M_{\Xi_{1r}}^{2}}
  \nonumber \\ &
  - \frac{
    4   \hat{\mu}^2_{\phi}
    (\kappa_{\Xi_1})^*_{r}
    (\kappa_{\Xi_1})_{r}}{
    M_{\Xi_{1r}}^{4}}
  + \frac{
    \hat{\mu}^2_{\phi}
    (\hat{g}^\phi_{\mathcal{B}_{1}})^*_{r}
    (\hat{g}^\phi_{\mathcal{B}_{1}})_{r}}{
    M_{\mathcal{B}_{1r}}^{2}}
  + \frac{
    \hat{\mu}^2_{\phi}
    (\hat{g}^\phi_{\mathcal{W}})^*_{r}
    (\hat{g}^\phi_{\mathcal{W}})_{r}}{
    2  M_{\mathcal{W}_r}^{2}}
  + \frac{
    \hat{\mu}^2_{\phi}
    (\hat{g}^\phi_{\mathcal{W}_{1}})^*_{r}
    (\hat{g}^\phi_{\mathcal{W}_{1}})_{r}}{
    4  M_{\mathcal{W}_{1r}}^{2}}
  \nonumber \\ &
  - \frac{
    \hat{\mu}^2_{\phi}
    g_2
    (g^W_{\mathcal{L}_1})_{sr}
    (\gamma_{\mathcal{L}_1})^*_{s}
    (\gamma_{\mathcal{L}_1})_{r}}{
    M_{\mathcal{L}_{1r}}^{2}
    M_{\mathcal{L}_{1s}}^{2}}
  + \frac{
    \hat{\mu}^2_{\phi}
    (h^{(1)}_{\mathcal{L}_1})_{rs}
    (\gamma_{\mathcal{L}_1})^*_{r}
    (\gamma_{\mathcal{L}_1})_{s}}{
    M_{\mathcal{L}_{1r}}^{2}
    M_{\mathcal{L}_{1s}}^{2}}
    \nonumber \\ &
  + \frac{
    2 \hat{\mu}^2_{\phi}
    \operatorname{Im}{\left(
        (\hat{g}^\phi_{\mathcal{W}})_{r}
      \right)}
    (\delta_{\mathcal{W}\Xi})_{rs}
    (\kappa_{\Xi})_{s}}{
    M_{\mathcal{W}_r}^{2}
    M_{\Xi_s}^{2}}
  + \frac{
    2   \hat{\mu}^2_{\phi}
    (\delta_{\mathcal{W}\Xi})_{ts}
    (\delta_{\mathcal{W}\Xi})_{tr}
    (\kappa_{\Xi})_{r}
    (\kappa_{\Xi})_{s}}{
    M_{\Xi_r}^{2}
    M_{\Xi_s}^{2}
    M_{\mathcal{W}_t}^{2}}
  \nonumber \\ &
  + \frac{
    2 \hat{\mu}^2_{\phi}
    \operatorname{Im}{\left(
        (\hat{g}^\phi_{\mathcal{W}_{1}})^*_{r}
        (\delta_{\mathcal{W}_1\Xi_1})_{rs}
        (\kappa_{\Xi_1})_{s}
      \right)}}{
    M_{\Xi_{1s}}^{2}
    M_{\mathcal{W}_{1r}}^{2}}
  + \frac{
    4 \hat{\mu}^2_{\phi}
    (\delta_{\mathcal{W}_1\Xi_1})^*_{st}
    (\delta_{\mathcal{W}_1\Xi_1})_{sr}
    (\kappa_{\Xi_1})_{r}
    (\kappa_{\Xi_1})^*_{t}}{
    M_{\Xi_{1r}}^{2}
    M_{\mathcal{W}_{1s}}^{2}
    M_{\Xi_{1t}}^{2}}
  \nonumber \\ &
   - \frac{
    2 \hat{\mu}^2_{\phi}
    \operatorname{Re}{\left(
        (g^{(2)}_{\mathcal{S}\mathcal{L}_1})_{rs}
        (\gamma_{\mathcal{L}_1})_{s}
      \right)}
    (\kappa_{\mathcal{S}})_{r}}{
    M_{\mathcal{S}_r}^{2}
    M_{\mathcal{L}_{1s}}^{2}}
  - \frac{
    \hat{\mu}^2_{\phi}
    (\varepsilon_{\mathcal{S}\mathcal{L}_1})_{rts}
    (\kappa_{\mathcal{S}})_{r}
    (\gamma_{\mathcal{L}_1})^*_{t}
    (\gamma_{\mathcal{L}_1})_{s}}{
    M_{\mathcal{S}_r}^{2}
    M_{\mathcal{L}_{1s}}^{2}
    M_{\mathcal{L}_{1t}}^{2}}
  \nonumber \\ &
  - \frac{
    2 \hat{\mu}^2_{\phi}
    \operatorname{Re}{\left(
        (\delta_{\mathcal{L}_1\varphi})_{rs}
        (\gamma_{\mathcal{L}_1})^*_{r}
        (\lambda_{\varphi})_{s}
      \right)}}{
    M_{\varphi_s}^{2}
    M_{\mathcal{L}_{1r}}^{2}}
  - \frac{
    2 \hat{\mu}^2_{\phi}
    \hat{\lambda}_{\phi}
    (\delta_{\mathcal{L}_1\varphi})^*_{ts}
    (\gamma_{\mathcal{L}_1})_{t}
    (\delta_{\mathcal{L}_1\varphi})_{rs}
    (\gamma_{\mathcal{L}_1})^*_{r}}{
    M_{\mathcal{L}_{1r}}^{2}
    M_{\varphi_s}^{2}
    M_{\mathcal{L}_{1t}}^{2}}
  \nonumber \\ &
  + \frac{
    2 \hat{\mu}^2_{\phi}
    \operatorname{Re}{\left(
        (g^{(2)}_{\Xi\mathcal{L}_1})_{rs}
        (\gamma_{\mathcal{L}_1})_{s}
      \right)}
    (\kappa_{\Xi})_{r}}{
    M_{\Xi_r}^{2}
    M_{\mathcal{L}_{1s}}^{2}}
  + \frac{
    \hat{\mu}^2_{\phi}
    (\varepsilon_{\Xi\mathcal{L}_1})_{srt}
    (\kappa_{\Xi})_{s}
    (\gamma_{\mathcal{L}_1})^*_{r}
    (\gamma_{\mathcal{L}_1})_{t}}{
    M_{\mathcal{L}_{1r}}^{2}
    M_{\Xi_s}^{2}
    M_{\mathcal{L}_{1t}}^{2}}
  \nonumber \\ &
  - \frac{
    4 \hat{\mu}^2_{\phi}
    \operatorname{Re}{\left(
        (g^{(1)}_{\Xi\mathcal{L}_1})^*_{rs}
        (\gamma_{\mathcal{L}_1})^*_{s}
      \right)}
    (\kappa_{\Xi})_{r}}{
    M_{\Xi_r}^{2}
    M_{\mathcal{L}_{1s}}^{2}}
  - \frac{
    4 \hat{\mu}^2_{\phi}
    \operatorname{Re}{\left(
        (g^{(1)}_{\Xi_1\mathcal{L}_1})^*_{rs}
        (\gamma_{\mathcal{L}_1})^*_{s}
        (\kappa_{\Xi_1})_{r}
      \right)}}{
    M_{\mathcal{L}_{1s}}^{2}
    M_{\Xi_{1r}}^{2}}
  \nonumber \\ &
  + \frac{
    2 \hat{\mu}^2_{\phi}
    \operatorname{Re}{\left(
        (\delta_{\mathcal{L}_1\varphi})_{rs}
        (\gamma_{\mathcal{L}_1})^*_{r}
        (\kappa_{\mathcal{S}\varphi})_{ts}
      \right)}
    (\kappa_{\mathcal{S}})_{t}}{
    M_{\varphi_s}^{2}
    M_{\mathcal{S}_t}^{2}
    M_{\mathcal{L}_{1r}}^{2}}
  + \frac{
    2 \hat{\mu}^2_{\phi}
    \operatorname{Re}{\left(
        (\delta_{\mathcal{L}_1\varphi})_{rs}
        (\gamma_{\mathcal{L}_1})^*_{r}
        (\kappa_{\Xi\varphi})_{ts}
      \right)}
    (\kappa_{\Xi})_{t}}{
    M_{\mathcal{L}_{1r}}^{2}
    M_{\Xi_t}^{2}
    M_{\varphi_s}^{2}}
  \nonumber \\ &
  + \frac{
    4 \hat{\mu}^2_{\phi}
    \operatorname{Re}{\left(
        (\delta_{\mathcal{L}_1\varphi})_{rs}
        (\gamma_{\mathcal{L}_1})^*_{r}
        (\kappa_{\Xi_1\varphi})^*_{ts}
        (\kappa_{\Xi_1})_{t}
      \right)}}{
    M_{\varphi_s}^{2}
    M_{\mathcal{L}_{1r}}^{2}
    M_{\Xi_{1t}}^{2}}
  \nonumber \\ &
  + \frac{1}{f} \Bigg\{
  - \frac{
    \hat{\mu}^2_{\phi}
    (\tilde{k}^{\phi}_{\mathcal{S}})_{r}
    (\kappa_{\mathcal{S}})_{r}}{
    M_{\mathcal{S}_r}^{2}}
  + \frac{
    \hat{\mu}^2_{\phi}
    (\tilde{k}^{\phi}_{\Xi})_{r}
    (\kappa_{\Xi})_{r}}{
    M_{\Xi_r}^{2}}
  + \frac{
    2 \hat{\mu}^2_{\phi}
    \operatorname{Re}{\left(
        (\tilde{\gamma}^{(3)}_{\mathcal{L}_1})_{r}
        (\gamma_{\mathcal{L}_1})^*_{r}
      \right)}}{
    M_{\mathcal{L}_{1r}}^{2}}
  \nonumber \\ &
  \qquad
  - \frac{
    2 \hat{\mu}^2_{\phi}
    \operatorname{Im}{\left(
        (\tilde{\gamma}^W_{\mathcal{L}_1})_{r}
        (\gamma_{\mathcal{L}_1})^*_{r}
      \right)}
    g_2}{
    M_{\mathcal{L}_{1r}}^{2}}
  \Bigg\}.
\end{align}

\newpage

\noindent These contributions can be absorbed into redefinitions of the SM Yukawa and quartic Higgs couplings: 
\begin{eqnarray}
  \hat{y}^{e,u,d}_{ij} & = & Z_\phi^{\frac {1}{2}} ( y^{e,u,d}_{ij} - \left(C_{y^{e,u,d}}\right)_{ij} ),
  \label{eq:y} \\
  \hat{\lambda}_\phi & = & Z_\phi^{2} (\lambda_\phi - C_{\phi 4}).
  \label{eq:lambda}
\end{eqnarray}
Due to the Higgs-field renormalization, the coefficient of the Higgs mass term is also redefined:
\beq
\hat{\mu}^2_\phi =  Z_\phi~\mu_\phi^2.
  \label{eq:mu}
\eeq
We remind the reader that the hatted couplings on the left-hand side of the last three equations are the coefficients of the corresponding operators---with the original Higgs-field normalization---in the SM part of $\LBSM$. The corresponding unhatted couplings are the coefficients of these operators---built with canonically-normalized fields---in $\mathcal{L}_{\mathrm{eff}}$.
Note that the right-hand sides depend linearly on the explicit hatted couplings on the left-hand sides. Solving this linear system is straightforward. 

In terms of the renormalized Higgs field and the redefined couplings $\mu^2_\phi$, $y^{e,u,d}$ and $\lambda_\phi$, all the heavy-field contributions appear in the Wilson coefficients of higher-dimensional operators. In order to keep our results as compact and clear as possible, we write the dimension-six operators in terms of the original, hatted couplings. They can be readily substituted by the solutions to eqs.~\refeq{eq:y}, \refeq{eq:lambda} and~\refeq{eq:mu} to get the expressions in terms of the redefined couplings. 
In practice, these expressions can be greatly simplified. Indeed, all the contributions to $C_{y^{e,u,d}}$, except the one inside $Z_\phi$,  and most of the contributions to $C_{\phi 4}$ are not $O(1)$ but carry an extra suppression $\mu^2_\phi/M^2$. For calculations to order $E^2/M^2$, with $E$ a low-energy scale, all these contributions can be neglected. In this approximation, the hatted couplings do not appear on the right-hand sides of eqs.~\refeq{eq:y}, \refeq{eq:lambda} and~\refeq{eq:mu}, which thus give explicitly their expressions in terms of the redefined ones.

\subsection{Dimension Five}
\label{sec:dim5}

The only dimension-five operator in the basis receives the following contributions: 
\begin{align}
  Z_\phi \left(C_5\right)_{ij}= & 
  - \frac{
    2   (\kappa_{\Xi_1})_{r}
    (y_{\Xi_1})^*_{rji}}{
    M_{\Xi_{1r}}^{2}}
  + \frac{
    (\lambda_N)_{rj}
    (\lambda_N)_{ri}}{
    2  M_{N_r}}
  + \frac{
    (\lambda_{\Sigma})_{rj}
    (\lambda_{\Sigma})_{ri}}{
    8  M_{\Sigma_r}}.
\end{align}

\subsection{Four-fermion Operators}
\label{sec:d64F}

\subsubsection{$\left(\overline{L}L\right)\left(\overline{L}L\right)$}
\label{sec:LLLL}

\begin{align}
  \left(C_{ll}\right)_{ijkl}= & 
  \frac{
    (y_{\mathcal{S}_1})^*_{rjl}
    (y_{\mathcal{S}_1})_{rik}}{
    M_{\mathcal{S}_{1r}}^{2}}
  + \frac{
    (y_{\Xi_1})_{rki}
    (y_{\Xi_1})^*_{rlj}}{
    M_{\Xi_{1r}}^{2}}
  - \frac{
    (g^l_{\mathcal{B}})_{rkl}
    (g^l_{\mathcal{B}})_{rij}}{
    2  M_{\mathcal{B}_r}^{2}}
  \nonumber \\ &
  - \frac{
    (g^l_{\mathcal{W}})_{rkj}
    (g^l_{\mathcal{W}})_{ril}}{
    4  M_{\mathcal{W}_r}^{2}}
  + \frac{
    (g^l_{\mathcal{W}})_{rkl}
    (g^l_{\mathcal{W}})_{rij}}{
    8  M_{\mathcal{W}_r}^{2}},
  \label{eq:Cll}
  \\[5mm]
  \left(C^{(1)}_{qq}\right)_{ijkl}= & 
  \frac{
    (y^{qq}_{\omega_1})_{rik}
    (y^{qq}_{\omega_1})^*_{rlj}}{
    2  M_{\omega_{1r}}^{2}}
  + \frac{
    3   (y^{qq}_{\zeta})_{rki}
    (y^{qq}_{\zeta})^*_{rlj}}{
    2  M_{\zeta_{r}}^{2}}
  + \frac{
    (y^{qq}_{\Omega_1})^*_{rik}
    (y^{qq}_{\Omega_1})_{rjl}}{
    4  M_{\Omega_{1r}}^{2}}
  + \frac{
    3   (y_{\Upsilon})_{rlj}
    (y_{\Upsilon})^*_{rki}}{
    4  M_{\Upsilon_{r}}^{2}}
  \nonumber \\ &
  - \frac{
    (g^q_{\mathcal{B}})_{rkl}
    (g^q_{\mathcal{B}})_{rij}}{
    2  M_{\mathcal{B}_r}^{2}}
  - \frac{
    (g^q_{\mathcal{G}})_{rkj}
    (g^q_{\mathcal{G}})_{ril}}{
    8  M_{\mathcal{G}_r}^{2}}
  + \frac{
    (g^q_{\mathcal{G}})_{rkl}
    (g^q_{\mathcal{G}})_{rij}}{
    12  M_{\mathcal{G}_r}^{2}}
  - \frac{
    3   (g_{\mathcal{H}})_{rkj}
    (g_{\mathcal{H}})_{ril}}{
    32  M_{\mathcal{H}_r}^{2}},
    \label{eq:Cqq1}
  \\[5mm]
  \left(C^{(3)}_{qq}\right)_{ijkl}= & 
  - \frac{
    (y^{qq}_{\omega_1})_{rki}
    (y^{qq}_{\omega_1})^*_{rjl}}{
    2  M_{\omega_{1r}}^{2}}
  - \frac{
    (y^{qq}_{\zeta})_{rki}
    (y^{qq}_{\zeta})^*_{rjl}}{
    2  M_{\zeta_{r}}^{2}}
  + \frac{
    (y^{qq}_{\Omega_1})^*_{rik}
    (y^{qq}_{\Omega_1})_{rlj}}{
    4  M_{\Omega_{1r}}^{2}}
  + \frac{
    (y_{\Upsilon})^*_{rki}
    (y_{\Upsilon})_{rjl}}{
    4  M_{\Upsilon_{r}}^{2}}
  \nonumber \\ &
  - \frac{
    (g^q_{\mathcal{W}})_{rkl}
    (g^q_{\mathcal{W}})_{rij}}{
    8  M_{\mathcal{W}_r}^{2}}
  - \frac{
    (g^q_{\mathcal{G}})_{rkj}
    (g^q_{\mathcal{G}})_{ril}}{
    8  M_{\mathcal{G}_r}^{2}}
  + \frac{
    (g_{\mathcal{H}})_{rkl}
    (g_{\mathcal{H}})_{rij}}{
    48  M_{\mathcal{H}_r}^{2}}
  + \frac{
    (g_{\mathcal{H}})_{rkj}
    (g_{\mathcal{H}})_{ril}}{
    32  M_{\mathcal{H}_r}^{2}},
  \label{eq:Cqq3}
  \\[5mm]
  \left(C^{(1)}_{lq}\right)_{ijkl}= & 
  \frac{
    (y^{ql}_{\omega_1})^*_{rki}
    (y^{ql}_{\omega_1})_{rlj}}{
    4  M_{\omega_{1r}}^{2}}
  + \frac{
    3   (y^{ql}_{\zeta})^*_{rki}
    (y^{ql}_{\zeta})_{rlj}}{
    4  M_{\zeta_{r}}^{2}}
  - \frac{
    (g^q_{\mathcal{B}})_{rkl}
    (g^l_{\mathcal{B}})_{rij}}{
    M_{\mathcal{B}_r}^{2}}
  \nonumber \\ &
  - \frac{
    (g^{lq}_{\mathcal{U}_2})^*_{rjk}
    (g^{lq}_{\mathcal{U}_2})_{ril}}{
    2  M_{\mathcal{U}_{2r}}^{2}}
  - \frac{
    3   (g_{\mathcal{X}})^*_{rjk}
    (g_{\mathcal{X}})_{ril}}{
    8  M_{\mathcal{X}_r}^{2}},
  \label{eq:Clq1}
  \\[5mm]
  \left(C^{(3)}_{lq}\right)_{ijkl}= & 
  - \frac{
    (y^{ql}_{\omega_1})^*_{rki}
    (y^{ql}_{\omega_1})_{rlj}}{
    4  M_{\omega_{1r}}^{2}}
  + \frac{
    (y^{ql}_{\zeta})^*_{rki}
    (y^{ql}_{\zeta})_{rlj}}{
    4  M_{\zeta_{r}}^{2}}
  - \frac{
    (g^q_{\mathcal{W}})_{rkl}
    (g^l_{\mathcal{W}})_{rij}}{
    4  M_{\mathcal{W}_r}^{2}}
  \nonumber \\ &
  - \frac{
    (g^{lq}_{\mathcal{U}_2})^*_{rjk}
    (g^{lq}_{\mathcal{U}_2})_{ril}}{
    2  M_{\mathcal{U}_{2r}}^{2}}
  + \frac{
    (g_{\mathcal{X}})^*_{rjk}
    (g_{\mathcal{X}})_{ril}}{
    8  M_{\mathcal{X}_r}^{2}}.
    \label{eq:Clq3}
\end{align}

\newpage

\subsubsection{$\left(\overline{R}R\right)\left(\overline{R}R\right)$}
\label{sec:RRRR}

\begin{align}
  \left(C_{ee}\right)_{ijkl}= & 
  \frac{
    (y_{\mathcal{S}_2})_{rki}
    (y_{\mathcal{S}_2})^*_{rlj}}{
    2  M_{\mathcal{S}_{2r}}^{2}}
  - \frac{
    (g^e_{\mathcal{B}})_{rkl}
    (g^e_{\mathcal{B}})_{rij}}{
    2  M_{\mathcal{B}_r}^{2}},
  \\[5mm]
  \left(C_{dd}\right)_{ijkl}= & 
  \frac{
    (y_{\omega_2})^*_{rlj}
    (y_{\omega_2})_{rki}}{
    M_{\omega_{2r}}^{2}}
  + \frac{
    (y_{\Omega_2})^*_{rik}
    (y_{\Omega_2})_{rjl}}{
    2  M_{\Omega_{2r}}^{2}}
  - \frac{
    (g^d_{\mathcal{B}})_{rkl}
    (g^d_{\mathcal{B}})_{rij}}{
    2  M_{\mathcal{B}_r}^{2}}
  \nonumber \\ &
  - \frac{
    (g^d_{\mathcal{G}})_{rkj}
    (g^d_{\mathcal{G}})_{ril}}{
    4  M_{\mathcal{G}_r}^{2}}
  + \frac{
    (g^d_{\mathcal{G}})_{rkl}
    (g^d_{\mathcal{G}})_{rij}}{
    12  M_{\mathcal{G}_r}^{2}},
  \\[5mm]
  \left(C_{uu}\right)_{ijkl}= & 
  \frac{
    (y^{uu}_{\omega_4})^*_{rlj}
    (y^{uu}_{\omega_4})_{rki}}{
    M_{\omega_{4r}}^{2}}
  + \frac{
    (y_{\Omega_4})^*_{rik}
    (y_{\Omega_4})_{rjl}}{
    2  M_{\Omega_{4r}}^{2}}
  - \frac{
    (g^u_{\mathcal{B}})_{rkl}
    (g^u_{\mathcal{B}})_{rij}}{
    2  M_{\mathcal{B}_r}^{2}}
  \nonumber \\ &
  - \frac{
    (g^u_{\mathcal{G}})_{rkj}
    (g^u_{\mathcal{G}})_{ril}}{
    4  M_{\mathcal{G}_r}^{2}}
  + \frac{
    (g^u_{\mathcal{G}})_{rkl}
    (g^u_{\mathcal{G}})_{rij}}{
    12  M_{\mathcal{G}_r}^{2}},
  \\[5mm]
  \left(C_{ed}\right)_{ijkl}= & 
  \frac{
    (y^{ed}_{\omega_4})^*_{rik}
    (y^{ed}_{\omega_4})_{rjl}}{
    2  M_{\omega_{4r}}^{2}}
  - \frac{
    (g^d_{\mathcal{B}})_{rkl}
    (g^e_{\mathcal{B}})_{rij}}{
    M_{\mathcal{B}_r}^{2}}
  - \frac{
    (g^{ed}_{\mathcal{U}_2})^*_{rjk}
    (g^{ed}_{\mathcal{U}_2})_{ril}}{
    M_{\mathcal{U}_{2r}}^{2}},
  \\[5mm]
  \left(C_{eu}\right)_{ijkl}= & 
  \frac{
    (y^{eu}_{\omega_1})^*_{rik}
    (y^{eu}_{\omega_1})_{rjl}}{
    2  M_{\omega_{1r}}^{2}}
  - \frac{
    (g^u_{\mathcal{B}})_{rkl}
    (g^e_{\mathcal{B}})_{rij}}{
    M_{\mathcal{B}_r}^{2}}
  - \frac{
    (g_{\mathcal{U}_5})^*_{rjk}
    (g_{\mathcal{U}_5})_{ril}}{
    M_{\mathcal{U}_{5r}}^{2}},
  \\[5mm]
  \left(C^{(1)}_{ud}\right)_{ijkl}= & 
  \frac{
    (y^{du}_{\omega_1})^*_{rlj}
    (y^{du}_{\omega_1})_{rki}}{
    3  M_{\omega_{1r}}^{2}}
  + \frac{
    (y^{ud}_{\Omega_1})^*_{rik}
    (y^{ud}_{\Omega_1})_{rjl}}{
    3  M_{\Omega_{1r}}^{2}}
  - \frac{
    (g^u_{\mathcal{B}})_{rij}
    (g^d_{\mathcal{B}})_{rkl}}{
    M_{\mathcal{B}_r}^{2}}
  \nonumber \\ &
  - \frac{
    (g^{du}_{\mathcal{B}_1})^*_{rli}
    (g^{du}_{\mathcal{B}_1})_{rkj}}{
    3  M_{\mathcal{B}_{1r}}^{2}}
  - \frac{
    4   (g_{\mathcal{G}_1})^*_{rli}
    (g_{\mathcal{G}_1})_{rkj}}{
    9  M_{\mathcal{G}_{1r}}^{2}},
  \\[5mm]
  \left(C^{(8)}_{ud}\right)_{ijkl}= & 
  - \frac{
    (y^{du}_{\omega_1})^*_{rlj}
    (y^{du}_{\omega_1})_{rki}}{
    M_{\omega_{1r}}^{2}}
  + \frac{
    (y^{ud}_{\Omega_1})^*_{rik}
    (y^{ud}_{\Omega_1})_{rjl}}{
    2  M_{\Omega_{1r}}^{2}}
  - \frac{
    (g^d_{\mathcal{G}})_{rkl}
    (g^u_{\mathcal{G}})_{rij}}{
    M_{\mathcal{G}_r}^{2}}
  \nonumber \\ &
  - \frac{
    2   (g^{du}_{\mathcal{B}_1})^*_{rli}
    (g^{du}_{\mathcal{B}_1})_{rkj}}{
    M_{\mathcal{B}_{1r}}^{2}}
  + \frac{
    (g_{\mathcal{G}_1})^*_{rli}
    (g_{\mathcal{G}_1})_{rkj}}{
    3  M_{\mathcal{G}_{1r}}^{2}}.
\end{align}

\newpage

\subsubsection{$\left(\overline{L}L\right)\left(\overline{R}R\right)$}
\label{sec:LLRR}

Recall that $\hat{y}^{e,u,d}$ are defined in equation \refeq{eq:y}.

\begin{align}
  \left(C_{le}\right)_{ijkl}= & 
  - \frac{
    (y^e_{\varphi})^*_{rli}
    (y^e_{\varphi})_{rkj}}{
    2  M_{\varphi_r}^{2}}
  - \frac{
    (g^e_{\mathcal{B}})_{rkl}
    (g^l_{\mathcal{B}})_{rij}}{
    M_{\mathcal{B}_r}^{2}}
  + \frac{
    (g_{\mathcal{L}_3})^*_{rki}
    (g_{\mathcal{L}_3})_{rlj}}{
    M_{\mathcal{L}_{3r}}^{2}}
  \nonumber \\ &
  - \frac{
    \hat{y}^{e*}_{li}
    (\delta_{\mathcal{L}_1\varphi})_{sr}
    (\gamma_{\mathcal{L}_1})^*_{s}
    (y^e_{\varphi})_{rkj}}{
    2  M_{\varphi_r}^{2}
    M_{\mathcal{L}_{1s}}^{2}}
  - \frac{
    \hat{y}^e_{kj}
    (\delta_{\mathcal{L}_1\varphi})^*_{sr}
    (\gamma_{\mathcal{L}_1})_{s}
    (y^e_{\varphi})^*_{rli}}{
    2  M_{\varphi_r}^{2}
    M_{\mathcal{L}_{1s}}^{2}}
  \nonumber \\ &
  - \frac{
    \hat{y}^e_{kj}
    \hat{y}^{e*}_{li}
    (\delta_{\mathcal{L}_1\varphi})^*_{ts}
    (\gamma_{\mathcal{L}_1})_{t}
    (\delta_{\mathcal{L}_1\varphi})_{rs}
    (\gamma_{\mathcal{L}_1})^*_{r}}{
    2  M_{\mathcal{L}_{1r}}^{2}
    M_{\varphi_s}^{2}
    M_{\mathcal{L}_{1t}}^{2}}
  \nonumber \\ &
  + \frac{1}{f} \Bigg\{
  \frac{
    \hat{y}^{e*}_{li}
    (\tilde{g}^{eDl}_{\mathcal{L}_1})_{rkj}
    (\gamma_{\mathcal{L}_1})^*_{r}}{
    4 M_{\mathcal{L}_{1r}}^{2}}
  + \frac{
    \hat{y}^{e*}_{li}
    (\tilde{g}^{Del}_{\mathcal{L}_1})_{rkj}
    (\gamma_{\mathcal{L}_1})^*_{r}}{
    4 M_{\mathcal{L}_{1r}}^{2}}
  \nonumber \\ &
  \qquad
  + \frac{
    \hat{y}^e_{kj}
    (\tilde{g}^{eDl}_{\mathcal{L}_1})^*_{rli}
    (\gamma_{\mathcal{L}_1})_{r}}{
    4 M_{\mathcal{L}_{1r}}^{2}}
  + \frac{
    \hat{y}^e_{kj}
    (\tilde{g}^{Del}_{\mathcal{L}_1})^*_{rli}
    (\gamma_{\mathcal{L}_1})_{r}}{
    4 M_{\mathcal{L}_{1r}}^{2}}
  \Bigg\},
  \\[7.5mm]
  \left(C_{ld}\right)_{ijkl}= & 
  - \frac{
    (y_{\Pi_1})^*_{rjk}
    (y_{\Pi_1})_{ril}}{
    2  M_{\Pi_{1r}}^{2}}
  - \frac{
    (g^d_{\mathcal{B}})_{rkl}
    (g^l_{\mathcal{B}})_{rij}}{
    M_{\mathcal{B}_r}^{2}}
  + \frac{
    (g^{dl}_{\mathcal{Q}_5})^*_{rki}
    (g^{dl}_{\mathcal{Q}_5})_{rlj}}{
    M_{\mathcal{Q}_{5r}}^{2}},
  \\[7.5mm]
  \left(C_{lu}\right)_{ijkl}= & 
  - \frac{
    (y^{lu}_{\Pi_7})^*_{rjk}
    (y^{lu}_{\Pi_7})_{ril}}{
    2  M_{\Pi_{7r}}^{2}}
  - \frac{
    (g^u_{\mathcal{B}})_{rkl}
    (g^l_{\mathcal{B}})_{rij}}{
    M_{\mathcal{B}_r}^{2}}
  + \frac{
    (g^{ul}_{\mathcal{Q}_1})^*_{rki}
    (g^{ul}_{\mathcal{Q}_1})_{rlj}}{
    M_{\mathcal{Q}_{1r}}^{2}},
  \\[7.5mm]
  \left(C_{qe}\right)_{ijkl}= & 
  - \frac{
    (y^{eq}_{\Pi_7})^*_{rli}
    (y^{eq}_{\Pi_7})_{rkj}}{
    2  M_{\Pi_{7r}}^{2}}
  - \frac{
    (g^e_{\mathcal{B}})_{rkl}
    (g^q_{\mathcal{B}})_{rij}}{
    M_{\mathcal{B}_r}^{2}}
  + \frac{
    (g^{eq}_{\mathcal{Q}_5})^*_{rki}
    (g^{eq}_{\mathcal{Q}_5})_{rlj}}{
    M_{\mathcal{Q}_{5r}}^{2}},
  \\[7.5mm]
  \left(C^{(1)}_{qu}\right)_{ijkl}= & 
  - \frac{
    (y^u_{\varphi})^*_{rjk}
    (y^u_{\varphi})_{ril}}{
    6  M_{\varphi_r}^{2}}
  - \frac{
    2   (y^{qu}_{\Phi})^*_{rjk}
    (y^{qu}_{\Phi})_{ril}}{
    9  M_{\Phi_{r}}^{2}}
  \nonumber \\ &
  - \frac{
    (g^u_{\mathcal{B}})_{rkl}
    (g^q_{\mathcal{B}})_{rij}}{
    M_{\mathcal{B}_r}^{2}}
  + \frac{
    2   (g^{uq}_{\mathcal{Q}_5})^*_{rlj}
    (g^{uq}_{\mathcal{Q}_5})_{rki}}{
    3  M_{\mathcal{Q}_{5r}}^{2}}
  + \frac{
    2   (g_{\mathcal{Y}_5})^*_{rlj}
    (g_{\mathcal{Y}_5})_{rki}}{
    3  M_{\mathcal{Y}_{5r}}^{2}}
  \nonumber \\ &
  + \frac{
    \hat{y}^u_{kj}
    (\delta_{\mathcal{L}_1\varphi})_{rs}
    (\gamma_{\mathcal{L}_1})^*_{r}
    (y^u_{\varphi})_{sil}}{
    6  M_{\mathcal{L}_{1r}}^{2}
    M_{\varphi_s}^{2}}
  + \frac{
    \hat{y}^{u*}_{li}
    (\delta_{\mathcal{L}_1\varphi})^*_{rs}
    (\gamma_{\mathcal{L}_1})_{r}
    (y^u_{\varphi})^*_{sjk}}{
    6  M_{\mathcal{L}_{1r}}^{2}
    M_{\varphi_s}^{2}}
  \nonumber \\ &
  - \frac{
    \hat{y}^u_{kj}
    \hat{y}^{u*}_{li}
    (\delta_{\mathcal{L}_1\varphi})^*_{ts}
    (\gamma_{\mathcal{L}_1})_{t}
    (\delta_{\mathcal{L}_1\varphi})_{rs}
    (\gamma_{\mathcal{L}_1})^*_{r}}{
    6  M_{\mathcal{L}_{1r}}^{2}
    M_{\varphi_s}^{2}
    M_{\mathcal{L}_{1t}}^{2}}
  \nonumber \\ &
  + \frac{1}{f} \Bigg\{
  - \frac{
    \hat{y}^u_{kj}
    (\tilde{g}^{qDu}_{\mathcal{L}_1})_{rli}
    (\gamma_{\mathcal{L}_1})^*_{r}}{
    12 M_{\mathcal{L}_{1r}}^{2}}
  - \frac{
    \hat{y}^u_{kj}
    (\tilde{g}^{Dqu}_{\mathcal{L}_1})_{rli}
    (\gamma_{\mathcal{L}_1})^*_{r}}{
    12 M_{\mathcal{L}_{1r}}^{2}}
  \nonumber \\ &
  \qquad
  - \frac{
    \hat{y}^{u*}_{li}
    (\tilde{g}^{qDu}_{\mathcal{L}_1})^*_{rkj}
    (\gamma_{\mathcal{L}_1})_{r}}{
    12 M_{\mathcal{L}_{1r}}^{2}}
  - \frac{
    \hat{y}^{u*}_{li}
    (\tilde{g}^{Dqu}_{\mathcal{L}_1})^*_{rkj}
    (\gamma_{\mathcal{L}_1})_{r}}{
    12 M_{\mathcal{L}_{1r}}^{2}}
  \Bigg\},
\end{align}

\begin{align}
  \left(C^{(8)}_{qu}\right)_{ijkl}= & 
  - \frac{
    (y^u_{\varphi})^*_{rjk}
    (y^u_{\varphi})_{ril}}{
    M_{\varphi_r}^{2}}
  + \frac{
    (y^{qu}_{\Phi})^*_{rjk}
    (y^{qu}_{\Phi})_{ril}}{
    6  M_{\Phi_{r}}^{2}}
  \nonumber \\ &
  - \frac{
    (g^u_{\mathcal{G}})_{rkl}
    (g^q_{\mathcal{G}})_{rij}}{
    M_{\mathcal{G}_r}^{2}}
  - \frac{
    2   (g^{uq}_{\mathcal{Q}_5})^*_{rlj}
    (g^{uq}_{\mathcal{Q}_5})_{rki}}{
    M_{\mathcal{Q}_{5r}}^{2}}
  + \frac{
    (g_{\mathcal{Y}_5})^*_{rlj}
    (g_{\mathcal{Y}_5})_{rki}}{
    M_{\mathcal{Y}_{5r}}^{2}}
  \nonumber \\ &
  + \frac{
    \hat{y}^u_{kj}
    (\delta_{\mathcal{L}_1\varphi})_{rs}
    (\gamma_{\mathcal{L}_1})^*_{r}
    (y^u_{\varphi})_{sil}}{
    M_{\mathcal{L}_{1r}}^{2}
    M_{\varphi_s}^{2}}
  + \frac{
    \hat{y}^{u*}_{li}
    (\delta_{\mathcal{L}_1\varphi})^*_{rs}
    (\gamma_{\mathcal{L}_1})_{r}
    (y^u_{\varphi})^*_{sjk}}{
    M_{\mathcal{L}_{1r}}^{2}
    M_{\varphi_s}^{2}}
  \nonumber \\ &
  - \frac{
    \hat{y}^u_{kj}
    \hat{y}^{u*}_{li}
    (\delta_{\mathcal{L}_1\varphi})^*_{ts}
    (\gamma_{\mathcal{L}_1})_{t}
    (\delta_{\mathcal{L}_1\varphi})_{rs}
    (\gamma_{\mathcal{L}_1})^*_{r}}{
    M_{\mathcal{L}_{1r}}^{2}
    M_{\varphi_s}^{2}
    M_{\mathcal{L}_{1t}}^{2}}
  \nonumber \\ &
  + \frac{1}{f} \Bigg\{
  - \frac{
    \hat{y}^u_{kj}
    (\tilde{g}^{qDu}_{\mathcal{L}_1})_{rli}
    (\gamma_{\mathcal{L}_1})^*_{r}}{
    2 M_{\mathcal{L}_{1r}}^{2}}
  - \frac{
    \hat{y}^u_{kj}
    (\tilde{g}^{Dqu}_{\mathcal{L}_1})_{rli}
    (\gamma_{\mathcal{L}_1})^*_{r}}{
    2 M_{\mathcal{L}_{1r}}^{2}}
  \nonumber \\ &
  \qquad
  - \frac{
    \hat{y}^{u*}_{li}
    (\tilde{g}^{qDu}_{\mathcal{L}_1})^*_{rkj}
    (\gamma_{\mathcal{L}_1})_{r}}{
    2 M_{\mathcal{L}_{1r}}^{2}}
  - \frac{
    \hat{y}^{u*}_{li}
    (\tilde{g}^{Dqu}_{\mathcal{L}_1})^*_{rkj}
    (\gamma_{\mathcal{L}_1})_{r}}{
    2 M_{\mathcal{L}_{1r}}^{2}}
  \Bigg\},
  \\[5mm]
  \left(C^{(1)}_{qd}\right)_{ijkl}= & 
  - \frac{
    (y^d_{\varphi})^*_{rli}
    (y^d_{\varphi})_{rkj}}{
    6  M_{\varphi_r}^{2}}
  - \frac{
    2   (y^{dq}_{\Phi})^*_{rli}
    (y^{dq}_{\Phi})_{rkj}}{
    9  M_{\Phi_{r}}^{2}}
  \nonumber \\ &
  - \frac{
    (g^d_{\mathcal{B}})_{rkl}
    (g^q_{\mathcal{B}})_{rij}}{
    M_{\mathcal{B}_r}^{2}}
  + \frac{
    2   (g^{dq}_{\mathcal{Q}_1})^*_{rlj}
    (g^{dq}_{\mathcal{Q}_1})_{rki}}{
    3  M_{\mathcal{Q}_{1r}}^{2}}
  + \frac{
    2   (g_{\mathcal{Y}_1})^*_{rlj}
    (g_{\mathcal{Y}_1})_{rki}}{
    3  M_{\mathcal{Y}_{1r}}^{2}}
  \nonumber \\ &
  - \frac{
    \hat{y}^{d*}_{li}
    (\delta_{\mathcal{L}_1\varphi})_{sr}
    (\gamma_{\mathcal{L}_1})^*_{s}
    (y^d_{\varphi})_{rkj}}{
    6  M_{\varphi_r}^{2}
    M_{\mathcal{L}_{1s}}^{2}}
  - \frac{
    \hat{y}^d_{kj}
    (\delta_{\mathcal{L}_1\varphi})^*_{sr}
    (\gamma_{\mathcal{L}_1})_{s}
    (y^d_{\varphi})^*_{rli}}{
    6  M_{\varphi_r}^{2}
    M_{\mathcal{L}_{1s}}^{2}}
  \nonumber \\ &
  - \frac{
    \hat{y}^d_{kj}
    \hat{y}^{d*}_{li}
    (\delta_{\mathcal{L}_1\varphi})^*_{ts}
    (\gamma_{\mathcal{L}_1})_{t}
    (\delta_{\mathcal{L}_1\varphi})_{rs}
    (\gamma_{\mathcal{L}_1})^*_{r}}{
    6  M_{\mathcal{L}_{1r}}^{2}
    M_{\varphi_s}^{2}
    M_{\mathcal{L}_{1t}}^{2}}
  \nonumber \\ &
  + \frac{1}{f} \Bigg\{
  \frac{
    \hat{y}^{d*}_{li}
    (\tilde{g}^{dDq}_{\mathcal{L}_1})_{rkj}
    (\gamma_{\mathcal{L}_1})^*_{r}}{
    12 M_{\mathcal{L}_{1r}}^{2}}
  + \frac{
    \hat{y}^{d*}_{li}
    (\tilde{g}^{Ddq}_{\mathcal{L}_1})_{rkj}
    (\gamma_{\mathcal{L}_1})^*_{r}}{
    12 M_{\mathcal{L}_{1r}}^{2}}
  \nonumber \\ &
  \qquad
  + \frac{
    \hat{y}^d_{kj}
    (\tilde{g}^{dDq}_{\mathcal{L}_1})^*_{rli}
    (\gamma_{\mathcal{L}_1})_{r}}{
    12 M_{\mathcal{L}_{1r}}^{2}}
  + \frac{
    \hat{y}^d_{kj}
    (\tilde{g}^{Ddq}_{\mathcal{L}_1})^*_{rli}
    (\gamma_{\mathcal{L}_1})_{r}}{
    12 M_{\mathcal{L}_{1r}}^{2}}
  \Bigg\},
  \\[5mm]
  \left(C^{(8)}_{qd}\right)_{ijkl}= & 
  - \frac{
    (y^d_{\varphi})^*_{rli}
    (y^d_{\varphi})_{rkj}}{
    M_{\varphi_r}^{2}}
  + \frac{
    (y^{dq}_{\Phi})^*_{rli}
    (y^{dq}_{\Phi})_{rkj}}{
    6  M_{\Phi_{r}}^{2}}
  \nonumber \\ &
  - \frac{
    (g^d_{\mathcal{G}})_{rkl}
    (g^q_{\mathcal{G}})_{rij}}{
    M_{\mathcal{G}_r}^{2}}
  - \frac{
    2   (g^{dq}_{\mathcal{Q}_1})^*_{rlj}
    (g^{dq}_{\mathcal{Q}_1})_{rki}}{
    M_{\mathcal{Q}_{1r}}^{2}}
  + \frac{
    (g_{\mathcal{Y}_1})^*_{rlj}
    (g_{\mathcal{Y}_1})_{rki}}{
    M_{\mathcal{Y}_{1r}}^{2}}
  \nonumber \\ &
  - \frac{
    \hat{y}^{d*}_{li}
    (\delta_{\mathcal{L}_1\varphi})_{sr}
    (\gamma_{\mathcal{L}_1})^*_{s}
    (y^d_{\varphi})_{rkj}}{
    M_{\varphi_r}^{2}
    M_{\mathcal{L}_{1s}}^{2}}
  - \frac{
    \hat{y}^d_{kj}
    (\delta_{\mathcal{L}_1\varphi})^*_{sr}
    (\gamma_{\mathcal{L}_1})_{s}
    (y^d_{\varphi})^*_{rli}}{
    M_{\varphi_r}^{2}
    M_{\mathcal{L}_{1s}}^{2}}
  \nonumber \\ &
  - \frac{
    \hat{y}^d_{kj}
    \hat{y}^{d*}_{li}
    (\delta_{\mathcal{L}_1\varphi})^*_{ts}
    (\gamma_{\mathcal{L}_1})_{t}
    (\delta_{\mathcal{L}_1\varphi})_{rs}
    (\gamma_{\mathcal{L}_1})^*_{r}}{
    M_{\mathcal{L}_{1r}}^{2}
    M_{\varphi_s}^{2}
    M_{\mathcal{L}_{1t}}^{2}}
  \nonumber \\ &
  + \frac{1}{f} \Bigg\{
  \frac{
    \hat{y}^{d*}_{li}
    (\tilde{g}^{dDq}_{\mathcal{L}_1})_{rkj}
    (\gamma_{\mathcal{L}_1})^*_{r}}{
    2 M_{\mathcal{L}_{1r}}^{2}}
  + \frac{
    \hat{y}^{d*}_{li}
    (\tilde{g}^{Ddq}_{\mathcal{L}_1})_{rkj}
    (\gamma_{\mathcal{L}_1})^*_{r}}{
    2 M_{\mathcal{L}_{1r}}^{2}}
  \nonumber \\ &
  \qquad
  + \frac{
    \hat{y}^d_{kj}
    (\tilde{g}^{dDq}_{\mathcal{L}_1})^*_{rli}
    (\gamma_{\mathcal{L}_1})_{r}}{
    2 M_{\mathcal{L}_{1r}}^{2}}
  + \frac{
    \hat{y}^d_{kj}
    (\tilde{g}^{Ddq}_{\mathcal{L}_1})^*_{rli}
    (\gamma_{\mathcal{L}_1})_{r}}{
    2 M_{\mathcal{L}_{1r}}^{2}}
  \Big\}.
\end{align}

\newpage

\subsubsection{$\left(\overline{L}R\right)\left(\overline{R}L\right)$ and
  $\left(\overline{L}R\right)\left(\overline{L}R\right)$}
\label{sec:LRRLnLRLR}
\vspace{-0.04cm}

Recall that $\hat{y}^{e,u,d}$ are defined in equation \refeq{eq:y}.
\begin{align}
  \left(C_{ledq}\right)_{ijkl}= & 
  \frac{
    (y^d_{\varphi})_{rkl}
    (y^e_{\varphi})^*_{rji}}{
    M_{\varphi_r}^{2}}
  + \frac{
    2   (g^{lq}_{\mathcal{U}_2})_{ril}
    (g^{ed}_{\mathcal{U}_2})^*_{rjk}}{
    M_{\mathcal{U}_{2r}}^{2}}
  - \frac{
    2   (g^{eq}_{\mathcal{Q}_5})_{rjl}
    (g^{dl}_{\mathcal{Q}_5})^*_{rki}}{
    M_{\mathcal{Q}_{5r}}^{2}}
  \nonumber \\ &
  + \frac{
    \hat{y}^{e*}_{ji}
    (\delta_{\mathcal{L}_1\varphi})_{sr}
    (\gamma_{\mathcal{L}_1})^*_{s}
    (y^d_{\varphi})_{rkl}}{
    M_{\varphi_r}^{2}
    M_{\mathcal{L}_{1s}}^{2}}
  + \frac{
    \hat{y}^d_{kl}
    (\delta_{\mathcal{L}_1\varphi})^*_{sr}
    (\gamma_{\mathcal{L}_1})_{s}
    (y^e_{\varphi})^*_{rji}}{
    M_{\varphi_r}^{2}
    M_{\mathcal{L}_{1s}}^{2}}
  \nonumber \\ &
  + \frac{
    \hat{y}^d_{kl}
    \hat{y}^{e*}_{ji}
    (\delta_{\mathcal{L}_1\varphi})^*_{ts}
    (\gamma_{\mathcal{L}_1})_{t}
    (\delta_{\mathcal{L}_1\varphi})_{rs}
    (\gamma_{\mathcal{L}_1})^*_{r}}{
    M_{\mathcal{L}_{1r}}^{2}
    M_{\varphi_s}^{2}
    M_{\mathcal{L}_{1t}}^{2}}
  \nonumber \\ &
  + \frac{1}{f} \Bigg\{
  - \frac{
    \hat{y}^{e*}_{ji}
    (\tilde{g}^{dDq}_{\mathcal{L}_1})_{rkl}
    (\gamma_{\mathcal{L}_1})^*_{r}}{
    2 M_{\mathcal{L}_{1r}}^{2}}
  - \frac{
    \hat{y}^{e*}_{ji}
    (\tilde{g}^{Ddq}_{\mathcal{L}_1})_{rkl}
    (\gamma_{\mathcal{L}_1})^*_{r}}{
    2 M_{\mathcal{L}_{1r}}^{2}}
  \nonumber \\ &
  \qquad
  - \frac{
    \hat{y}^d_{kl}
    (\tilde{g}^{eDl}_{\mathcal{L}_1})^*_{rji}
    (\gamma_{\mathcal{L}_1})_{r}}{
    2 M_{\mathcal{L}_{1r}}^{2}}
  - \frac{
    \hat{y}^d_{kl}
    (\tilde{g}^{Del}_{\mathcal{L}_1})^*_{rji}
    (\gamma_{\mathcal{L}_1})_{r}}{
    2 M_{\mathcal{L}_{1r}}^{2}}
  \Bigg\},
  \\[4mm]
  \left(C^{(1)}_{quqd}\right)_{ijkl}= & 
  - \frac{
    (y^u_{\varphi})_{rij}
    (y^d_{\varphi})^*_{rlk}}{
    M_{\varphi_r}^{2}}
  + \frac{
    4   (y^{qq}_{\omega_1})_{rki}
    (y^{du}_{\omega_1})^*_{rlj}}{
    3  M_{\omega_{1r}}^{2}}
  + \frac{
    4   (y^{qq}_{\Omega_1})^*_{rki}
    (y^{ud}_{\Omega_1})_{rjl}}{
    3  M_{\Omega_{1r}}^{2}}
  \nonumber \\ &
  - \frac{
    \hat{y}^{d*}_{lk}
    (\delta_{\mathcal{L}_1\varphi})_{sr}
    (\gamma_{\mathcal{L}_1})^*_{s}
    (y^u_{\varphi})_{rij}}{
    M_{\varphi_r}^{2}
    M_{\mathcal{L}_{1s}}^{2}}
  + \frac{
    \hat{y}^{u*}_{ji}
    (\delta_{\mathcal{L}_1\varphi})^*_{sr}
    (\gamma_{\mathcal{L}_1})_{s}
    (y^d_{\varphi})^*_{rlk}}{
    M_{\varphi_r}^{2}
    M_{\mathcal{L}_{1s}}^{2}}
  \nonumber \\ &
  + \frac{
    \hat{y}^{u*}_{ji}
    \hat{y}^{d*}_{lk}
    (\delta_{\mathcal{L}_1\varphi})^*_{ts}
    (\gamma_{\mathcal{L}_1})_{t}
    (\delta_{\mathcal{L}_1\varphi})_{rs}
    (\gamma_{\mathcal{L}_1})^*_{r}}{
    M_{\mathcal{L}_{1r}}^{2}
    M_{\varphi_s}^{2}
    M_{\mathcal{L}_{1t}}^{2}}
  \nonumber \\ &
  + \frac{1}{f} \Bigg\{
  \frac{
    \hat{y}^{d*}_{lk}
    (\tilde{g}^{qDu}_{\mathcal{L}_1})_{rji}
    (\gamma_{\mathcal{L}_1})^*_{r}}{
    2 M_{\mathcal{L}_{1r}}^{2}}
  + \frac{
    \hat{y}^{d*}_{lk}
    (\tilde{g}^{Dqu}_{\mathcal{L}_1})_{rji}
    (\gamma_{\mathcal{L}_1})^*_{r}}{
    2 M_{\mathcal{L}_{1r}}^{2}}
  \nonumber \\ &
  \qquad
  - \frac{
    \hat{y}^{u*}_{ji}
    (\tilde{g}^{dDq}_{\mathcal{L}_1})^*_{rlk}
    (\gamma_{\mathcal{L}_1})_{r}}{
    2 M_{\mathcal{L}_{1r}}^{2}}
  - \frac{
    \hat{y}^{u*}_{ji}
    (\tilde{g}^{Ddq}_{\mathcal{L}_1})^*_{rlk}
    (\gamma_{\mathcal{L}_1})_{r}}{
    2 M_{\mathcal{L}_{1r}}^{2}}
  \Bigg\},
  \\[4mm]
  \left(C^{(8)}_{quqd}\right)_{ijkl}= & 
  - \frac{
    4 (y^{qq}_{\omega_1})_{rki}
    (y^{du}_{\omega_1})^*_{rlj}}{
    M_{\omega_{1r}}^{2}}
  + \frac{
    2 (y^{qq}_{\Omega_1})^*_{rki}
    (y^{ud}_{\Omega_1})_{rjl}}{
    M_{\Omega_{1r}}^{2}}
  - \frac{
    (y^{dq}_{\Phi})^*_{rlk}
    (y^{qu}_{\Phi})_{rij}}{
    M_{\Phi_{r}}^{2}},
  \\[4mm]
  \left(C^{(1)}_{lequ}\right)_{ijkl}= & 
  \frac{
    (y^u_{\varphi})_{rkl}
    (y^e_{\varphi})^*_{rji}}{
    M_{\varphi_r}^{2}}
  + \frac{
    (y^{eu}_{\omega_1})_{rjl}
    (y^{ql}_{\omega_1})^*_{rki}}{
    2  M_{\omega_{1r}}^{2}}
  + \frac{
    (y^{eq}_{\Pi_7})^*_{rjk}
    (y^{lu}_{\Pi_7})_{ril}}{
    2  M_{\Pi_{7r}}^{2}}
  \nonumber \\ &
  + \frac{
    \hat{y}^{e*}_{ji}
    (\delta_{\mathcal{L}_1\varphi})_{sr}
    (\gamma_{\mathcal{L}_1})^*_{s}
    (y^u_{\varphi})_{rkl}}{
    M_{\varphi_r}^{2}
    M_{\mathcal{L}_{1s}}^{2}}
  - \frac{
    \hat{y}^{u*}_{lk}
    (\delta_{\mathcal{L}_1\varphi})^*_{sr}
    (\gamma_{\mathcal{L}_1})_{s}
    (y^e_{\varphi})^*_{rji}}{
    M_{\varphi_r}^{2}
    M_{\mathcal{L}_{1s}}^{2}}
  \nonumber \\ &
  - \frac{
    \hat{y}^{u*}_{lk}
    \hat{y}^{e*}_{ji}
    (\delta_{\mathcal{L}_1\varphi})^*_{ts}
    (\gamma_{\mathcal{L}_1})_{t}
    (\delta_{\mathcal{L}_1\varphi})_{rs}
    (\gamma_{\mathcal{L}_1})^*_{r}}{
    M_{\mathcal{L}_{1r}}^{2}
    M_{\varphi_s}^{2}
    M_{\mathcal{L}_{1t}}^{2}}
  \nonumber \\ &
  + \frac{1}{f} \Bigg\{
  - \frac{
    \hat{y}^{e*}_{ji}
    (\tilde{g}^{qDu}_{\mathcal{L}_1})_{rlk}
    (\gamma_{\mathcal{L}_1})^*_{r}}{
    2 M_{\mathcal{L}_{1r}}^{2}}
  - \frac{
    \hat{y}^{e*}_{ji}
    (\tilde{g}^{Dqu}_{\mathcal{L}_1})_{rlk}
    (\gamma_{\mathcal{L}_1})^*_{r}}{
    2 M_{\mathcal{L}_{1r}}^{2}}
  \nonumber \\ &
  \qquad
  + \frac{
    \hat{y}^{u*}_{lk}
    (\tilde{g}^{eDl}_{\mathcal{L}_1})^*_{rji}
    (\gamma_{\mathcal{L}_1})_{r}}{
    2 M_{\mathcal{L}_{1r}}^{2}}
  + \frac{
    \hat{y}^{u*}_{lk}
    (\tilde{g}^{Del}_{\mathcal{L}_1})^*_{rji}
    (\gamma_{\mathcal{L}_1})_{r}}{
    2 M_{\mathcal{L}_{1r}}^{2}}
  \Bigg\},
\end{align}

\newpage

\begin{align}
  \left(C^{(3)}_{lequ}\right)_{ijkl}= & 
  - \frac{
    (y^{eu}_{\omega_1})_{rjl}
    (y^{ql}_{\omega_1})^*_{rki}}{
    8  M_{\omega_{1r}}^{2}}
  + \frac{
    (y^{eq}_{\Pi_7})^*_{rjk}
    (y^{lu}_{\Pi_7})_{ril}}{
    8  M_{\Pi_{7r}}^{2}}.
\end{align}

\subsubsection{B-violating}
\label{sec:Bviolating}

\begin{align}
  \left(C_{duq}\right)_{ijkl}= & 
  \frac{
    (y^{du}_{\omega_1})^*_{rij}
    (y^{ql}_{\omega_1})_{rkl}}{
    M_{\omega_{1r}}^{2}}
  + \frac{
    2   (g^{dq}_{\mathcal{Q}_1})^*_{rik}
    (g^{ul}_{\mathcal{Q}_1})_{rjl}}{
    M_{\mathcal{Q}_{1r}}^{2}}
  - \frac{
    2   (g^{uq}_{\mathcal{Q}_5})^*_{rjk}
    (g^{dl}_{\mathcal{Q}_5})_{ril}}{
    M_{\mathcal{Q}_{5r}}^{2}},
  \\[5mm]
  \left(C_{qqu}\right)_{ijkl}= & 
  \frac{
    (y^{eu}_{\omega_1})_{rlk}
    (y^{qq}_{\omega_1})^*_{rij}}{
    M_{\omega_{1r}}^{2}}
  - \frac{
    2   (g^{uq}_{\mathcal{Q}_5})^*_{rki}
    (g^{eq}_{\mathcal{Q}_5})_{rlj}}{
    M_{\mathcal{Q}_{5r}}^{2}},
  \\[5mm]
  \left(C_{qqq}\right)_{ijkl}= & 
  \frac{
    2   (y^{qq}_{\omega_1})^*_{rij}
    (y^{ql}_{\omega_1})_{rkl}}{
    M_{\omega_{1r}}^{2}}
  - \frac{
    2   (y^{qq}_{\zeta})^*_{rij}
    (y^{ql}_{\zeta})_{rkl}}{
    M_{\zeta_{r}}^{2}},
    \label{eq:Cqqq}
  \\[5mm]
  \left(C_{duu}\right)_{ijkl}= & 
  \frac{
    (y^{du}_{\omega_1})^*_{rij}
    (y^{eu}_{\omega_1})_{rlk}}{
    M_{\omega_{1r}}^{2}}
  - \frac{
    2   (y^{uu}_{\omega_4})^*_{rjk}
    (y^{ed}_{\omega_4})_{rli}}{
    M_{\omega_{4r}}^{2}}.
\end{align}

\subsection{Bosonic Operators}
\label{sec:d6Bosonic}

\subsubsection{$\phi^6$ and $\phi^4 D^2$}
\label{phi6nphi4D2}

Recall that $\hat{g}^\phi_V$ contains contributions from $\mathcal{L}_1$ (see
equations \refeq{eq:ghat_first}--\refeq{eq:ghat_last}) and that
$\hat{\lambda}_\phi$ is defined in equation \refeq{eq:lambda}.

Due to the length of the contributions to the coefficient of the ${\cal O}_{\phi}$ operator we have separated them as follows:
\begin{equation}
 Z_\phi^{3}~\! C_\phi = C^{\mathrm{S}}_\phi + C^{\mathrm{V}}_\phi + C^{\mathrm{SV}}_\phi,
\end{equation}
where $C^{\mathrm{S}}_\phi$, $C^{\mathrm{V}}_\phi$, and $C^{\mathrm{SV}}_\phi$ are given below.

\begin{align}
  C^{\mathrm{V}}_\phi = & \;
  - \frac{
    2 \hat{\lambda}_{\phi}
    (\hat{g}^\phi_{\mathcal{B}_1})^*_{r}
    (\hat{g}^\phi_{\mathcal{B}_1})_{r}}{
    M_{\mathcal{B}_{1r}}^{2}}
  - \frac{ \hat{\lambda}_{\phi}
    (\hat{g}^\phi_{\mathcal{W}})^*_{r}
    (\hat{g}^\phi_{\mathcal{W}})_{r}}{
    M_{\mathcal{W}_r}^{2}}
  - \frac{
    \hat{\lambda}_{\phi}
    (\hat{g}^\phi_{\mathcal{W}_1})^*_{r}
    (\hat{g}^\phi_{\mathcal{W}_1})_{r}}{
    2 M_{\mathcal{W}_{1r}}^{2}}
  \nonumber \\ &
  + \frac{
    2 g_2
    \hat{\lambda}_{\phi}
    (g^W_{\mathcal{L}_1})_{sr}
    (\gamma_{\mathcal{L}_1})^*_{s}
    (\gamma_{\mathcal{L}_1})_{r}}{
    M_{\mathcal{L}_{1r}}^{2}
    M_{\mathcal{L}_{1s}}^{2}}
  - \frac{
    2 \hat{\lambda}_{\phi}
    (h^{(1)}_{\mathcal{L}_1})_{rs}
    (\gamma_{\mathcal{L}_1})^*_{r}
    (\gamma_{\mathcal{L}_1})_{s}}{
    M_{\mathcal{L}_{1r}}^{2}
    M_{\mathcal{L}_{1s}}^{2}}
  \nonumber \\ &
  + \frac{1}{f}
  \Bigg\{
  - \frac{
    4 \hat{\lambda}_{\phi} \operatorname{Re}{\left(
        (\tilde{\gamma}^{(3)}_{\mathcal{L}_1})_{r}
        (\gamma_{\mathcal{L}_1})^*_{r}
      \right)}}{
    M_{\mathcal{L}_{1r}}^{2}}
  + \frac{
    4 \hat{\lambda}_{\phi} \operatorname{Im}{\left(
        (\tilde{\gamma}^W_{\mathcal{L}_1})_{r}
        (\gamma_{\mathcal{L}_1})^*_{r}
      \right)}
    g_2}{
    M_{\mathcal{L}_{1r}}^{2}}
  \Bigg\},
\end{align}

\begin{align}
  C^{\mathrm{S}}_\phi = & 
  - \frac{
    (\lambda_{\mathcal{S}})_{rs}
    (\kappa_{\mathcal{S}})_{r}
    (\kappa_{\mathcal{S}})_{s}}{
    M_{\mathcal{S}_r}^{2}
    M_{\mathcal{S}_s}^{2}}
  + \frac{
    (\kappa_{\mathcal{S}_3})_{rts}
    (\kappa_{\mathcal{S}})_{r}
    (\kappa_{\mathcal{S}})_{t}
    (\kappa_{\mathcal{S}})_{s}}{
    M_{\mathcal{S}_r}^{2}
    M_{\mathcal{S}_s}^{2}
    M_{\mathcal{S}_t}^{2}}
  + \frac{
    (\lambda_{\varphi})^*_{r}
    (\lambda_{\varphi})_{r}}{
    M_{\varphi_r}^{2}}
  + \frac{
    4 \hat{\lambda}_{\phi}
    (\kappa_{\Xi})_{r}
    (\kappa_{\Xi})_{r}}{
    M_{\Xi_r}^{4}}
  \nonumber \\ &
  - \frac{
    (\lambda_{\Xi})_{s}
    (\kappa_{\Xi})_{s}
    (\kappa_{\Xi})_{r}}{
    M_{\Xi_r}^{2}
    M_{\Xi_s}^{2}}
  + \frac{
    8 \hat{\lambda}_{\phi}
    (\kappa_{\Xi_1})^*_{r}
    (\kappa_{\Xi_1})_{r}}{
    M_{\Xi_{1r}}^{4}}
  - \frac{
    2 (\lambda_{\Xi_1})_{rs}
    (\kappa_{\Xi_1})^*_{r}
    (\kappa_{\Xi_1})_{s}}{
    M_{\Xi_{1r}}^{2}
    M_{\Xi_{1s}}^{2}}
  \nonumber \\ &
  + \frac{
    \sqrt{2}
    (\lambda'_{\Xi_1})_{rs}
    (\kappa_{\Xi_1})^*_{r}
    (\kappa_{\Xi_1})_{s}}{
    M_{\Xi_{1r}}^{2}
    M_{\Xi_{1s}}^{2}}
  + \frac{
    (\lambda_{\Theta_1})^*_{r}
    (\lambda_{\Theta_1})_{r}}{
    6 M_{\Theta_{1r}}^{2}}
  + \frac{
    (\lambda_{\Theta_3})^*_{r}
    (\lambda_{\Theta_3})_{r}}{
    2 M_{\Theta_{3r}}^{2}}
  \nonumber \\ &
  - \frac{
    2 \operatorname{Re}{\left(
        (\kappa_{\mathcal{S}\varphi})_{rs}
        (\lambda_{\varphi})^*_{s}
      \right)}
    (\kappa_{\mathcal{S}})_{r}}{
    M_{\mathcal{S}_r}^{2}
    M_{\varphi_s}^{2}}
  + \frac{
    (\kappa_{\mathcal{S}\varphi})^*_{rt}
    (\kappa_{\mathcal{S}})_{r}
    (\kappa_{\mathcal{S}\varphi})_{st}
    (\kappa_{\mathcal{S}})_{s}}{
    M_{\mathcal{S}_r}^{2}
    M_{\mathcal{S}_s}^{2}
    M_{\varphi_t}^{2}}
  \nonumber \\ &
  - \frac{
    (\lambda_{\mathcal{S}\Xi})_{sr}
    (\kappa_{\mathcal{S}})_{s}
    (\kappa_{\Xi})_{r}}{
    M_{\Xi_r}^{2}
    M_{\mathcal{S}_s}^{2}}
  + \frac{
    (\kappa_{\mathcal{S}\Xi})_{tsr}
    (\kappa_{\mathcal{S}})_{t}
    (\kappa_{\Xi})_{s}
    (\kappa_{\Xi})_{r}}{
    M_{\Xi_r}^{2}
    M_{\Xi_s}^{2}
    M_{\mathcal{S}_t}^{2}}
  \nonumber \\ &
  - \frac{
    2 \operatorname{Re}{\left(
        (\kappa_{\Xi\varphi})_{rs}
        (\lambda_{\varphi})^*_{s}
      \right)}
    (\kappa_{\Xi})_{r}}{
    M_{\Xi_r}^{2}
    M_{\varphi_s}^{2}}
  + \frac{
    (\kappa_{\Xi\varphi})^*_{tr}
    (\kappa_{\Xi})_{t}
    (\kappa_{\Xi\varphi})_{sr}
    (\kappa_{\Xi})_{s}}{
    M_{\varphi_r}^{2}
    M_{\Xi_s}^{2}
    M_{\Xi_t}^{2}}
  \nonumber \\ &
  - \frac{
    4 \operatorname{Re}{\left(
        (\lambda_{\mathcal{S}\Xi_1})_{rs}
        (\kappa_{\Xi_1})^*_{s}
      \right)}
    (\kappa_{\mathcal{S}})_{r}}{
    M_{\Xi_{1s}}^{2}
    M_{\mathcal{S}_r}^{2}}
  + \frac{
    2 (\kappa_{\mathcal{S}\Xi_1})_{trs}
    (\kappa_{\mathcal{S}})_{t}
    (\kappa_{\Xi_1})^*_{r}
    (\kappa_{\Xi_1})_{s}}{
    M_{\Xi_{1r}}^{2}
    M_{\Xi_{1s}}^{2}
    M_{\mathcal{S}_t}^{2}}
  \nonumber \\ &
  - \frac{
    2 \sqrt{2} \operatorname{Re}{\left(
        (\lambda_{\Xi_1\Xi})_{rs}
        (\kappa_{\Xi_1})^*_{r}
      \right)}
    (\kappa_{\Xi})_{s}}{
    M_{\Xi_s}^{2}
    M_{\Xi_{1r}}^{2}}
  - \frac{
    \sqrt{2}
    (\kappa_{\Xi\Xi_1})_{srt}
    (\kappa_{\Xi})_{s}
    (\kappa_{\Xi_1})^*_{r}
    (\kappa_{\Xi_1})_{t}}{
    M_{\Xi_{1r}}^{2}
    M_{\Xi_s}^{2}
    M_{\Xi_{1t}}^{2}}
  \nonumber \\ &
  - \frac{
    4 \operatorname{Re}{\left(
        (\kappa_{\Xi_1\varphi})^*_{rs}
        (\kappa_{\Xi_1})_{r}
        (\lambda_{\varphi})^*_{s}
      \right)}}{
    M_{\Xi_{1r}}^{2}
    M_{\varphi_s}^{2}}
  + \frac{
    4   (\kappa_{\Xi_1\varphi})^*_{st}
    (\kappa_{\Xi_1})_{s}
    (\kappa_{\Xi_1\varphi})_{rt}
    (\kappa_{\Xi_1})^*_{r}}{
    M_{\Xi_{1r}}^{2}
    M_{\Xi_{1s}}^{2}
    M_{\varphi_t}^{2}}
  \nonumber \\ &
  - \frac{
    \operatorname{Re}{\left(
        (\kappa_{\Xi\Theta_1})_{rs}
        (\lambda_{\Theta_1})^*_{s}
      \right)}
    (\kappa_{\Xi})_{r}}{
    3 M_{\Theta_{1s}}^{2}
    M_{\Xi_r}^{2}}
  + \frac{
    (\kappa_{\Xi\Theta_1})^*_{rs}
    (\kappa_{\Xi})_{r}
    (\kappa_{\Xi\Theta_1})_{ts}
    (\kappa_{\Xi})_{t}}{
    6 M_{\Xi_r}^{2}
    M_{\Theta_{1s}}^{2}
    M_{\Xi_t}^{2}}
  \nonumber \\ &
  - \frac{
    \operatorname{Re}{\left(
        (\kappa_{\Xi_1\Theta_1})_{rs}
        (\kappa_{\Xi_1})^*_{r}
        (\lambda_{\Theta_1})^*_{s}
      \right)}}{
    3 M_{\Xi_{1r}}^{2}
    M_{\Theta_{1s}}^{2}}
  + \frac{
    (\kappa_{\Xi_1\Theta_1})^*_{tr}
    (\kappa_{\Xi_1})_{t}
    (\kappa_{\Xi_1\Theta_1})_{sr}
    (\kappa_{\Xi_1})^*_{s}}{
    6 M_{\Theta_{1r}}^{2}
    M_{\Xi_{1s}}^{2}
    M_{\Xi_{1t}}^{2}}
  \nonumber \\ &
  - \frac{
    \operatorname{Re}{\left(
        (\kappa_{\Xi_1\Theta_3})_{rs}
        (\kappa_{\Xi_1})^*_{r}
        (\lambda_{\Theta_3})^*_{s}
      \right)}}{
    M_{\Xi_{1r}}^{2}
    M_{\Theta_{3s}}^{2}}
  + \frac{
    (\kappa_{\Xi_1\Theta_3})^*_{tr}
    (\kappa_{\Xi_1})_{t}
    (\kappa_{\Xi_1\Theta_3})_{sr}
    (\kappa_{\Xi_1})^*_{s}}{
    2 M_{\Theta_{3r}}^{2}
    M_{\Xi_{1s}}^{2}
    M_{\Xi_{1t}}^{2}}
  \nonumber \\ &
  + \frac{
    2 \operatorname{Re}{\left(
        (\kappa_{\Xi\varphi})^*_{rs}
        (\kappa_{\mathcal{S}\varphi})_{ts}
      \right)}
    (\kappa_{\Xi})_{r}
    (\kappa_{\mathcal{S}})_{t}}{
    M_{\Xi_r}^{2}
    M_{\mathcal{S}_t}^{2}
    M_{\varphi_s}^{2}}
  + \frac{
    4 \operatorname{Re}{\left(
        (\kappa_{\Xi_1\varphi})_{rs}
        (\kappa_{\Xi_1})^*_{r}
        (\kappa_{\mathcal{S}\varphi})_{ts}
      \right)}
    (\kappa_{\mathcal{S}})_{t}}{
    M_{\varphi_s}^{2}
    M_{\Xi_{1r}}^{2}
    M_{\mathcal{S}_t}^{2}}
  \nonumber \\ &
  + \frac{
    4 \operatorname{Re}{\left(
        (\kappa_{\Xi_1\varphi})^*_{rs}
        (\kappa_{\Xi_1})_{r}
        (\kappa_{\Xi\varphi})^*_{ts}
      \right)}
    (\kappa_{\Xi})_{t}}{
    M_{\varphi_s}^{2}
    M_{\Xi_{1r}}^{2}
    M_{\Xi_t}^{2}}
  + \frac{
    \operatorname{Re}{\left(
        (\kappa_{\Xi_1\Theta_1})^*_{rs}
        (\kappa_{\Xi_1})_{r}
        (\kappa_{\Xi\Theta_1})_{ts}
      \right)}
    (\kappa_{\Xi})_{t}}{
    3 M_{\Theta_{1s}}^{2}
    M_{\Xi_{1r}}^{2}
    M_{\Xi_t}^{2}}
  \nonumber \\ &
  + \frac{1}{f} \Bigg\{
  \frac{
    2   \hat{\lambda}_{\phi}
    (\tilde{k}^{\phi}_{\mathcal{S}})_{r}
    (\kappa_{\mathcal{S}})_{r}}{
    M_{\mathcal{S}_r}^{2}}
  + \frac{
    (\tilde{\lambda}_{\mathcal{S}})_{r}
    (\kappa_{\mathcal{S}})_{r}}{
    M_{\mathcal{S}_r}^{2}}
  - \frac{
    2 \hat{\lambda}_{\phi}
    (\tilde{k}^{\phi}_{\Xi})_{r}
    (\kappa_{\Xi})_{r}}{
    M_{\Xi_r}^{2}}
  + \frac{
    (\tilde{\lambda}_{\Xi})_{r}
    (\kappa_{\Xi})_{r}}{
    M_{\Xi_r}^{2}}
  \nonumber \\ &
  \qquad
  + \frac{
    4 \operatorname{Re}{\left(
        (\tilde{\lambda}_{\Xi_1})_{r}
        (\kappa_{\Xi_1})^*_{r}
      \right)}}{
    M_{\Xi_{1r}}^{2}}
  \Bigg\},
\end{align}

\newpage

\begin{align}
  \frac{C^{\mathrm{SV}}_\phi}{\hat{\lambda}_{\phi}} = & \;
  - \frac{
    4 \operatorname{Im}{\left(
        (\hat{g}^\phi_{\mathcal{W}})_{r}
      \right)}
    (\delta_{\mathcal{W}\Xi})_{rs}
    (\kappa_{\Xi})_{s}}{
    M_{\mathcal{W}_r}^{2}
    M_{\Xi_s}^{2}}
  - \frac{
    4 (\delta_{\mathcal{W}\Xi})_{ts}
    (\delta_{\mathcal{W}\Xi})_{tr}
    (\kappa_{\Xi})_{r}
    (\kappa_{\Xi})_{s}}{
    M_{\Xi_r}^{2}
    M_{\Xi_s}^{2}
    M_{\mathcal{W}_t}^{2}}
  \nonumber \\ &
  - \frac{
    4 \operatorname{Im}{\left(
        (\hat{g}^\phi_{\mathcal{W}_1})^*_{r}
        (\delta_{\mathcal{W}_1\Xi_1})_{rs}
        (\kappa_{\Xi_1})_{s}
      \right)}}{
    M_{\Xi_{1s}}^{2}
    M_{\mathcal{W}_{1r}}^{2}}
  - \frac{
    8 (\delta_{\mathcal{W}_1\Xi_1})^*_{st}
    (\delta_{\mathcal{W}_1\Xi_1})_{sr}
    (\kappa_{\Xi_1})_{r}
    (\kappa_{\Xi_1})^*_{t}}{
    M_{\Xi_{1r}}^{2}
    M_{\mathcal{W}_{1s}}^{2}
    M_{\Xi_{1t}}^{2}}
  \nonumber \\ &
  + \frac{
    4 \operatorname{Re}{\left(
        (g^{(2)}_{\mathcal{S}\mathcal{L}_1})_{rs}
        (\gamma_{\mathcal{L}_1})_{s}
      \right)}
    (\kappa_{\mathcal{S}})_{r}}{
    M_{\mathcal{S}_r}^{2}
    M_{\mathcal{L}_{1s}}^{2}}
  + \frac{
    2 (\varepsilon_{\mathcal{S}\mathcal{L}_1})_{rts}
    (\kappa_{\mathcal{S}})_{r}
    (\gamma_{\mathcal{L}_1})^*_{t}
    (\gamma_{\mathcal{L}_1})_{s}}{
    M_{\mathcal{S}_r}^{2}
    M_{\mathcal{L}_{1s}}^{2}
    M_{\mathcal{L}_{1t}}^{2}}
  \nonumber \\ &
   + \frac{
    4 \operatorname{Re}{\left(
        (\delta_{\mathcal{L}_1\varphi})_{rs}
        (\gamma_{\mathcal{L}_1})^*_{r}
        (\lambda_{\varphi})_{s}
      \right)}}{
    M_{\varphi_s}^{2}
    M_{\mathcal{L}_{1r}}^{2}}
  + \frac{
    4 \hat{\lambda}_{\phi}
    (\delta_{\mathcal{L}_1\varphi})^*_{ts}
    (\gamma_{\mathcal{L}_1})_{t}
    (\delta_{\mathcal{L}_1\varphi})_{rs}
    (\gamma_{\mathcal{L}_1})^*_{r}}{
    M_{\mathcal{L}_{1r}}^{2}
    M_{\varphi_s}^{2}
    M_{\mathcal{L}_{1t}}^{2}}
  \nonumber \\ &
  - \frac{
    4 \operatorname{Re}{\left(
        (g^{(2)}_{\Xi\mathcal{L}_1})_{rs}
        (\gamma_{\mathcal{L}_1})_{s}
      \right)}
    (\kappa_{\Xi})_{r}}{
    M_{\Xi_r}^{2}
    M_{\mathcal{L}_{1s}}^{2}}
  - \frac{
    2 (\varepsilon_{\Xi\mathcal{L}_1})_{srt}
    (\kappa_{\Xi})_{s}
    (\gamma_{\mathcal{L}_1})^*_{r}
    (\gamma_{\mathcal{L}_1})_{t}}{
    M_{\mathcal{L}_{1r}}^{2}
    M_{\Xi_s}^{2}
    M_{\mathcal{L}_{1t}}^{2}}
  \nonumber \\ &
  + \frac{
    8 \operatorname{Re}{\left(
        (g^{(1)}_{\Xi\mathcal{L}_1})^*_{rs}
        (\gamma_{\mathcal{L}_1})^*_{s}
      \right)}
    (\kappa_{\Xi})_{r}}{
    M_{\Xi_r}^{2}
    M_{\mathcal{L}_{1s}}^{2}}
  + \frac{
    8 \operatorname{Re}{\left(
        (g^{(1)}_{\Xi_1\mathcal{L}_1})^*_{rs}
        (\gamma_{\mathcal{L}_1})^*_{s}
        (\kappa_{\Xi_1})_{r}
      \right)}}{
    M_{\mathcal{L}_{1s}}^{2}
    M_{\Xi_{1r}}^{2}}
  \nonumber \\ &
  - \frac{
    4 \operatorname{Re}{\left(
        (\delta_{\mathcal{L}_1\varphi})_{rs}
        (\gamma_{\mathcal{L}_1})^*_{r}
        (\kappa_{\mathcal{S}\varphi})_{ts}
      \right)}
    (\kappa_{\mathcal{S}})_{t}}{
    M_{\varphi_s}^{2}
    M_{\mathcal{S}_t}^{2}
    M_{\mathcal{L}_{1r}}^{2}}
  - \frac{
    4 \operatorname{Re}{\left(
        (\delta_{\mathcal{L}_1\varphi})_{rs}
        (\gamma_{\mathcal{L}_1})^*_{r}
        (\kappa_{\Xi\varphi})_{ts}
      \right)}
    (\kappa_{\Xi})_{t}}{
    M_{\mathcal{L}_{1r}}^{2}
    M_{\Xi_t}^{2}
    M_{\varphi_s}^{2}}
  \nonumber \\ &
  - \frac{
    8 \operatorname{Re}{\left(
        (\delta_{\mathcal{L}_1\varphi})_{rs}
        (\gamma_{\mathcal{L}_1})^*_{r}
        (\kappa_{\Xi_1\varphi})^*_{ts}
        (\kappa_{\Xi_1})_{t}
      \right)}}{
    M_{\varphi_s}^{2}
    M_{\mathcal{L}_{1r}}^{2}
    M_{\Xi_{1t}}^{2}},
\end{align}

\newpage

\begin{align}
  Z_\phi^{2}~\!C_{\phi D} = & 
  - \frac{
    2 (\kappa_{\Xi})_{r}
    (\kappa_{\Xi})_{r}}{
    M_{\Xi_r}^{4}}
  + \frac{
    4 (\kappa_{\Xi_1})^*_{r}
    (\kappa_{\Xi_1})_{r}}{
    M_{\Xi_{1r}}^{4}}
  - \frac{
    \operatorname{Re}{\left(
        (\hat{g}^\phi_{\mathcal{B}})_{r}
        (\hat{g}^\phi_{\mathcal{B}})_{r}
      \right)}}{
    M_{\mathcal{B}_r}^{2}}
  - \frac{
    (\hat{g}^\phi_{\mathcal{B}})^*_{r}
    (\hat{g}^\phi_{\mathcal{B}})_{r}}{
    M_{\mathcal{B}_r}^{2}}
  \nonumber \\ &
  + \frac{
    (\hat{g}^\phi_{\mathcal{B}_1})^*_{r}
    (\hat{g}^\phi_{\mathcal{B}_1})_{r}}{
    M_{\mathcal{B}_{1r}}^{2}}
  - \frac{
    \operatorname{Re}{\left(
        (\hat{g}^\phi_{\mathcal{W}})_{r}
        (\hat{g}^\phi_{\mathcal{W}})_{r}
      \right)}}{
    4 M_{\mathcal{W}_r}^{2}}
  + \frac{
    (\hat{g}^\phi_{\mathcal{W}})^*_{r}
    (\hat{g}^\phi_{\mathcal{W}})_{r}}{
    4 M_{\mathcal{W}_r}^{2}}
  - \frac{
    (\hat{g}^\phi_{\mathcal{W}_1})^*_{r}
    (\hat{g}^\phi_{\mathcal{W}_1})_{r}}{
    4 M_{\mathcal{W}_{1r}}^{2}}
  \nonumber \\ &
  + \frac{
    g_1
    (g^B_{\mathcal{L}_1})_{rs}
    (\gamma_{\mathcal{L}_1})^*_{r}
    (\gamma_{\mathcal{L}_1})_{s}}{
    M_{\mathcal{L}_{1r}}^{2}
    M_{\mathcal{L}_{1s}}^{2}}
  - \frac{
    (h^{(2)}_{\mathcal{L}_1})_{rs}
    (\gamma_{\mathcal{L}_1})^*_{r}
    (\gamma_{\mathcal{L}_1})_{s}}{
    M_{\mathcal{L}_{1r}}^{2}
    M_{\mathcal{L}_{1s}}^{2}}
  + \frac{
    2 \operatorname{Re}{\left(
        (h^{(3)}_{\mathcal{L}_1})_{rs}
        (\gamma_{\mathcal{L}_1})^*_{r}
        (\gamma_{\mathcal{L}_1})^*_{s}
      \right)}}{
    M_{\mathcal{L}_{1r}}^{2}
    M_{\mathcal{L}_{1s}}^{2}}
  \nonumber \\ &
    + \frac{
    2 \operatorname{Im}{\left(
        (\hat{g}^\phi_{\mathcal{W}})_{r}
      \right)}
    (\delta_{\mathcal{W}\Xi})_{rs}
    (\kappa_{\Xi})_{s}}{
    M_{\mathcal{W}_r}^{2}
    M_{\Xi_s}^{2}}
  + \frac{
    2 (\delta_{\mathcal{W}\Xi})_{ts}
    (\delta_{\mathcal{W}\Xi})_{tr}
    (\kappa_{\Xi})_{r}
    (\kappa_{\Xi})_{s}}{
    M_{\Xi_r}^{2}
    M_{\Xi_s}^{2}
    M_{\mathcal{W}_t}^{2}}
  \nonumber \\ &
  - \frac{
    2 \operatorname{Im}{\left(
        (\hat{g}^\phi_{\mathcal{W}_1})^*_{r}
        (\delta_{\mathcal{W}_1\Xi_1})_{rs}
        (\kappa_{\Xi_1})_{s}
      \right)}}{
    M_{\mathcal{W}_{1r}}^{2}
    M_{\Xi_{1s}}^{2}}
  - \frac{
    4 (\delta_{\mathcal{W}_1\Xi_1})^*_{tr}
    (\delta_{\mathcal{W}_1\Xi_1})_{ts}
    (\kappa_{\Xi_1})_{s}
    (\kappa_{\Xi_1})^*_{r}}{
    M_{\Xi_{1r}}^{2}
    M_{\Xi_{1s}}^{2}
    M_{\mathcal{W}_{1t}}^{2}}
  \nonumber \\ &
  - \frac{
    4 \operatorname{Re}{\left(
        (g^{(1)}_{\Xi\mathcal{L}_1})^*_{rs}
        (\gamma_{\mathcal{L}_1})^*_{s}
      \right)}
    (\kappa_{\Xi})_{r}}{
    M_{\Xi_r}^{2}
    M_{\mathcal{L}_{1s}}^{2}}
  + \frac{
    4 \operatorname{Re}{\left(
        (g^{(2)}_{\Xi\mathcal{L}_1})_{rs}
        (\gamma_{\mathcal{L}_1})_{s}
      \right)}
    (\kappa_{\Xi})_{r}}{
    M_{\Xi_r}^{2}
    M_{\mathcal{L}_{1s}}^{2}}
  \nonumber \\ &
  + \frac{
    2 (\varepsilon_{\Xi\mathcal{L}_1})_{srt}
    (\kappa_{\Xi})_{s}
    (\gamma_{\mathcal{L}_1})^*_{r}
    (\gamma_{\mathcal{L}_1})_{t}}{
    M_{\mathcal{L}_{1r}}^{2}
    M_{\Xi_s}^{2}
    M_{\mathcal{L}_{1t}}^{2}}
   + \frac{
    4 \operatorname{Re}{\left(
        (g^{(1)}_{\Xi_1\mathcal{L}_1})^*_{rs}
        (\gamma_{\mathcal{L}_1})^*_{s}
        (\kappa_{\Xi_1})_{r}
      \right)}}{
    M_{\mathcal{L}_{1s}}^{2}
    M_{\Xi_{1r}}^{2}}
  \nonumber \\ &
  - \frac{
    4 \operatorname{Re}{\left(
        (g^{(2)}_{\Xi_1\mathcal{L}_1})_{rs}
        (\kappa_{\Xi_1})^*_{r}
        (\gamma_{\mathcal{L}_1})_{s}
      \right)}}{
    M_{\mathcal{L}_{1s}}^{2}
    M_{\Xi_{1r}}^{2}}
  - \frac{
    4 \operatorname{Re}{\left(
        (\varepsilon_{\Xi_1\mathcal{L}_1})^*_{rst}
        (\kappa_{\Xi_1})^*_{r}
        (\gamma_{\mathcal{L}_1})_{t}
        (\gamma_{\mathcal{L}_1})_{s}
      \right)}}{
    M_{\mathcal{L}_{1s}}^{2}
    M_{\mathcal{L}_{1t}}^{2}
    M_{\Xi_{1r}}^{2}}
  \nonumber \\ &
  + \frac{1}{f}
  \Bigg\{
  \frac{
    2 (\tilde{k}^{\phi}_{\Xi})_{r}
    (\kappa_{\Xi})_{r}}{
    M_{\Xi_r}^{2}}
  - \frac{
    4 \operatorname{Re}{\left(
        (\tilde{k}_{\Xi_1})_{r}
        (\kappa_{\Xi_1})^*_{r}
      \right)}}{
    M_{\Xi_{1r}}^{2}}
  - \frac{
    2 \operatorname{Re}{\left(
        (\tilde{\gamma}^{(1)}_{\mathcal{L}_1})_{r}
        (\gamma_{\mathcal{L}_1})^*_{r}
      \right)}}{
    M_{\mathcal{L}_{1r}}^{2}}
  \nonumber \\ &
  \qquad
  + \frac{
    2 \operatorname{Re}{\left(
        (\tilde{\gamma}^{(2)}_{\mathcal{L}_1})_{r}
        (\gamma_{\mathcal{L}_1})^*_{r}
      \right)}}{
    M_{\mathcal{L}_{1r}}^{2}}
  + \frac{
    2 \operatorname{Im}{\left(
        (\tilde{\gamma}^B_{\mathcal{L}_1})_{r}
        (\gamma_{\mathcal{L}_1})^*_{r}
      \right)}
    g_1}{
    M_{\mathcal{L}_{1r}}^{2}}
  \Bigg\},
\end{align}

\newpage

\begin{align}
  Z_\phi^{2}~\!C_{\phi\square}= & 
  - \frac{
    (\kappa_{\mathcal{S}})_{r}
    (\kappa_{\mathcal{S}})_{r}}{
    2 M_{\mathcal{S}_r}^{4}}
  + \frac{
    (\kappa_{\Xi})_{r}
    (\kappa_{\Xi})_{r}}{
    2 M_{\Xi_r}^{4}}
  + \frac{
    2 (\kappa_{\Xi_1})^*_{r}
    (\kappa_{\Xi_1})_{r}}{
    M_{\Xi_{1r}}^{4}}
  \nonumber \\ &
  - \frac{
    \operatorname{Re}{\left(
        (\hat{g}^\phi_{\mathcal{B}})_{r}
        (\hat{g}^\phi_{\mathcal{B}})_{r}
      \right)}}{
    2 M_{\mathcal{B}_r}^{2}}
  - \frac{
    \operatorname{Re}{\left(
        (\hat{g}^\phi_{\mathcal{W}})_{r}
        (\hat{g}^\phi_{\mathcal{W}})_{r}
      \right)}}{
    8 M_{\mathcal{W}_r}^{2}}
  - \frac{
    (\hat{g}^\phi_{\mathcal{B}_1})^*_{r}
    (\hat{g}^\phi_{\mathcal{B}_1})_{r}}{
    2 M_{\mathcal{B}_{1r}}^{2}}
  - \frac{
    (\hat{g}^\phi_{\mathcal{W}})^*_{r}
    (\hat{g}^\phi_{\mathcal{W}})_{r}}{
    4 M_{\mathcal{W}_r}^{2}}
  \nonumber \\ &
  - \frac{
    (\hat{g}^\phi_{\mathcal{W}_1})^*_{r}
    (\hat{g}^\phi_{\mathcal{W}_1})_{r}}{
    8 M_{\mathcal{W}_{1r}}^{2}}
  + \frac{
    g_1
    (g^B_{\mathcal{L}_1})_{rs}
    (\gamma_{\mathcal{L}_1})^*_{r}
    (\gamma_{\mathcal{L}_1})_{s}}{
    4 M_{\mathcal{L}_{1r}}^{2}
    M_{\mathcal{L}_{1s}}^{2}}
  + \frac{
    3 g_2
    (g^W_{\mathcal{L}_1})_{sr}
    (\gamma_{\mathcal{L}_1})^*_{s}
    (\gamma_{\mathcal{L}_1})_{r}}{
    4 M_{\mathcal{L}_{1r}}^{2}
    M_{\mathcal{L}_{1s}}^{2}}
  \nonumber \\ &
  - \frac{
    (h^{(1)}_{\mathcal{L}_1})_{rs}
    (\gamma_{\mathcal{L}_1})^*_{r}
    (\gamma_{\mathcal{L}_1})_{s}}{
    2 M_{\mathcal{L}_{1r}}^{2}
    M_{\mathcal{L}_{1s}}^{2}}
  + \frac{
    \operatorname{Re}{\left(
        (h^{(3)}_{\mathcal{L}_1})_{rs}
        (\gamma_{\mathcal{L}_1})^*_{r}
        (\gamma_{\mathcal{L}_1})^*_{s}
      \right)}}{
    M_{\mathcal{L}_{1r}}^{2}
    M_{\mathcal{L}_{1s}}^{2}}
    \nonumber \\ &
  + \frac{
    \operatorname{Im}{\left(
        (\hat{g}^\phi_{\mathcal{B}})_{r}
      \right)}
    (\delta_{\mathcal{B}\mathcal{S}})_{rs}
    (\kappa_{\mathcal{S}})_{s}}{
    M_{\mathcal{B}_r}^{2}
    M_{\mathcal{S}_s}^{2}}
  + \frac{
    (\delta_{\mathcal{B}\mathcal{S}})_{rt}
    (\delta_{\mathcal{B}\mathcal{S}})_{rs}
    (\kappa_{\mathcal{S}})_{s}
    (\kappa_{\mathcal{S}})_{t}}{
    2 M_{\mathcal{B}_r}^{2}
    M_{\mathcal{S}_s}^{2}
    M_{\mathcal{S}_t}^{2}}
  \nonumber \\ &
  - \frac{
    \operatorname{Im}{\left(
        (\hat{g}^\phi_{\mathcal{W}})_{r}
      \right)}
    (\delta_{\mathcal{W}\Xi})_{rs}
    (\kappa_{\Xi})_{s}}{
    2 M_{\mathcal{W}_r}^{2}
    M_{\Xi_s}^{2}}
  - \frac{
    (\delta_{\mathcal{W}\Xi})_{ts}
    (\delta_{\mathcal{W}\Xi})_{tr}
    (\kappa_{\Xi})_{r}
    (\kappa_{\Xi})_{s}}{
    2 M_{\Xi_r}^{2}
    M_{\Xi_s}^{2}
    M_{\mathcal{W}_t}^{2}}
  \nonumber \\ &
  - \frac{
    \operatorname{Im}{\left(
        (\hat{g}^\phi_{\mathcal{W}_1})^*_{r}
        (\delta_{\mathcal{W}_1\Xi_1})_{rs}
        (\kappa_{\Xi_1})_{s}
      \right)}}{
    M_{\Xi_{1s}}^{2}
    M_{\mathcal{W}_{1r}}^{2}}
  - \frac{
    2 (\delta_{\mathcal{W}_1\Xi_1})^*_{st}
    (\delta_{\mathcal{W}_1\Xi_1})_{sr}
    (\kappa_{\Xi_1})_{r}
    (\kappa_{\Xi_1})^*_{t}}{
    M_{\Xi_{1r}}^{2}
    M_{\mathcal{W}_{1s}}^{2}
    M_{\Xi_{1t}}^{2}}
  \nonumber \\ &
  - \frac{
    \operatorname{Re}{\left(
        (g^{(1)}_{\mathcal{S}\mathcal{L}_1})^*_{rs}
        (\gamma_{\mathcal{L}_1})^*_{s}
      \right)}
    (\kappa_{\mathcal{S}})_{r}}{
    M_{\mathcal{S}_r}^{2}
    M_{\mathcal{L}_{1s}}^{2}}
  + \frac{
    \operatorname{Re}{\left(
        (g^{(2)}_{\mathcal{S}\mathcal{L}_1})_{rs}
        (\gamma_{\mathcal{L}_1})_{s}
      \right)}
    (\kappa_{\mathcal{S}})_{r}}{
    M_{\mathcal{S}_r}^{2}
    M_{\mathcal{L}_{1s}}^{2}}
  \nonumber \\ &
  + \frac{
    (\varepsilon_{\mathcal{S}\mathcal{L}_1})_{rts}
    (\kappa_{\mathcal{S}})_{r}
    (\gamma_{\mathcal{L}_1})^*_{t}
    (\gamma_{\mathcal{L}_1})_{s}}{
    2 M_{\mathcal{S}_r}^{2}
    M_{\mathcal{L}_{1s}}^{2}
    M_{\mathcal{L}_{1t}}^{2}}
  + \frac{
    \operatorname{Re}{\left(
        (g^{(1)}_{\Xi\mathcal{L}_1})^*_{rs}
        (\gamma_{\mathcal{L}_1})^*_{s}
      \right)}
    (\kappa_{\Xi})_{r}}{
    M_{\Xi_r}^{2}
    M_{\mathcal{L}_{1s}}^{2}}
  \nonumber \\ &
  - \frac{
    \operatorname{Re}{\left(
        (g^{(2)}_{\Xi\mathcal{L}_1})_{rs}
        (\gamma_{\mathcal{L}_1})_{s}
      \right)}
    (\kappa_{\Xi})_{r}}{
    M_{\Xi_r}^{2}
    M_{\mathcal{L}_{1s}}^{2}}
  - \frac{
    (\varepsilon_{\Xi\mathcal{L}_1})_{srt}
    (\kappa_{\Xi})_{s}
    (\gamma_{\mathcal{L}_1})^*_{r}
    (\gamma_{\mathcal{L}_1})_{t}}{
    2 M_{\mathcal{L}_{1r}}^{2}
    M_{\Xi_s}^{2}
    M_{\mathcal{L}_{1t}}^{2}}
  \nonumber \\ &
  + \frac{
    2 \operatorname{Re}{\left(
        (g^{(1)}_{\Xi_1\mathcal{L}_1})^*_{rs}
        (\gamma_{\mathcal{L}_1})^*_{s}
        (\kappa_{\Xi_1})_{r}
      \right)}}{
    M_{\mathcal{L}_{1s}}^{2}
    M_{\Xi_{1r}}^{2}}
  - \frac{
    2 \operatorname{Re}{\left(
        (g^{(2)}_{\Xi_1\mathcal{L}_1})_{rs}
        (\kappa_{\Xi_1})^*_{r}
        (\gamma_{\mathcal{L}_1})_{s}
      \right)}}{
    M_{\mathcal{L}_{1s}}^{2}
    M_{\Xi_{1r}}^{2}}
  \nonumber \\ &
  - \frac{
    2 \operatorname{Re}{\left(
        (\varepsilon_{\Xi_1\mathcal{L}_1})^*_{rst}
        (\kappa_{\Xi_1})^*_{r}
        (\gamma_{\mathcal{L}_1})_{t}
        (\gamma_{\mathcal{L}_1})_{s}
      \right)}}{
    M_{\mathcal{L}_{1s}}^{2}
    M_{\mathcal{L}_{1t}}^{2}
    M_{\Xi_{1r}}^{2}}
  \nonumber \\ &
  + \frac{1}{f}
  \Bigg\{
  \frac{
    (\tilde{k}^{\phi}_{\mathcal{S}})_{r}
    (\kappa_{\mathcal{S}})_{r}}{
    2 M_{\mathcal{S}_r}^{2}}
  - \frac{
    (\tilde{k}^{\phi}_{\Xi})_{r}
    (\kappa_{\Xi})_{r}}{
    2 M_{\Xi_r}^{2}}
  - \frac{
    2 \operatorname{Re}{\left(
        (\tilde{k}_{\Xi_1})_{r}
        (\kappa_{\Xi_1})^*_{r}
      \right)}}{
    M_{\Xi_{1r}}^{2}}
  + \frac{
    \operatorname{Re}{\left(
        (\tilde{\gamma}^{(2)}_{\mathcal{L}_1})_{r}
        (\gamma_{\mathcal{L}_1})^*_{r}
      \right)}}{
    M_{\mathcal{L}_{1r}}^{2}}
  \nonumber \\ &
  \qquad
  - \frac{
    \operatorname{Re}{\left(
        (\tilde{\gamma}^{(3)}_{\mathcal{L}_1})_{r}
        (\gamma_{\mathcal{L}_1})^*_{r}
      \right)}}{
    M_{\mathcal{L}_{1r}}^{2}}
  + \frac{
    \operatorname{Im}{\left(
        (\tilde{\gamma}^B_{\mathcal{L}_1})_{r}
        (\gamma_{\mathcal{L}_1})^*_{r}
      \right)}
    g_1}{
    2 M_{\mathcal{L}_{1r}}^{2}}
  + \frac{
    3 \operatorname{Im}{\left(
        (\tilde{\gamma}^W_{\mathcal{L}_1})_{r}
        (\gamma_{\mathcal{L}_1})^*_{r}
      \right)}
    g_2}{
    2 M_{\mathcal{L}_{1r}}^{2}}
  \Bigg\}.
\end{align}

\newpage

\subsubsection{$X^2 \phi^2$}
\label{sec:F2phi2}

\begin{align}
  Z_\phi~\!C_{\phi B} = & \;
  - \frac{
    (g_1)^2
    (\gamma_{\mathcal{L}_1})^*_{r}
    (\gamma_{\mathcal{L}_1})_{r}}{
    8  M_{\mathcal{L}_{1r}}^{4}}
  - \frac{
    g_1
    (g^B_{\mathcal{L}_1})_{rs}
    (\gamma_{\mathcal{L}_1})^*_{r}
    (\gamma_{\mathcal{L}_1})_{s}}{
    4  M_{\mathcal{L}_{1r}}^{2}
    M_{\mathcal{L}_{1s}}^{2}}
  \nonumber \\ &
  + \frac{1}{f} \left\{
    \frac{
      (\tilde{k}^B_{\mathcal{S}})_{r}
      (\kappa_{\mathcal{S}})_{r}}{
      M_{\mathcal{S}_r}^{2}}
    - \frac{
      \operatorname{Im}{\left(
          (\tilde{\gamma}^B_{\mathcal{L}_1})_{r}
          (\gamma_{\mathcal{L}_1})^*_{r}
        \right)}
      g_1}{
      2 M_{\mathcal{L}_{1r}}^{2}}
  \right\},
  \\[5mm]
  Z_\phi~\!C_{\phi\tilde{B}} = & \;
  - \frac{
    g_1
    (g^{\tilde{B}}_{\mathcal{L}_1})_{rs}
    (\gamma_{\mathcal{L}_1})^*_{r}
    (\gamma_{\mathcal{L}_1})_{s}}{
    4  M_{\mathcal{L}_{1r}}^{2}
    M_{\mathcal{L}_{1s}}^{2}}
  + \frac{1}{f} \left\{
    \frac{
      (\tilde{k}^{\tilde{B}}_{\mathcal{S}})_{r}
      (\kappa_{\mathcal{S}})_{r}}{
      M_{\mathcal{S}_r}^{2}}
    - \frac{
      \operatorname{Im}{\left(
          (\tilde{\gamma}^{\tilde{B}}_{\mathcal{L}_1})_{r}
          (\gamma_{\mathcal{L}_1})^*_{r}
        \right)}
      g_1}{
      2 M_{\mathcal{L}_{1r}}^{2}}
  \right\},
  \\[5mm]
  Z_\phi~\!C_{\phi W} = & \;
  - \frac{
    (g_2)^2
    (\gamma_{\mathcal{L}_1})^*_{r}
    (\gamma_{\mathcal{L}_1})_{r}}{
    8  M_{\mathcal{L}_{1r}}^{4}}
  - \frac{
    g_2
    (g^W_{\mathcal{L}_1})_{rs}
    (\gamma_{\mathcal{L}_1})^*_{r}
    (\gamma_{\mathcal{L}_1})_{s}}{
    4  M_{\mathcal{L}_{1r}}^{2}
    M_{\mathcal{L}_{1s}}^{2}}
  \nonumber \\ &
  + \frac{1}{f} \left\{
    \frac{
      (\tilde{k}^W_{\mathcal{S}})_{r}
      (\kappa_{\mathcal{S}})_{r}}{
      M_{\mathcal{S}_r}^{2}}
    - \frac{
      \operatorname{Im}{\left(
          (\tilde{\gamma}^W_{\mathcal{L}_1})_{r}
          (\gamma_{\mathcal{L}_1})^*_{r}
        \right)}
      g_2}{
      2 M_{\mathcal{L}_{1r}}^{2}}
  \right\},
  \\[5mm]
  Z_\phi~\!C_{\phi\tilde{W}} = & \;
  - \frac{
    g_2
    (g^{\tilde{W}}_{\mathcal{L}_1})_{rs}
    (\gamma_{\mathcal{L}_1})^*_{r}
    (\gamma_{\mathcal{L}_1})_{s}}{
    4  M_{\mathcal{L}_{1r}}^{2}
    M_{\mathcal{L}_{1s}}^{2}}
  + \frac{1}{f} \left\{
    \frac{
      (\tilde{k}^{\tilde{W}}_{\mathcal{S}})_{r}
      (\kappa_{\mathcal{S}})_{r}}{
      M_{\mathcal{S}_r}^{2}}
    - \frac{
      \operatorname{Im}{\left(
          (\tilde{\gamma}^{\tilde{W}}_{\mathcal{L}_1})_{r}
          (\gamma_{\mathcal{L}_1})^*_{r}
        \right)}
      g_2}{
      2 M_{\mathcal{L}_{1r}}^{2}}
  \right\},
  \\[5mm]
  Z_\phi~\!C_{\phi WB} = & \;
  - \frac{
    g_1
    g_2
    (\gamma_{\mathcal{L}_1})^*_{r}
    (\gamma_{\mathcal{L}_1})_{r}}{
    4  M_{\mathcal{L}_{1r}}^{4}}
  - \frac{
    g_2
    (g^B_{\mathcal{L}_1})_{rs}
    (\gamma_{\mathcal{L}_1})^*_{r}
    (\gamma_{\mathcal{L}_1})_{s}}{
    4  M_{\mathcal{L}_{1r}}^{2}
    M_{\mathcal{L}_{1s}}^{2}}
  - \frac{
    g_1
    (g^W_{\mathcal{L}_1})_{rs}
    (\gamma_{\mathcal{L}_1})^*_{r}
    (\gamma_{\mathcal{L}_1})_{s}}{
    4  M_{\mathcal{L}_{1r}}^{2}
    M_{\mathcal{L}_{1s}}^{2}}
  \nonumber \\ &
  + \frac{1}{f} \left\{
    \frac{
      (\tilde{k}^{WB}_{\Xi})_{r}
      (\kappa_{\Xi})_{r}}{
      M_{\Xi_r}^{2}}
    - \frac{
      \operatorname{Im}{\left(
          (\tilde{\gamma}^B_{\mathcal{L}_1})_{r}
          (\gamma_{\mathcal{L}_1})^*_{r}
        \right)}
      g_2}{
      2 M_{\mathcal{L}_{1r}}^{2}}
    - \frac{
      \operatorname{Im}{\left(
          (\tilde{\gamma}^W_{\mathcal{L}_1})_{r}
          (\gamma_{\mathcal{L}_1})^*_{r}
        \right)}
      g_1}{
      2 M_{\mathcal{L}_{1r}}^{2}}
  \right\},
  \\[5mm]
  Z_\phi~\!C_{\phi W\tilde{B}} = & \;
  - \frac{
    g_2
    (g^{\tilde{B}}_{\mathcal{L}_1})_{rs}
    (\gamma_{\mathcal{L}_1})^*_{r}
    (\gamma_{\mathcal{L}_1})_{s}}{
    4  M_{\mathcal{L}_{1r}}^{2}
    M_{\mathcal{L}_{1s}}^{2}}
  - \frac{
    g_1
    (g^{\tilde{W}}_{\mathcal{L}_1})_{rs}
    (\gamma_{\mathcal{L}_1})^*_{r}
    (\gamma_{\mathcal{L}_1})_{s}}{
    4  M_{\mathcal{L}_{1r}}^{2}
    M_{\mathcal{L}_{1s}}^{2}}
  \nonumber \\ &
  + \frac{1}{f} \left\{
    \frac{
      (\tilde{k}^{W\tilde{B}}_{\Xi})_{r}
      (\kappa_{\Xi})_{r}}{
      M_{\Xi_r}^{2}}
    - \frac{
      \operatorname{Im}{\left(
          (\tilde{\gamma}^{\tilde{B}}_{\mathcal{L}_1})_{r}
          (\gamma_{\mathcal{L}_1})^*_{r}
        \right)}
      g_2}{
      2 M_{\mathcal{L}_{1r}}^{2}}
    - \frac{
      \operatorname{Im}{\left(
          (\tilde{\gamma}^{\tilde{W}}_{\mathcal{L}_1})_{r}
          (\gamma_{\mathcal{L}_1})^*_{r}
        \right)}
      g_1}{
      2 M_{\mathcal{L}_{1r}}^{2}}
  \right\},
  \\[5mm]
  Z_\phi~\!C_{\phi G} = & \;
  \frac{1}{f}
  \frac{
    (\tilde{k}^G_{\mathcal{S}})_{r}
    (\kappa_{\mathcal{S}})_{r}}{
    M_{\mathcal{S}_r}^{2}},
  \\[5mm]
  Z_\phi~\!C_{\phi\tilde{G}} = & \;
  \frac{1}{f}
  \frac{
    (\tilde{k}^{\tilde{G}}_{\mathcal{S}})_{r}
    (\kappa_{\mathcal{S}})_{r}}{
    M_{\mathcal{S}_r}^{2}}.
\end{align}

\newpage

\subsection{Operators with Bosons and Fermions}
\label{sec:d6BosonFermion}

There are three types of operators coupling bosonic and fermionic fields: the operators of the form $\psi^2 \phi^3$ represent couplings between scalars and fermions only, while those of the form $X \psi^2 \phi$ and $\psi^2 D \phi^2$ contain covariant interactions between the SM scalar, fermions and gauge fields.

\subsubsection{$\psi^2 \phi^3$}
\label{sec:psi2phi3}

Due to the length of the contributions to the coefficients of the different $\psi^2 \phi^3$ operators 
(${\cal O}_{e\phi}$, ${\cal O}_{d\phi}$ and ${\cal O}_{u\phi}$), we have separated them as follows:
\begin{align}
  Z_\phi^{\frac{3}{2}}~\!\left(C_{e\phi}\right)_{ij} = & \;
  \hat{y}^{e*}_{ji} a + b^e_{ij} + c^e_{ij}, \\
  Z_\phi^{\frac{3}{2}}~\!\left(C_{d\phi}\right)_{ij} = & \;
  \hat{y}^{d*}_{ji} a + b^d_{ij} + c^d_{ij}, \\
  Z_\phi^{\frac{3}{2}}~\!\left(C_{u\phi}\right)_{ij} = & \;
  \hat{y}^{u*}_{ji} a^* + b^u_{ij} + c^u_{ij},
\end{align}
where the coefficients $a$, $b^\psi_{ij}$ and $c^\psi_{ij}$ are defined below (equations
\refeq{eq:a}--\refeq{eq:cu}). (The coefficients $b^\psi_{ij}$ and $c^\psi_{ij}$ refer to the contributions from only one type of particle and mixed contributions, respectively.)\\

Recall also that $\hat{g}^\phi_V$ contains contributions from $\mathcal{L}_1$ (see
equations \refeq{eq:ghat_first}--\refeq{eq:ghat_last}) and that
$\hat{y}^{e,d,u}_{ji}$ and $\hat{\lambda}_\phi$ are defined in equations
\refeq{eq:y} and \refeq{eq:lambda}.

\newpage

\begin{align}
  a = & \;
  \frac{
    (\kappa_{\Xi})_{r}
    (\kappa_{\Xi})_{r}}{
    M_{\Xi_r}^{4}}
  + \frac{
    2 (\kappa_{\Xi_1})^*_{r}
    (\kappa_{\Xi_1})_{r}}{
    M_{\Xi_{1r}}^{4}}
  - \frac{
    i  \operatorname{Im}{\left(
        (\hat{g}^\phi_{\mathcal{B}})_{r}
        (\hat{g}^\phi_{\mathcal{B}})_{r}
      \right)}}{
    2  M_{\mathcal{B}_r}^{2}}
  - \frac{
    (\hat{g}^\phi_{\mathcal{B}_1})^*_{r}
    (\hat{g}^\phi_{\mathcal{B}_1})_{r}}{
    2  M_{\mathcal{B}_{1r}}^{2}}
  \nonumber \\ &
  - \frac{
    i  \operatorname{Im}{\left(
        (\hat{g}^\phi_{\mathcal{W}})_{r}
        (\hat{g}^\phi_{\mathcal{W}})_{r}
      \right)}}{
    8  M_{\mathcal{W}_r}^{2}}
  - \frac{
    (\hat{g}^\phi_{\mathcal{W}})^*_{r}
    (\hat{g}^\phi_{\mathcal{W}})_{r}}{
    4  M_{\mathcal{W}_r}^{2}}
  - \frac{
    (\hat{g}^\phi_{\mathcal{W}_1})^*_{r}
    (\hat{g}^\phi_{\mathcal{W}_1})_{r}}{
    8  M_{\mathcal{W}_{1r}}^{2}}
  \nonumber \\ &
  + \frac{
    g_2
    (g^W_{\mathcal{L}_1})_{sr}
    (\gamma_{\mathcal{L}_1})^*_{s}
    (\gamma_{\mathcal{L}_1})_{r}}{
    2  M_{\mathcal{L}_{1r}}^{2}
    M_{\mathcal{L}_{1s}}^{2}}
  - \frac{
    (h^{(1)}_{\mathcal{L}_1})_{rs}
    (\gamma_{\mathcal{L}_1})^*_{r}
    (\gamma_{\mathcal{L}_1})_{s}}{
    2  M_{\mathcal{L}_{1r}}^{2}
    M_{\mathcal{L}_{1s}}^{2}}
  - \frac{
    i  \operatorname{Im}{\left(
        (h^{(3)}_{\mathcal{L}_1})_{rs}
        (\gamma_{\mathcal{L}_1})^*_{r}
        (\gamma_{\mathcal{L}_1})^*_{s}
      \right)}}{
    M_{\mathcal{L}_{1r}}^{2}
    M_{\mathcal{L}_{1s}}^{2}}
  \nonumber \\ &
    - \frac{
    i  \operatorname{Re}{\left(
        (\hat{g}^\phi_{\mathcal{B}})_{r}
      \right)}
    (\delta_{\mathcal{B}\mathcal{S}})_{rs}
    (\kappa_{\mathcal{S}})_{s}}{
    M_{\mathcal{B}_r}^{2}
    M_{\mathcal{S}_s}^{2}}
  \nonumber \\ &
  + \frac{
    i \left(
      (\hat{g}^\phi_{\mathcal{W}})_{r}
      - 3 (\hat{g}^\phi_{\mathcal{W}})^*_{s}
    \right)
    (\delta_{\mathcal{W}\Xi})_{rs}
    (\kappa_{\Xi})_{s}}{
    4  M_{\mathcal{W}_r}^{2}
    M_{\Xi_s}^{2}}
  - \frac{
    (\delta_{\mathcal{W}\Xi})_{ts}
    (\delta_{\mathcal{W}\Xi})_{tr}
    (\kappa_{\Xi})_{r}
    (\kappa_{\Xi})_{s}}{
    M_{\Xi_r}^{2}
    M_{\Xi_s}^{2}
    M_{\mathcal{W}_t}^{2}}
  \nonumber \\ &
  - \frac{
    \operatorname{Im}{\left(
        (\hat{g}^\phi_{\mathcal{W}_1})^*_{r}
        (\delta_{\mathcal{W}_1\Xi_1})_{rs}
        (\kappa_{\Xi_1})_{s}
      \right)}}{
    M_{\Xi_{1s}}^{2}
    M_{\mathcal{W}_{1r}}^{2}}
  - \frac{
    2 (\delta_{\mathcal{W}_1\Xi_1})^*_{st}
    (\delta_{\mathcal{W}_1\Xi_1})_{sr}
    (\kappa_{\Xi_1})_{r}
    (\kappa_{\Xi_1})^*_{t}}{
    M_{\Xi_{1r}}^{2}
    M_{\mathcal{W}_{1s}}^{2}
    M_{\Xi_{1t}}^{2}}
  \nonumber \\ &
  + \frac{
    i  \operatorname{Im}{\left(
        (g^{(1)}_{\mathcal{S} \mathcal{L}_1})^*_{rs}
        (\gamma_{\mathcal{L}_1})^*_{s}
      \right)}
    (\kappa_{\mathcal{S}})_{r}}{
    M_{\mathcal{S}_r}^{2}
    M_{\mathcal{L}_{1s}}^{2}}
  + \frac{
    \operatorname{Re}{\left(
        (g^{(2)}_{\mathcal{S}\mathcal{L}_1})_{rs}
        (\gamma_{\mathcal{L}_1})_{s}
      \right)}
    (\kappa_{\mathcal{S}})_{r}}{
    M_{\mathcal{S}_r}^{2}
    M_{\mathcal{L}_{1s}}^{2}}
  \nonumber \\ &
  + \frac{
    (\varepsilon_{\mathcal{S}\mathcal{L}_1})_{rts}
    (\kappa_{\mathcal{S}})_{r}
    (\gamma_{\mathcal{L}_1})^*_{t}
    (\gamma_{\mathcal{L}_1})_{s}}{
    2  M_{\mathcal{S}_r}^{2}
    M_{\mathcal{L}_{1s}}^{2}
    M_{\mathcal{L}_{1t}}^{2}}
  + \frac{
    (\delta_{\mathcal{L}_1\varphi})_{sr}
    (\gamma_{\mathcal{L}_1})^*_{s}
    (\lambda_{\varphi})_{r}}{
    M_{\varphi_r}^{2}
    M_{\mathcal{L}_{1s}}^{2}}
  \nonumber \\ &
  + \frac{
    2 \hat{\lambda}_{\phi}
    (\delta_{\mathcal{L}_1\varphi})^*_{ts}
    (\gamma_{\mathcal{L}_1})_{t}
    (\delta_{\mathcal{L}_1\varphi})_{rs}
    (\gamma_{\mathcal{L}_1})^*_{r}}{
    M_{\mathcal{L}_{1r}}^{2}
    M_{\varphi_s}^{2}
    M_{\mathcal{L}_{1t}}^{2}}
  + \frac{
    \left(
      (g^{(1)}_{\Xi\mathcal{L}_1})_{rs}
      (\gamma_{\mathcal{L}_1})_{s}
      + 3 (g^{(1)}_{\Xi\mathcal{L}_1})^*_{rs}
      (\gamma_{\mathcal{L}_1})^*_{s}
    \right)
    (\kappa_{\Xi})_{r}}{
    2  M_{\Xi_r}^{2}
    M_{\mathcal{L}_{1s}}^{2}}
  \nonumber \\ &
  - \frac{
    \operatorname{Re}{\left(
        (g^{(2)}_{\Xi\mathcal{L}_1})_{rs}
        (\gamma_{\mathcal{L}_1})_{s}
      \right)}
    (\kappa_{\Xi})_{r}}{
    M_{\Xi_r}^{2}
    M_{\mathcal{L}_{1s}}^{2}}
  - \frac{
    (\varepsilon_{\Xi\mathcal{L}_1})_{srt}
    (\kappa_{\Xi})_{s}
    (\gamma_{\mathcal{L}_1})^*_{r}
    (\gamma_{\mathcal{L}_1})_{t}}{
    2  M_{\mathcal{L}_{1r}}^{2}
    M_{\Xi_s}^{2}
    M_{\mathcal{L}_{1t}}^{2}}
  \nonumber \\ &
  + \frac{
    2 \operatorname{Re}{\left(
        (g^{(1)}_{\Xi_1\mathcal{L}_1})^*_{rs}
        (\gamma_{\mathcal{L}_1})^*_{s}
        (\kappa_{\Xi_1})_{r}
      \right)}}{
    M_{\mathcal{L}_{1s}}^{2}
    M_{\Xi_{1r}}^{2}}
  - \frac{
    2  i  \operatorname{Im}{\left(
        (g^{(2)}_{\Xi_1\mathcal{L}_1})_{rs}
        (\kappa_{\Xi_1})^*_{r}
        (\gamma_{\mathcal{L}_1})_{s}
      \right)}}{
    M_{\mathcal{L}_{1s}}^{2}
    M_{\Xi_{1r}}^{2}}
  \nonumber \\ &
  - \frac{
    2  i  \operatorname{Im}{\left(
        (\varepsilon_{\Xi_1\mathcal{L}_1})^*_{rst}
        (\kappa_{\Xi_1})^*_{r}
        (\gamma_{\mathcal{L}_1})_{t}
        (\gamma_{\mathcal{L}_1})_{s}
      \right)}}{
    M_{\mathcal{L}_{1s}}^{2}
    M_{\mathcal{L}_{1t}}^{2}
    M_{\Xi_{1r}}^{2}}
  \nonumber \\ &
  - \frac{
    (\delta_{\mathcal{L}_1\varphi})_{tr}
    (\gamma_{\mathcal{L}_1})^*_{t}
    (\kappa_{\mathcal{S}\varphi})_{sr}
    (\kappa_{\mathcal{S}})_{s}}{
    M_{\varphi_r}^{2}
    M_{\mathcal{S}_s}^{2}
    M_{\mathcal{L}_{1t}}^{2}}
  - \frac{
    (\delta_{\mathcal{L}_1\varphi})_{rt}
    (\gamma_{\mathcal{L}_1})^*_{r}
    (\kappa_{\Xi\varphi})_{st}
    (\kappa_{\Xi})_{s}}{
    M_{\mathcal{L}_{1r}}^{2}
    M_{\Xi_s}^{2}
    M_{\varphi_t}^{2}}
  \nonumber \\ &
  - \frac{
    2 (\delta_{\mathcal{L}_1\varphi})_{sr}
    (\gamma_{\mathcal{L}_1})^*_{s}
    (\kappa_{\Xi_1\varphi})^*_{tr}
    (\kappa_{\Xi_1})_{t}}{
    M_{\varphi_r}^{2}
    M_{\mathcal{L}_{1s}}^{2}
    M_{\Xi_{1t}}^{2}}
  \nonumber \\ &
  + \frac{1}{f} \Bigg\{
  \frac{
    (\tilde{k}^{\phi}_{\mathcal{S}})_{r}
    (\kappa_{\mathcal{S}})_{r}}{
    2  M_{\mathcal{S}_r}^{2}}
  - \frac{
    (\tilde{k}^{\phi}_{\Xi})_{r}
    (\kappa_{\Xi})_{r}}{
    2 M_{\Xi_r}^{2}}
  - \frac{
    2  i  \operatorname{Im}{\left(
        (\tilde{k}_{\Xi_1})_{r}
        (\kappa_{\Xi_1})^*_{r}
      \right)}}{
    M_{\Xi_{1r}}^{2}}
  - \frac{
    i  \operatorname{Im}{\left(
        (\tilde{\gamma}^{(2)}_{\mathcal{L}_1})_{r}
        (\gamma_{\mathcal{L}_1})^*_{r}
      \right)}}{
    M_{\mathcal{L}_{1r}}^{2}}
  \nonumber \\ &
  \qquad
  - \frac{
    \operatorname{Re}{\left(
        (\tilde{\gamma}^{(3)}_{\mathcal{L}_1})_{r}
        (\gamma_{\mathcal{L}_1})^*_{r}
      \right)}}{
    M_{\mathcal{L}_{1r}}^{2}}
  + \frac{
    \operatorname{Im}{\left(
        (\tilde{\gamma}^W_{\mathcal{L}_1})_{r}
        (\gamma_{\mathcal{L}_1})^*_{r}
      \right)}
    g_2}{
    M_{\mathcal{L}_{1r}}^{2}}
  \Bigg\},
  \label{eq:a}
\end{align}

\newpage

\begin{align}
  b^e_{ij} = &
  \frac{
    (\lambda_{\varphi})_{r}
    (y^e_{\varphi})^*_{rji}}{
    M_{\varphi_r}^{2}}
  + \frac{
    \hat{y}^{e*}_{jk}
    (\lambda_E)_{rk}
    (\lambda_E)^*_{ri}}{
    2  M_{E_r}^{2}}
  + \frac{
    \hat{y}^{e*}_{ki}
    (\lambda_{\Delta_1})_{rj}
    (\lambda_{\Delta_1})^*_{rk}}{
    2  M_{\Delta_{1r}}^{2}}
  \nonumber \\ &
  + \frac{
    \hat{y}^{e*}_{ki}
    (\lambda_{\Delta_3})_{rj}
    (\lambda_{\Delta_3})^*_{rk}}{
    2  M_{\Delta_{3r}}^{2}}
  + \frac{
    \hat{y}^{e*}_{jk}
    (\lambda_{\Sigma})^*_{ri}
    (\lambda_{\Sigma})_{rk}}{
    4  M_{\Sigma_r}^{2}}
  + \frac{
    \hat{y}^{e*}_{jk}
    (\lambda_{\Sigma_1})_{rk}
    (\lambda_{\Sigma_1})^*_{ri}}{
    8  M_{\Sigma_{1r}}^{2}}
  \nonumber \\ &
  + \frac{
    i \hat{y}^{e*}_{jk}
    \operatorname{Im}{\left(
        (\hat{g}^\phi_{\mathcal{B}})_{r}
      \right)}
    (g^{l}_{\mathcal{B}})_{rik}}{
    M_{\mathcal{B}_r}^{2}}
  - \frac{
    i \hat{y}^{e*}_{ki}
    \operatorname{Im}{\left(
        (\hat{g}^\phi_{\mathcal{B}})_{r}
      \right)}
    (g^e_{\mathcal{B}})_{rkj}}{
    M_{\mathcal{B}_r}^{2}}
  + \frac{
    i \hat{y}^{e*}_{jk}
    \operatorname{Im}{\left(
        (\hat{g}^\phi_{\mathcal{W}})_{r}
      \right)}
    (g^{l}_{\mathcal{W}})_{rik}}{
    4 M_{\mathcal{W}_r}^{2}}
  \nonumber \\ &
  + \frac{1}{f} \Bigg\{
  \frac{
    (\tilde{y}^e_{\mathcal{S}})^*_{rji}
    (\kappa_{\mathcal{S}})_{r}}{
    M_{\mathcal{S}_r}^{2}}
  + \frac{
    (\tilde{y}^e_{\Xi})^*_{rji}
    (\kappa_{\Xi})_{r}}{
    M_{\Xi_r}^{2}}
  + \frac{
    2 (\tilde{y}^e_{\Xi_1})^*_{rji}
    (\kappa_{\Xi_1})_{r}}{
    M_{\Xi_{1r}}^2}
  + \frac{
    i  \hat{y}^{e*}_{jk}
    (\tilde{\lambda}^l_{E})_{rk}
    (\lambda_E)^*_{ri}}{
    2 M_{E_r}}
  \nonumber \\ &
  \qquad
  + \frac{
    (\tilde{\lambda}^e_{E})_{rj}
    (\lambda_E)^*_{ri}}{
    M_{E_r}}
  + \frac{
    i  \hat{y}^{e*}_{jk}
    (\tilde{\lambda}^l_{E})^*_{ri}
    (\lambda_E)_{rk}}{
    2 M_{E_r}}
  - \frac{
    i  \hat{y}^{e*}_{ki}
    (\tilde{\lambda}^e_{\Delta_1})^*_{rk}
    (\lambda_{\Delta_1})_{rj}}{
    2 M_{\Delta_{1r}}}
  \nonumber \\ &
  \qquad
  + \frac{
    (\tilde{\lambda}^l_{\Delta_1})^*_{ri}
    (\lambda_{\Delta_1})_{rj}}{
    M_{\Delta_{1r}}}
  + \frac{
    (\tilde{\lambda}^{l\prime}_{\Delta_1})^*_{ri}
    (\lambda_{\Delta_1})_{rj}}{
    M_{\Delta_{1r}}}
  - \frac{
    i  \hat{y}^{e*}_{ki}
    (\tilde{\lambda}^e_{\Delta_1})_{rj}
    (\lambda_{\Delta_1})^*_{rk}}{
    2 M_{\Delta_{1r}}}
  \nonumber \\ &
  \qquad
  - \frac{
    i  \hat{y}^{e*}_{ki}
    (\tilde{\lambda}^e_{\Delta_3})^*_{rk}
    (\lambda_{\Delta_3})_{rj}}{
    2 M_{\Delta_{3r}}}
  + \frac{
    (\tilde{\lambda}^l_{\Delta_3})^*_{ri}
    (\lambda_{\Delta_3})_{rj}}{
    M_{\Delta_{3r}}}
  - \frac{
    i  \hat{y}^{e*}_{ki}
    (\tilde{\lambda}^e_{\Delta_3})_{rj}
    (\lambda_{\Delta_3})^*_{rk}}{
    2 M_{\Delta_{3r}}}
  \nonumber \\ &
  \qquad
  + \frac{
    i  \hat{y}^{e*}_{jk}
    (\tilde{\lambda}^l_{\Sigma})^*_{ri}
    (\lambda_{\Sigma})_{rk}}{
    2 M_{\Sigma_r}}
  + \frac{
    i  \hat{y}^{e*}_{jk}
    (\tilde{\lambda}^l_{\Sigma})_{rk}
    (\lambda_{\Sigma})^*_{ri}}{
    2 M_{\Sigma_r}}
  + \frac{
    (\tilde{\lambda}^e_{\Sigma})_{rj}
    (\lambda_{\Sigma})^*_{ri}}{
    M_{\Sigma_r}}
  \nonumber \\ &
  \qquad
  + \frac{
    i  \hat{y}^{e*}_{jk}
    (\tilde{\lambda}^l_{\Sigma_1})_{rk}
    (\lambda_{\Sigma_1})^*_{ri}}{
    4 M_{\Sigma_{1r}}}
  + \frac{
    (\tilde{\lambda}^e_{\Sigma_1})_{rj}
    (\lambda_{\Sigma_1})^*_{ri}}{
    2 M_{\Sigma_{1r}}}
  + \frac{
    i  \hat{y}^{e*}_{jk}
    (\tilde{\lambda}^l_{\Sigma_1})^*_{ri}
    (\lambda_{\Sigma_1})_{rk}}{
    4 M_{\Sigma_{1r}}}
  \nonumber \\ &
  \qquad
  + \frac{
    \hat{y}^{e*}_{jk}
    \hat{y}^e_{lk}
    (\tilde{g}^{eDl}_{\mathcal{L}_1})^*_{rli}
    (\gamma_{\mathcal{L}_1})_{r}}{
    4 M_{\mathcal{L}_{1r}}^{2}}
  - \frac{
    \hat{y}^{e*}_{ki}
    \hat{y}^e_{kl}
    (\tilde{g}^{eDl}_{\mathcal{L}_1})^*_{rjl}
    (\gamma_{\mathcal{L}_1})_{r}}{
    4 M_{\mathcal{L}_{1r}}^{2}}
  - \frac{
    \hat{\lambda}_{\phi}
    (\tilde{g}^{eDl}_{\mathcal{L}_1})^*_{rji}
    (\gamma_{\mathcal{L}_1})_{r}}{
    M_{\mathcal{L}_{1r}}^{2}}
  \nonumber \\ &
  \qquad
  - \frac{
    \hat{y}^{e*}_{jk}
    \hat{y}^e_{lk}
    (\tilde{g}^{Del}_{\mathcal{L}_1})^*_{rli}
    (\gamma_{\mathcal{L}_1})_{r}}{
    4 M_{\mathcal{L}_{1r}}^{2}}
  + \frac{
    \hat{y}^{e*}_{ki}
    \hat{y}^e_{kl}
    (\tilde{g}^{Del}_{\mathcal{L}_1})^*_{rjl}
    (\gamma_{\mathcal{L}_1})_{r}}{
    4 M_{\mathcal{L}_{1r}}^{2}}
  - \frac{
    \hat{\lambda}_{\phi}
    (\tilde{g}^{Del}_{\mathcal{L}_1})^*_{rji}
    (\gamma_{\mathcal{L}_1})_{r}}{
    M_{\mathcal{L}_{1r}}^{2}}
  \nonumber \\ &
  \qquad
  + \frac{
    i  \hat{y}^{e*}_{ki}
    (\tilde{g}^{e}_{\mathcal{L}_1})_{rkj}
    (\gamma_{\mathcal{L}_1})^*_{r}}{
    2 M_{\mathcal{L}_{1r}}^{2}}
  - \frac{
    i  \hat{y}^{e*}_{jk}
    (\tilde{g}^{l}_{\mathcal{L}_1})_{rik}
    (\gamma_{\mathcal{L}_1})^*_{r}}{
    2 M_{\mathcal{L}_{1r}}^{2}}
  - \frac{
    i  \hat{y}^{e*}_{jk}
    (\tilde{g}^{l\prime}_{\mathcal{L}_1})_{rik}
    (\gamma_{\mathcal{L}_1})^*_{r}}{
    2 M_{\mathcal{L}_{1r}}^{2}}
  \nonumber \\ &
  \qquad
  + \frac{
    i  \hat{y}^{e*}_{ki}
    (\tilde{g}^{e}_{\mathcal{L}_1})^*_{rjk}
    (\gamma_{\mathcal{L}_1})_{r}}{
    2 M_{\mathcal{L}_{1r}}^{2}}
  - \frac{
    i  \hat{y}^{e*}_{jk}
    (\tilde{g}^{l}_{\mathcal{L}_1})^*_{rki}
    (\gamma_{\mathcal{L}_1})_{r}}{
    2 M_{\mathcal{L}_{1r}}^{2}}
  - \frac{
    i  \hat{y}^{e*}_{jk}
    (\tilde{g}^{l\prime}_{\mathcal{L}_1})^*_{rki}
    (\gamma_{\mathcal{L}_1})_{r}}{
    2 M_{\mathcal{L}_{1r}}^{2}}
  \Bigg\},
\end{align}

\newpage

\begin{align}
  c^e_{ij} = &
  - \frac{
    (\kappa_{\mathcal{S}\varphi})_{rs}
    (\kappa_{\mathcal{S}})_{r}
    (y^e_{\varphi})^*_{sji}}{
    M_{\mathcal{S}_r}^{2}
    M_{\varphi_s}^{2}}
  - \frac{
    (\kappa_{\Xi\varphi})_{sr}
    (\kappa_{\Xi})_{s}
    (y^e_{\varphi})^*_{rji}}{
    M_{\varphi_r}^{2}
    M_{\Xi_s}^{2}}
  \nonumber \\ &
  - \frac{
    2   (\kappa_{\Xi_1\varphi})^*_{sr}
    (\kappa_{\Xi_1})_{s}
    (y^e_{\varphi})^*_{rji}}{
    M_{\varphi_r}^{2}
    M_{\Xi_{1s}}^{2}}
  - \frac{
    (\lambda_{E \Delta_1})_{rs}
    (\lambda_E)^*_{ri}
    (\lambda_{\Delta_1})_{sj}}{
    M_{E_r}
    M_{\Delta_{1s}}}
  \nonumber \\ &
  - \frac{
    (\lambda_{E \Delta_3})_{sr}
    (\lambda_E)^*_{si}
    (\lambda_{\Delta_3})_{rj}}{
    M_{\Delta_{3r}}
    M_{E_s}}
  - \frac{
    (\lambda_{\Sigma})^*_{si}
    (\lambda_{\Sigma \Delta_1})_{sr}
    (\lambda_{\Delta_1})_{rj}}{
    2  M_{\Delta_{1r}}
    M_{\Sigma_s}}
  \nonumber \\ &
  - \frac{
    (\lambda_{\Sigma_1 \Delta_1})_{rs}
    (\lambda_{\Sigma_1})^*_{ri}
    (\lambda_{\Delta_1})_{sj}}{
    4  M_{\Sigma_{1r}}
    M_{\Delta_{1s}}}
  + \frac{
    (\lambda_{\Sigma_1 \Delta_3})_{sr}
    (\lambda_{\Sigma_1})^*_{si}
    (\lambda_{\Delta_3})_{rj}}{
    4  M_{\Delta_{3r}}
    M_{\Sigma_{1s}}}
  \nonumber \\ &
  - \frac{
    (w_{\mathcal{S}E})_{rsj}
    (\kappa_{\mathcal{S}})_{r}
    (\lambda_E)^*_{si}}{
    M_{\mathcal{S}_r}^{2}
    M_{E_s}}
  - \frac{
    (w_{\mathcal{S}\Delta_1})^*_{rsi}
    (\kappa_{\mathcal{S}})_{r}
    (\lambda_{\Delta_1})_{sj}}{
    M_{\mathcal{S}_r}^{2}
    M_{\Delta_{1s}}}
  \nonumber \\ &
  - \frac{
    (w_{\Xi \Delta_3})^*_{rsi}
    (\kappa_{\Xi})_{r}
    (\lambda_{\Delta_1})_{sj}}{
    M_{\Xi_r}^{2}
    M_{\Delta_{1s}}}
   - \frac{
    (w_{\Xi \Sigma_1})_{srj}
    (\kappa_{\Xi})_{s}
    (\lambda_{\Sigma_1})^*_{ri}}{
    2  M_{\Sigma_{1r}}
    M_{\Xi_s}^{2}}
  \nonumber \\ &
  - \frac{
    2   (w_{\Xi_1 \Delta_3})^*_{rsi}
    (\kappa_{\Xi_1})_{r}
    (\lambda_{\Delta_3})_{sj}}{
    M_{\Xi_{1r}}^{2}
    M_{\Delta_{3s}}}
  - \frac{
    (\lambda_{\Sigma})^*_{si}
    (w_{\Xi_1 \Sigma})^*_{rsj}
    (\kappa_{\Xi_1})_{r}}{
    M_{\Xi_{1r}}^{2}
    M_{\Sigma_s}}
  \nonumber \\ &
  + \frac{
    i  \hat{y}^{e*}_{jk}
    (z_{E\mathcal{L}_1})_{rsk}
    (\lambda_E)^*_{ri}
    (\gamma_{\mathcal{L}_1})^*_{s}}{
    2  M_{E_r}
    M_{\mathcal{L}_{1s}}^{2}}
  + \frac{
    i  \hat{y}^{e*}_{jk}
    (z_{E\mathcal{L}_1})^*_{rsi}
    (\gamma_{\mathcal{L}_1})_{s}
    (\lambda_E)_{rk}}{
    2  M_{E_r}
    M_{\mathcal{L}_{1s}}^{2}}
  \nonumber \\ &
  - \frac{
    i  \hat{y}^{e*}_{ki}
    (z_{\Delta_1\mathcal{L}_1})^*_{rsk}
    (\gamma_{\mathcal{L}_1})^*_{s}
    (\lambda_{\Delta_1})_{rj}}{
    2  M_{\Delta_{1r}}
    M_{\mathcal{L}_{1s}}^{2}}
  - \frac{
    i  \hat{y}^{e*}_{ki}
    (z_{\Delta_1\mathcal{L}_1})_{rsj}
    (\lambda_{\Delta_1})^*_{rk}
    (\gamma_{\mathcal{L}_1})_{s}}{
    2  M_{\Delta_{1r}}
    M_{\mathcal{L}_{1s}}^{2}}
  \nonumber \\ &
  - \frac{
    i  \hat{y}^{e*}_{ki}
    (z_{\Delta_3\mathcal{L}_1})^*_{srk}
    (\gamma_{\mathcal{L}_1})_{r}
    (\lambda_{\Delta_3})_{sj}}{
    2  M_{\mathcal{L}_{1r}}^{2}
    M_{\Delta_{3s}}}
  - \frac{
    i  \hat{y}^{e*}_{ki}
    (z_{\Delta_3\mathcal{L}_1})_{srj}
    (\lambda_{\Delta_3})^*_{sk}
    (\gamma_{\mathcal{L}_1})^*_{r}}{
    2  M_{\mathcal{L}_{1r}}^{2}
    M_{\Delta_{3s}}}
  \nonumber \\ &
  + \frac{
    i  \hat{y}^{e*}_{jk}
    (\lambda_{\Sigma})^*_{ri}
    (z_{\Sigma\mathcal{L}_1})_{rsk}
    (\gamma_{\mathcal{L}_1})_{s}}{
    2  M_{\Sigma_r}
    M_{\mathcal{L}_{1s}}^{2}}
  + \frac{
    i  \hat{y}^{e*}_{jk}
    (\lambda_{\Sigma})_{sk}
    (z_{\Sigma\mathcal{L}_1})^*_{sri}
    (\gamma_{\mathcal{L}_1})^*_{r}}{
    2  M_{\mathcal{L}_{1r}}^{2}
    M_{\Sigma_s}}
  \nonumber \\ &
  + \frac{
    i  \hat{y}^{e*}_{jk}
    (z_{\Sigma_1\mathcal{L}_1})_{srk}
    (\lambda_{\Sigma_1})^*_{si}
    (\gamma_{\mathcal{L}_1})^*_{r}}{
    4  M_{\mathcal{L}_{1r}}^{2}
    M_{\Sigma_{1s}}}
  + \frac{
    i  \hat{y}^{e*}_{jk}
    (z_{\Sigma_1\mathcal{L}_1})^*_{rsi}
    (\gamma_{\mathcal{L}_1})_{s}
    (\lambda_{\Sigma_1})_{rk}}{
    4  M_{\Sigma_{1r}}
    M_{\mathcal{L}_{1s}}^{2}}
  \nonumber \\ &
  + \frac{
    i \hat{y}^{e*}_{jk}
    (\delta_{\mathcal{B}\mathcal{S}})_{rs}
    (g^{l}_{\mathcal{B}})_{rik}
    (\kappa_{\mathcal{S}})_{s}}{
    M_{\mathcal{B}_r}^{2}
    M_{\mathcal{S}_s}^{2}}
  - \frac{
    i  \hat{y}^{e*}_{ki}
    (\delta_{\mathcal{B}\mathcal{S}})_{rs}
    (g^e_{\mathcal{B}})_{rkj}
    (\kappa_{\mathcal{S}})_{s}}{
    M_{\mathcal{B}_r}^{2}
    M_{\mathcal{S}_s}^{2}}
  \nonumber \\ &
  + \frac{
    i  \hat{y}^{e*}_{jk}
    (\delta_{\mathcal{W}\Xi})_{sr}
    (g^{l}_{\mathcal{W}})_{sik}
    (\kappa_{\Xi})_{r}}{
    2  M_{\Xi_r}^{2}
    M_{\mathcal{W}_s}^{2}}
  + \frac{
    2   \hat{\lambda}_{\phi}
    (\delta_{\mathcal{L}_1\varphi})^*_{sr}
    (\gamma_{\mathcal{L}_1})_{s}
    (y^e_{\varphi})^*_{rji}}{
    M_{\varphi_r}^{2}
    M_{\mathcal{L}_{1s}}^{2}},
\end{align}

\newpage

\begin{align}
  b^d_{ij} = &
  \frac{
    (\lambda_{\varphi})_{r}
    (y^d_{\varphi})^*_{rji}}{
    M_{\varphi_r}^{2}}
  + \frac{
    \hat{y}^{d*}_{jk}
    (\lambda_D)_{rk}
    (\lambda_D)^*_{ri}}{
    2  M_{D_r}^{2}}
  + \frac{
    \hat{y}^{d*}_{ki}
    (\lambda^d_{Q_1})_{rj}
    (\lambda^d_{Q_1})^*_{rk}}{
    2  M_{Q_{1r}}^{2}}
  \nonumber \\ &
  + \frac{
    \hat{y}^{d*}_{ki}
    (\lambda_{Q_5})_{rj}
    (\lambda_{Q_5})^*_{rk}}{
    2  M_{Q_{5r}}^{2}}
  + \frac{
    \hat{y}^{d*}_{jk}
    (\lambda_{T_1})_{rk}
    (\lambda_{T_1})^*_{ri}}{
    8  M_{T_{1r}}^{2}}
  + \frac{
    \hat{y}^{d*}_{jk}
    (\lambda_{T_2})_{rk}
    (\lambda_{T_2})^*_{ri}}{
    4  M_{T_{2r}}^{2}}
  \nonumber \\ &
  + \frac{
    i \hat{y}^{d*}_{jk}
    \operatorname{Im}{\left(
        (\hat{g}^\phi_{\mathcal{B}})_{r}
      \right)}
    (g^q_{\mathcal{B}})_{rik}}{
    M_{\mathcal{B}_r}^{2}}
  - \frac{
    i \hat{y}^{d*}_{ki}
    \operatorname{Im}{\left(
        (\hat{g}^\phi_{\mathcal{B}})_{r}
      \right)}
    (g^d_{\mathcal{B}})_{rkj}}{
    M_{\mathcal{B}_r}^{2}}
  + \frac{
      i \hat{y}^{d*}_{jk}
      \operatorname{Im}{\left(
          (\hat{g}^\phi_{\mathcal{W}})_{r}
        \right)}
      (g^{q}_{\mathcal{W}})_{rik}}{
      4 M_{\mathcal{W}_r}^{2}}
  \nonumber \\ &
  + \frac{1}{f} \Bigg\{
  \frac{
    (\tilde{y}^d_{\mathcal{S}})^*_{rji}
    (\kappa_{\mathcal{S}})_{r}}{
    M_{\mathcal{S}_r}^{2}}
  + \frac{
    (\tilde{y}^d_{\Xi})^*_{rji}
    (\kappa_{\Xi})_{r}}{
    M_{\Xi_r}^{2}}
  + \frac{
    2   (\tilde{y}^d_{\Xi_1})^*_{rji}
    (\kappa_{\Xi_1})_{r}}{
    M_{\Xi_{1r}}^{2}}
  + \frac{
    i  \hat{y}^{d*}_{jk}
    (\tilde{\lambda}^q_{D})_{rk}
    (\lambda_D)^*_{ri}}{
    2 M_{D_r}}
  \nonumber \\ &
  \qquad
  + \frac{
    (\tilde{\lambda}^d_{D})_{rj}
    (\lambda_D)^*_{ri}}{
    M_{D_r}}
  + \frac{
    i  \hat{y}^{d*}_{jk}
    (\tilde{\lambda}^q_{D})^*_{ri}
    (\lambda_D)_{rk}}{
    2 M_{D_r}}
  - \frac{
    i  \hat{y}^{d*}_{ki}
    (\tilde{\lambda}^d_{Q_1})^*_{rk}
    (\lambda^d_{Q_1})_{rj}}{
    2 M_{Q_{1r}}}
  \nonumber \\ &
  \qquad
  + \frac{
    (\tilde{\lambda}^q_{Q_1})^*_{ri}
    (\lambda^d_{Q_1})_{rj}}{
    M_{Q_{1r}}}
  + \frac{
    (\tilde{\lambda}^{q\prime}_{Q_1})^*_{ri}
    (\lambda^d_{Q_1})_{rj}}{
    M_{Q_{1r}}}
  - \frac{
    i  \hat{y}^{d*}_{ki}
    (\tilde{\lambda}^d_{Q_1})_{rj}
    (\lambda^d_{Q_1})^*_{rk}}{
    2 M_{Q_{1r}}}
  \nonumber \\ &
  \qquad
  - \frac{
    i  \hat{y}^{d*}_{ki}
    (\tilde{\lambda}^d_{Q_5})^*_{rk}
    (\lambda_{Q_5})_{rj}}{
    2 M_{Q_{5r}}}
  + \frac{
    (\tilde{\lambda}^q_{Q_5})^*_{ri}
    (\lambda_{Q_5})_{rj}}{
    M_{Q_{5r}}}
  - \frac{
    i  \hat{y}^{d*}_{ki}
    (\tilde{\lambda}^d_{Q_5})_{rj}
    (\lambda_{Q_5})^*_{rk}}{
    2 M_{Q_{5r}}}
  \nonumber \\ &
  \qquad
  + \frac{
    i  \hat{y}^{d*}_{jk}
    (\tilde{\lambda}^q_{T_1})_{rk}
    (\lambda_{T_1})^*_{ri}}{
    4 M_{T_{1r}}}
  + \frac{
    (\tilde{\lambda}^d_{T_1})_{rj}
    (\lambda_{T_1})^*_{ri}}{
    2 M_{T_{1r}}}
  + \frac{
    i  \hat{y}^{d*}_{jk}
    (\tilde{\lambda}^q_{T_1})^*_{ri}
    (\lambda_{T_1})_{rk}}{
    4 M_{T_{1r}}}
  \nonumber \\ &
  \qquad
  + \frac{
    i  \hat{y}^{d*}_{jk}
    (\tilde{\lambda}^q_{T_2})_{rk}
    (\lambda_{T_2})^*_{ri}}{
    2 M_{T_{2r}}}
  + \frac{
    (\tilde{\lambda}^d_{T_2})_{rj}
    (\lambda_{T_2})^*_{ri}}{
    M_{T_{2r}}}
    + \frac{
    i  \hat{y}^{d*}_{jk}
    (\tilde{\lambda}^q_{T_2})^*_{ri}
    (\lambda_{T_2})_{rk}}{
    2 M_{T_{2r}}}
  \nonumber \\ &
  \qquad
  + \frac{
    \hat{y}^{d*}_{jk}
    \hat{y}^d_{lk}
    (\tilde{g}^{dDq}_{\mathcal{L}_1})^*_{rli}
    (\gamma_{\mathcal{L}_1})_{r}}{
    4 M_{\mathcal{L}_{1r}}^{2}}
  - \frac{
    \hat{y}^{d*}_{ki}
    \hat{y}^d_{kl}
    (\tilde{g}^{dDq}_{\mathcal{L}_1})^*_{rjl}
    (\gamma_{\mathcal{L}_1})_{r}}{
    4 M_{\mathcal{L}_{1r}}^{2}}
  - \frac{
    \hat{\lambda}_{\phi}
    (\tilde{g}^{dDq}_{\mathcal{L}_1})^*_{rji}
    (\gamma_{\mathcal{L}_1})_{r}}{
    M_{\mathcal{L}_{1r}}^{2}}
  \nonumber \\ &
  \qquad
  - \frac{
    \hat{y}^{d*}_{jk}
    \hat{y}^d_{lk}
    (\tilde{g}^{Ddq}_{\mathcal{L}_1})^*_{rli}
    (\gamma_{\mathcal{L}_1})_{r}}{
    4 M_{\mathcal{L}_{1r}}^{2}}
  + \frac{
    \hat{y}^{d*}_{ki}
    \hat{y}^d_{kl}
    (\tilde{g}^{Ddq}_{\mathcal{L}_1})^*_{rjl}
    (\gamma_{\mathcal{L}_1})_{r}}{
    4 M_{\mathcal{L}_{1r}}^{2}}
  - \frac{
    \hat{\lambda}_{\phi}
    (\tilde{g}^{Ddq}_{\mathcal{L}_1})^*_{rji}
    (\gamma_{\mathcal{L}_1})_{r}}{
    M_{\mathcal{L}_{1r}}^{2}}
  \nonumber \\ &
  \qquad
  + \frac{
    i  \hat{y}^{d*}_{ki}
    (\tilde{g}^{d}_{\mathcal{L}_1})_{rkj}
    (\gamma_{\mathcal{L}_1})^*_{r}}{
    2 M_{\mathcal{L}_{1r}}^{2}}
  - \frac{
    i  \hat{y}^{d*}_{jk}
    (\tilde{g}^{q}_{\mathcal{L}_1})_{rik}
    (\gamma_{\mathcal{L}_1})^*_{r}}{
    2 M_{\mathcal{L}_{1r}}^{2}}
  - \frac{
    i  \hat{y}^{d*}_{jk}
    (\tilde{g}^{q\prime}_{\mathcal{L}_1})_{rik}
    (\gamma_{\mathcal{L}_1})^*_{r}}{
    2 M_{\mathcal{L}_{1r}}^{2}}
  \nonumber \\ &
  \qquad
  + \frac{
    i  \hat{y}^{d*}_{ki}
    (\tilde{g}^{d}_{\mathcal{L}_1})^*_{rjk}
    (\gamma_{\mathcal{L}_1})_{r}}{
    2 M_{\mathcal{L}_{1r}}^{2}}
  - \frac{
    i  \hat{y}^{d*}_{jk}
    (\tilde{g}^{q}_{\mathcal{L}_1})^*_{rki}
    (\gamma_{\mathcal{L}_1})_{r}}{
    2 M_{\mathcal{L}_{1r}}^{2}}
  - \frac{
    i  \hat{y}^{d*}_{jk}
    (\tilde{g}^{q\prime}_{\mathcal{L}_1})^*_{rki}
    (\gamma_{\mathcal{L}_1})_{r}}{
    2 M_{\mathcal{L}_{1r}}^{2}}
  \Bigg\},
\end{align}

\newpage

\begin{align}
  c^d_{ij} = & \;
  - \frac{
    (\kappa_{\mathcal{S}\varphi})_{rs}
    (\kappa_{\mathcal{S}})_{r}
    (y^d_{\varphi})^*_{sji}}{
    M_{\mathcal{S}_r}^{2}
    M_{\varphi_s}^{2}}
  - \frac{
    (\kappa_{\Xi\varphi})_{rs}
    (\kappa_{\Xi})_{r}
    (y^d_{\varphi})^*_{sji}}{
    M_{\Xi_r}^{2}
    M_{\varphi_s}^{2}}
  \nonumber \\ &
  - \frac{
    2   (\kappa_{\Xi_1\varphi})^*_{sr}
    (\kappa_{\Xi_1})_{s}
    (y^d_{\varphi})^*_{rji}}{
    M_{\varphi_r}^{2}
    M_{\Xi_{1s}}^{2}}
  - \frac{
    (\lambda_{D Q_1})_{sr}
    (\lambda_D)^*_{si}
    (\lambda^d_{Q_1})_{rj}}{
    M_{Q_{1r}}
    M_{D_s}}
  \nonumber \\ &
  - \frac{
    (\lambda_{D Q_5})_{rs}
    (\lambda_D)^*_{ri}
    (\lambda_{Q_5})_{sj}}{
    M_{D_r}
    M_{Q_{5s}}}
  - \frac{
    (\lambda_{T_1 Q_1})_{rs}
    (\lambda_{T_1})^*_{ri}
    (\lambda^d_{Q_1})_{sj}}{
    4  M_{T_{1r}}
    M_{Q_{1s}}}
  \nonumber \\ &
  - \frac{
    (\lambda_{T_2 Q_1})_{sr}
    (\lambda_{T_2})^*_{si}
    (\lambda^d_{Q_1})_{rj}}{
    2  M_{Q_{1r}}
    M_{T_{2s}}}
  + \frac{
    (\lambda_{T_1 Q_5})_{rs}
    (\lambda_{T_1})^*_{ri}
    (\lambda_{Q_5})_{sj}}{
    4  M_{T_{1r}}
    M_{Q_{5s}}}
  \nonumber \\ &
  - \frac{
    (w_{\mathcal{S}D})_{rsj}
    (\kappa_{\mathcal{S}})_{r}
    (\lambda_D)^*_{si}}{
    M_{\mathcal{S}_r}^{2}
    M_{D_s}}
  - \frac{
    (w_{\mathcal{S}Q_1})^*_{rsi}
    (\kappa_{\mathcal{S}})_{r}
    (\lambda^d_{Q_1})_{sj}}{
    M_{\mathcal{S}_r}^{2}
    M_{Q_{1s}}}
  \nonumber \\ &
  - \frac{
    (w_{\Xi Q_7})^*_{sri}
    (\kappa_{\Xi})_{s}
    (\lambda^d_{Q_1})_{rj}}{
    M_{Q_{1r}}
    M_{\Xi_s}^{2}}
  - \frac{
    (w_{\Xi T_1})_{rsj}
    (\kappa_{\Xi})_{r}
    (\lambda_{T_1})^*_{si}}{
    2  M_{\Xi_r}^{2}
    M_{T_{1s}}}
  \nonumber \\ &
  - \frac{
    2 (w_{\Xi_1 Q_5})^*_{rsi}
    (\kappa_{\Xi_1})_{r}
    (\lambda_{Q_5})_{sj}}{
    M_{\Xi_{1r}}^{2}
    M_{Q_{5s}}}
  - \frac{
    (w_{\Xi_1 T_2})_{rsj}
    (\kappa_{\Xi_1})_{r}
    (\lambda_{T_2})^*_{si}}{
    M_{\Xi_{1r}}^{2}
    M_{T_{2s}}}
  \nonumber \\ &
  + \frac{
    i  \hat{y}^{d*}_{jk}
    (z_{D\mathcal{L}_1})_{rsk}
    (\lambda_D)^*_{ri}
    (\gamma_{\mathcal{L}_1})^*_{s}}{
    2  M_{D_r}
    M_{\mathcal{L}_{1s}}^{2}}
  + \frac{
    i  \hat{y}^{d*}_{jk}
    (z_{D\mathcal{L}_1})^*_{rsi}
    (\gamma_{\mathcal{L}_1})_{s}
    (\lambda_D)_{rk}}{
    2  M_{D_r}
    M_{\mathcal{L}_{1s}}^{2}}
  \nonumber \\ &
  - \frac{
    i  \hat{y}^{d*}_{ki}
    (z^d_{Q_1\mathcal{L}_1})^*_{rsk}
    (\gamma_{\mathcal{L}_1})^*_{s}
    (\lambda^d_{Q_1})_{rj}}{
    2  M_{Q_{1r}}
    M_{\mathcal{L}_{1s}}^{2}}
  - \frac{
    i  \hat{y}^{d*}_{ki}
    (z^d_{Q_1\mathcal{L}_1})_{rsj}
    (\lambda^d_{Q_1})^*_{rk}
    (\gamma_{\mathcal{L}_1})_{s}}{
    2  M_{Q_{1r}}
    M_{\mathcal{L}_{1s}}^{2}}
  \nonumber \\ &
  - \frac{
    i  \hat{y}^{d*}_{ki}
    (z_{Q_5\mathcal{L}_1})^*_{rsk}
    (\gamma_{\mathcal{L}_1})_{s}
    (\lambda_{Q_5})_{rj}}{
    2  M_{Q_{5r}}
    M_{\mathcal{L}_{1s}}^{2}}
  - \frac{
    i  \hat{y}^{d*}_{ki}
    (z_{Q_5\mathcal{L}_1})_{rsj}
    (\lambda_{Q_5})^*_{rk}
    (\gamma_{\mathcal{L}_1})^*_{s}}{
    2  M_{Q_{5r}}
    M_{\mathcal{L}_{1s}}^{2}}
  \nonumber \\ &
  + \frac{
    i  \hat{y}^{d*}_{jk}
    (z_{T_2\mathcal{L}_1})_{rsk}
    (\lambda_{T_2})^*_{ri}
    (\gamma_{\mathcal{L}_1})_{s}}{
    2  M_{T_{2r}}
    M_{\mathcal{L}_{1s}}^{2}}
  + \frac{
    i  \hat{y}^{d*}_{jk}
    (z_{T_2\mathcal{L}_1})^*_{sri}
    (\gamma_{\mathcal{L}_1})^*_{r}
    (\lambda_{T_2})_{sk}}{
    2  M_{\mathcal{L}_{1r}}^{2}
    M_{T_{2s}}}
  \nonumber \\ &
  + \frac{
    i  \hat{y}^{d*}_{jk}
    (z_{T_1\mathcal{L}_1})_{rsk}
    (\lambda_{T_1})^*_{ri}
    (\gamma_{\mathcal{L}_1})^*_{s}}{
    4  M_{T_{1r}}
    M_{\mathcal{L}_{1s}}^{2}}
  + \frac{
    i  \hat{y}^{d*}_{jk}
    (z_{T_1\mathcal{L}_1})^*_{sri}
    (\gamma_{\mathcal{L}_1})_{r}
    (\lambda_{T_1})_{sk}}{
    4  M_{\mathcal{L}_{1r}}^{2}
    M_{T_{1s}}}
  \nonumber \\ &
   + \frac{
    i \hat{y}^{d*}_{jk}
    (\delta_{\mathcal{B}\mathcal{S}})_{rs}
    (g^{q}_{\mathcal{B}})_{rik}
    (\kappa_{\mathcal{S}})_{s}}{
    M_{\mathcal{B}_r}^{2}
    M_{\mathcal{S}_s}^{2}}
  - \frac{
    i  \hat{y}^{d*}_{ki}
    (\delta_{\mathcal{B}\mathcal{S}})_{rs}
    (g^d_{\mathcal{B}})_{rkj}
    (\kappa_{\mathcal{S}})_{s}}{
    M_{\mathcal{B}_r}^{2}
    M_{\mathcal{S}_s}^{2}}
  \nonumber \\ &
  + \frac{
    i  \hat{y}^{d*}_{jk}
    (\delta_{\mathcal{W}\Xi})_{sr}
    (g^{q}_{\mathcal{W}})_{sik}
    (\kappa_{\Xi})_{r}}{
    2  M_{\Xi_r}^{2}
    M_{\mathcal{W}_s}^{2}}
  + \frac{
    2   \hat{\lambda}_{\phi}
    (\delta_{\mathcal{L}_1\varphi})^*_{sr}
    (\gamma_{\mathcal{L}_1})_{s}
    (y^d_{\varphi})^*_{rji}}{
    M_{\varphi_r}^{2}
    M_{\mathcal{L}_{1s}}^{2}},
\end{align}

\newpage

\begin{align}
  b^u_{ij} = & \;
  - \frac{
    (\lambda_{\varphi})^*_{r}
    (y^u_{\varphi})_{rij}}{
    M_{\varphi_r}^{2}}
  + \frac{
    \hat{y}^{u*}_{jk}
    (\lambda_U)_{rk}
    (\lambda_U)^*_{ri}}{
    2  M_{U_r}^{2}}
  + \frac{
    \hat{y}^{u*}_{ki}
    (\lambda^u_{Q_1})_{rj}
    (\lambda^u_{Q_1})^*_{rk}}{
    2  M_{Q_{1r}}^{2}}
  \nonumber \\ &
  + \frac{
    \hat{y}^{u*}_{ki}
    (\lambda_{Q_7})_{rj}
    (\lambda_{Q_7})^*_{rk}}{
    2  M_{Q_{7r}}^{2}}
  + \frac{
    \hat{y}^{u*}_{jk}
    (\lambda_{T_1})_{rk}
    (\lambda_{T_1})^*_{ri}}{
    4  M_{T_{1r}}^{2}}
  + \frac{
    \hat{y}^{u*}_{jk}
    (\lambda_{T_2})_{rk}
    (\lambda_{T_2})^*_{ri}}{
    8  M_{T_{2r}}^{2}}
  \nonumber \\ &
  + \frac{
    i \hat{y}^{u*}_{jk}
    \operatorname{Im}{\left(
        (\hat{g}^\phi_{\mathcal{B}})_{r}
      \right)}
    (g^{q}_{\mathcal{B}})_{rik}}{
    M_{\mathcal{B}_r}^{2}}
  - \frac{
    i \hat{y}^{u*}_{ki}
    \operatorname{Im}{\left(
        (\hat{g}^\phi_{\mathcal{B}})_{r}
      \right)}
    (g^u_{\mathcal{B}})_{rkj}}{
    M_{\mathcal{B}_r}^{2}}
  - \frac{
    i \hat{y}^{u*}_{jk}
    \operatorname{Im}{\left(
        (\hat{g}^\phi_{\mathcal{W}})_{r}
      \right)}
    (g^{q}_{\mathcal{W}})_{rik}}{
    4 M_{\mathcal{W}_r}^{2}}
  \nonumber \\ &
  + \frac{1}{f} \Bigg\{
  \frac{
    (\tilde{y}^u_{\mathcal{S}})^*_{rji}
    (\kappa_{\mathcal{S}})_{r}}{
    M_{\mathcal{S}_r}^{2}}
  - \frac{
    (\tilde{y}^u_{\Xi})^*_{rji}
    (\kappa_{\Xi})_{r}}{
    M_{\Xi_r}^{2}}
  + \frac{
    2 (\tilde{y}^u_{\Xi_1})_{rji}
    (\kappa_{\Xi_1})^*_{r}}{
    M_{\Xi_{1r}}^{2}}
  \nonumber \\ &
  \qquad
  + \frac{
    i \hat{y}^{u*}_{jk}
    (\tilde{\lambda}^q_{U})_{rk}
    (\lambda_U)^*_{ri}}{
    2 M_{U_r}}
  + \frac{
    (\tilde{\lambda}^u_{U})_{rj}
    (\lambda_U)^*_{ri}}{
    M_{U_r}}
  + \frac{
    i \hat{y}^{u*}_{jk}
    (\tilde{\lambda}^q_{U})^*_{ri}
    (\lambda_U)_{rk}}{
    2 M_{U_r}}
  \nonumber \\ &
  \qquad
  - \frac{
    i  \hat{y}^{u*}_{ki}
    (\tilde{\lambda}^u_{Q_1})^*_{rk}
    (\lambda^u_{Q_1})_{rj}}{
    2 M_{Q_{1r}}}
  + \frac{
    (\tilde{\lambda}^{q\prime}_{Q_1})^*_{ri}
    (\lambda^u_{Q_1})_{rj}}{
    M_{Q_{1r}}}
  - \frac{
    i  \hat{y}^{u*}_{ki}
    (\tilde{\lambda}^u_{Q_1})_{rj}
    (\lambda^u_{Q_1})^*_{rk}}{
    2 M_{Q_{1r}}}
  \nonumber \\ &
  \qquad
  - \frac{
    i  \hat{y}^{u*}_{ki}
    (\tilde{\lambda}^u_{Q_7})^*_{rk}
    (\lambda_{Q_7})_{rj}}{
    2 M_{Q_{7r}}}
  + \frac{
    (\tilde{\lambda}^q_{Q_7})^*_{ri}
    (\lambda_{Q_7})_{rj}}{
    M_{Q_{7r}}}
  - \frac{
    i  \hat{y}^{u*}_{ki}
    (\tilde{\lambda}^u_{Q_7})_{rj}
    (\lambda_{Q_7})^*_{rk}}{
    2 M_{Q_{7r}}}
  \nonumber \\ &
  \qquad
  + \frac{
    i  \hat{y}^{u*}_{jk}
    (\tilde{\lambda}^q_{T_1})_{rk}
    (\lambda_{T_1})^*_{ri}}{
    2 M_{T_{1r}}}
  + \frac{
    (\tilde{\lambda}^u_{T_1})_{rj}
    (\lambda_{T_1})^*_{ri}}{
    M_{T_{1r}}}
  + \frac{
    i  \hat{y}^{u*}_{jk}
    (\tilde{\lambda}^q_{T_1})^*_{ri}
    (\lambda_{T_1})_{rk}}{
    2 M_{T_{1r}}}
  \nonumber \\ &
  \qquad
  + \frac{
    i  \hat{y}^{u*}_{jk}
    (\tilde{\lambda}^q_{T_2})_{rk}
    (\lambda_{T_2})^*_{ri}}{
    4 M_{T_{2r}}}
  - \frac{
    (\tilde{\lambda}^u_{T_2})_{rj}
    (\lambda_{T_2})^*_{ri}}{
    2 M_{T_{2r}}}
  + \frac{
    i  \hat{y}^{u*}_{jk}
    (\tilde{\lambda}^q_{T_2})^*_{ri}
    (\lambda_{T_2})_{rk}}{
    4 M_{T_{2r}}}
  \nonumber \\ &
  \qquad
  - \frac{
    \hat{y}^{u*}_{jk}
    \hat{y}^u_{lk}
    (\tilde{g}^{qDu}_{\mathcal{L}_1})_{rli}
    (\gamma_{\mathcal{L}_1})^*_{r}}{
    4 M_{\mathcal{L}_{1r}}^{2}}
  + \frac{
    \hat{y}^{u*}_{ki}
    \hat{y}^u_{kl}
    (\tilde{g}^{qDu}_{\mathcal{L}_1})_{rjl}
    (\gamma_{\mathcal{L}_1})^*_{r}}{
    4 M_{\mathcal{L}_{1r}}^{2}}
  + \frac{
    \hat{\lambda}_{\phi}
    (\tilde{g}^{qDu}_{\mathcal{L}_1})_{rji}
    (\gamma_{\mathcal{L}_1})^*_{r}}{
    M_{\mathcal{L}_{1r}}^{2}}
  \nonumber \\ &
  \qquad
  + \frac{
    \hat{y}^{u*}_{jk}
    \hat{y}^u_{lk}
    (\tilde{g}^{Dqu}_{\mathcal{L}_1})_{rli}
    (\gamma_{\mathcal{L}_1})^*_{r}}{
    4 M_{\mathcal{L}_{1r}}^{2}}
  - \frac{
    \hat{y}^{u*}_{ki}
    \hat{y}^u_{kl}
    (\tilde{g}^{Dqu}_{\mathcal{L}_1})_{rjl}
    (\gamma_{\mathcal{L}_1})^*_{r}}{
    4 M_{\mathcal{L}_{1r}}^{2}}
  + \frac{
    \hat{\lambda}_{\phi}
    (\tilde{g}^{Dqu}_{\mathcal{L}_1})_{rji}
    (\gamma_{\mathcal{L}_1})^*_{r}}{
    M_{\mathcal{L}_{1r}}^{2}}
  \nonumber \\ &
  \qquad
  + \frac{
    i  \hat{y}^{u*}_{ki}
    (\tilde{g}^{u}_{\mathcal{L}_1})_{rkj}
    (\gamma_{\mathcal{L}_1})^*_{r}}{
    2 M_{\mathcal{L}_{1r}}^{2}}
  - \frac{
    i  \hat{y}^{u*}_{jk}
    (\tilde{g}^{q}_{\mathcal{L}_1})_{rik}
    (\gamma_{\mathcal{L}_1})^*_{r}}{
    2 M_{\mathcal{L}_{1r}}^{2}}
  + \frac{
    i  \hat{y}^{u*}_{jk}
    (\tilde{g}^{q\prime}_{\mathcal{L}_1})_{rik}
    (\gamma_{\mathcal{L}_1})^*_{r}}{
    2 M_{\mathcal{L}_{1r}}^{2}}
  \nonumber \\ &
  \qquad
  + \frac{
    i  \hat{y}^{u*}_{ki}
    (\tilde{g}^{u}_{\mathcal{L}_1})^*_{rjk}
    (\gamma_{\mathcal{L}_1})_{r}}{
    2 M_{\mathcal{L}_{1r}}^{2}}
  - \frac{
    i  \hat{y}^{u*}_{jk}
    (\tilde{g}^{q}_{\mathcal{L}_1})^*_{rki}
    (\gamma_{\mathcal{L}_1})_{r}}{
    2 M_{\mathcal{L}_{1r}}^{2}}
  + \frac{
    i  \hat{y}^{u*}_{jk}
    (\tilde{g}^{q\prime}_{\mathcal{L}_1})^*_{rki}
    (\gamma_{\mathcal{L}_1})_{r}}{
    2 M_{\mathcal{L}_{1r}}^{2}}
  \Bigg\},
\end{align}

\newpage

\begin{align}
  c^u_{ij} = & \;
  \frac{
    (\kappa_{\mathcal{S}\varphi})^*_{rs}
    (\kappa_{\mathcal{S}})_{r}
    (y^u_{\varphi})_{sij}}{
    M_{\mathcal{S}_r}^{2}
    M_{\varphi_s}^{2}}
  + \frac{
    (\kappa_{\Xi\varphi})^*_{sr}
    (\kappa_{\Xi})_{s}
    (y^u_{\varphi})_{rij}}{
    M_{\varphi_r}^{2}
    M_{\Xi_s}^{2}}
  \nonumber \\ &
  + \frac{
    2   (\kappa_{\Xi_1\varphi})_{rs}
    (\kappa_{\Xi_1})^*_{r}
    (y^u_{\varphi})_{sij}}{
    M_{\Xi_{1r}}^{2}
    M_{\varphi_s}^{2}}
  - \frac{
    (\lambda_{U Q_1})_{rs}
    (\lambda_U)^*_{ri}
    (\lambda^u_{Q_1})_{sj}}{
    M_{U_r}
    M_{Q_{1s}}}
  \nonumber \\ &
  - \frac{
    (\lambda_{U Q_7})_{rs}
    (\lambda_U)^*_{ri}
    (\lambda_{Q_7})_{sj}}{
    M_{U_r}
    M_{Q_{7s}}}
  - \frac{
    (\lambda_{T_1 Q_1})_{sr}
    (\lambda_{T_1})^*_{si}
    (\lambda^u_{Q_1})_{rj}}{
    2  M_{Q_{1r}}
    M_{T_{1s}}}
  \nonumber \\ &
    - \frac{
    (\lambda_{T_2 Q_1})_{sr}
    (\lambda_{T_2})^*_{si}
    (\lambda^u_{Q_1})_{rj}}{
    4  M_{Q_{1r}}
    M_{T_{2s}}}
  + \frac{
    (\lambda_{T_2 Q_7})_{sr}
    (\lambda_{T_2})^*_{si}
    (\lambda_{Q_7})_{rj}}{
    4  M_{Q_{7r}}
    M_{T_{2s}}}
  \nonumber \\ &
  - \frac{
    (w_{\mathcal{S}U})_{rsj}
    (\kappa_{\mathcal{S}})_{r}
    (\lambda_U)^*_{si}}{
    M_{\mathcal{S}_r}^{2}
    M_{U_s}}
  - \frac{
    (w_{\mathcal{S}Q_1})^*_{rsi}
    (\kappa_{\mathcal{S}})_{r}
    (\lambda^u_{Q_1})_{sj}}{
    M_{\mathcal{S}_r}^{2}
    M_{Q_{1s}}}
  \nonumber \\ &
  + \frac{
    (w_{\Xi T_2})_{srj}
    (\kappa_{\Xi})_{s}
    (\lambda_{T_2})^*_{ri}}{
    2  M_{T_{2r}}
    M_{\Xi_s}^{2}}
  + \frac{
    (w_{\Xi Q_7})^*_{rsi}
    (\kappa_{\Xi})_{r}
    (\lambda^u_{Q_1})_{sj}}{
    M_{\Xi_r}^{2}
    M_{Q_{1s}}}
  \nonumber \\ &
  - \frac{
    (w_{\Xi_1 T_1})_{srj}
    (\kappa_{\Xi_1})^*_{s}
    (\lambda_{T_1})^*_{ri}}{
    M_{T_{1r}}
    M_{\Xi_{1s}}^{2}}
  - \frac{
    2   (w_{\Xi_1 Q_7})^*_{rsi}
    (\kappa_{\Xi_1})^*_{r}
    (\lambda_{Q_7})_{sj}}{
    M_{\Xi_{1r}}^{2}
    M_{Q_{7s}}}
  \nonumber \\ &
  + \frac{
    i  \hat{y}^{u*}_{jk}
    (z_{U\mathcal{L}_1})_{rsk}
    (\lambda_U)^*_{ri}
    (\gamma_{\mathcal{L}_1})_{s}}{
    2  M_{U_r}
    M_{\mathcal{L}_{1s}}^{2}}
  + \frac{
    i  \hat{y}^{u*}_{jk}
    (z_{U\mathcal{L}_1})^*_{rsi}
    (\gamma_{\mathcal{L}_1})^*_{s}
    (\lambda_U)_{rk}}{
    2  M_{U_r}
    M_{\mathcal{L}_{1s}}^{2}}
  \nonumber \\ &
  - \frac{
    i  \hat{y}^{u*}_{ki}
    (z^u_{Q_1\mathcal{L}_1})^*_{srk}
    (\gamma_{\mathcal{L}_1})_{r}
    (\lambda^u_{Q_1})_{sj}}{
    2  M_{\mathcal{L}_{1r}}^{2}
    M_{Q_{1s}}}
  - \frac{
    i  \hat{y}^{u*}_{ki}
    (z^u_{Q_1\mathcal{L}_1})_{srj}
    (\lambda^u_{Q_1})^*_{sk}
    (\gamma_{\mathcal{L}_1})^*_{r}}{
    2  M_{\mathcal{L}_{1r}}^{2}
    M_{Q_{1s}}}
  \nonumber \\ &
  - \frac{
    i  \hat{y}^{u*}_{ki}
    (z_{Q_7\mathcal{L}_1})^*_{rsk}
    (\gamma_{\mathcal{L}_1})^*_{s}
    (\lambda_{Q_7})_{rj}}{
    2  M_{Q_{7r}}
    M_{\mathcal{L}_{1s}}^{2}}
  - \frac{
    i  \hat{y}^{u*}_{ki}
    (z_{Q_7\mathcal{L}_1})_{rsj}
    (\lambda_{Q_7})^*_{rk}
    (\gamma_{\mathcal{L}_1})_{s}}{
    2  M_{Q_{7r}}
    M_{\mathcal{L}_{1s}}^{2}}
  \nonumber \\ &
  + \frac{
    i  \hat{y}^{u*}_{jk}
    (z_{T_1\mathcal{L}_1})_{rsk}
    (\lambda_{T_1})^*_{ri}
    (\gamma_{\mathcal{L}_1})^*_{s}}{
    2  M_{T_{1r}}
    M_{\mathcal{L}_{1s}}^{2}}
  + \frac{
    i  \hat{y}^{u*}_{jk}
    (z_{T_1\mathcal{L}_1})^*_{sri}
    (\gamma_{\mathcal{L}_1})_{r}
    (\lambda_{T_1})_{sk}}{
    2  M_{\mathcal{L}_{1r}}^{2}
    M_{T_{1s}}}
  \nonumber \\ &
  + \frac{
    i  \hat{y}^{u*}_{jk}
    (z_{T_2\mathcal{L}_1})^*_{sri}
    (\gamma_{\mathcal{L}_1})^*_{r}
    (\lambda_{T_2})_{sk}}{
    4  M_{\mathcal{L}_{1r}}^{2}
    M_{T_{2s}}}
  + \frac{
    i  \hat{y}^{u*}_{jk}
    (z_{T_2\mathcal{L}_1})_{rsk}
    (\lambda_{T_2})^*_{ri}
    (\gamma_{\mathcal{L}_1})_{s}}{
    4  M_{T_{2r}}
    M_{\mathcal{L}_{1s}}^{2}}
  \nonumber \\ &
  + \frac{
    i \hat{y}^{u*}_{jk}
    (\delta_{\mathcal{B}\mathcal{S}})_{rs}
    (g^{q}_{\mathcal{B}})_{rik}
    (\kappa_{\mathcal{S}})_{s}}{
    M_{\mathcal{B}_r}^{2}
    M_{\mathcal{S}_s}^{2}}
  - \frac{
    i  \hat{y}^{u*}_{ki}
    (\delta_{\mathcal{B}\mathcal{S}})_{rs}
    (g^u_{\mathcal{B}})_{rkj}
    (\kappa_{\mathcal{S}})_{s}}{
    M_{\mathcal{B}_r}^{2}
    M_{\mathcal{S}_s}^{2}}
  \nonumber \\ &
  - \frac{
    i  \hat{y}^{u*}_{jk}
    (\delta_{\mathcal{W}\Xi})_{sr}
    (g^{q}_{\mathcal{W}})_{sik}
    (\kappa_{\Xi})_{r}}{
    2  M_{\Xi_r}^{2}
    M_{\mathcal{W}_s}^{2}}
  - \frac{
    2   \hat{\lambda}_{\phi}
    (\delta_{\mathcal{L}_1\varphi})_{sr}
    (\gamma_{\mathcal{L}_1})^*_{s}
    (y^u_{\varphi})_{rij}}{
    M_{\varphi_r}^{2}
    M_{\mathcal{L}_{1s}}^{2}}.
  \label{eq:cu}
\end{align}

\newpage

\subsubsection{$X \psi^2 \phi$}
\label{sec:Xpsi2phi}

\begin{align}
  Z_\phi^{\frac{1}{2}}~\!\left(C_{eB}\right)_{ij}= &
  \frac{1}{f} \left\{
    \frac{
      (\tilde{\lambda}^B_{E})_{rj}
      (\lambda_E)^*_{ri}}{
      M_{E_r}}
    + \frac{
      (\tilde{\lambda}^B_{\Delta_1})^*_{ri}
      (\lambda_{\Delta_1})_{rj}}{
      M_{\Delta_{1r}}}
    - \frac{
      g_1
      (\tilde{g}^{eDl}_{\mathcal{L}_1})^*_{rji}
      (\gamma_{\mathcal{L}_1})_{r}}{
      8 M_{\mathcal{L}_{1r}}^{2}}
    + \frac{
      g_1
      (\tilde{g}^{Del}_{\mathcal{L}_1})^*_{rji}
      (\gamma_{\mathcal{L}_1})_{r}}{
      8 M_{\mathcal{L}_{1r}}^{2}}
  \right\},
  \\
  &\nonumber\\
  Z_\phi^{\frac{1}{2}}~\!\left(C_{eW}\right)_{ij}= &
  \frac{1}{f} \left\{
    \frac{
      (\tilde{\lambda}^W_{\Delta_1})^*_{ri}
      (\lambda_{\Delta_1})_{rj}}{
      M_{\Delta_{1r}}}
    + \frac{
      (\tilde{\lambda}^W_{\Sigma_1})_{rj}
      (\lambda_{\Sigma_1})^*_{ri}}{
      2 M_{\Sigma_{1r}}}
    - \frac{
      g_2
      (\tilde{g}^{eDl}_{\mathcal{L}_1})^*_{rji}
      (\gamma_{\mathcal{L}_1})_{r}}{
      8 M_{\mathcal{L}_{1r}}^{2}}
    + \frac{
      g_2
      (\tilde{g}^{Del}_{\mathcal{L}_1})^*_{rji}
      (\gamma_{\mathcal{L}_1})_{r}}{
      8 M_{\mathcal{L}_{1r}}^{2}}
  \right\},
  \\
  &\nonumber\\
  Z_\phi^{\frac{1}{2}}~\!\left(C_{dB}\right)_{ij}= &
  \frac{1}{f} \left\{
    \frac{
      (\tilde{\lambda}^B_{D})_{rj}
      (\lambda_D)^*_{ri}}{
      M_{D_r}}
    + \frac{
      (\tilde{\lambda}^B_{Q_1})^*_{ri}
      (\lambda^d_{Q_1})_{rj}}{
      M_{Q_{1r}}}
    - \frac{
      g_1
      (\tilde{g}^{dDq}_{\mathcal{L}_1})^*_{rji}
      (\gamma_{\mathcal{L}_1})_{r}}{
      8 M_{\mathcal{L}_{1r}}^{2}}
    + \frac{
      g_1
      (\tilde{g}^{Ddq}_{\mathcal{L}_1})^*_{rji}
      (\gamma_{\mathcal{L}_1})_{r}}{
      8 M_{\mathcal{L}_{1r}}^{2}}
  \right\},
  \\
  &\nonumber\\
  Z_\phi^{\frac{1}{2}}~\!\left(C_{dW}\right)_{ij}= &
  \frac{1}{f} \left\{
    \frac{
      (\tilde{\lambda}^W_{Q_1})^*_{ri}
      (\lambda^d_{Q_1})_{rj}}{
      M_{Q_{1r}}}
    + \frac{
      (\tilde{\lambda}^W_{T_1})_{rj}
      (\lambda_{T_1})^*_{ri}}{
      2 M_{T_{1r}}}
    - \frac{
      g_2
      (\tilde{g}^{dDq}_{\mathcal{L}_1})^*_{rji}
      (\gamma_{\mathcal{L}_1})_{r}}{
      8 M_{\mathcal{L}_{1r}}^{2}}
    + \frac{
      g_2
      (\tilde{g}^{Ddq}_{\mathcal{L}_1})^*_{rji}
      (\gamma_{\mathcal{L}_1})_{r}}{
      8 M_{\mathcal{L}_{1r}}^{2}}
  \right\},
  \\
  &\nonumber\\
  Z_\phi^{\frac{1}{2}}~\!\left(C_{dG}\right)_{ij}= &
  \frac{1}{f} \left\{
    \frac{
      (\tilde{\lambda}^G_{D})_{rj}
      (\lambda_D)^*_{ri}}{
      M_{D_r}}
    + \frac{
      (\tilde{\lambda}^G_{Q_1})^*_{ri}
      (\lambda^d_{Q_1})_{rj}}{
      M_{Q_{1r}}}
  \right\},
  \\
  &\nonumber\\
  Z_\phi^{\frac{1}{2}}~\!\left(C_{uB}\right)_{ij}= &
  \frac{1}{f} \left\{
    \frac{
      (\tilde{\lambda}^B_{U})_{rj}
      (\lambda_U)^*_{ri}}{
      M_{U_r}}
    + \frac{
      (\tilde{\lambda}^B_{Q_1})^*_{ri}
      (\lambda^u_{Q_1})_{rj}}{
      M_{Q_{1r}}}
    + \frac{
      g_1
      (\tilde{g}^{qDu}_{\mathcal{L}_1})_{rji}
      (\gamma_{\mathcal{L}_1})^*_{r}}{
      8 M_{\mathcal{L}_{1r}}^{2}}
    - \frac{
      g_1
      (\tilde{g}^{Dqu}_{\mathcal{L}_1})_{rji}
      (\gamma_{\mathcal{L}_1})^*_{r}}{
      8 M_{\mathcal{L}_{1r}}^{2}}
  \right\},
  \\
  &\nonumber\\
  Z_\phi^{\frac{1}{2}}~\!\left(C_{uW}\right)_{ij}= &
  \frac{1}{f} \left\{
    \frac{
      (\tilde{\lambda}^W_{Q_1})^*_{ri}
      (\lambda^u_{Q_1})_{rj}}{
      M_{Q_{1r}}}
    + \frac{
      (\tilde{\lambda}^W_{T_2})_{rj}
      (\lambda_{T_2})^*_{ri}}{
      2 M_{T_{2r}}}
    + \frac{
      g_2
      (\tilde{g}^{qDu}_{\mathcal{L}_1})_{rji}
      (\gamma_{\mathcal{L}_1})^*_{r}}{
      8 M_{\mathcal{L}_{1r}}^{2}}
    - \frac{
      g_2
      (\tilde{g}^{Dqu}_{\mathcal{L}_1})_{rji}
      (\gamma_{\mathcal{L}_1})^*_{r}}{
      8 M_{\mathcal{L}_{1r}}^{2}}
  \right\},
  \\
  &\nonumber\\
  Z_\phi^{\frac{1}{2}}~\!\left(C_{uG}\right)_{ij}= &
  \frac{1}{f} \left\{
    \frac{
      (\tilde{\lambda}^G_{U})_{rj}
      (\lambda_U)^*_{ri}}{
      M_{U_r}}
    + \frac{
      (\tilde{\lambda}^G_{Q_1})^*_{ri}
      (\lambda^u_{Q_1})_{rj}}{
      M_{Q_{1r}}}
  \right\}.
\end{align}

\newpage

\subsubsection{$\psi^2 \phi^2 D$}
\label{sec:psi2phi2D}

Recall that $\hat{g}^\phi_V$ contains contributions from $\mathcal{L}_1$ (see
equations \refeq{eq:ghat_first}--\refeq{eq:ghat_last}) and that
$\hat{y}^{e,u,d}$ are defined in equation \refeq{eq:y}.

\begin{align}
  Z_\phi~\!\left(C^{(1)}_{\phi l}\right)_{ij}= &
  \frac{
    (\lambda_N)^*_{ri}
    (\lambda_N)_{rj}}{
    4  M_{N_r}^{2}}
  - \frac{
    (\lambda_E)_{rj}
    (\lambda_E)^*_{ri}}{
    4  M_{E_r}^{2}}
  + \frac{
    3   (\lambda_{\Sigma})^*_{ri}
    (\lambda_{\Sigma})_{rj}}{
    16  M_{\Sigma_r}^{2}}
  - \frac{
    3   (\lambda_{\Sigma_1})_{rj}
    (\lambda_{\Sigma_1})^*_{ri}}{
    16  M_{\Sigma_{1r}}^{2}}
  \nonumber \\ &
  - \frac{
    \operatorname{Re}{\left(
        (\hat{g}^\phi_{\mathcal{B}})_{r}
      \right)}
    (g^l_{\mathcal{B}})_{rij}}{
    M_{\mathcal{B}_r}^{2}}
  - \frac{
    g_1
    \delta_{ij}
    (g^B_{\mathcal{L}_1})_{rs}
    (\gamma_{\mathcal{L}_1})^*_{r}
    (\gamma_{\mathcal{L}_1})_{s}}{
    4  M_{\mathcal{L}_{1r}}^{2}
    M_{\mathcal{L}_{1s}}^{2}}
  \nonumber \\ &
  + \frac{
    i  (\lambda_N)_{rj}
    (z_{N\mathcal{L}_1})^*_{rsi}
    (\gamma_{\mathcal{L}_1})^*_{s}}{
    4  M_{N_r}
    M_{\mathcal{L}_{1s}}^{2}}
  - \frac{
    i  (\lambda_N)^*_{ri}
    (z_{N\mathcal{L}_1})_{rsj}
    (\gamma_{\mathcal{L}_1})_{s}}{
    4  M_{N_r}
    M_{\mathcal{L}_{1s}}^{2}}
  \nonumber \\ &
  - \frac{
    i  (z_{E\mathcal{L}_1})^*_{rsi}
    (\gamma_{\mathcal{L}_1})_{s}
    (\lambda_E)_{rj}}{
    4  M_{E_r}
    M_{\mathcal{L}_{1s}}^{2}}
  + \frac{
    i  (z_{E\mathcal{L}_1})_{rsj}
    (\lambda_E)^*_{ri}
    (\gamma_{\mathcal{L}_1})^*_{s}}{
    4  M_{E_r}
    M_{\mathcal{L}_{1s}}^{2}}
  \nonumber \\ &
  + \frac{
    3  i  (\lambda_{\Sigma})_{sj}
    (z_{\Sigma\mathcal{L}_1})^*_{sri}
    (\gamma_{\mathcal{L}_1})^*_{r}}{
    8  M_{\mathcal{L}_{1r}}^{2}
    M_{\Sigma_s}}
  - \frac{
    3  i  (\lambda_{\Sigma})^*_{ri}
    (z_{\Sigma\mathcal{L}_1})_{rsj}
    (\gamma_{\mathcal{L}_1})_{s}}{
    8  M_{\Sigma_r}
    M_{\mathcal{L}_{1s}}^{2}}
  \nonumber \\ &
  + \frac{
    3  i  (z_{\Sigma_1\mathcal{L}_1})_{srj}
    (\lambda_{\Sigma_1})^*_{si}
    (\gamma_{\mathcal{L}_1})^*_{r}}{
    8  M_{\mathcal{L}_{1r}}^{2}
    M_{\Sigma_{1s}}}
  - \frac{
    3  i  (z_{\Sigma_1\mathcal{L}_1})^*_{rsi}
    (\gamma_{\mathcal{L}_1})_{s}
    (\lambda_{\Sigma_1})_{rj}}{
    8  M_{\Sigma_{1r}}
    M_{\mathcal{L}_{1s}}^{2}}
  \nonumber \\ &
  + \frac{1}{f} \Bigg\{
  \frac{
    i  (\tilde{\lambda}_{N})^*_{ri}
    (\lambda_N)_{rj}}{
    4 M_{N_r}}  
  - \frac{
    i  (\tilde{\lambda}_{N})_{rj}
    (\lambda_N)^*_{ri}}{
    4  M_{N_r}}
  \nonumber \\ &
  \qquad
  + \frac{
    i  (\tilde{\lambda}^l_{E})_{rj}
    (\lambda_E)^*_{ri}}{
    4 M_{E_r}}
  - \frac{
    i  (\tilde{\lambda}^l_{E})^*_{ri}
    (\lambda_E)_{rj}}{
    4 M_{E_r}}
  \nonumber \\ &
  \qquad
  + \frac{
    3  i  (\tilde{\lambda}^l_{\Sigma})^*_{ri}
    (\lambda_{\Sigma})_{rj}}{
    8 M_{\Sigma_r}}
  - \frac{
    3  i  (\tilde{\lambda}^l_{\Sigma})_{rj}
    (\lambda_{\Sigma_0})^*_{ri}}{
    8 M_{\Sigma_r}}
  \nonumber \\ &
  \qquad
  + \frac{
    3  i  (\tilde{\lambda}^l_{\Sigma_1})_{rj}
    (\lambda_{\Sigma_1})^*_{ri}}{
    8 M_{\Sigma_{1r}}}
  - \frac{
    3  i  (\tilde{\lambda}^l_{\Sigma_1})^*_{ri}
    (\lambda_{\Sigma_1})_{rj}}{
    8 M_{\Sigma_{1r}}}
  \nonumber \\ &
  \qquad
  - \frac{
    \hat{y}^{e*}_{ki}
    (\tilde{g}^{eDl}_{\mathcal{L}_1})_{rkj}
    (\gamma_{\mathcal{L}_1})^*_{r}}{
    8 M_{\mathcal{L}_{1r}}^{2}}
  - \frac{
    \hat{y}^e_{kj}
    (\tilde{g}^{eDl}_{\mathcal{L}_1})^*_{rki}
    (\gamma_{\mathcal{L}_1})_{r}}{
    8 M_{\mathcal{L}_{1r}}^{2}}
  \nonumber \\ &
  \qquad
  + \frac{
    \hat{y}^{e*}_{ki}
    (\tilde{g}^{Del}_{\mathcal{L}_1})_{rkj}
    (\gamma_{\mathcal{L}_1})^*_{r}}{
    8 M_{\mathcal{L}_{1r}}^{2}}
  + \frac{
    \hat{y}^e_{kj}
    (\tilde{g}^{Del}_{\mathcal{L}_1})^*_{rki}
    (\gamma_{\mathcal{L}_1})_{r}}{
    8 M_{\mathcal{L}_{1r}}^{2}}
  \nonumber \\ &
  \qquad
  - \frac{
    i  (\tilde{g}^{l}_{\mathcal{L}_1})_{rij}
    (\gamma_{\mathcal{L}_1})^*_{r}}{
    2 M_{\mathcal{L}_{1r}}^{2}}
  + \frac{
    i  (\tilde{g}^{l}_{\mathcal{L}_1})^*_{rji}
    (\gamma_{\mathcal{L}_1})_{r}}{
    2 M_{\mathcal{L}_{1r}}^{2}}
  \nonumber \\ &
  \qquad
    - \frac{
    \operatorname{Im}{\left(
        (\tilde{\gamma}^B_{\mathcal{L}_1})_{r}
        (\gamma_{\mathcal{L}_1})^*_{r}
      \right)}
    g_1
    \delta_{ij}}{
    2 M_{\mathcal{L}_{1r}}^{2}}
  \Bigg\},
\end{align}

\newpage

\begin{align}
  Z_\phi~\!\left(C^{(3)}_{\phi l}\right)_{ij}= &
   - \frac{
    (\lambda_N)^*_{ri}
    (\lambda_N)_{rj}}{
    4  M_{N_r}^{2}}
  - \frac{
    (\lambda_E)_{rj}
    (\lambda_E)^*_{ri}}{
    4  M_{E_r}^{2}}
  + \frac{
    (\lambda_{\Sigma})^*_{ri}
    (\lambda_{\Sigma})_{rj}}{
    16  M_{\Sigma_r}^{2}}
  + \frac{
    (\lambda_{\Sigma_1})_{rj}
    (\lambda_{\Sigma_1})^*_{ri}}{
    16  M_{\Sigma_{1r}}^{2}}
  \nonumber \\ &
  - \frac{
    \operatorname{Re}{\left(
        (\hat{g}^\phi_{\mathcal{W}})_{r}
      \right)}
    (g^l_{\mathcal{W}})_{rij}}{
    4  M_{\mathcal{W}_r}^{2}}
  + \frac{
    g_2
    \delta_{ij}
    (g^W_{\mathcal{L}_1})_{rs}
    (\gamma_{\mathcal{L}_1})^*_{r}
    (\gamma_{\mathcal{L}_1})_{s}}{
    4  M_{\mathcal{L}_{1r}}^{2}
    M_{\mathcal{L}_{1s}}^{2}}
 \nonumber \\ &
  - \frac{
    i  (\lambda_N)_{rj}
    (z_{N\mathcal{L}_1})^*_{rsi}
    (\gamma_{\mathcal{L}_1})^*_{s}}{
    4  M_{N_r}
    M_{\mathcal{L}_{1s}}^{2}}
  + \frac{
    i  (\lambda_N)^*_{ri}
    (z_{N\mathcal{L}_1})_{rsj}
    (\gamma_{\mathcal{L}_1})_{s}}{
    4  M_{N_r}
    M_{\mathcal{L}_{1s}}^{2}}
  \nonumber \\ &
  + \frac{
    i  (z_{E\mathcal{L}_1})_{rsj}
    (\lambda_E)^*_{ri}
    (\gamma_{\mathcal{L}_1})^*_{s}}{
    4  M_{E_r}
    M_{\mathcal{L}_{1s}}^{2}}
  - \frac{
    i  (z_{E\mathcal{L}_1})^*_{rsi}
    (\gamma_{\mathcal{L}_1})_{s}
    (\lambda_E)_{rj}}{
    4  M_{E_r}
    M_{\mathcal{L}_{1s}}^{2}}
  \nonumber \\ &
  + \frac{
    i  (\lambda_{\Sigma})_{sj}
    (z_{\Sigma\mathcal{L}_1})^*_{sri}
    (\gamma_{\mathcal{L}_1})^*_{r}}{
    8  M_{\mathcal{L}_{1r}}^{2}
    M_{\Sigma_s}}
  - \frac{
    i  (\lambda_{\Sigma})^*_{ri}
    (z_{\Sigma\mathcal{L}_1})_{rsj}
    (\gamma_{\mathcal{L}_1})_{s}}{
    8  M_{\Sigma_r}
    M_{\mathcal{L}_{1s}}^{2}}
  \nonumber \\ &
  - \frac{
    i  (z_{\Sigma_1\mathcal{L}_1})_{srj}
    (\lambda_{\Sigma_1})^*_{si}
    (\gamma_{\mathcal{L}_1})^*_{r}}{
    8  M_{\mathcal{L}_{1r}}^{2}
    M_{\Sigma_{1s}}}
  + \frac{
    i  (z_{\Sigma_1\mathcal{L}_1})^*_{rsi}
    (\gamma_{\mathcal{L}_1})_{s}
    (\lambda_{\Sigma_1})_{rj}}{
    8  M_{\Sigma_{1r}}
    M_{\mathcal{L}_{1s}}^{2}}
  \nonumber \\ &
  + \frac{1}{f} \Bigg\{
   - \frac{
    i  (\tilde{\lambda}_{N})^*_{ri}
    (\lambda_N)_{rj}}{
    4 M_{N_r}}
  + \frac{
    i  (\tilde{\lambda}_{N})_{rj}
    (\lambda_N)^*_{ri}}{
    4  M_{N_r}}
  \nonumber \\ &
  \qquad
  + \frac{
    i  (\tilde{\lambda}^l_{E})_{rj}
    (\lambda_E)^*_{ri}}{
    4 M_{E_r}}
  - \frac{
    i  (\tilde{\lambda}^l_{E})^*_{ri}
    (\lambda_E)_{rj}}{
    4 M_{E_r}}
  \nonumber \\ &
  \qquad
  + \frac{
    i  (\tilde{\lambda}^l_{\Sigma})^*_{ri}
    (\lambda_{\Sigma})_{rj}}{
    8 M_{\Sigma_r}}
  - \frac{
    i  (\tilde{\lambda}^l_{\Sigma})_{rj}
    (\lambda_{\Sigma})^*_{ri}}{
    8 M_{\Sigma_r}}
  \nonumber \\ &
  \qquad
  - \frac{
    i  (\tilde{\lambda}^l_{\Sigma_1})_{rj}
    (\lambda_{\Sigma_1})^*_{ri}}{
    8 M_{\Sigma_{1r}}}
  + \frac{
    i  (\tilde{\lambda}^l_{\Sigma_1})^*_{ri}
    (\lambda_{\Sigma_1})_{rj}}{
    8 M_{\Sigma_{1r}}}
  \nonumber \\ &
  \qquad
  - \frac{
    \hat{y}^{e*}_{ki}
    (\tilde{g}^{eDl}_{\mathcal{L}_1})_{rkj}
    (\gamma_{\mathcal{L}_1})^*_{r}}{
    8 M_{\mathcal{L}_{1r}}^{2}}
  - \frac{
    \hat{y}^e_{kj}
    (\tilde{g}^{eDl}_{\mathcal{L}_1})^*_{rki}
    (\gamma_{\mathcal{L}_1})_{r}}{
    8 M_{\mathcal{L}_{1r}}^{2}}
  \nonumber \\ &
  \qquad
  + \frac{
    \hat{y}^{e*}_{ki}
    (\tilde{g}^{Del}_{\mathcal{L}_1})_{rkj}
    (\gamma_{\mathcal{L}_1})^*_{r}}{
    8 M_{\mathcal{L}_{1r}}^{2}}
  + \frac{
    \hat{y}^e_{kj}
    (\tilde{g}^{Del}_{\mathcal{L}_1})^*_{rki}
    (\gamma_{\mathcal{L}_1})_{r}}{
    8 M_{\mathcal{L}_{1r}}^{2}}
  \nonumber \\ &
  \qquad
  - \frac{
    i  (\tilde{g}^{l\prime}_{\mathcal{L}_1})_{rij}
    (\gamma_{\mathcal{L}_1})^*_{r}}{
    2 M_{\mathcal{L}_{1r}}^{2}}
  + \frac{
    i  (\tilde{g}^{l\prime}_{\mathcal{L}_1})^*_{rji}
    (\gamma_{\mathcal{L}_1})_{r}}{
    2 M_{\mathcal{L}_{1r}}^{2}}
  \nonumber \\ &
  \qquad
  + \frac{
    \operatorname{Im}{\left(
        (\tilde{\gamma}^W_{\mathcal{L}_1})_{r}
        (\gamma_{\mathcal{L}_1})^*_{r}
      \right)}
    g_2
    \delta_{ij}}{
    2 M_{\mathcal{L}_{1r}}^{2}}
  \Bigg\},
\end{align}

\newpage

\begin{align}
  Z_\phi~\!\left(C^{(1)}_{\phi q}\right)_{ij}= &
  \frac{
    (\lambda_U)_{rj}
    (\lambda_U)^*_{ri}}{
    4  M_{U_r}^{2}}
  - \frac{
    (\lambda_D)_{rj}
    (\lambda_D)^*_{ri}}{
    4  M_{D_r}^{2}}
  - \frac{
    3   (\lambda_{T_1})_{rj}
    (\lambda_{T_1})^*_{ri}}{
    16  M_{T_{1r}}^{2}}
  + \frac{
    3   (\lambda_{T_2})_{rj}
    (\lambda_{T_2})^*_{ri}}{
    16  M_{T_{2r}}^{2}}
  \nonumber \\ &
  - \frac{
    \operatorname{Re}{\left(
        (\hat{g}^\phi_{\mathcal{B}})_{r}
      \right)}
    (g^q_{\mathcal{B}})_{rij}}{
    M_{\mathcal{B}_r}^{2}}
  + \frac{
    g_1
    \delta_{ij}
    (g^B_{\mathcal{L}_1})_{rs}
    (\gamma_{\mathcal{L}_1})^*_{r}
    (\gamma_{\mathcal{L}_1})_{s}}{
    12  M_{\mathcal{L}_{1r}}^{2}
    M_{\mathcal{L}_{1s}}^{2}}
  \nonumber \\ &
  - \frac{
    i  (z_{U\mathcal{L}_1})_{rsj}
    (\lambda_U)^*_{ri}
    (\gamma_{\mathcal{L}_1})_{s}}{
    4  M_{U_r}
    M_{\mathcal{L}_{1s}}^{2}}
  + \frac{
    i  (z_{U\mathcal{L}_1})^*_{rsi}
    (\gamma_{\mathcal{L}_1})^*_{s}
    (\lambda_U)_{rj}}{
    4  M_{U_r}
    M_{\mathcal{L}_{1s}}^{2}}
  \nonumber \\ &
  + \frac{
    i  (z_{D\mathcal{L}_1})_{rsj}
    (\lambda_D)^*_{ri}
    (\gamma_{\mathcal{L}_1})^*_{s}}{
    4  M_{D_r}
    M_{\mathcal{L}_{1s}}^{2}}
  - \frac{
    i  (z_{D\mathcal{L}_1})^*_{rsi}
    (\gamma_{\mathcal{L}_1})_{s}
    (\lambda_D)_{rj}}{
    4  M_{D_r}
    M_{\mathcal{L}_{1s}}^{2}}
  \nonumber \\ &
    + \frac{
    3  i  (z_{T_1\mathcal{L}_1})_{rsj}
    (\lambda_{T_1})^*_{ri}
    (\gamma_{\mathcal{L}_1})^*_{s}}{
    8  M_{T_{1r}}
    M_{\mathcal{L}_{1s}}^{2}}
  - \frac{
    3  i  (z_{T_1\mathcal{L}_1})^*_{sri}
    (\gamma_{\mathcal{L}_1})_{r}
    (\lambda_{T_1})_{sj}}{
    8  M_{\mathcal{L}_{1r}}^{2}
    M_{T_{1s}}}
  \nonumber \\ &
    - \frac{
    3  i  (z_{T_2\mathcal{L}_1})_{rsj}
    (\lambda_{T_2})^*_{ri}
    (\gamma_{\mathcal{L}_1})_{s}}{
    8  M_{T_{2r}}
    M_{\mathcal{L}_{1s}}^{2}}
  + \frac{
    3  i  (z_{T_2\mathcal{L}_1})^*_{sri}
    (\gamma_{\mathcal{L}_1})^*_{r}
    (\lambda_{T_2})_{sj}}{
    8  M_{\mathcal{L}_{1r}}^{2}
    M_{T_{2s}}}
  \nonumber \\ &
  + \frac{1}{f} \Bigg\{
  - \frac{
    i  (\tilde{\lambda}^q_{U})_{rj}
    (\lambda_U)^*_{ri}}{
    4 M_{U_r}}
  + \frac{
    i  (\tilde{\lambda}^q_{U})^*_{ri}
    (\lambda_U)_{rj}}{
    4  M_{U_r}}
  \nonumber \\ &
  \qquad
  + \frac{
    i  (\tilde{\lambda}^q_{D})_{rj}
    (\lambda_D)^*_{ri}}{
    4 M_{D_r}}
  - \frac{
    i  (\tilde{\lambda}^q_{D})^*_{ri}
    (\lambda_D)_{rj}}{
    4 M_{D_r}}
  \nonumber \\ &
  \qquad
  + \frac{
    3  i  (\tilde{\lambda}^q_{T_1})_{rj}
    (\lambda_{T_1})^*_{ri}}{
    8 M_{T_{1r}}}
  - \frac{
    3  i  (\tilde{\lambda}^q_{T_1})^*_{ri}
    (\lambda_{T_1})_{rj}}{
    8 M_{T_{1r}}}
  \nonumber \\ &
  \qquad
  - \frac{
    3  i  (\tilde{\lambda}^q_{T_2})_{rj}
    (\lambda_{T_2})^*_{ri}}{
    8 M_{T_{2r}}}
  + \frac{
    3  i  (\tilde{\lambda}^q_{T_2})^*_{ri}
    (\lambda_{T_2})_{rj}}{
    8 M_{T_{2r}}}
  \nonumber \\ &
  \qquad
  - \frac{
    \hat{y}^{d*}_{ki}
    (\tilde{g}^{dDq}_{\mathcal{L}_1})_{rkj}
    (\gamma_{\mathcal{L}_1})^*_{r}}{
    8 M_{\mathcal{L}_{1r}}^{2}}
  - \frac{
    \hat{y}^d_{kj}
    (\tilde{g}^{dDq}_{\mathcal{L}_1})^*_{rki}
    (\gamma_{\mathcal{L}_1})_{r}}{
    8 M_{\mathcal{L}_{1r}}^{2}}
  \nonumber \\ &
  \qquad
  + \frac{
    \hat{y}^{d*}_{ki}
    (\tilde{g}^{Ddq}_{\mathcal{L}_1})_{rkj}
    (\gamma_{\mathcal{L}_1})^*_{r}}{
    8 M_{\mathcal{L}_{1r}}^{2}}
  + \frac{
    \hat{y}^d_{kj}
    (\tilde{g}^{Ddq}_{\mathcal{L}_1})^*_{rki}
    (\gamma_{\mathcal{L}_1})_{r}}{
    8 M_{\mathcal{L}_{1r}}^{2}}
  \nonumber \\ &
  \qquad
  - \frac{
    \hat{y}^u_{kj}
    (\tilde{g}^{qDu}_{\mathcal{L}_1})_{rki}
    (\gamma_{\mathcal{L}_1})^*_{r}}{
    8 M_{\mathcal{L}_{1r}}^{2}}
  - \frac{
    \hat{y}^{u*}_{ki}
    (\tilde{g}^{qDu}_{\mathcal{L}_1})^*_{rkj}
    (\gamma_{\mathcal{L}_1})_{r}}{
    8 M_{\mathcal{L}_{1r}}^{2}}
  \nonumber \\ &
  \qquad
  + \frac{
    \hat{y}^u_{kj}
    (\tilde{g}^{Dqu}_{\mathcal{L}_1})_{rki}
    (\gamma_{\mathcal{L}_1})^*_{r}}{
    8 M_{\mathcal{L}_{1r}}^{2}}
  + \frac{
    \hat{y}^{u*}_{ki}
    (\tilde{g}^{Dqu}_{\mathcal{L}_1})^*_{rkj}
    (\gamma_{\mathcal{L}_1})_{r}}{
    8 M_{\mathcal{L}_{1r}}^{2}}
  \nonumber \\ &
  \qquad
  - \frac{
    i  (\tilde{g}^{q}_{\mathcal{L}_1})_{rij}
    (\gamma_{\mathcal{L}_1})^*_{r}}{
    2 M_{\mathcal{L}_{1r}}^{2}}
  + \frac{
    i  (\tilde{g}^{q}_{\mathcal{L}_1})^*_{rji}
    (\gamma_{\mathcal{L}_1})_{r}}{
    2 M_{\mathcal{L}_{1r}}^{2}}
  \nonumber \\ &
  \qquad
  + \frac{
    \operatorname{Im}{\left(
        (\tilde{\gamma}^B_{\mathcal{L}_1})_{r}
        (\gamma_{\mathcal{L}_1})^*_{r}
      \right)}
    g_1
    \delta_{ij}}{
    6
    M_{\mathcal{L}_{1r}}^{2}}
  \Bigg\},
\end{align}

\newpage

\begin{align}
  Z_\phi~\!\left(C^{(3)}_{\phi q}\right)_{ij}= &
  - \frac{
    (\lambda_U)_{rj}
    (\lambda_U)^*_{ri}}{
    4  M_{U_r}^{2}}
  - \frac{
    (\lambda_D)_{rj}
    (\lambda_D)^*_{ri}}{
    4  M_{D_r}^{2}}
  + \frac{
    (\lambda_{T_1})_{rj}
    (\lambda_{T_1})^*_{ri}}{
    16  M_{T_{1r}}^{2}}
  + \frac{
    (\lambda_{T_2})_{rj}
    (\lambda_{T_2})^*_{ri}}{
    16  M_{T_{2r}}^{2}}
  \nonumber \\ &
  - \frac{
    \operatorname{Re}{\left(
        (\hat{g}^\phi_{\mathcal{W}})_{r}
      \right)}
    (g^q_{\mathcal{W}})_{rij}}{
    4  M_{\mathcal{W}_r}^{2}}
  + \frac{
    g_2
    \delta_{ij}
    (g^W_{\mathcal{L}_1})_{rs}
    (\gamma_{\mathcal{L}_1})^*_{r}
    (\gamma_{\mathcal{L}_1})_{s}}{
    4  M_{\mathcal{L}_{1r}}^{2}
    M_{\mathcal{L}_{1s}}^{2}}
  \nonumber \\ &
  + \frac{
    i  (z_{U\mathcal{L}_1})_{rsj}
    (\lambda_U)^*_{ri}
    (\gamma_{\mathcal{L}_1})_{s}}{
    4  M_{U_r}
    M_{\mathcal{L}_{1s}}^{2}}
  - \frac{
    i  (z_{U\mathcal{L}_1})^*_{rsi}
    (\gamma_{\mathcal{L}_1})^*_{s}
    (\lambda_U)_{rj}}{
    4  M_{U_r}
    M_{\mathcal{L}_{1s}}^{2}}
  \nonumber \\ &
  + \frac{
    i  (z_{D\mathcal{L}_1})_{rsj}
    (\lambda_D)^*_{ri}
    (\gamma_{\mathcal{L}_1})^*_{s}}{
    4  M_{D_r}
    M_{\mathcal{L}_{1s}}^{2}}
  - \frac{
    i  (z_{D\mathcal{L}_1})^*_{rsi}
    (\gamma_{\mathcal{L}_1})_{s}
    (\lambda_D)_{rj}}{
    4  M_{D_r}
    M_{\mathcal{L}_{1s}}^{2}}
  \nonumber \\ &
  - \frac{
    i  (z_{T_1\mathcal{L}_1})_{rsj}
    (\lambda_{T_1})^*_{ri}
    (\gamma_{\mathcal{L}_1})^*_{s}}{
    8  M_{T_{1r}}
    M_{\mathcal{L}_{1s}}^{2}}
  + \frac{
    i  (z_{T_1\mathcal{L}_1})^*_{sri}
    (\gamma_{\mathcal{L}_1})_{r}
    (\lambda_{T_1})_{sj}}{
    8  M_{\mathcal{L}_{1r}}^{2}
    M_{T_{1s}}}
  \nonumber \\ &
  - \frac{
    i  (z_{T_2\mathcal{L}_1})_{rsj}
    (\lambda_{T_2})^*_{ri}
    (\gamma_{\mathcal{L}_1})_{s}}{
    8  M_{T_{2r}}
    M_{\mathcal{L}_{1s}}^{2}}
  + \frac{
    i  (z_{T_2\mathcal{L}_1})^*_{sri}
    (\gamma_{\mathcal{L}_1})^*_{r}
    (\lambda_{T_2})_{sj}}{
    8  M_{\mathcal{L}_{1r}}^{2}
    M_{T_{2s}}}
  \nonumber \\ &
  + \frac{1}{f} \Bigg\{
  \frac{
    i  (\tilde{\lambda}^q_{U})_{rj}
    (\lambda_U)^*_{ri}}{
    4 M_{U_r}}
  - \frac{
    i  (\tilde{\lambda}^q_{U})^*_{ri}
    (\lambda_U)_{rj}}{
    4  M_{U_r}}
  \nonumber \\ &
  \qquad
  + \frac{
    i  (\tilde{\lambda}^q_{D})_{rj}
    (\lambda_D)^*_{ri}}{
    4 M_{D_r}}
  - \frac{
    i  (\tilde{\lambda}^q_{D})^*_{ri}
    (\lambda_D)_{rj}}{
    4 M_{D_r}}
  \nonumber \\ &
  \qquad
  - \frac{
    i  (\tilde{\lambda}^q_{T_2})_{rj}
    (\lambda_{T_2})^*_{ri}}{
    8 M_{T_{2r}}}
  + \frac{
    i  (\tilde{\lambda}^q_{T_2})^*_{ri}
    (\lambda_{T_2})_{rj}}{
    8 M_{T_{2r}}}
  \nonumber \\ &
  \qquad
  - \frac{
    i  (\tilde{\lambda}^q_{T_1})_{rj}
    (\lambda_{T_1})^*_{ri}}{
    8 M_{T_{1r}}}
  + \frac{
    i  (\tilde{\lambda}^q_{T_1})^*_{ri}
    (\lambda_{T_1})_{rj}}{
    8 M_{T_{1r}}}
  \nonumber \\ &
  \qquad
  - \frac{
    \hat{y}^{d*}_{ki}
    (\tilde{g}^{dDq}_{\mathcal{L}_1})_{rkj}
    (\gamma_{\mathcal{L}_1})^*_{r}}{
    8 M_{\mathcal{L}_{1r}}^{2}}
  - \frac{
    \hat{y}^d_{kj}
    (\tilde{g}^{dDq}_{\mathcal{L}_1})^*_{rki}
    (\gamma_{\mathcal{L}_1})_{r}}{
    8 M_{\mathcal{L}_{1r}}^{2}}
  \nonumber \\ &
  \qquad
  + \frac{
    \hat{y}^{d*}_{ki}
    (\tilde{g}^{Ddq}_{\mathcal{L}_1})_{rkj}
    (\gamma_{\mathcal{L}_1})^*_{r}}{
    8 M_{\mathcal{L}_{1r}}^{2}}
  + \frac{
    \hat{y}^d_{kj}
    (\tilde{g}^{Ddq}_{\mathcal{L}_1})^*_{rki}
    (\gamma_{\mathcal{L}_1})_{r}}{
    8 M_{\mathcal{L}_{1r}}^{2}}
  \nonumber \\ &
  \qquad
  + \frac{
    \hat{y}^u_{kj}
    (\tilde{g}^{qDu}_{\mathcal{L}_1})_{rki}
    (\gamma_{\mathcal{L}_1})^*_{r}}{
    8 M_{\mathcal{L}_{1r}}^{2}}
  + \frac{
    \hat{y}^{u*}_{ki}
    (\tilde{g}^{qDu}_{\mathcal{L}_1})^*_{rkj}
    (\gamma_{\mathcal{L}_1})_{r}}{
    8 M_{\mathcal{L}_{1r}}^{2}}
  \nonumber \\ &
  \qquad
  - \frac{
    \hat{y}^u_{kj}
    (\tilde{g}^{Dqu}_{\mathcal{L}_1})_{rki}
    (\gamma_{\mathcal{L}_1})^*_{r}}{
    8 M_{\mathcal{L}_{1r}}^{2}}
  - \frac{
    \hat{y}^{u*}_{ki}
    (\tilde{g}^{Dqu}_{\mathcal{L}_1})^*_{rkj}
    (\gamma_{\mathcal{L}_1})_{r}}{
    8 M_{\mathcal{L}_{1r}}^{2}}
  \nonumber \\ &
  \qquad
  - \frac{
    i  (\tilde{g}^{q\prime}_{\mathcal{L}_1})_{rij}
    (\gamma_{\mathcal{L}_1})^*_{r}}{
    2 M_{\mathcal{L}_{1r}}^{2}}
  + \frac{
    i  (\tilde{g}^{q\prime}_{\mathcal{L}_1})^*_{rji}
    (\gamma_{\mathcal{L}_1})_{r}}{
    2 M_{\mathcal{L}_{1r}}^{2}}
  \nonumber \\ &
  \qquad
  + \frac{
    \operatorname{Im}{\left(
        (\tilde{\gamma}^W_{\mathcal{L}_1})_{r}
        (\gamma_{\mathcal{L}_1})^*_{r}
      \right)}
    g_2
    \delta_{ij}}{
    2 M_{\mathcal{L}_{1r}}^{2}}
  \Bigg\},
\end{align}

\newpage

\begin{align}
  Z_\phi~\!\left(C_{\phi e}\right)_{ij}= &
  \frac{
    (\lambda_{\Delta_1})_{rj}
    (\lambda_{\Delta_1})^*_{ri}}{
    2  M_{\Delta_{1r}}^{2}}
  - \frac{
    (\lambda_{\Delta_3})_{rj}
    (\lambda_{\Delta_3})^*_{ri}}{
    2  M_{\Delta_{3r}}^{2}}
  \nonumber \\ &
  - \frac{
    \operatorname{Re}{\left(
        (\hat{g}^\phi_{\mathcal{B}})_{r}
      \right)}
    (g^e_{\mathcal{B}})_{rij}}{
    M_{\mathcal{B}_r}^{2}}
  - \frac{
    g_1
    \delta_{ij}
    (g^B_{\mathcal{L}_1})_{rs}
    (\gamma_{\mathcal{L}_1})^*_{r}
    (\gamma_{\mathcal{L}_1})_{s}}{
    2  M_{\mathcal{L}_{1r}}^{2}
    M_{\mathcal{L}_{1s}}^{2}}
  \nonumber \\ &
  + \frac{
    i  (z_{\Delta_1\mathcal{L}_1})^*_{rsi}
    (\gamma_{\mathcal{L}_1})^*_{s}
    (\lambda_{\Delta_1})_{rj}}{
    2  M_{\Delta_{1r}}
    M_{\mathcal{L}_{1s}}^{2}}
  - \frac{
    i  (z_{\Delta_1\mathcal{L}_1})_{rsj}
    (\lambda_{\Delta_1})^*_{ri}
    (\gamma_{\mathcal{L}_1})_{s}}{
    2  M_{\Delta_{1r}}
    M_{\mathcal{L}_{1s}}^{2}}
  \nonumber \\ &
  - \frac{
    i  (z_{\Delta_3\mathcal{L}_1})^*_{sri}
    (\gamma_{\mathcal{L}_1})_{r}
    (\lambda_{\Delta_3})_{sj}}{
    2  M_{\mathcal{L}_{1r}}^{2}
    M_{\Delta_{3s}}}
  + \frac{
    i  (z_{\Delta_3\mathcal{L}_1})_{srj}
    (\lambda_{\Delta_3})^*_{si}
    (\gamma_{\mathcal{L}_1})^*_{r}}{
    2  M_{\mathcal{L}_{1r}}^{2}
    M_{\Delta_{3s}}}
  \nonumber \\ &
  + \frac{1}{f} \Bigg\{
  \frac{
    i  (\tilde{\lambda}^e_{\Delta_1})^*_{ri}
    (\lambda_{\Delta_1})_{rj}}{
    2 M_{\Delta_{1r}}}
  - \frac{
    i  (\tilde{\lambda}^e_{\Delta_1})_{rj}
    (\lambda_{\Delta_1})^*_{ri}}{
    2 M_{\Delta_{1r}}}
  \nonumber \\ &
  \qquad
  - \frac{
    i  (\tilde{\lambda}^e_{\Delta_3})^*_{ri}
    (\lambda_{\Delta_3})_{rj}}{
    2 M_{\Delta_{3r}}}
  + \frac{
    i  (\tilde{\lambda}^e_{\Delta_3})_{rj}
    (\lambda_{\Delta_3})^*_{ri}}{
    2 M_{\Delta_{3r}}}
  \nonumber \\ &
  \qquad
  - \frac{
    \hat{y}^{e*}_{jk}
    (\tilde{g}^{eDl}_{\mathcal{L}_1})_{rik}
    (\gamma_{\mathcal{L}_1})^*_{r}}{
    4 M_{\mathcal{L}_{1r}}^{2}}
  - \frac{
    \hat{y}^e_{ik}
    (\tilde{g}^{eDl}_{\mathcal{L}_1})^*_{rjk}
    (\gamma_{\mathcal{L}_1})_{r}}{
    4 M_{\mathcal{L}_{1r}}^{2}}
  \nonumber \\ &
  \qquad
  + \frac{
    \hat{y}^{e*}_{jk}
    (\tilde{g}^{Del}_{\mathcal{L}_1})_{rik}
    (\gamma_{\mathcal{L}_1})^*_{r}}{
    4 M_{\mathcal{L}_{1r}}^{2}}
  + \frac{
    \hat{y}^e_{ik}
    (\tilde{g}^{Del}_{\mathcal{L}_1})^*_{rjk}
    (\gamma_{\mathcal{L}_1})_{r}}{
    4 M_{\mathcal{L}_{1r}}^{2}}
  \nonumber \\ &
  \qquad
  - \frac{
    i  (\tilde{g}^{e}_{\mathcal{L}_1})_{rij}
    (\gamma_{\mathcal{L}_1})^*_{r}}{
    2 M_{\mathcal{L}_{1r}}^{2}}
  + \frac{
    i  (\tilde{g}^{e}_{\mathcal{L}_1})^*_{rji}
    (\gamma_{\mathcal{L}_1})_{r}}{
    2 M_{\mathcal{L}_{1r}}^{2}}
  \nonumber \\ &
  \qquad
  - \frac{
    \operatorname{Im}{\left(
        (\tilde{\gamma}^B_{\mathcal{L}_1})_{r}
        (\gamma_{\mathcal{L}_1})^*_{r}
      \right)}
    g_1
    \delta_{ij}}{
    M_{\mathcal{L}_{1r}}^{2}}
  \Bigg\},
\end{align}

\newpage

\begin{align}
  Z_\phi~\!\left(C_{\phi d}\right)_{ij}= &
  \frac{
    (\lambda^d_{Q_1})_{rj}
    (\lambda^d_{Q_1})^*_{ri}}{
    2  M_{Q_{1r}}^{2}}
  - \frac{
    (\lambda_{Q_5})_{rj}
    (\lambda_{Q_5})^*_{ri}}{
    2  M_{Q_{5r}}^{2}}
  \nonumber \\ &
  - \frac{
    \operatorname{Re}{\left(
        (\hat{g}^\phi_{\mathcal{B}})_{r}
      \right)}
    (g^d_{\mathcal{B}})_{rij}}{
    M_{\mathcal{B}_r}^{2}}
  - \frac{
    g_1
    \delta_{ij}
    (g^B_{\mathcal{L}_1})_{rs}
    (\gamma_{\mathcal{L}_1})^*_{r}
    (\gamma_{\mathcal{L}_1})_{s}}{
    6  M_{\mathcal{L}_{1r}}^{2}
    M_{\mathcal{L}_{1s}}^{2}}
  \nonumber \\ &
  + \frac{
    i  (z^d_{Q_1\mathcal{L}_1})^*_{rsi}
    (\gamma_{\mathcal{L}_1})^*_{s}
    (\lambda^d_{Q_1})_{rj}}{
    2  M_{Q_{1r}}
    M_{\mathcal{L}_{1s}}^{2}}
  - \frac{
    i  (z^d_{Q_1\mathcal{L}_1})_{rsj}
    (\lambda^d_{Q_1})^*_{ri}
    (\gamma_{\mathcal{L}_1})_{s}}{
    2  M_{Q_{1r}}
    M_{\mathcal{L}_{1s}}^{2}}
  \nonumber \\ &
  - \frac{
    i  (z_{Q_5\mathcal{L}_1})^*_{rsi}
    (\gamma_{\mathcal{L}_1})_{s}
    (\lambda_{Q_5})_{rj}}{
    2  M_{Q_{5r}}
    M_{\mathcal{L}_{1s}}^{2}}
  + \frac{
    i  (z_{Q_5\mathcal{L}_1})_{rsj}
    (\lambda_{Q_5})^*_{ri}
    (\gamma_{\mathcal{L}_1})^*_{s}}{
    2  M_{Q_{5r}}
    M_{\mathcal{L}_{1s}}^{2}}
  \nonumber \\ &
  + \frac{1}{f} \Bigg\{
  \frac{
    i  (\tilde{\lambda}^d_{Q_1})^*_{ri}
    (\lambda^d_{Q_1})_{rj}}{
    2 M_{Q_{1r}}}
  - \frac{
    i  (\tilde{\lambda}^d_{Q_1})_{rj}
    (\lambda^d_{Q_1})^*_{ri}}{
    2 M_{Q_{1r}}}
  \nonumber \\ &
  \qquad
  - \frac{
    i  (\tilde{\lambda}^d_{Q_5})^*_{ri}
    (\lambda_{Q_5})_{rj}}{
    2 M_{Q_{5r}}}
  + \frac{
    i  (\tilde{\lambda}^d_{Q_5})_{rj}
    (\lambda_{Q_5})^*_{ri}}{
    2 M_{Q_{5r}}}
  \nonumber \\ &
  \qquad
  - \frac{
    \hat{y}^{d*}_{jk}
    (\tilde{g}^{dDq}_{\mathcal{L}_1})_{rik}
    (\gamma_{\mathcal{L}_1})^*_{r}}{
    4 M_{\mathcal{L}_{1r}}^{2}}
  - \frac{
    \hat{y}^d_{ik}
    (\tilde{g}^{dDq}_{\mathcal{L}_1})^*_{rjk}
    (\gamma_{\mathcal{L}_1})_{r}}{
    4 M_{\mathcal{L}_{1r}}^{2}}
  \nonumber \\ &
  \qquad
  + \frac{
    \hat{y}^{d*}_{jk}
    (\tilde{g}^{Ddq}_{\mathcal{L}_1})_{rik}
    (\gamma_{\mathcal{L}_1})^*_{r}}{
    4 M_{\mathcal{L}_{1r}}^{2}}
  + \frac{
    \hat{y}^d_{ik}
    (\tilde{g}^{Ddq}_{\mathcal{L}_1})^*_{rjk}
    (\gamma_{\mathcal{L}_1})_{r}}{
    4 M_{\mathcal{L}_{1r}}^{2}}
  \nonumber \\ &
  \qquad
  - \frac{
    i  (\tilde{g}^{d}_{\mathcal{L}_1})_{rij}
    (\gamma_{\mathcal{L}_1})^*_{r}}{
    2 M_{\mathcal{L}_{1r}}^{2}}
  + \frac{
    i  (\tilde{g}^{d}_{\mathcal{L}_1})^*_{rji}
    (\gamma_{\mathcal{L}_1})_{r}}{
    2 M_{\mathcal{L}_{1r}}^{2}}
  \nonumber \\ &
  \qquad
  - \frac{
    \operatorname{Im}{\left(
        (\tilde{\gamma}^B_{\mathcal{L}_1})_{r}
        (\gamma_{\mathcal{L}_1})^*_{r}
      \right)}
    g_1
    \delta_{ij}}{
    3 M_{\mathcal{L}_{1r}}^{2}}
  \Bigg\},
\end{align}

\newpage

\begin{align}
  Z_\phi~\!\left(C_{\phi u}\right)_{ij}= &
  - \frac{
    (\lambda^u_{Q_1})_{rj}
    (\lambda^u_{Q_1})^*_{ri}}{
    2  M_{Q_{1r}}^{2}}
  + \frac{
    (\lambda_{Q_7})_{rj}
    (\lambda_{Q_7})^*_{ri}}{
    2  M_{Q_{7r}}^{2}}
  \nonumber \\ &
  - \frac{
    \operatorname{Re}{\left(
        (\hat{g}^\phi_{\mathcal{B}})_{r}
      \right)}
    (g^u_{\mathcal{B}})_{rij}}{
    M_{\mathcal{B}_r}^{2}}
  + \frac{
    g_1
    \delta_{ij}
    (g^B_{\mathcal{L}_1})_{rs}
    (\gamma_{\mathcal{L}_1})^*_{r}
    (\gamma_{\mathcal{L}_1})_{s}}{
    3  M_{\mathcal{L}_{1r}}^{2}
    M_{\mathcal{L}_{1s}}^{2}}
  \nonumber \\ &
  - \frac{
    i  (z^u_{Q_1\mathcal{L}_1})^*_{sri}
    (\gamma_{\mathcal{L}_1})_{r}
    (\lambda^u_{Q_1})_{sj}}{
    2  M_{\mathcal{L}_{1r}}^{2}
    M_{Q_{1s}}}
  + \frac{
    i  (z^u_{Q_1\mathcal{L}_1})_{srj}
    (\lambda^u_{Q_1})^*_{si}
    (\gamma_{\mathcal{L}_1})^*_{r}}{
    2  M_{\mathcal{L}_{1r}}^{2}
    M_{Q_{1s}}}
  \nonumber \\ &
  + \frac{
    i  (z_{Q_7\mathcal{L}_1})^*_{rsi}
    (\gamma_{\mathcal{L}_1})^*_{s}
    (\lambda_{Q_7})_{rj}}{
    2  M_{Q_{7r}}
    M_{\mathcal{L}_{1s}}^{2}}
  - \frac{
    i  (z_{Q_7\mathcal{L}_1})_{rsj}
    (\lambda_{Q_7})^*_{ri}
    (\gamma_{\mathcal{L}_1})_{s}}{
    2  M_{Q_{7r}}
    M_{\mathcal{L}_{1s}}^{2}}
  \nonumber \\ &
  + \frac{1}{f} \Bigg\{
  - \frac{
    i  (\tilde{\lambda}^u_{Q_1})^*_{ri}
    (\lambda^u_{Q_1})_{rj}}{
    2 M_{Q_{1r}}}
  + \frac{
    i  (\tilde{\lambda}^u_{Q_1})_{rj}
    (\lambda^u_{Q_1})^*_{ri}}{
    2 M_{Q_{1r}}}
  \nonumber \\ &
  \qquad
  + \frac{
    i  (\tilde{\lambda}^u_{Q_7})^*_{ri}
    (\lambda_{Q_7})_{rj}}{
    2 M_{Q_{7r}}}
  - \frac{
    i  (\tilde{\lambda}^u_{Q_7})_{rj}
    (\lambda_{Q_7})^*_{ri}}{
    2 M_{Q_{7r}}}
  \nonumber \\ &
  \qquad
  - \frac{
    \hat{y}^u_{ik}
    (\tilde{g}^{qDu}_{\mathcal{L}_1})_{rjk}
    (\gamma_{\mathcal{L}_1})^*_{r}}{
    4 M_{\mathcal{L}_{1r}}^{2}}
  - \frac{
    \hat{y}^{u*}_{jk}
    (\tilde{g}^{qDu}_{\mathcal{L}_1})^*_{rik}
    (\gamma_{\mathcal{L}_1})_{r}}{
    4 M_{\mathcal{L}_{1r}}^{2}}
  \nonumber \\ &
  \qquad
  + \frac{
    \hat{y}^u_{ik}
    (\tilde{g}^{Dqu}_{\mathcal{L}_1})_{rjk}
    (\gamma_{\mathcal{L}_1})^*_{r}}{
    4 M_{\mathcal{L}_{1r}}^{2}}
  + \frac{
    \hat{y}^{u*}_{jk}
    (\tilde{g}^{Dqu}_{\mathcal{L}_1})^*_{rik}
    (\gamma_{\mathcal{L}_1})_{r}}{
    4  M_{\mathcal{L}_{1r}}^{2}}
  \nonumber \\ &
  \qquad
  - \frac{
    i  (\tilde{g}^{u}_{\mathcal{L}_1})_{rij}
    (\gamma_{\mathcal{L}_1})^*_{r}}{
    2 M_{\mathcal{L}_{1r}}^{2}}
  + \frac{
    i  (\tilde{g}^{u}_{\mathcal{L}_1})^*_{rji}
    (\gamma_{\mathcal{L}_1})_{r}}{
    2 M_{\mathcal{L}_{1r}}^{2}}
  \nonumber \\ &
  \qquad
  + \frac{
    2   \operatorname{Im}{\left(
        (\tilde{\gamma}^B_{\mathcal{L}_1})_{r}
        (\gamma_{\mathcal{L}_1})^*_{r}
      \right)}
    g_1
    \delta_{ij}}{
    3 M_{\mathcal{L}_{1r}}^{2}}
  \Bigg\},
\end{align}

\begin{align}
    Z_\phi~\!\left(C_{\phi ud}\right)_{ij}= &
  \frac{
    (\lambda^d_{Q_1})_{rj}
    (\lambda^u_{Q_1})^*_{ri}}{
    M_{Q_{1r}}^{2}}
  - \frac{
    (\hat{g}^\phi_{\mathcal{B}_1})_{r}
    (g^{du}_{\mathcal{B}_1})^*_{rji}}{
    M_{\mathcal{B}_{1r}}^{2}}
  \nonumber \\ &
  + \frac{
    i  (z^u_{Q_1\mathcal{L}_1})^*_{rsi}
    (\gamma_{\mathcal{L}_1})_{s}
    (\lambda^d_{Q_1})_{rj}}{
    M_{Q_{1r}}
    M_{\mathcal{L}_{1s}}^{2}}
  - \frac{
    i  (z^d_{Q_1\mathcal{L}_1})_{rsj}
    (\lambda^u_{Q_1})^*_{ri}
    (\gamma_{\mathcal{L}_1})_{s}}{
    M_{Q_{1r}}
    M_{\mathcal{L}_{1s}}^{2}}
  \nonumber \\ &
  + \frac{1}{f} \Bigg\{
  \frac{
    i  (\tilde{\lambda}^u_{Q_1})^*_{ri}
    (\lambda^d_{Q_1})_{rj}}{
    M_{Q_{1r}}}
  - \frac{
    i  (\tilde{\lambda}^d_{Q_1})_{rj}
    (\lambda^u_{Q_1})^*_{ri}}{
    M_{Q_{1r}}}
  + \frac{
    i  (\tilde{g}^{du}_{\mathcal{L}_1})^*_{rji}
    (\gamma_{\mathcal{L}_1})_{r}}{
    M_{\mathcal{L}_{1r}}^{2}}
  \nonumber \\ &
  \qquad
  - \frac{
    \hat{y}^u_{ik}
    (\tilde{g}^{dDq}_{\mathcal{L}_1})^*_{rjk}
    (\gamma_{\mathcal{L}_1})_{r}}{
    2 M_{\mathcal{L}_{1r}}^{2}}
  + \frac{
    \hat{y}^u_{ik}
    (\tilde{g}^{Ddq}_{\mathcal{L}_1})^*_{rjk}
    (\gamma_{\mathcal{L}_1})_{r}}{
    2 M_{\mathcal{L}_{1r}}^{2}}
  \nonumber \\ &
  \qquad
  + \frac{
    \hat{y}^{d*}_{jk}
    (\tilde{g}^{qDu}_{\mathcal{L}_1})^*_{rik}
    (\gamma_{\mathcal{L}_1})_{r}}{
    2 M_{\mathcal{L}_{1r}}^{2}}
  - \frac{
    \hat{y}^{d*}_{jk}
    (\tilde{g}^{Dqu}_{\mathcal{L}_1})^*_{rik}
    (\gamma_{\mathcal{L}_1})_{r}}{
    2 M_{\mathcal{L}_{1r}}^{2}}
  \Bigg\}.
\end{align}

\newpage
\bibliography{BSM_EFT_Dict}{}
\bibliographystyle{JHEP-2}

\end{document}